\documentclass[twocolumn,times]{aastex631}
\usepackage{threeparttablex}
\usepackage{amsmath}

\shorttitle{IR-X correlations}
\shortauthors{Spinoglio, Fern\'andez-Ontiveros \& Malkan}
\usepackage{subfigure}
\usepackage{url}
\colorlet{juan}{Fuchsia!50!WildStrawberry}

\graphicspath{{./}{figures/}}

\newcommand{\ergs}{erg~s$^{-1}$}

\newcommand{\neii}{[Ne {\small II}] 12.81 $\mu$m}

\newcommand{\nev}{[NeV]}
\newcommand{\nevp}{[NeV] 14.32 $\mu$m}
\newcommand{\nevs}{[NeV] 24.32 $\mu$m}
\newcommand{\neiii}{[Ne {\small III}] 15.55 $\mu$m}
\newcommand{\oiv}{[O {\small IV}]}
\newcommand{\oivp}{[O {\small IV}] 25.89 $\mu$m}
\newcommand{\oiii}{[O {\small III}] 5007 \AA}

\begin{document}

\title{The high-ionization IR fine-structure lines as bolometric indicators of the AGN power:\\ study of the complete 12$\mu$m AGN sample}


\author[0000-0001-8840-1551]{Luigi Spinoglio}
\affiliation{Istituto di Astrofisica e Planetologia Spaziali (INAF--IAPS), Via Fosso del Cavaliere 100, I--00133 Roma, Italy}

\author[0000-0001-9490-899X]{Juan Antonio Fern\'andez-Ontiveros}
\affiliation{Centro de Estudios de F\'isica del Cosmos de Arag\'on (CEFCA), Plaza San Juan 1, E--44001, Teruel, Spain}
\affiliation{Istituto di Astrofisica e Planetologia Spaziali (INAF--IAPS), Via Fosso del Cavaliere 100, I--00133 Roma, Italy}

\author[0000-0001-6919-1237]{Matthew A. Malkan}
\affiliation{Department of Physics and Astronomy, UCLA, Los Angeles, CA 90095-1547, USA}



\begin{abstract}

The high-ionization mid-IR lines, excited in the Narrow Line Regions (NLR) of Active Galactic Nuclei (AGN), barely affected by stellar excitation and dust extinction, trace the AGN bolometric power. We used the complete 12$\rm{\micron}$ sample of Seyfert galaxies, for which 100/116 objects have reliable 2-10keV observations. The [NeV] and [OIV] mid-IR lines linearly correlate with several AGN bolometric indicators (intrinsic 2–10keV and observed 14–195keV X-ray emission, compact nuclear 12$\rm{\micron}$ emission, [OIII] $\lambda$5007\AA \  line emission), both in terms of flux and luminosity. No evidence of systematic differences in these correlations is found among the Seyfert populations, including type 1 and type 2, and Compton thick and thin AGN. Nevertheless, we find that a sequence of high-to-low Eddington ratio together with strong-to-weak line excitation (traced by the [OIV]/[Ne II] line ratio) encompasses from type 1 through type 2 AGN to Low Ionization Nuclear Emission Line Region (LINER) galaxies, showing intrinsic differences in these three AGN populations. A positive correlation between the Black Hole Accretion Rate (BHAR) and the Star Formation Rate (SFR) is found, but no correlation between the specific SFR (sSFR) and the ratio BHAR/M$_{BH}$, simply reflecting the fact that the more massive is a galaxy, the more it is forming stars and feeding its central black hole.  The JWST telescope, just beginning operations, will allow large samples of AGN to be observed in these lines in the nearby Universe (z$<$0.9).

\end{abstract}



\section{Introduction} \label{sec:intro}

A full understanding of galaxy evolution still needs both observational and theoretical efforts. While our theoretical understanding of complex baryonic processes is advancing \citep[see, e.g.][]{wechsler2018}, there still remains a large gap in comparing theory with currently limited  observations. Rest-frame  optical and UV data are used to probe star formation in distant galaxies, but are greatly affected by dust. X-ray data are sensitive tracers of black hole activity, but are often limited by sensitivity and resolution. In the most obscured regions, such as those at the centers of forming galaxies, all but the highest energy X-rays can be absorbed. It has become evident during the last few decades that the key phases of galaxy evolution -- the birth and growth of stars and the growth of supermassive black holes (SMBH) -- take place within regions enshrouded in dust, largely hidden from view at optical and UV wavelengths \citep{madau2014} and suffering from the effects of obscuration at X-ray energies (see, e.g. Compton thick objects with hydrogen column densities in excess of 10$^{24}$ cm$^{-2}$; \citealt{bassani1999,delmoro2016}). The observed evolution of star formation and black hole growth over cosmic time appear remarkably similar, with a peak around z$\sim$2 and then decreasing by a factor of 30 in the last 10 Gyrs. At this peak, an epoch known as cosmic noon, nearly 90\% of the energy emitted by young stars emerges in the infrared from dust, re-radiating the energy absorbed from young stars and growing supermassive black holes \citep{goto2019}. The infrared spectrum therefore provides a critical window through which we can understand the crucial hidden part of galaxy evolution, and measure both the star formation rate (SFR) and the black hole accretion rate (BHAR) in individual galaxies over much of cosmic time.  

This work aims to establish what are the best IR tracers for black hole accretion, and for star formation (to be able to separate the two components), in the local Universe. These can then be used at intermediate redshift by \textit{JWST} (z$<$0.9) and have the potential to be used at high redshift (z$\lesssim$3.5) by the upcoming IR space telescopes of the next decade(s), for studying galaxy evolution. A suitable laboratory to study the interplay of black hole accretion and star formation, which is fundamental to the evolution of galaxies, is given in the local Universe  by the class of the Seyfert galaxies, which are powered by both black hole accretion in their nuclei and also by star formation, either circum-nuclear ( at physical scales up to $\sim$200 pc for circumnuclear disks) and/or in the galaxy disk (at scales of $\sim$1-10 kpc). 

The mid-infrared (mid-IR) spectra of Seyfert galaxies are very rich in narrow emission lines that can be powered by both the presence of active galactic nuclei (AGN) and star formation occurring in circumnuclear regions and in the galactic disk (see, e.g. \citealt{tommasin2008,tommasin2010}). 
The potential of the IR line diagnostics was first demonstrated by \citet{spinoglio1992}. Unlike optical spectroscopy, mid-IR spectroscopy is much less affected by dust extinction, as can be quantified by the fact that 50 mag of extinction in the V-band corresponds to only 1 mag at 24$\mu$m \citep[e.g., ][]{genzel1998}. Thus, infrared observations are the only practical tool for spectroscopically studying dust-obscured galaxies and AGN. The brightest AGN features in the mid-IR range are the high-ionization forbidden emission lines from [NeV] at 14.3 and 24.3$\mu$m and [OIV] at 25.9$\mu$m. Given its high ionization potential (IP = 97\,eV), Ne$^{4+}$ is an unambiguous tracers of the AGN narrow-line region (NLR), whereas O$^{3+}$ (IP = 54.9\,eV) is dominated by the AGN contribution, with an equivalent width of about a factor ten higher compared to Starburst galaxies \citep{sturm2002,melendez2008,tommasin2008,tommasin2010}. The star-formation tracers include the strong PAH emission features (at, e.g., 11.3$\mu$m), the mid-IR pure rotational lines of H$_2$ and the low-ionization fine-structure lines (e.g. [NeII]12.8$\mu$m, with IP=21.6eV, [SIII]18 and 33$\mu$m, IP = 23.3eV, and [SiII]34.8$\mu$m, IP = 16.3eV), as well as the ubiquitous [CII]158$\mu$m line ( IP = 11.3eV) \citep[see, e.g.][]{delooze2011,delooze2014}. We refer to \citet{mordini2021,mordini2022} for a detailed recent calibration of the fine-structure lines luminosities against the total IR luminosities.

On the other hand, the hard X-ray emission in the $2$--$10\, \rm{keV}$ band is a direct measure of the current (instantaneous) accretion luminosity for all AGNs except the extreme Compton thick objects. For Compton-thin sources ($\rm{N_H} \lesssim 10^{24}\, \rm{cm^{-2}}$) 
a correction can be applied to recover a reliable intrinsic luminosity \citep[see, e.g.,][]{ricci2015}.
The higher energy continuum (e.g. $14$--$195\, \rm{keV}$) can be used in some cases to measure the nuclear continuum in Compton-thick sources.
The relationship between the accretion luminosity, as derived from the X-rays, and the NLR luminosity, measured by the high-ionization mid-IR fine structure lines of [NeV] and [OIV], can shed light on the differences between the three main types of active galaxies. The NLR emission, especially if traced by long-wavelength transitions, is thought to be essentially isotropic, because the spatial scale of the NLR ($100$--$500\, \rm{pc}$) is much larger than the predicted size of the obscuring dust tori. These latter are expected to be of the order of a few parsecs in size \citep[see, e.g.][]{jaffe2004,elitzur2006,tristram2007,pott2010}.

Previous studies have found intrinsic differences in the X-ray to NLR line ratios between Seyfert 1 and 2, ascribed to obscuration affecting the observed hard X-ray fluxes (up to the 14--195\,keV band) and the optical NLR emission in type 2 nuclei \citep{melendez2008,rigby2009,lamassa2010}. However, this is in contrast with the indistinguishable [OIV] luminosity distributions of both Seyfert classes \citep{Diamond-Stanic2009,tommasin2010}.

In this work, instead of using X-ray or optically selected samples of Seyfert galaxies, we use the complete $12\, \rm{\micron}$ selected sample of Seyfert galaxies \citep{rush1993}, exploiting the large set of  multi-frequency data available, to better understand the nature of the ``pure'' Seyfert 2's and how they are related to type 1's. For example, are they intrinsically the same, and can they be related through a simple unification, such as viewing angle? Or are AGN intrinsically different? Furthermore, we have studied the relation between the host galaxy properties and those of the AGN, comparing these galaxies with the so-called Main-Sequence of star forming galaxies \citep{elbaz2011}.

Taking advantage of the complete and virtually unbiased galaxy sample that we use, we have studied: {\it (i)} the correlation between the bolometric luminosity as derived from X-rays observations and that one derived from the high-ionization \oivp~ line; {\it (ii)} the differences in the accretion process among the various AGN spectroscopic classes; {\it (iii)} the relation between the AGN properties (BHAR, Eddington ratio) and those of the host galaxy (M$\star$, SFR, sSFR).

The paper is organized as follows: section \ref{sec:sample} describes the sample of AGN used; section \ref{sec:results} gives the major results of this paper and is divided into the following parts: \ref{sec:corr} gives the correlations between the X-ray emission and the mid-IR fine structure lines; \ref{sec:corr2} shows the correlations between the X-ray emission and both the \oiii~ optical line and the 12$\mu$m nuclear continuum; \ref{sec:bol} shows how the X-ray and mid-IR determinations of the AGN bolometric luminosity agree; \ref{sec:led} shows the LED diagram for our sample; \ref{sec:edd} shows the Eddington ratio as a function of the bolometric luminosity; \ref{sec:oxygen} shows the correlation between the \oivp~ and \oiii~ line fluxes and luminosities and finally \ref{sec:sfr&BH} gives the correlation between SFR and BHAR. Finally in section \ref{sec:discuss} the various results are compared with previous studies.

\section{The sample}\label{sec:sample}

Our study is based on the $12\, \rm{\micron}$ Seyfert galaxy sample
(53 Seyfert 1's and 63 Seyfert 2's) 
(hereafter 12MSG) extracted from the Infrared Astronomical Satellite (IRAS) $12\, \rm{\micron}$ galaxy sample \citep{rush1993}.  
The chosen sample is essentially a bolometric flux-limited survey outside the galactic plane, and therefore largely unbiased, given the empirical fact that galaxies emit an approximately constant fraction of their total bolometric luminosity at $12\, \rm{\micron}$. This fraction is $\sim 15\%$ for AGNs \citep{spinoglio1989} and $7\%$ for normal and starburst galaxies, independent of star formation activity \citep{spinoglio1995}.

The multi-frequency observations that we use in our analysis include the hard X-rays observations in the bands $2$--$10\, \rm{keV}$ and $14$--$195\, \rm{keV}$ from \citet{ricci2017} and \citet{oh2018}, which include the broadband X-ray ($0.3$--$150\, \rm{keV}$) characteristics of the BAT-detected AGNs, obtained by combining XMM-Newton, Swift/XRT, ASCA, Chandra, and Suzaku observations in the soft X-ray band, in addition to the Swift-BAT AGN 70 months catalog. We also use the optical spectroscopy from \citet{malkan2017} and the 12$\mu$m nuclear fluxes from \citet{asmus2014}, as well as the {\it Spitzer} spectroscopic data \citep{tommasin2008, wu2009, gallimore2010, tommasin2010} and the {\it Herschel} PACS spectra \citep{fernandez2016}.

Using the optical spectropolarimetric results that have been extensively obtained for the $12\, \rm{\micron}$ selected Seyfert sample \citep[][and references therein]{tran2003}, \citet{tommasin2010} have proposed a classification of Seyfert galaxies, dividing them into the three classes: Seyfert type 1's (hereafter Sy1), Hidden Broad Line Seyfert type 2's (hereafter HBL), showing the presence of broad lines in polarized light, and ``pure'' Seyfert type 2's (hereafter Sy2), without any sign of detected broad emission lines. The infrared spectral behavior of the Sy1 and HBL is indistinguishable, implying that they indeed belong to the larger class of ``type 1'' AGNs. Sy2 are not a homogeneous class--some of them have properties similar to type 1's, and may include a fraction of undetected type 1 nuclei, while others are more similar to non-Seyfert galaxies, either starburst or Low Ionization Nuclear Emission Regions (LINERs) \citep[see, e.g.][]{ho1997}.

The 12MSG sample has complete observational coverage at virtually every wavelength, from radio to hard X-rays \citep[see][for a review up to 2008]{tommasin2008} and \citet{wu2009,baum2010,tommasin2010,gallimore2010,lamassa2010,brightman2011,pereira2013,spinoglio2015,theios2016,gruppioni2016,gomez2017,malkan2017,lacaria2019}.

Table \ref{tab:sample0} lists the galaxies of the full AGN sample selected from the revised 12$\mu$m sample of galaxies \citep{rush1993} with the galaxy coordinates, redshift and activity type based on optical spectroscopy.
Table \ref{tab:sample1} gives for each galaxy the 
X-ray data information, where available, including the derived hydrogen column density, whether the galaxy is Compton thick (CT), the SWIFT identification, where available, the (2-10)KeV intrinsic, i.e. absorption corrected, X-ray flux and luminosity, the (14-195)KeV X-ray flux and luminosity \citep{oh2018, ricci2017,cusumano2010} and the bolometric luminosity as derived from the (2-10)keV intrinsic luminosity, and the bolometric correction from \citet{lusso2012}.
Table \ref{tab:CT} lists the Compton thick objects of our sample with the observed and corrected (2-10)keV fluxes and luminosities and the bolometric luminosities as derived from these X-ray fluxes.

Table \ref{tab:sample2} gives the fluxes of the emission lines considered in this study: the [NeII]12.8$\mu$m, [NeV]14.32$\mu$m, [NeII]15.5$\mu$m, [NeV]24.32$\mu$m and [OIV]25.89$\mu$m mid-IR fine structure lines from \citet{tommasin2008,tommasin2010}, the [CII]158$\mu$m line from \citet{fernandez2016} and the [OIII]5007\AA \ optical line flux from \citet{malkan2017}. We include the 12$\mu$m nuclear flux density from \citet{asmus2014}. 

The {\it Spitzer} IRS spectrograph apertures used for the fine structure lines were 4.7" $\times$  11.3" for the short wavelength lines (\neii, \neiii \ and \nevp) and 11.1" $\times$ 22.3" for the long wavelength lines (\nevs, \oivp). Given the average redshift of the 12MSG active galaxies of z=0.028 \citep{rush1993}, which corresponds to a mean distance of 124.5 Mpc, ($H_0 = 67.4\,\rm{km\,s^{-1}\,Mpc^{-1}}$) the short and long wavelength apertures of the Spitzer IRS spectrograph correspond to a linear area of $\sim 1\,\rm{kpc}$ and 5\,kpc in diameter, respectively. Therefore the NLRs of our sample of AGN are --\,on average\,-- fully covered by the mid-IR spectroscopic observations.

 Since the goal of our spectroscopy is to measure line emission produced by the AGN, our optical line measurements typically refer to the central several arcseconds of the galaxy. The smallest apertures used in \citet{malkan2017} were the SDSS \citep{york2000} fibers (3" diameter) while the largest apertures (typically rectangular slits) were roughly twice that size  (a few kpc). The line emission from these relatively small regions is strongly dominated by the NLR powered by the AGN, with only minor contributions from \textsc{H\,ii} regions in the disk of the host spiral galaxy \citep{theios2016,xia2018}.
Conversely, for the closest Seyfert galaxies, it is possible that extended NLR emission from the AGN could extend beyond 1\,kpc from the nucleus. Fortunately even in the most extreme known cases the [OIII] line emission powered by the AGN is compact, nearly all of it confined within $\sim 700\, \rm{pc}$ from the central nucleus \citep{schmitt1996}.

Table \ref{tab:sample3} gives the K-band photometry of each of the observed galaxies: the estimated total and nuclear K-band flux densities, derived from the 2MASS Extended Source Catalog, the estimated stellar mass of the galaxy, the bolometric luminosity as derived from the [OIV]25.89$\mu$m, using the calibration of \citet{mordini2021} with the bolometric correction of \citet{spinoglio1995}, and the supermassive black hole mass, as derived mostly from stellar velocity dispersion observations that can be found in \citet[and references therein]{fernandez2021}. The galaxy flux has been derived starting from the total 2MASS XSC flux (called ``$K\_mag\_ext'"$, which uses the extrapolation of the surface brightness model fit to the data, integrated over the entire object). From this total luminosity, we then subtract the non-stellar nuclear component. We estimate this from either high spatial resolution K-band observations (Keck/OSIRIS with adaptive optics) \citep{muller-sanchez2018} or by using the ``peak" H and K magnitudes given in the 2MASS XSC. 
For 19 AGN (see Table \ref{tab:sample3}), for which HST observations have been used to compute the R-band absolute magnitudes of the host galaxy \citep{kim2017}, through image decomposition using the code GALFIT \citep{peng2002}, we have adopted the stellar masses as derived from the relation between R-absolute magnitude and stellar mass of \citet{mahajan2018}, who used a sample of 428 galaxies in the redshift range 0.002 $<$ z $<$ 0.02, with MAGPHYS \citep{dacunha2008} fits of the 21-band photometry from \citet{driver2016}.

 In each table the AGN have been divided in four classes: Seyfert type 1 galaxies, Hidden Broad Line Region galaxies (hereafter HBL), Seyfert type 2 galaxies and LINER galaxies (hereafter LIN). The inclusion of the HBL galaxies follows the work by \citet{tommasin2010}, who found that the mid-IR spectroscopic properties of the HBL are indistinguishable from those of the type 1 Seyfert galaxies, and the optical spectropolarimetry that has been collected primarily by \citet{tran2003}.

\section{Results}\label{sec:results}

\subsection{The X-ray/mid-IR lines correlation}\label{sec:corr}

In this study we have considered that, in order to establish a physical link between two quantities (either emission line or continuum strengths) a significant correlation has to be found between their fluxes, because correlations based only on the luminosities are not reliable enough. This is because the luminosity correlations are {\it boosted} by the distance squared factor, which may produce a spurious trend from uncorrelated fluxes. Nevertheless, our results will be mostly compared with luminosity correlations in the literature, since few studies discuss the flux correlations.

To study the relationship between the bolometric continuum of the AGN and the luminosity emitted from the NLR and powered by the accretion activity (hereafter, X--mid-IR relation), we analyzed the hard X-ray $2$--$10\, \rm{keV}$ intrinsic (i.e. absorption-corrected) flux and luminosity as a function of the three high-ionization mid-IR fine-structure NLR emission lines ([NeV] 14.32 $\mu$m, [NeV] 24.32 $\mu$m and [OIV] 25.89 $\mu$m. 
The latter requirement guarantees that most of the energy carried by these lines originates from accretion processes rather than star formation, while the mid-IR wavelength region allows us to neglect dust extinction of the line fluxes. For comparison, we also included in the analysis the lower ionization lines of [NeIII]15.5 $\mu$m (IP = 40.9eV) and [NeII]12.8 $\mu$m. The former is excited by both the AGN and the star-forming component of the galaxy and the latter mostly by the star-forming component \citep{spinoglio1992}.

 \begin{figure*}
  \includegraphics[width=0.5\textwidth]{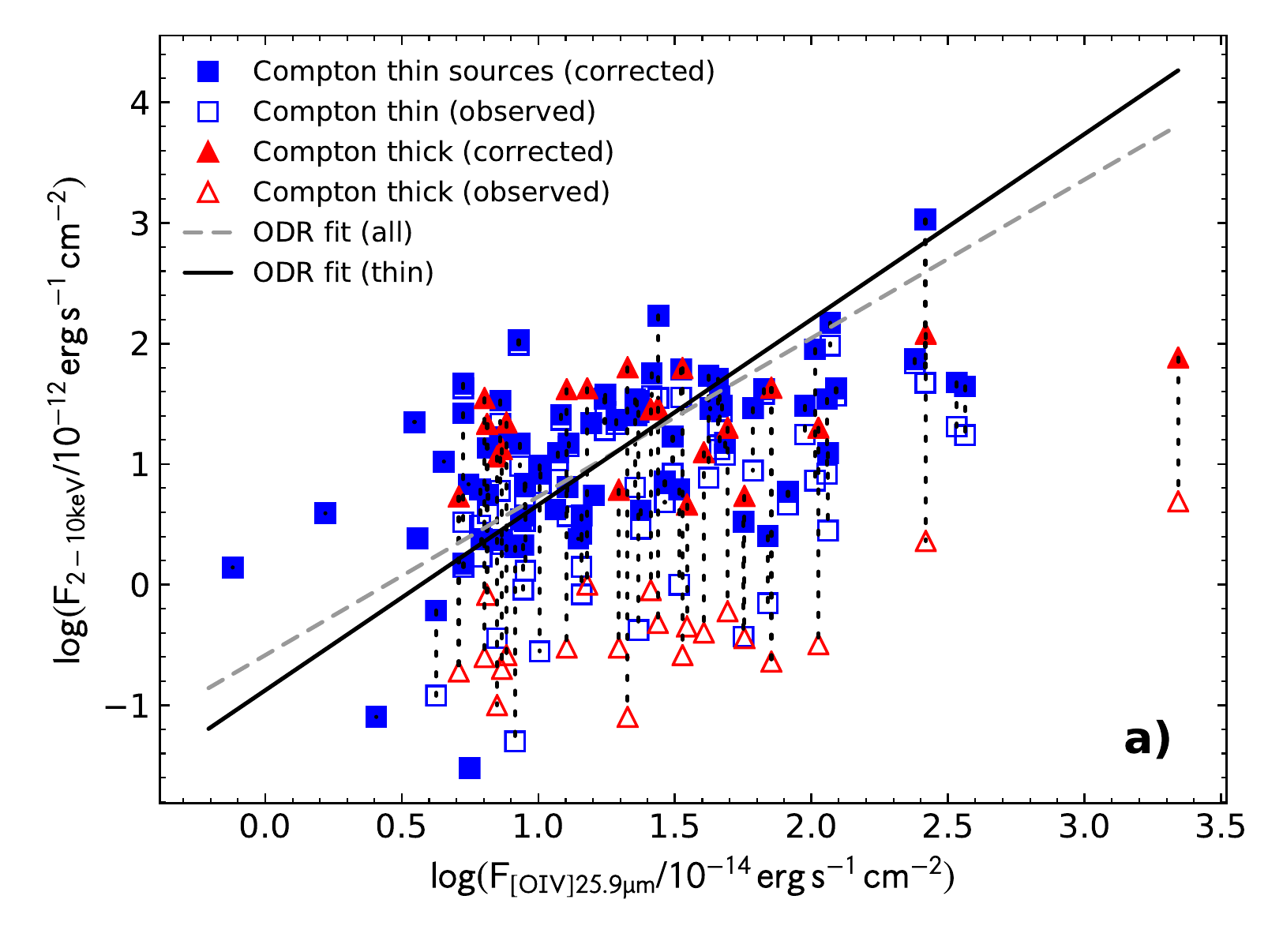}
  \includegraphics[width=0.5\textwidth]{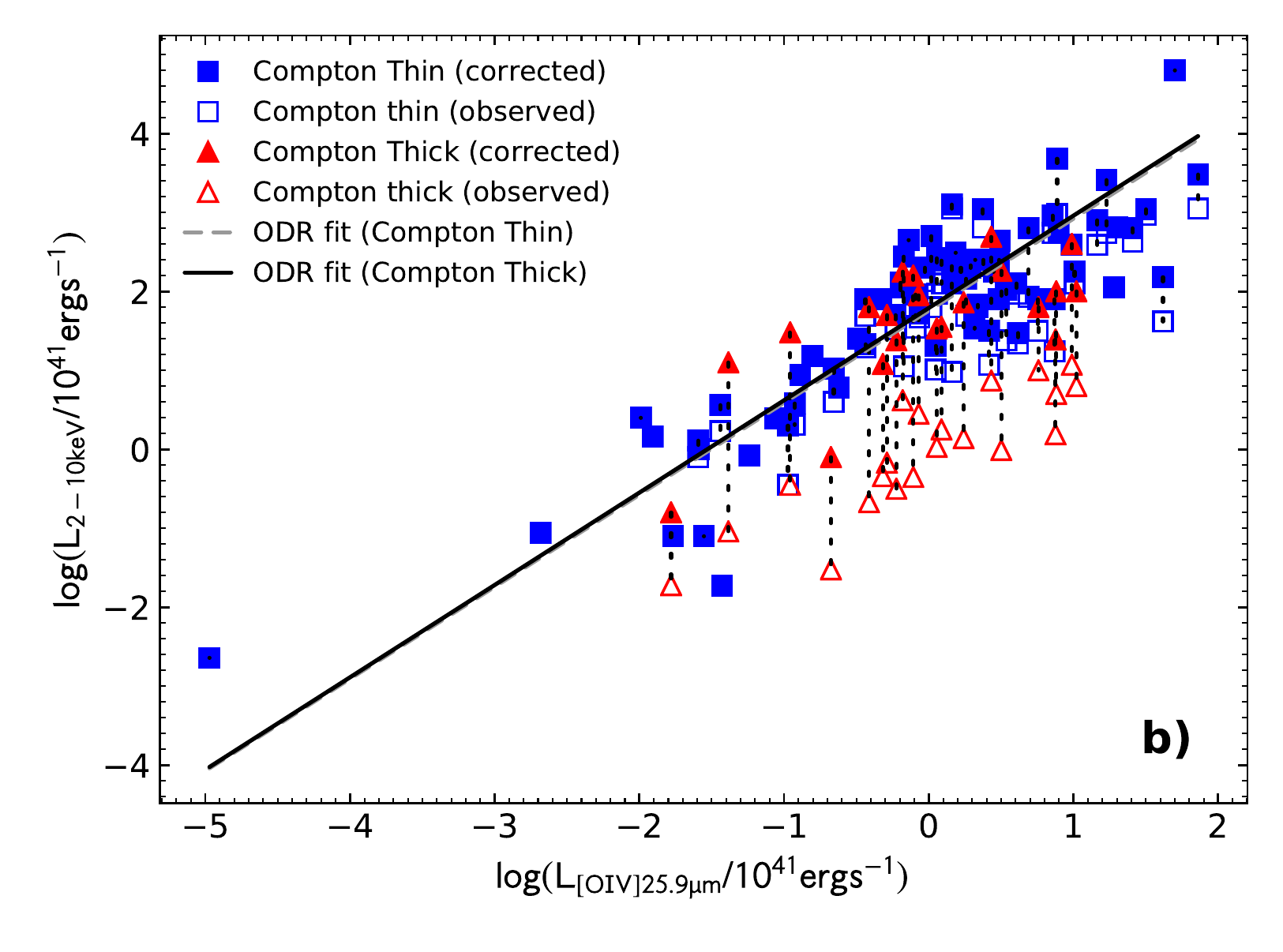}\\
    \includegraphics[width=0.5\textwidth]{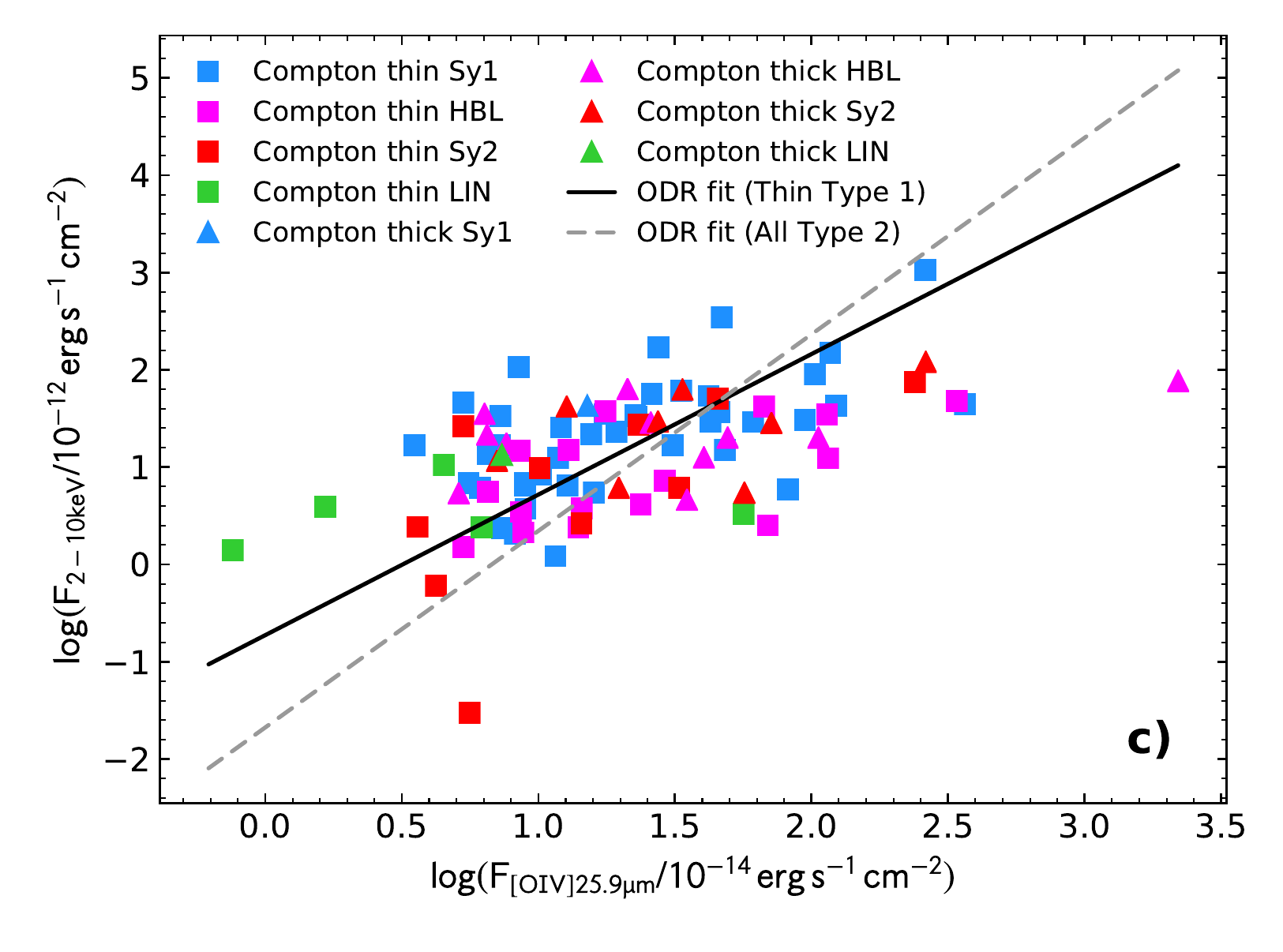}
  \includegraphics[width=0.5\textwidth]{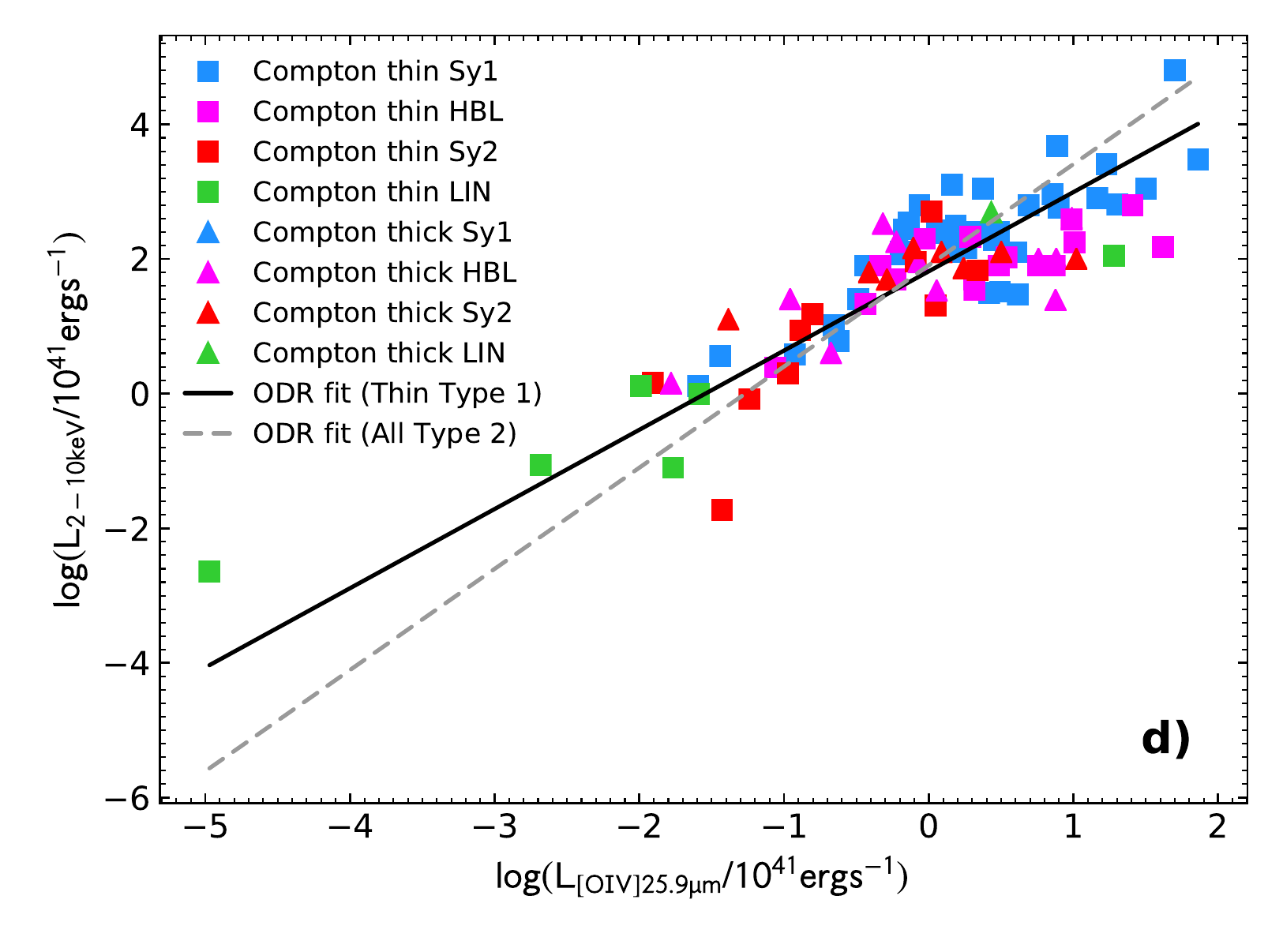}
  \caption{{\bf a: upper left} X-ray (2--10) keV intrinsic flux as a function of the \oiv\ 25.9$\mu$m line flux for all sources with reliable hard-X data and an optical spectroscopic classification. The blue squares indicate the Compton-thin AGN, while Compton- thick objects are denoted by red triangles. The solid lines represent the linear regression fit with all objects, while the dotted line shows the fit to only Compton-thin objects (see Table \ref{tab:cor}). {\bf b: upper right} X-ray (2-10) keV intrinsic luminosity vs. \oiv\ line luminosity (both in logarithmic units of \ergs). {\bf c: lower left} Same as {\bf a}, but with the objects color coded: Seyfert types 1 (Sy1): blue; Hidden Broad Line Region Galaxies (HBL): magenta; Seyfert types 2 (Sy2): red; LINERS (LIN): green. The solid line shows the fit for all Compton-thin Type 1 AGN, which include both Seyfert 1's and HBL, while the broken line gives the bet fit for all the Type 2 objects, i.e. the {\it pure} Seyfert 2's, both Compton- thick and thin (see Table \ref{tab:cor}). {\bf d: lower right} Same as {\bf b}, but with the objects color coded, according to panel {\bf c}. The fits are similar to those in panel {\bf c}, but between luminosities (see Table \ref{tab:cor}).
  }
  \label{fig:X-OIVF&L}
  \end{figure*}
  
\begin{figure*}
  \includegraphics[width=0.5\textwidth]{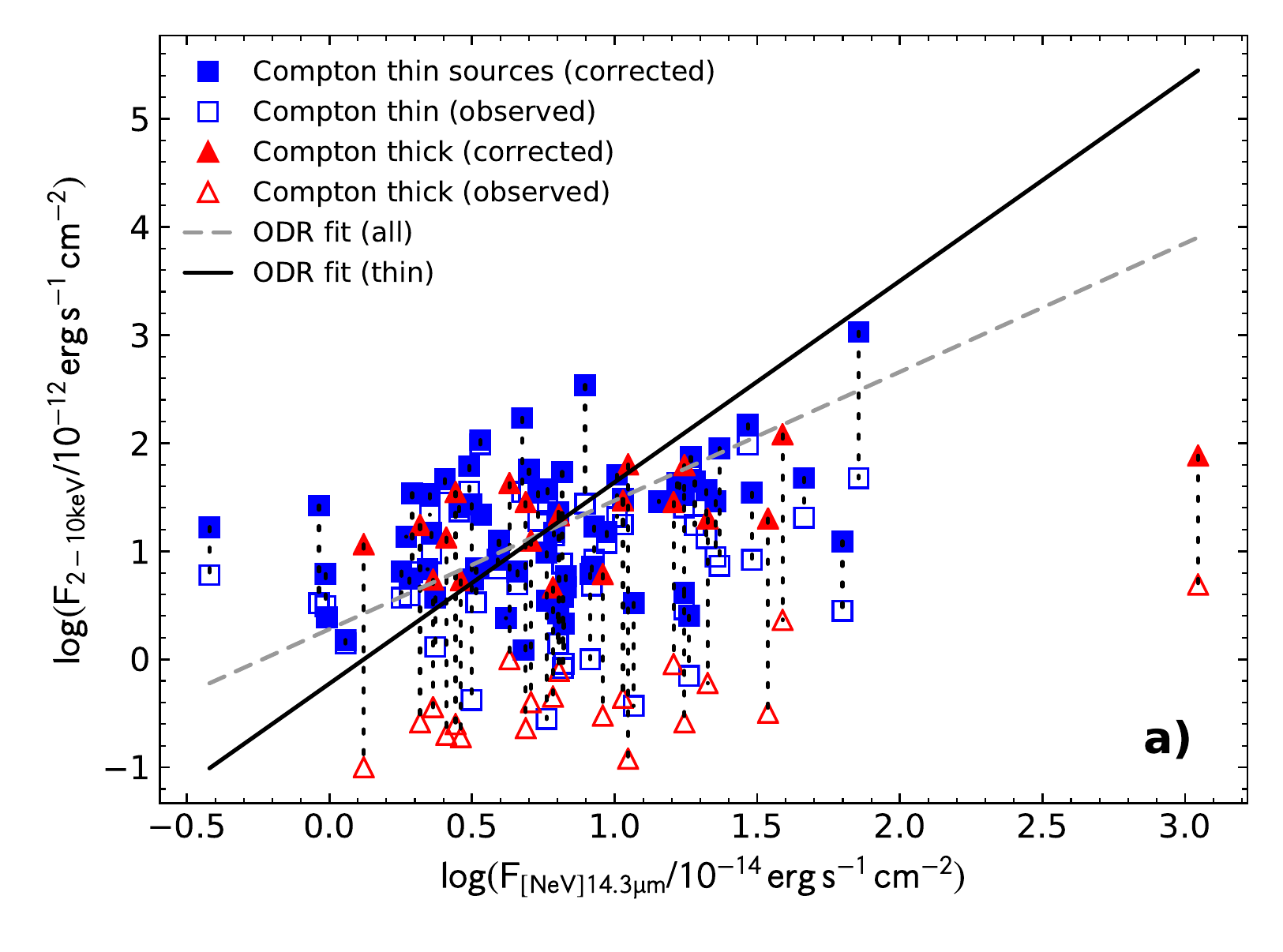}
  \includegraphics[width=0.5\textwidth]{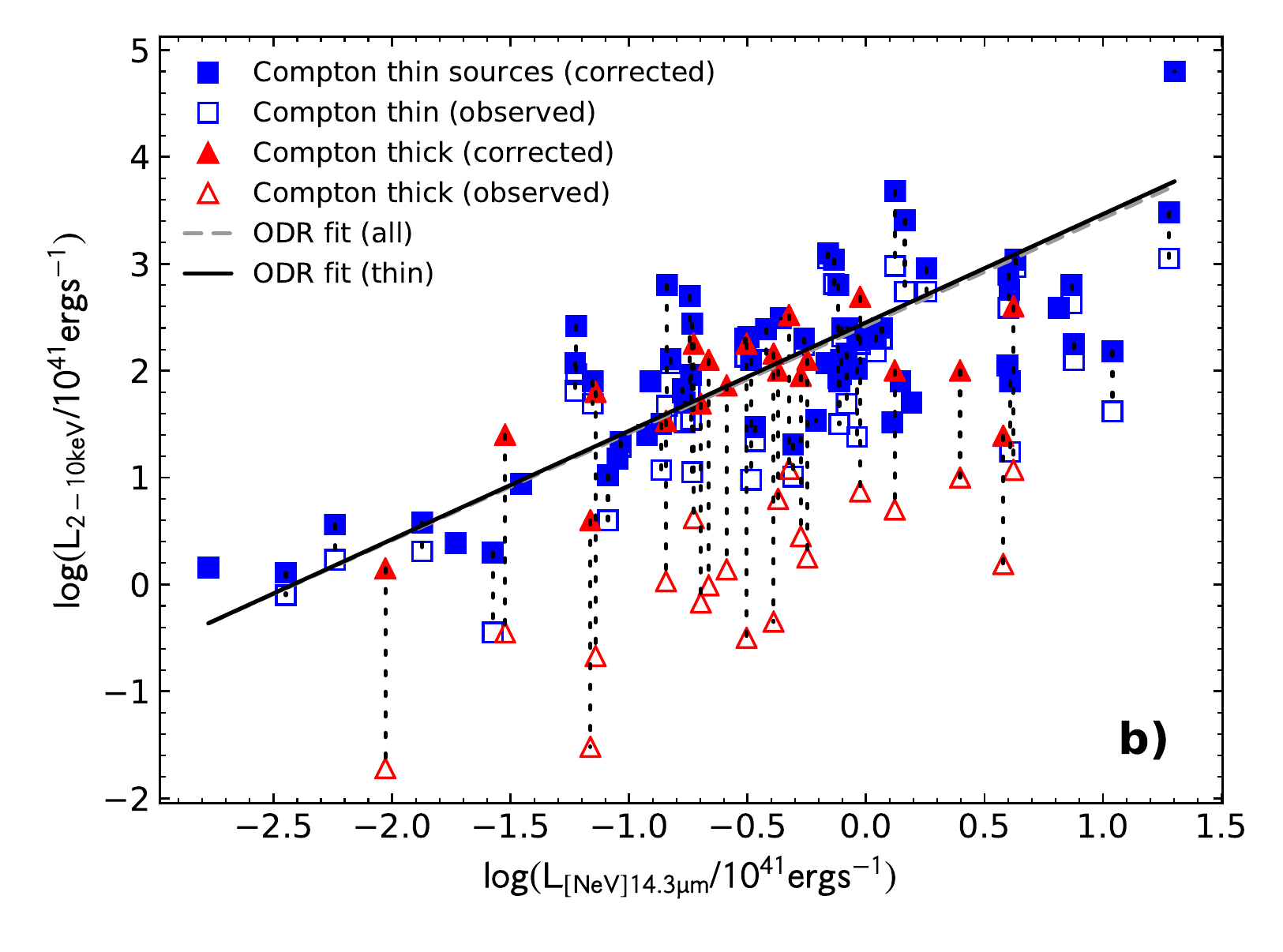}
  \caption{{\bf a: left} X-ray (2--10) keV intrinsic flux as a function of the \nev\ 14.3$\mu$m line flux for all sources with reliable hard-X data and an optical spectroscopic classification. The blue squares indicate the Compton-thin AGN, while Compton-thick objects are denoted by red triangles. The solid lines represent the linear regression fit with all objects, while the dotted line the fit with only Compton-thin objects. {\bf b: right} X-ray (2-10) keV intrinsic luminosity vs. \nev\ line luminosity (both in logarithmic units of \ergs).}  \label{fig:X-NeV14F&L}
  \end{figure*}
  
  \begin{figure*}
  \includegraphics[width=0.5\textwidth]{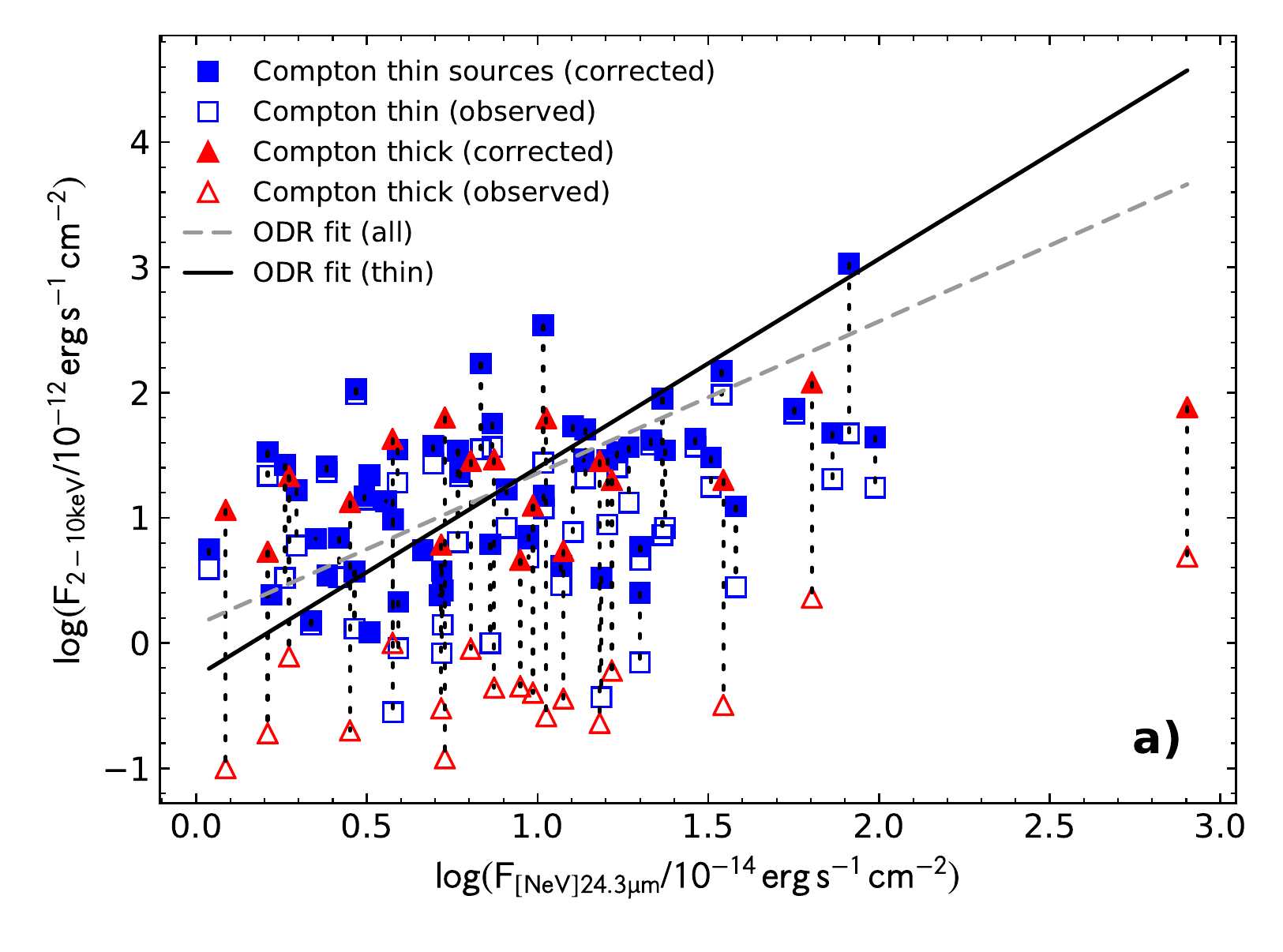}
  \includegraphics[width=0.5\textwidth]{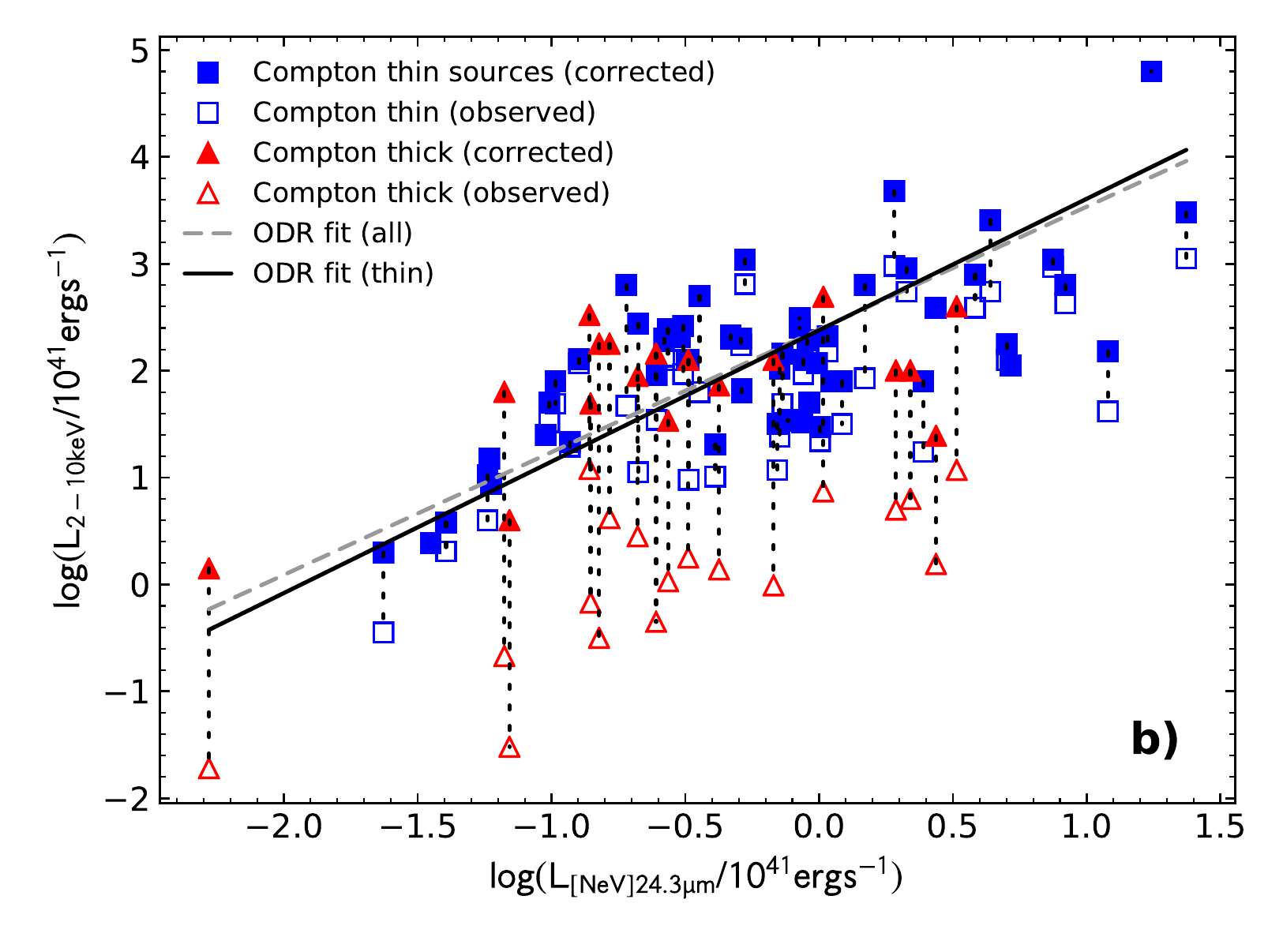}
  \caption{{\bf a: left} X-ray (2--10) keV intrinsic flux as a function of the \nev\ 24.3$\mu$m line flux for all sources with reliable hard-X data and an optical spectroscopic classification. The blue squares indicate the Compton-thin AGN, while Compton-thick objects are denoted by red triangles. The solid lines represent the linear regression fit with all objects, while the dotted line the fit with only Compton-thin objects. {\bf b: right} X-ray (2-10) keV intrinsic luminosity vs. \nev\ 24.3$\mu$m line luminosity (both in logarithmic units of \ergs). 
  }
  \label{fig:X-NeV24F&L}
  \end{figure*}

To quantify the statistical significance of these relations, we used a orthogonal distance regression (ODR) fit. The results are presented in Table \ref{tab:cor}, where the Pearson linear correlation coefficients are shown for the flux and luminosity relations for each of the five considered mid-IR fine structure lines, together with the linear regression best-fit parameters with relative uncertainties. All correlations have been computed for the whole population of AGN and for the sub-sample of Compton-thin objects (C-thin). For the particular case of the [OIV]25.9$\mu$m line, because of the higher statistics compared to the other lines, also the correlations of the sub-sample of Type 1 objects, which include all the Compton-thin Seyfert type 1 and HBL galaxies, have been computed and compared to the Type 2 objects, which include ''{\it pure}" Seyfert type 2 galaxies (see Table \ref{tab:cor}). 

 \begin{figure*}
  \includegraphics[width=0.5\textwidth]{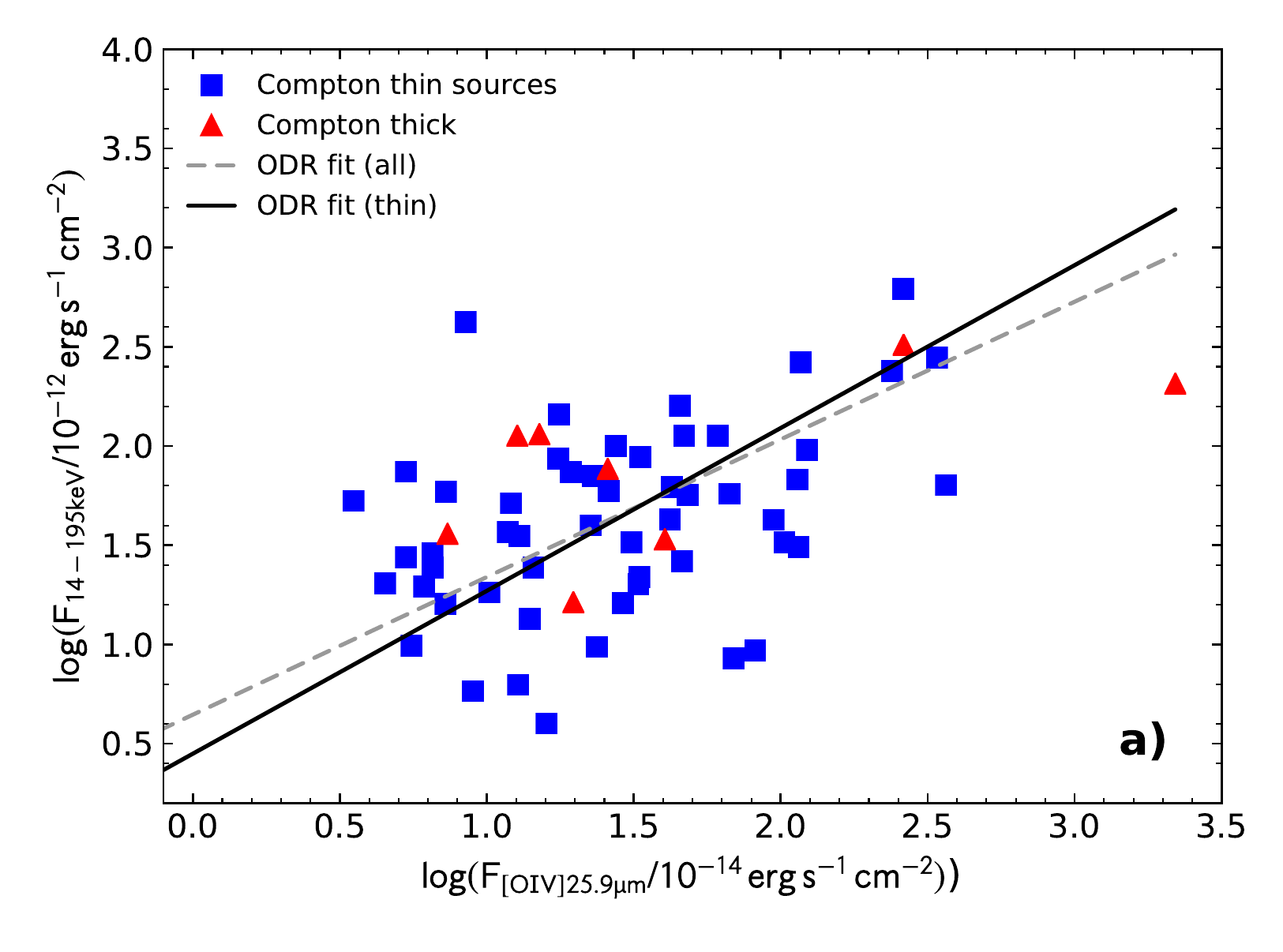}
  \includegraphics[width=0.5\textwidth]{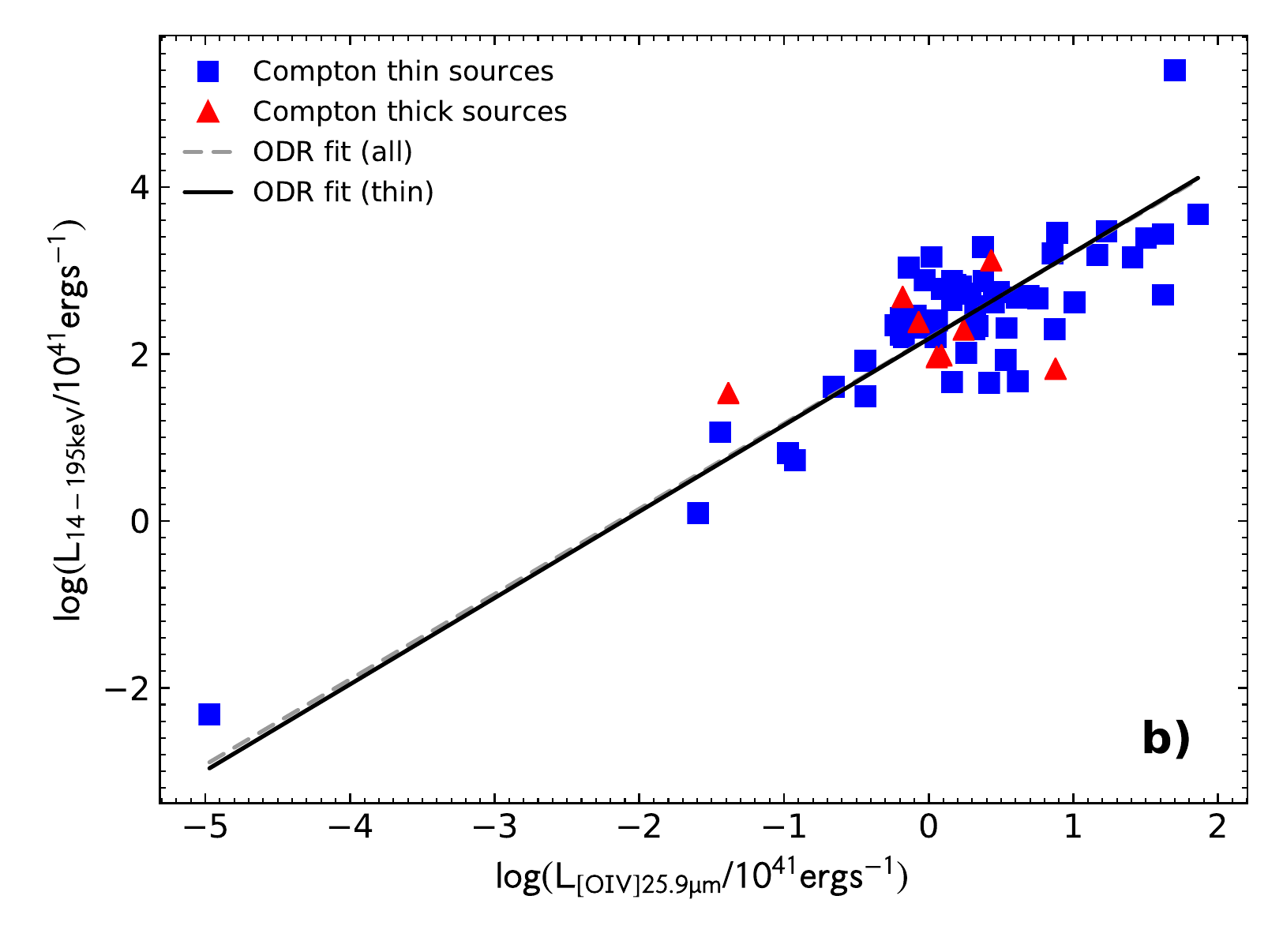}\\
    \includegraphics[width=0.5\textwidth]{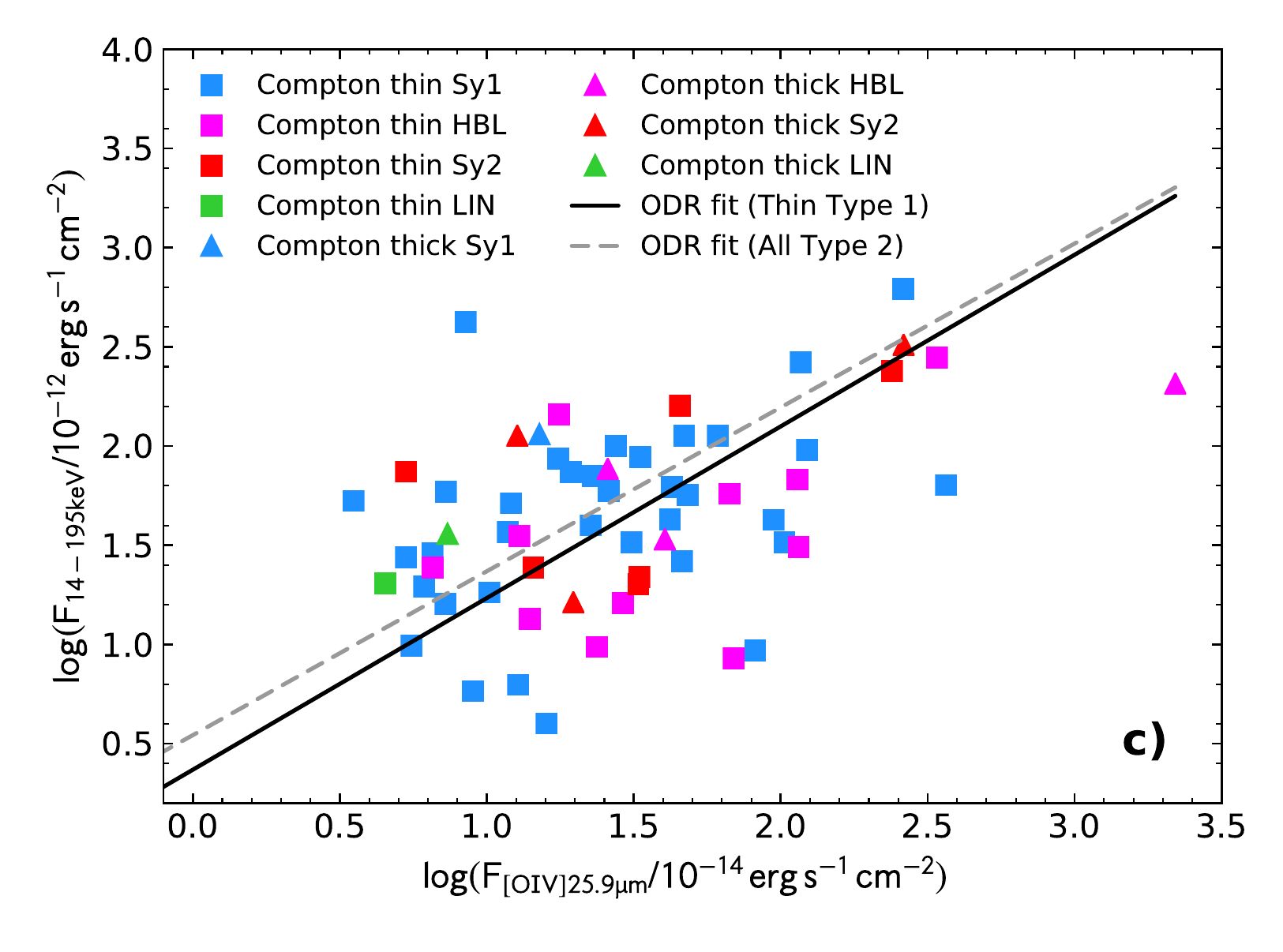}
  \includegraphics[width=0.5\textwidth]{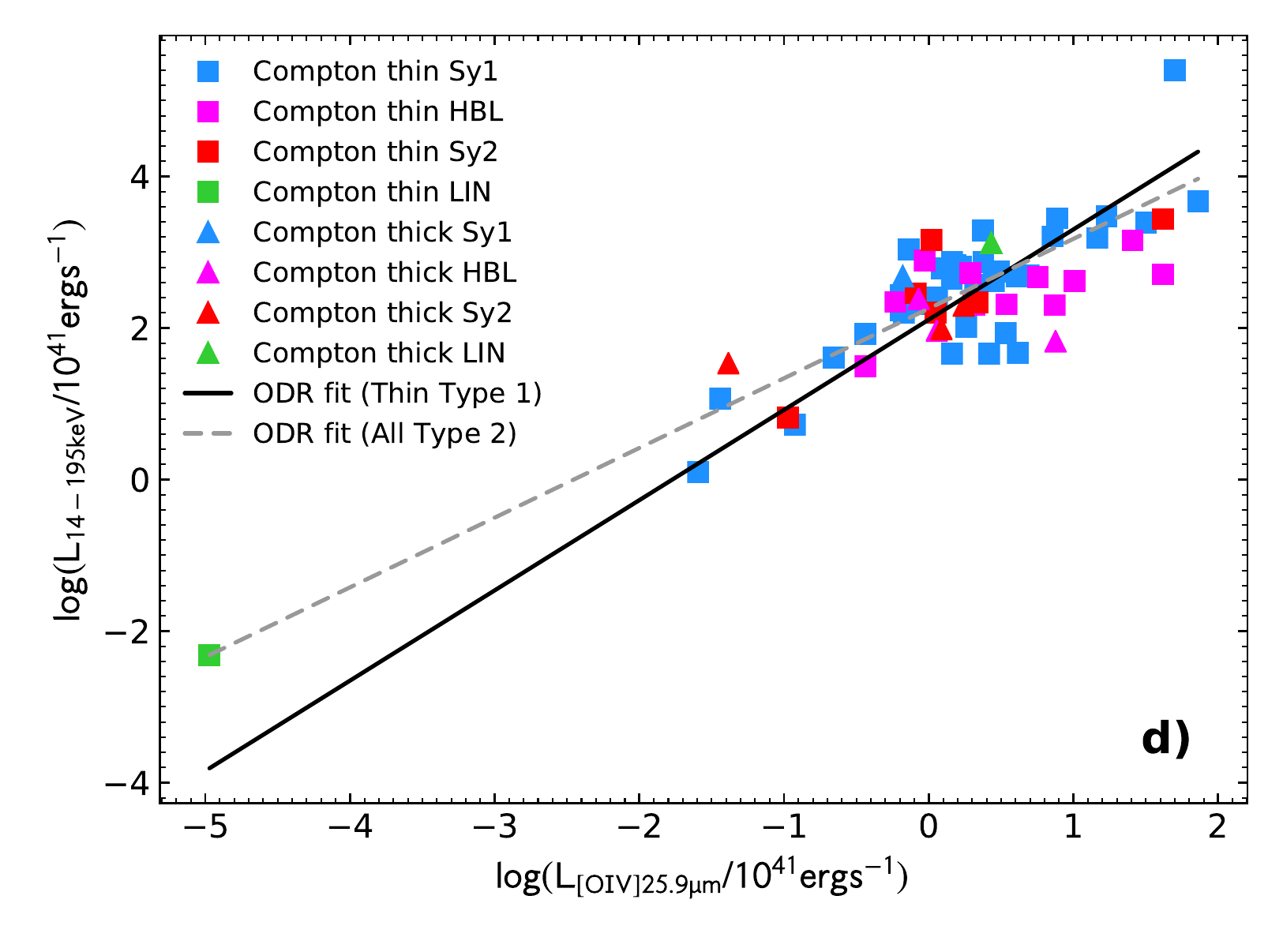}
  \caption{{\bf a: left} X-ray (14-195) keV intrinsic flux as a function of the \oiv\ 25.9$\mu$m line flux for all sources with reliable hard-X data and an optical spectroscopic classification. The blue squares indicate the Compton-thin AGN, while Compton-thick objects are denoted by red triangles. The solid lines represent the linear regression fit with all objects, while the dotted line the fit with only Compton-thin objects . {\bf b: right} X-ray (14-195) keV intrinsic luminosity vs. \oiv\ line luminosity (both in logarithmic units of \ergs). 
 {\bf c: lower left} Same as {\bf a}, but with the objects color coded: Seyfert types 1 (Sy1): blue; Hidden Broad Line Region Galaxies (HBL): magenta; Seyfert types 2 (Sy2): red; LINERS (LIN): green. 
  {\bf d: lower right} Same as {\bf b}, but with the objects color coded, according to panel {\bf c}. }
  \label{fig:HX-OIVF&L}
  \end{figure*}

\begin{figure*}
  \includegraphics[width=0.5\textwidth]{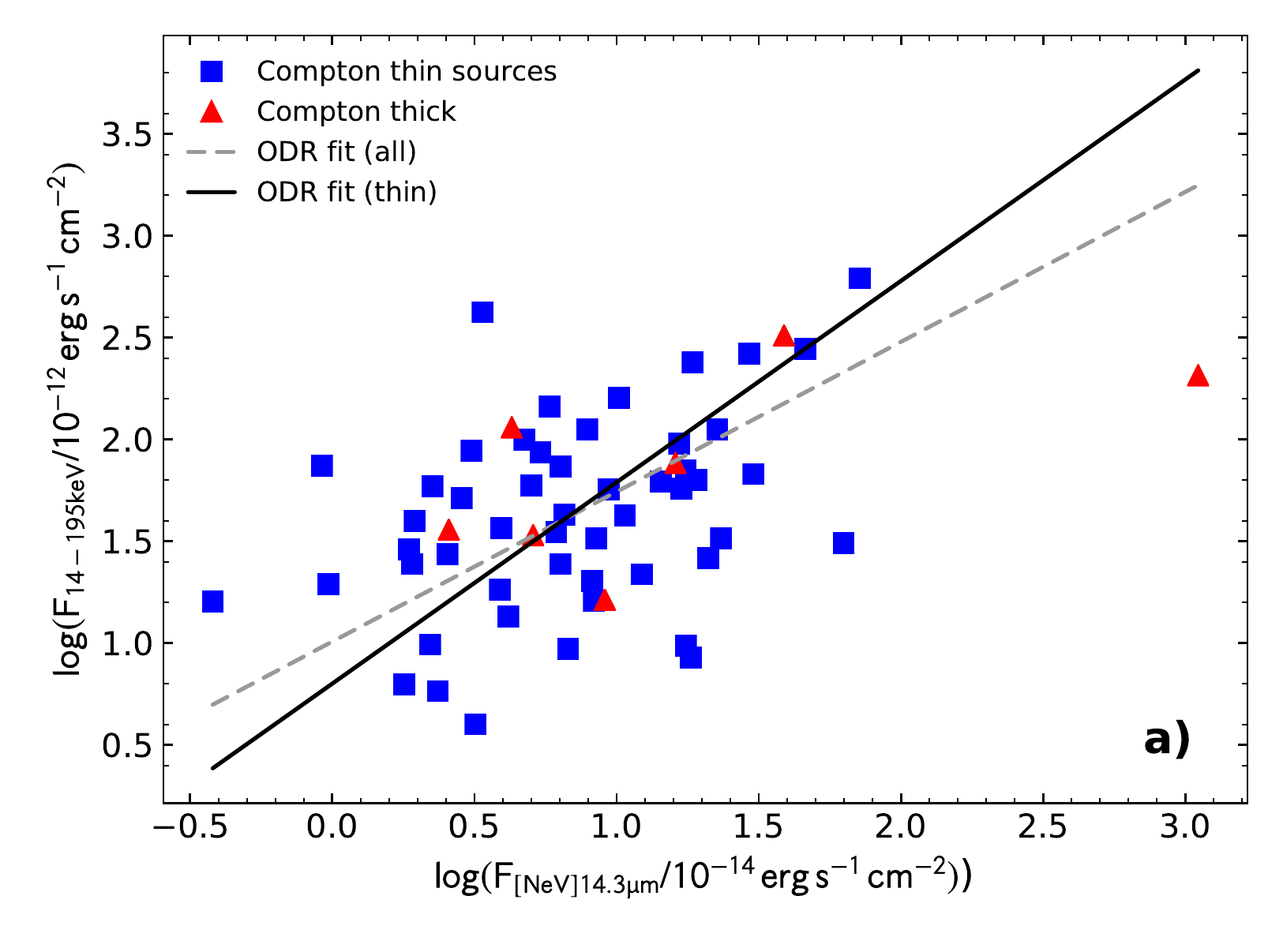}
  \includegraphics[width=0.5\textwidth]{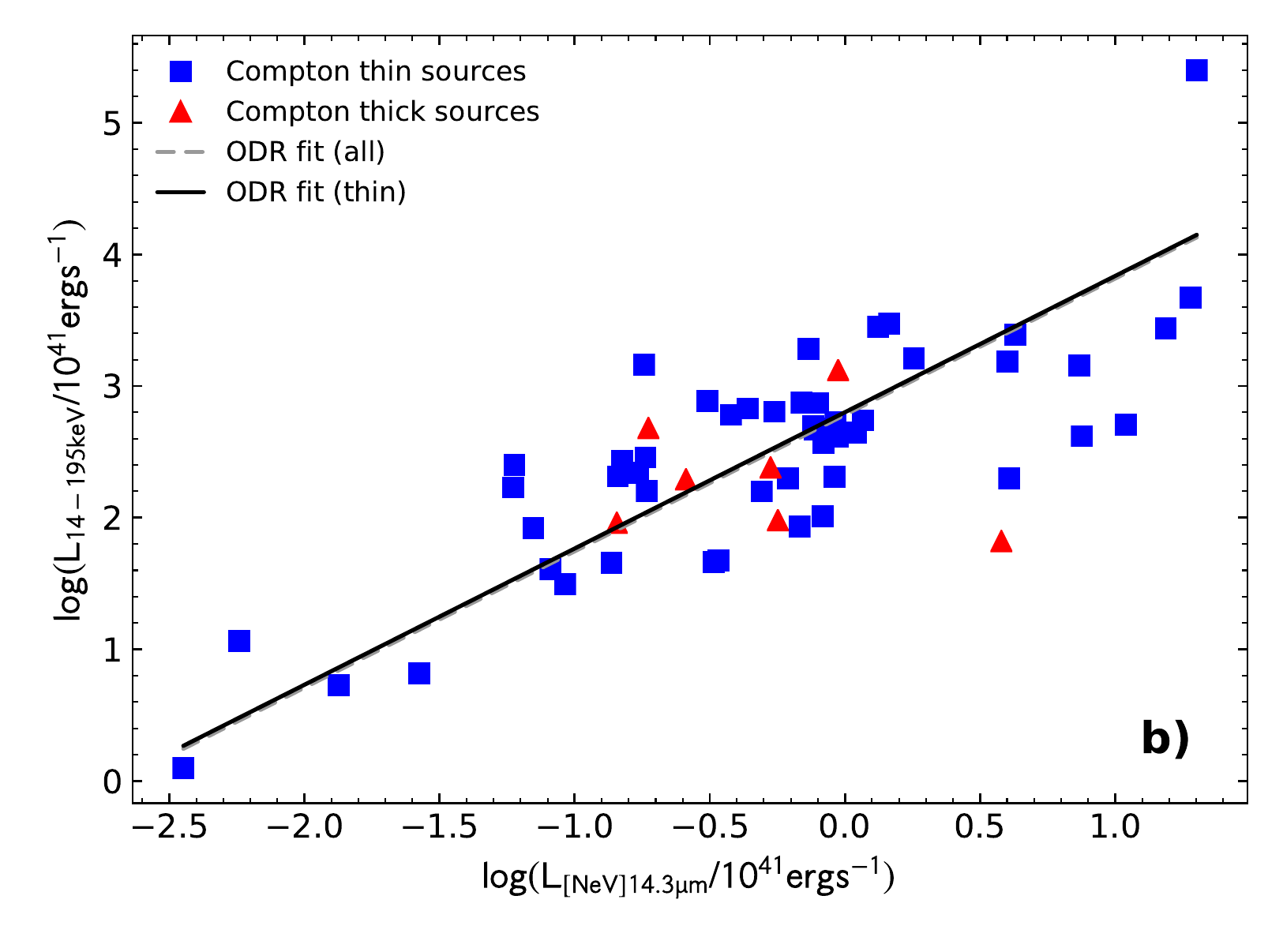}
  \caption{{\bf a: left} X-ray (14-195) keV intrinsic flux as a function of the \nev\ 14.3$\mu$m line flux for all sources with reliable hard-X data and an optical spectroscopic classification. The blue squares indicate the Compton-thin AGN, while Compton-thick objects are denoted by red triangles. The solid lines represent the linear regression fit with all objects, while the dotted line the fit with only Compton-thin objects. {\bf b: right} X-ray (14-195) keV intrinsic luminosity vs. \nev\ 14.3$\mu$m line luminosity (both in logarithmic units of \ergs). 
  }
  \label{fig:HX-NeV14F&L}
  \end{figure*}
  
  \begin{figure*}
  \includegraphics[width=0.5\textwidth]{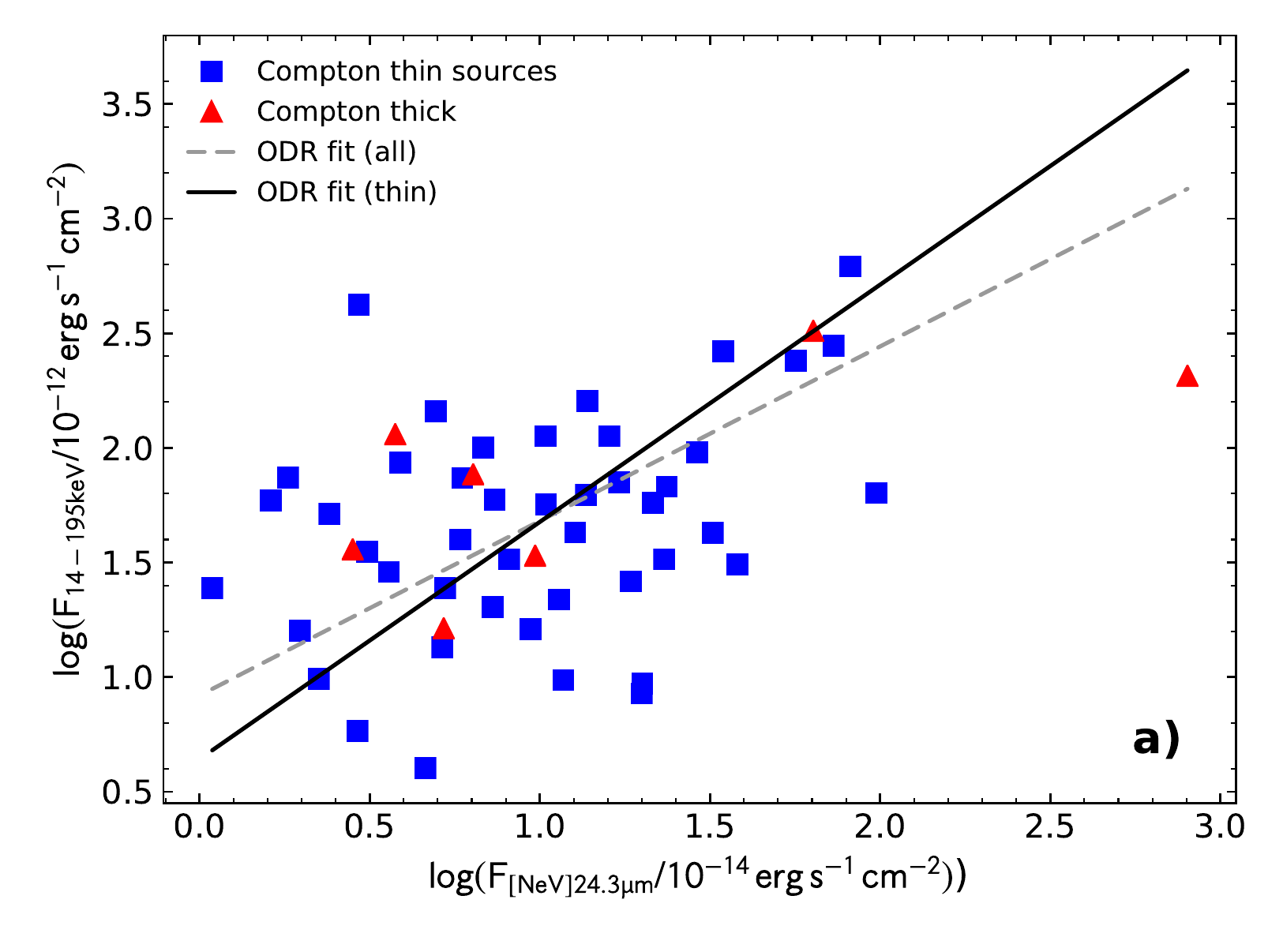}
  \includegraphics[width=0.5\textwidth]{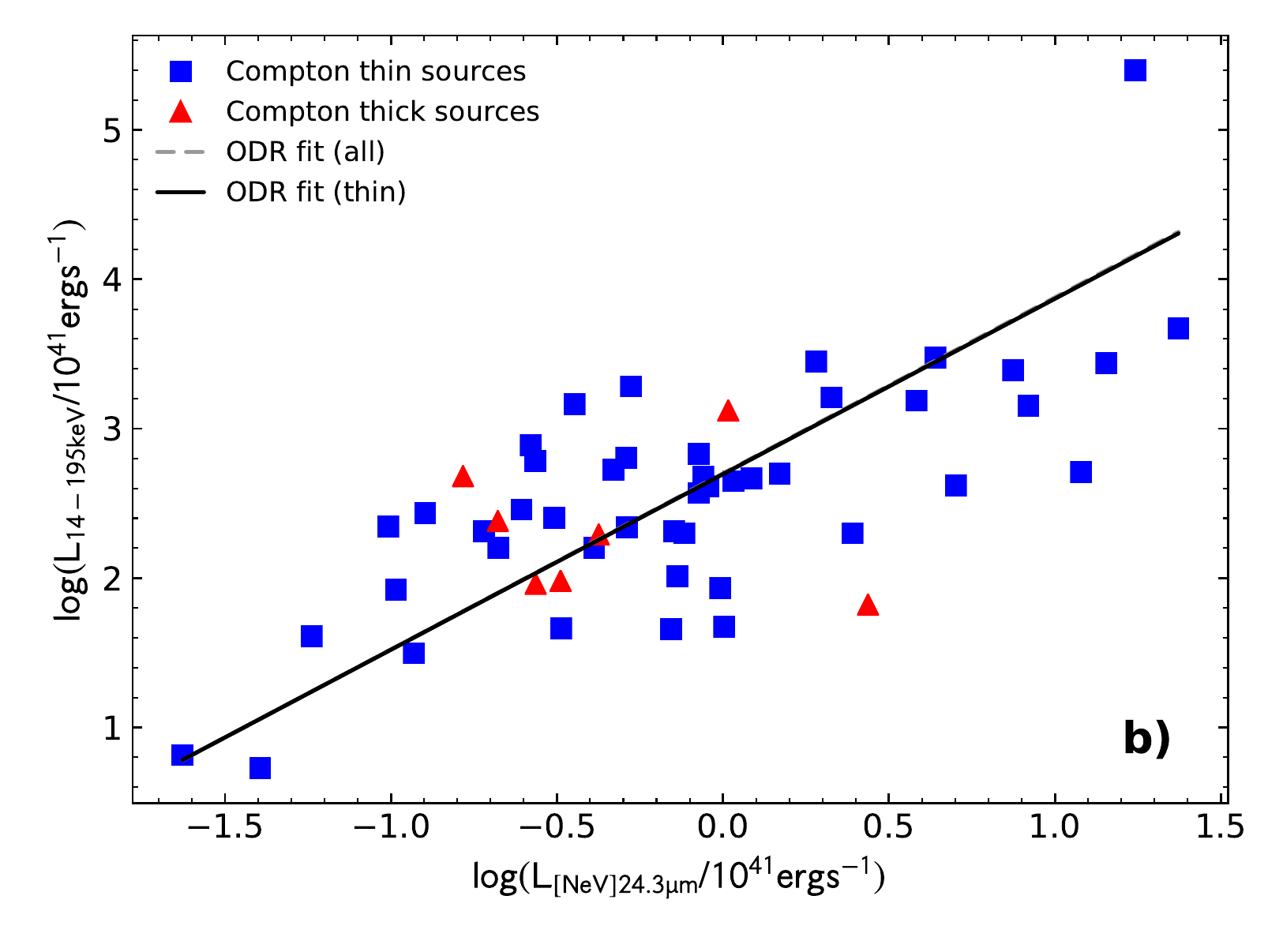}
  \caption{{\bf a: left} X-ray (14-195) keV intrinsic flux as a function of the \nev\ 24.3$\mu$m line flux for all sources with reliable hard-X data and an optical spectroscopic classification. The blue squares indicate the Compton-thin AGN, while Compton-thick objects are denoted by red triangles. The solid lines represent the linear regression fit with all objects, while the dotted line the fit with only Compton-thin objects. {\bf b: right} X-ray (14-195) keV intrinsic luminosity vs. \nev\ 24.3$\mu$m line luminosity (both in logarithmic units of \ergs). 
  }
  \label{fig:HX-NeV24F&L}
  \end{figure*}  

In Figure \ref{fig:X-OIVF&L} the hard X-ray ($2$--$10)\, \rm{keV}$ observed (open symbols) and intrinsic (i.e. absorption-corrected) fluxes (panel {\bf a}) and luminosities (panel {\bf b}) of both the Compton-thin and the Compton-thick AGN are plotted as a function of the flux and luminosity, respectively, of the \oiv\ 25.9$\mu$m line for the subsample of the 12MSG with reliable hard X-ray data obtained from the literature. The observed and corrected X-ray fluxes and luminosities are connected in the diagrams by a broken line, which indicates the correction applied either by fitting the X-ray spectra or by extrapolating the X-ray ($14$--$195)\, \rm{keV}$ flux and luminosity to the ($2$--$10)\, \rm{keV}$ band \citep[see, e.g.,][]{ricci2017}. In the same Fig. \ref{fig:X-OIVF&L}, panels {\bf c} and {\bf d} show the intrinsic fluxes and luminosities, respectively, but with the color coded types of AGN, which are divided in the four classes of Seyfert 1's (Sy1), Hidden Broad Line region galaxies (HBL), Seyfert 2's (Sy2) and LINER  (LIN).

Figures \ref{fig:X-NeV14F&L} and \ref{fig:X-NeV24F&L} show the ($2$--$10)\, \rm{keV}$ observed and intrinsic fluxes (panel {\bf a}) and luminosities (panel {\bf b}) versus that of the \nev\ 14.3$\mu$m line and \nev\ 24.3$\mu$m line, respectively.

It is clear from the three figures \ref{fig:X-OIVF&L}, \ref{fig:X-NeV14F&L} and \ref{fig:X-NeV24F&L} that to derive a reliable relation between the X-ray and mid-IR line fluxes and luminosities it is mandatory to correct the X-ray data, as the application of absorption corrections reduces the scatter of the data points by a large amount. We do not even try to compute correlations using the observed ($2$--$10)\, \rm{keV}$ X-ray fluxes.

To demonstrate that the higher ionization lines of \nev~and \oiv~are the best tracers of AGN emission, we also show in Fig. \ref{fig:X-NeIIIF&L} and \ref{fig:X-NeIIF&L} in the Appendix the same plots between the ($2$--$10)\, \rm{keV}$ observed and intrinsic fluxes (panel {\bf a}) and luminosities (panel {\bf b}) and the \neiii~and \neii~lines.

Examining the results of the orthogonal distance regression (ODR) fits, shown in Table \ref{tab:cor}, it is clear that the best flux correlation is obtained for the \oiv~25.9$\mu$m line, followed by the \nev 24.3$\mu$m and \nev 14.3$\mu$m. The slope of the correlation for these three mid-IR fine structure lines is of the order of $B=1.2-1.4$ and becomes relatively steeper $B=1.6-1.8$ when considering only the Compton-thin AGN, instead of the total observed population. The \neiii~ line has a relatively high value of the Pearson coefficient (R$\sim$0.47, n=100), however the slope is abnormally high ($B=2.7-3.5$). To the contrary, no correlation is apparent between the \neii~ line and the X-ray flux. 

In analogy with Fig.\ref{fig:X-OIVF&L}, Fig.\ref{fig:HX-OIVF&L} shows the higher energy X-ray ($14$--$195)\, \rm{keV}$ fluxes (panel {\bf a}) and luminosities (panel {\bf b}) of both the Compton-thin and the Compton-thick AGN as a function of the flux and luminosity, respectively, of the \oiv\ 25.9$\mu$m line. In the same Fig. \ref{fig:HX-OIVF&L}, panels {\bf c} and  {\bf d} show the intrinsic fluxes and luminosities, respectively, but with the color coded types of AGN.
In the following figures, i.e. Fig. \ref{fig:HX-NeV14F&L}, \ref{fig:HX-NeV24F&L} and Fig. \ref{fig:HX-NeIIIF&L}, \ref{fig:HX-NeIIF&L} in the Appendix, the X-ray ($14$--$195)\, \rm{keV}$ flux and luminosity is plotted as a function of the \nev 14.3$\mu$m, \nev 24.3$\mu$m, \neii~ and \neiii~ fluxes and luminosities, respectively.

The ODR fits results, given in Table \ref{tab:cor}, show that for the \oiv~ and \nev~ lines the slope is significantly flatter than unity ($B \sim 0.7$) if all AGN are considered, but it becomes close to unity ($B=0.8-1.0$) when only Compton-thin objects are included. This is also true for the lower-ionization \neiii~ line and in most cases the correlations are relatively strong (i.e. with a $P_{null}$ probability below 10$^{-2}$). The case of \neii~, as already seen for the lower energy X-ray band of ($2$--$10)\, \rm{keV}$, shows the weakest correlation.

Considering the correlations between the luminosities of either ($2$--$10)\, \rm{keV}$ and ($14$--$195)\, \rm{keV}$ bands and the mid-IR fine-structure line luminosities, as can be seen from either the figures and Table \ref{tab:cor}, these are always very strong and are almost linear (with a slope of  $B=1.0-1.3$). This is not surprising, because luminosity-luminosity correlations are boosted by the 4$\pi$R$^2$ factor, where R is the distance of galaxy,  and therefore they do not necessarily mean that the observed physical quantities are indeed correlated. We consider here that a real physical correlation must appear also between the observed (or corrected, in the case of the ($2$--$10)\, \rm{keV}$ X-ray band) fluxes and the considered emission line.


\subsection{The AGN bolometric luminosities as a function of the \oiv~ and \nev~ line luminosities}\label{sec:bollum_lines}

\begin{figure*}
  \includegraphics[width=0.33\textwidth]{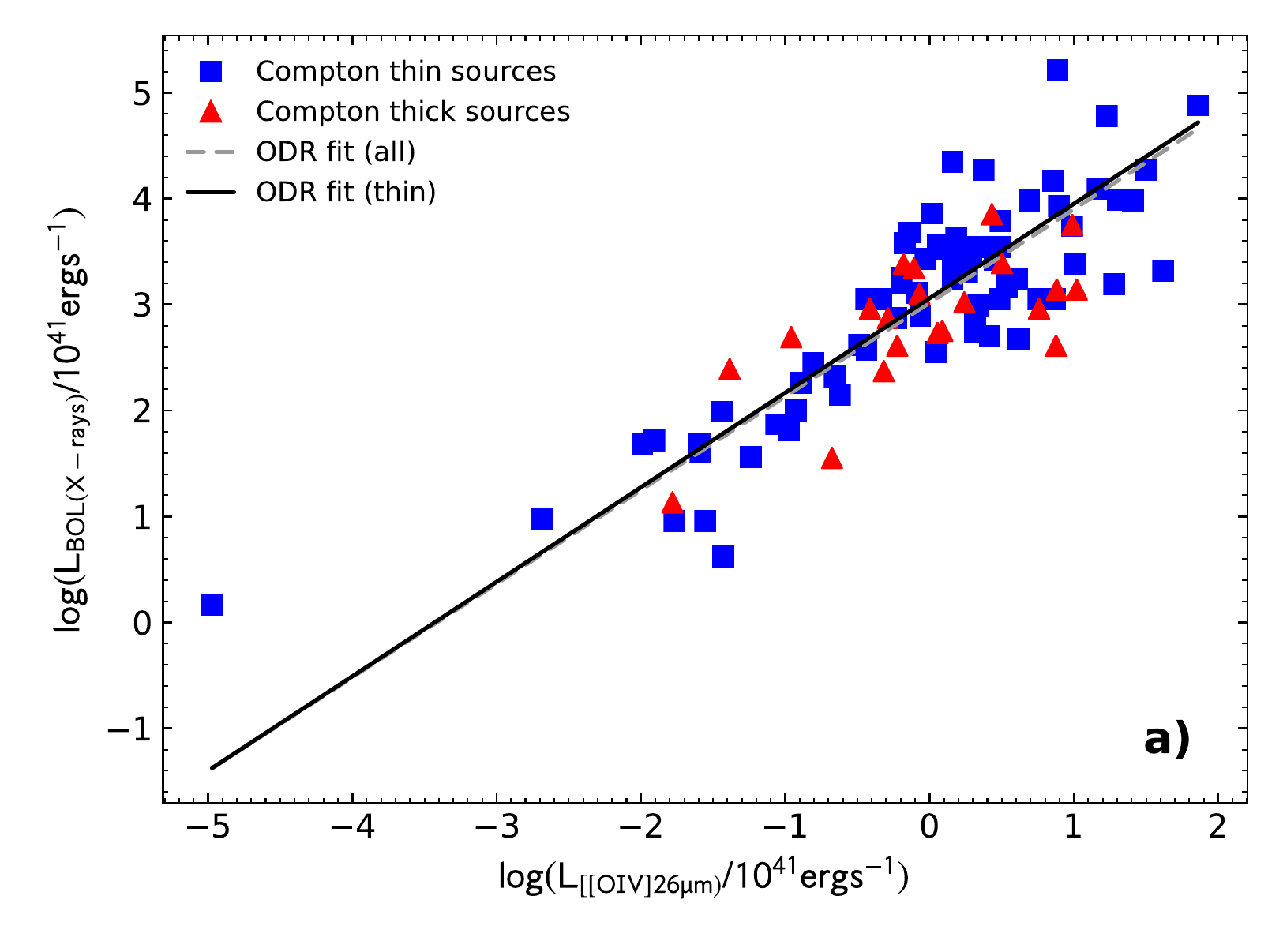}
  \includegraphics[width=0.33\textwidth]{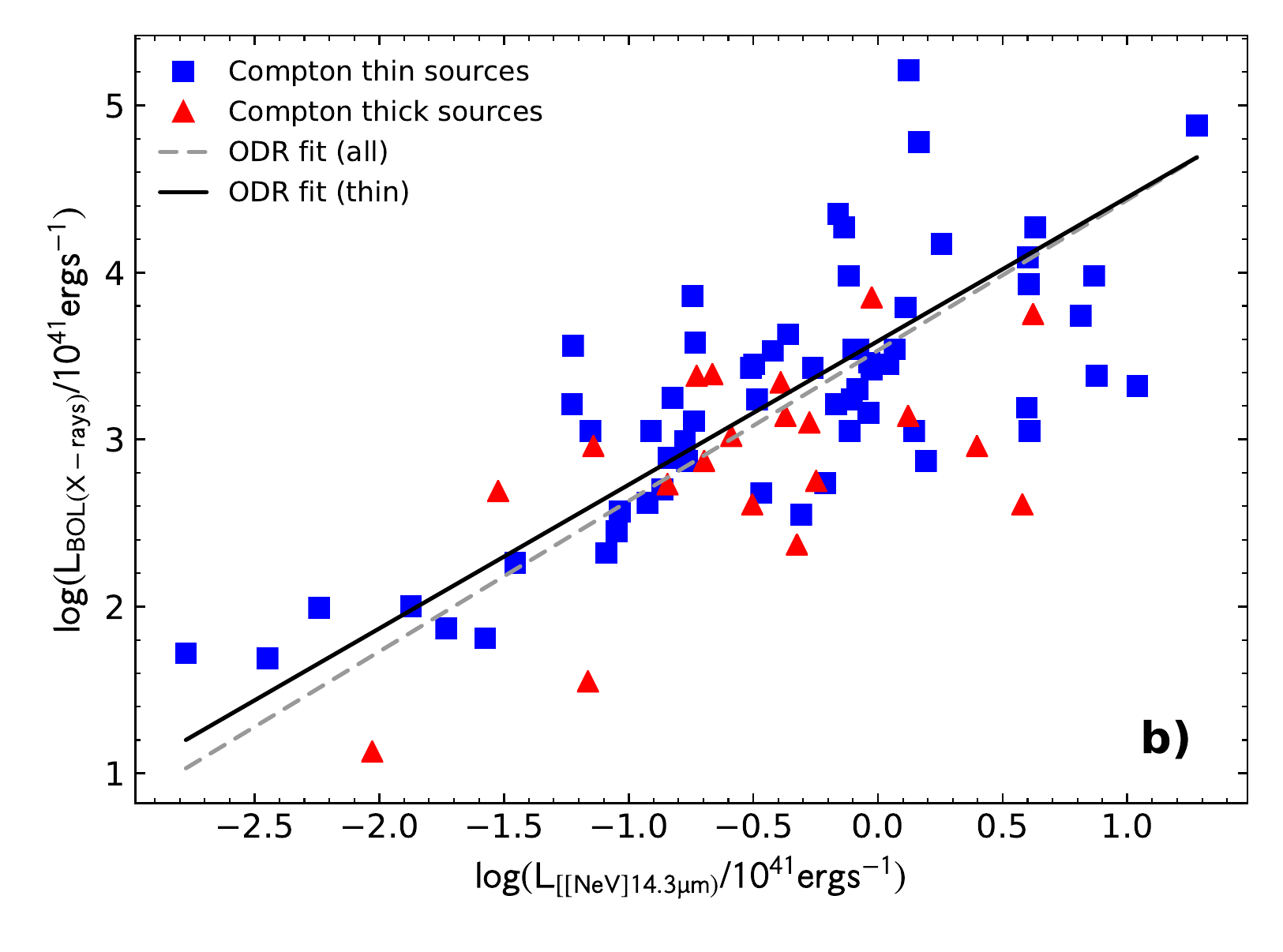}
  \includegraphics[width=0.33\textwidth]{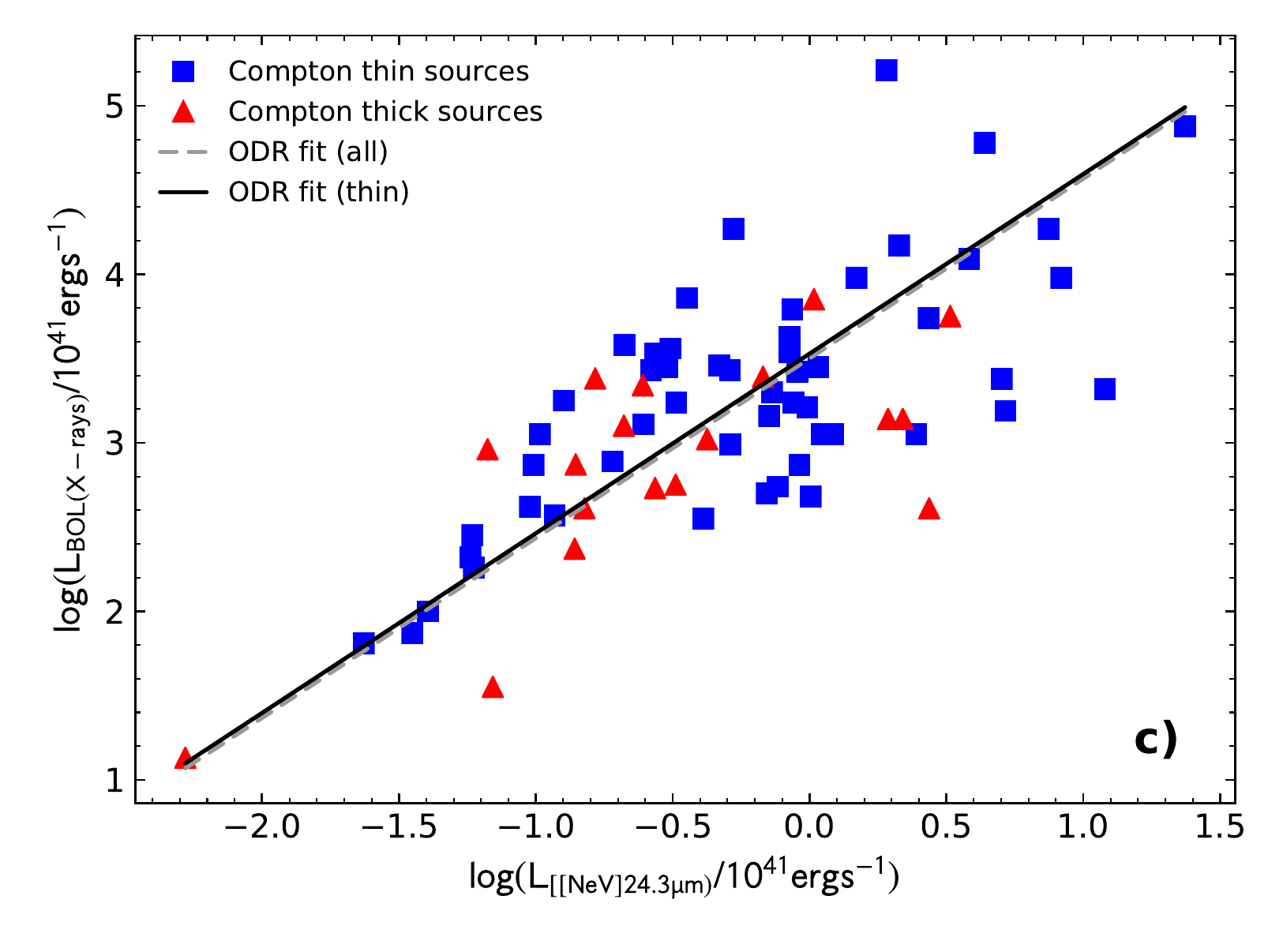}
  \caption{{\bf a: left} AGN bolometric luminosity as a function of the \oivp~ line luminosity. The blue squares indicate the Compton-thin AGN, while Compton-thick objects are denoted by red triangles. The solid lines represent the linear regression fit with all objects, while the dotted line the fit with only Compton-thin objects. {\bf b: center} AGN bolometric luminosity as a function of the \nevp~ line luminosity. 
  {\bf c: right} AGN bolometric luminosity as a function of the \nevs~ line luminosity.
  }
  \label{fig:lbol_lines}
  \end{figure*}

We have computed the AGN bolometric luminosities, using the \citet{lusso2012} bolometric correction and the corrected (2-10)keV luminosities, as a function of the luminosities of the three mid-IR high ionization lines of \oivp, \nevp~ and \nevs. The correlations are shown in Fig. \ref{fig:lbol_lines}({\bf a}, {\bf b}, and {\bf c}) and are tabulated in Table \ref{tab:cor2}.

The following formulae can be used when one wants to compute the AGN bolometric luminosity from the luminosities of these lines:
\begin{equation}
\begin{split}
    \log\left(\frac{\rm L_{BOL}}{\rm 10^{41}\,erg\,s^{-1}}\right)=(3.02 \pm 0.06)\\
    +(0.88 \pm 0.06)\log\left(\frac{L_{\rm [OIV]25.9{\mu}m}}{\rm 10^{41}\,erg\,s^{-1}}\right),
    \end{split}
\end{equation}

\begin{equation}
\begin{split}
    \log\left(\frac{\rm L_{BOL}}{\rm 10^{41}\,erg\,s^{-1}}\right)=(3.53 \pm 0.07)\\
    +(0.90 \pm 0.08)\log\left(\frac{\rm L_{[NeV]14.3{\mu}m}}{\rm 10^{41}\,erg\,s^{-1}}\right),
    \end{split}
\end{equation}

\begin{equation}
\begin{split}
    \log\left(\frac{\rm L_{BOL}}{\rm 10^{41}\,erg\,s^{-1}}\right)=(3.50 \pm 0.07)\\
    +(1.06 \pm 0.10)\log\left(\frac{\rm L_{[NeV]24.3{\mu}m}}{\rm 10^{41}\,erg\,s^{-1}}\right),
    \end{split}
\end{equation}

The RMS scatter in these correlations is of the order of $\sim$0.3 dex, meaning that an emission line measurement can predict the AGN bolometric luminosity to an accuracy of about a factor two. 

\subsection{The X-ray correlation with other AGN strength indicators}\label{sec:corr2}

In order to complete the picture of the observables correlated with the strength of an active nucleus in a galaxy, 
we have used the optical spectra of the 12$\mu$m AGN sample published in \citet{malkan2017} to include 
the [OIII]$\lambda$5007\AA \ line, and the observations of the 12$\mu$m nuclear flux given in \citet{asmus2015}. These two
fluxes are in Table \ref{tab:sample2}, in columns (9) and (10), respectively.

Figures  \ref{fig:X-OIII5007F&L} and  \ref{fig:HX-OIII5007F&L} in the Appendix show the correlations between fluxes and luminosities of the ($2$--$10)\, \rm{keV}$ X-ray band and of the ($14$--$195)\, \rm{keV}$ X-ray band fluxes and luminosities, respectively, with the [OIII]$\lambda$5007\AA \ line flux and luminosity (panels (a) and (b), respectively). As can be seen from these figures and quantified in Table \ref{tab:cor2}, the flux correlation between the [OIII]$\lambda$5007\AA \ line and the ($2$--$10)\, \rm{keV}$ X-ray band is quite strong (P$_{null}$ $\sim$ 10$^{-10}$), while the one with the ($14$--$195)\, \rm{keV}$ X-ray band is very weak (being P$_{null}$ $\sim$ 10$^{-2}$) (see Fig.\ref{fig:HX-OIII5007F&L}(a)). 
This is partly due to the scarcity of the number of AGN that have 14-195keV data together with [OIII]$\lambda$5007\AA \ line fluxes. Additionally the extremely hard X-ray band may present an intrinsically larger scatter due to a more uncertain proportionality to the intrinsic strength of the AGN, compared to the 2-10keV X-ray band \citep[see, e.g. fig.\,13 in][]{ricci2017}. The 
Luminosity correlations, as already discussed in section \ref{sec:corr}, are always strong, with probabilities P$_{null}$ $\sim$ 10$^{-10}$ or even lower. The normalization of the [OIII] correlation with corrected ($2$--$10)\, \rm{keV}$ X-ray luminosity is very similar to that found recently by \citet{saade2022}. However, they identified a few Seyfert 2 galaxies fall well below the correlation perhaps because their hard X-ray emission has turned off for the last several decades.

Figures \ref{fig:X-12umF&L} and \ref{fig:HX-12umF&L} in the Appendix show the correlations between fluxes and luminosities of the ($2$--$10)\, \rm{keV}$ X-ray band and of the ($14$--$195)\, \rm{keV}$ X-ray band, respectively, with the nuclear flux and luminosity measured at 12$\mu$m. The correlations between the ($2$--$10)\, \rm{keV}$ X-ray band flux and the 12$\mu$m nuclear flux is relatively strong (P$_{null}$ $\sim$ 10$^{-4}$ for the whole AGN sample and P$_{null}$ $\sim$ 10$^{-3}$ for the Compton- thin AGN), while at higher X-ray energies, the correlation becomes weaker (P$_{null}$ $\sim$ 10$^{-3}$ for the whole AGN sample and P$_{null}$ $\sim$ 10$^{-2}$ for the Compton-thin AGN). This is again due to the small number of AGN that have both 14-195keV data and 12$\mu$m nuclear fluxes and to the intrinsic scatter of the 14-195keV emission.

 \begin{figure*}
   \includegraphics[width=0.5\textwidth]{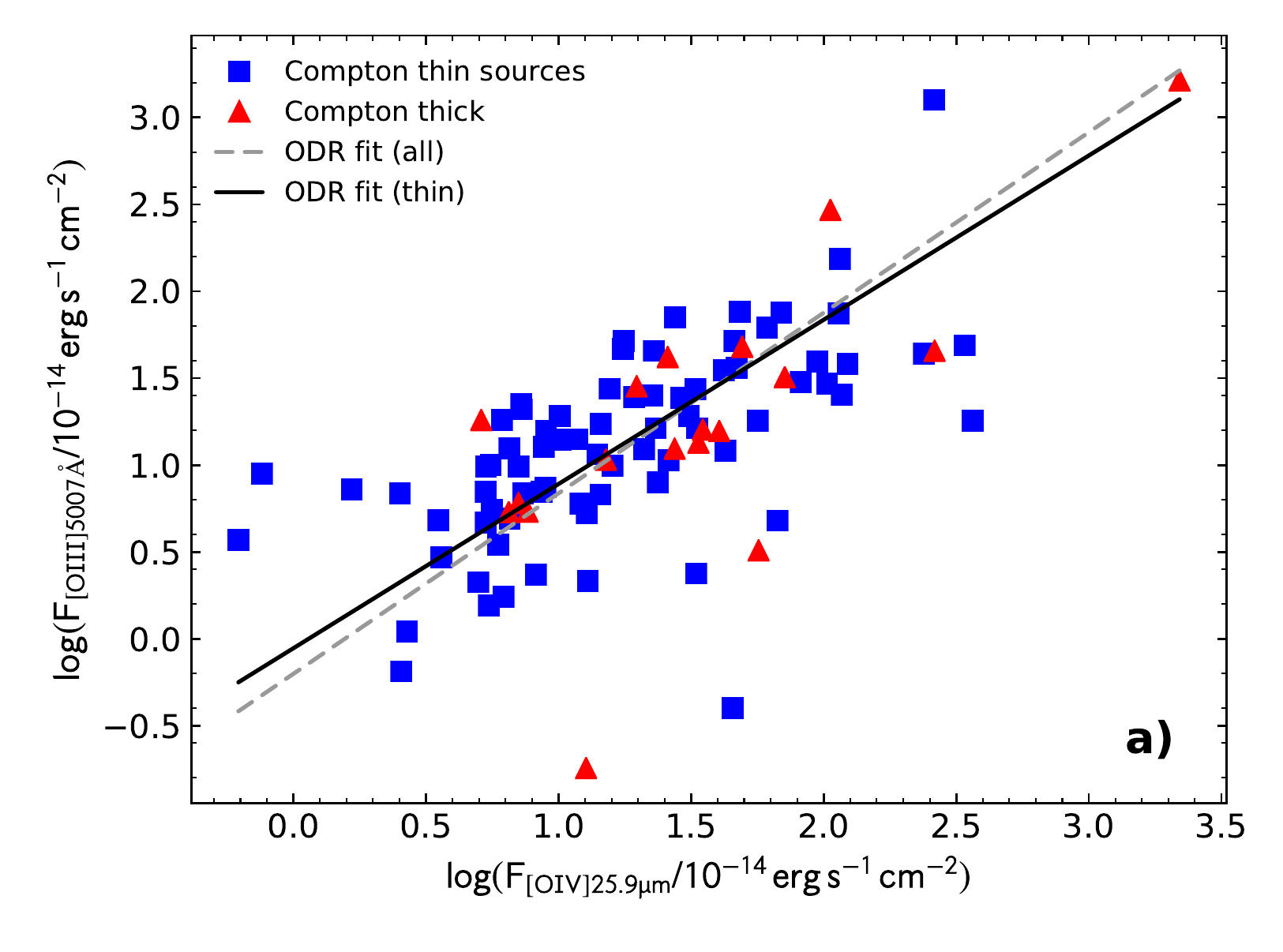}
      \includegraphics[width=0.5\textwidth]{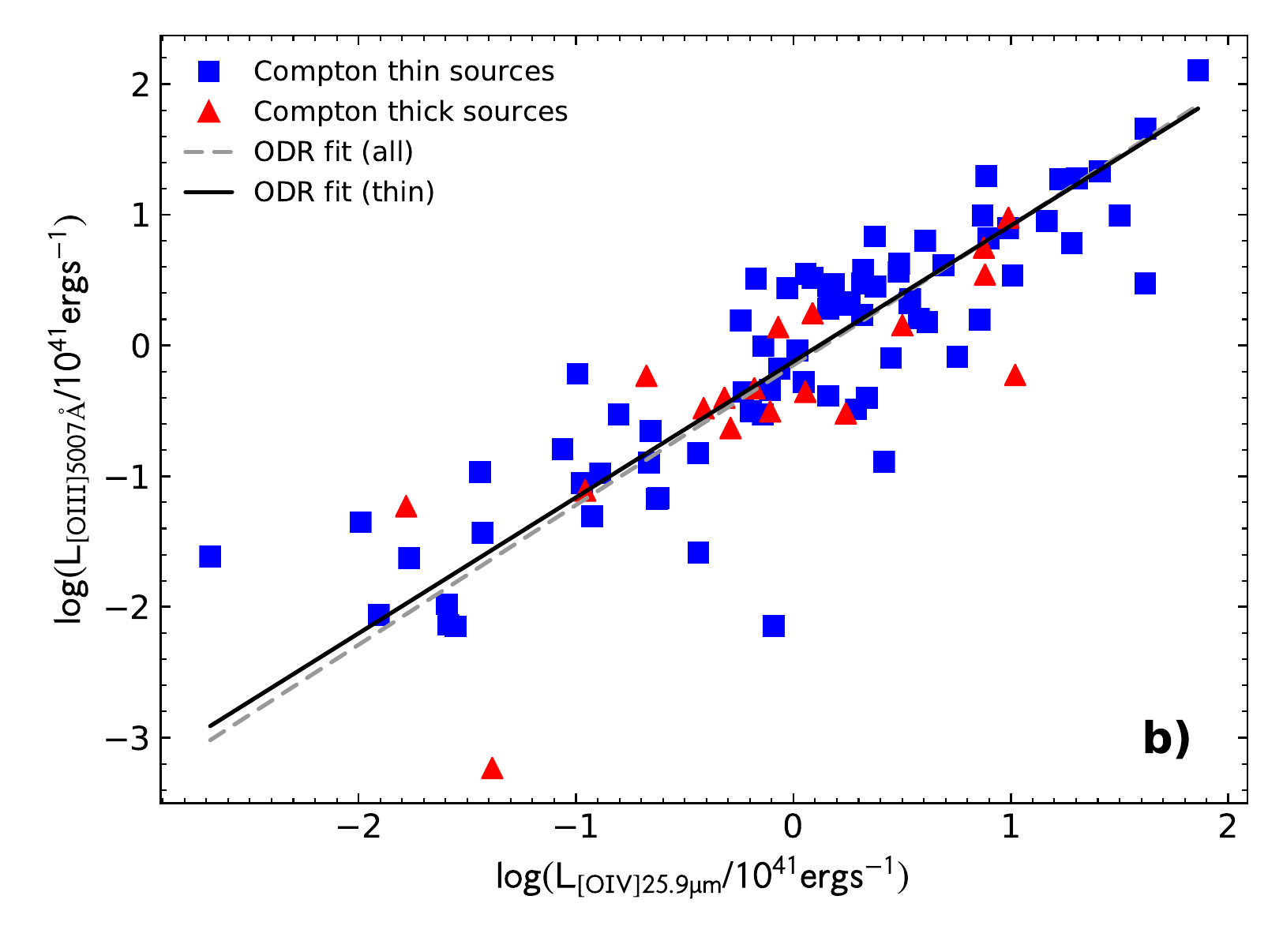}\\
  \includegraphics[width=0.5\textwidth]{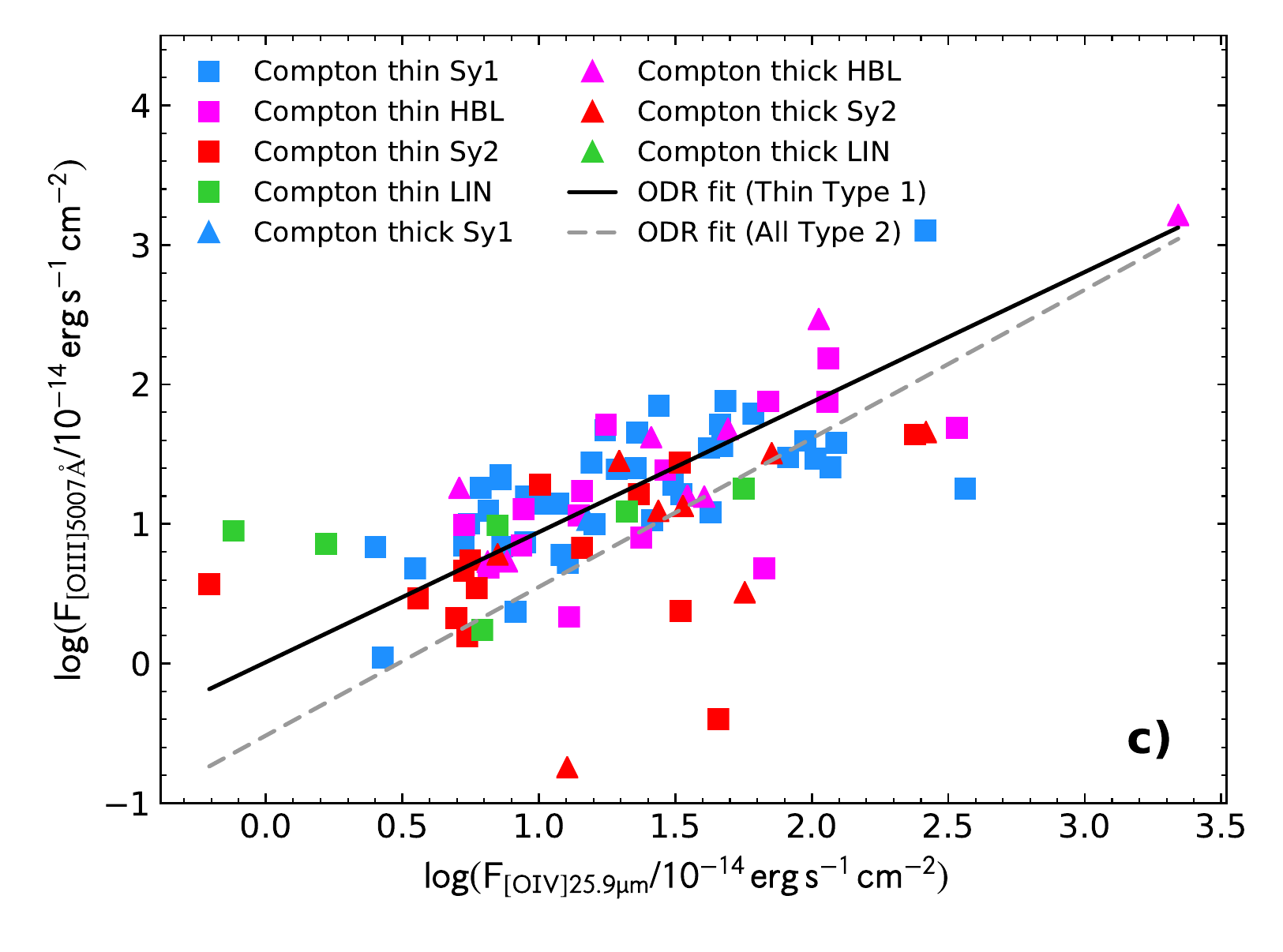}
  \includegraphics[width=0.5\textwidth]{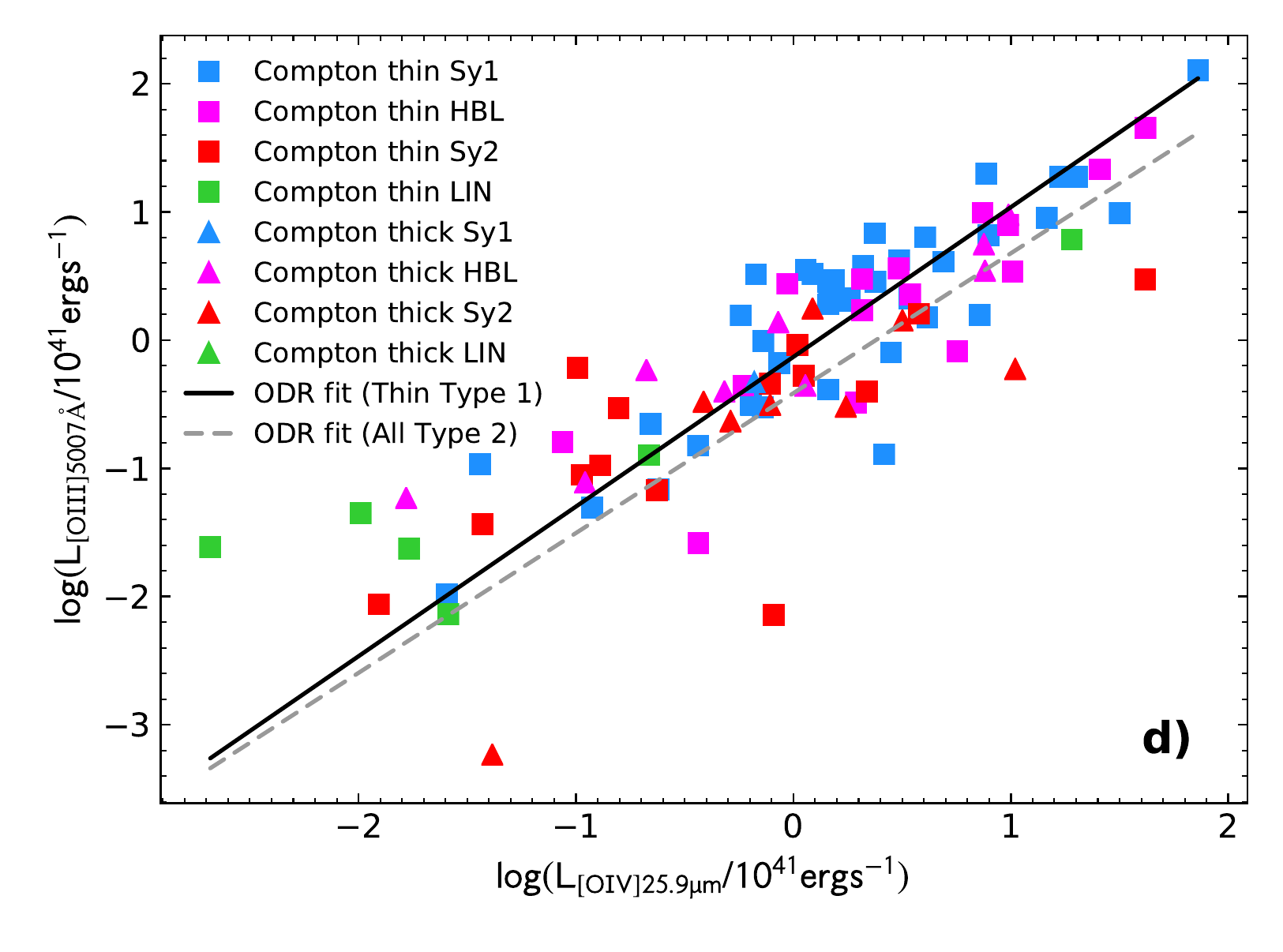}\\
  \caption{{\bf a: upper left} \oiii~ line flux as a function of the \oiv\ 25.9$\mu$m line flux for all sources with reliable optical spectroscopic classification. Blue squares indicate Compton-thin objects, while red triangles Compton-thick. The solid line shows the fit for all Compton-thin AGN, while the broken line the fit for all objects. 
  {\bf b: upper right}  \oiii~ line luminosity vs. \oiv\ line luminosity (both in logarithmic units of \ergs). The solid line shows the fit for all Compton thin AGN, while the broken line the fit for all objects. 
  {\bf c: lower left} Same as {\bf a:}, but with the objects color coded: Seyfert types 1 (Sy1): blue; Hidden Broad Line Region Galaxies (HBL): magenta; Seyfert types 2 (Sy2): red; LINERS (LIN): green. The solid line shows the fit for all Compton thin Type 1 AGN, which include both Seyfert 1's and HBL, while the broken line gives the bet fit for all the Type 2 objects, i.e. the {\it pure} Seyfert 2's, both Compton thick and thin (see Table \ref{tab:cor2})
   {\bf b: lower right} Same as {\bf b:}, but with the objects color coded. All correlations are reported in Table \ref{tab:cor2}
  }
  \label{fig:OIII-OIVF&L}
  \end{figure*}

\subsection{Flux and luminosity correlations between \oivp~ and \oiii~ }\label{sec:oxygen}

We show in Fig. \ref{fig:OIII-OIVF&L}(a) \& (c) the correlation between the line fluxes of the \oiii~ line and the \oivp~ line, and in Fig. \ref{fig:OIII-OIVF&L}(b) \& (d) the correlation between their luminosities, either considering all objects together and only Compton thin AGN and also considering all type 1's together (including Seyfert 1's and HBL) and all types 2. We do not find any significant difference in the correlation slopes (see Table \ref{tab:cor2}) between either Compton thin or CT AGN and between type 1 and type 2 objects.

\citet{melendez2008} computed also the correlation between the \oiii~ line luminosity and the \oivp~ line luminosity in Seyfert galaxies and find a steeper correlation for Seyfert 2's (B$\sim$1.8-2) compared to Seyfert 1's (B$\sim$0.9).
They also found that $L_{\rm [OIII]}$ correlates better with $L_{\rm 2-10 keV}$ than $L_{\rm [OIV]}$ and suggest that this can be due to extinction affecting both [\textsc{O\,iii}] and $2$--$10\, \rm{keV}$, but not [\textsc{O\,iv}]. However, as already noticed, we better trust the flux correlations than the luminosity correlations: for our sample, the flux correlations are in agreement within the errors, with only a slightly higher Pearson coefficient in the flux correlation for the [\textsc{O\,iii}] ($\sim 0.68$) (see Table \ref{tab:cor2}) when compared to that of the [\textsc{O\,iv}] ($\sim 0.58$) (see Table \ref{tab:cor}). The [\textsc{O\,iv}] luminosity correlation has a slightly higher slope coefficient ($0.88$) when compared to that of the [\textsc{O\,iii}] ($0.80$).
We consider that the results of \citet{melendez2008} are not valid in general for the classes of AGN considered, because their smaller sample (40 galaxies only) might be biased toward high X-ray brightness \citep{rigby2009}.

\subsection{Bolometric luminosity: X-ray versus IR method}\label{sec:bol}

\begin{figure*}
  \includegraphics[width=0.5\textwidth]{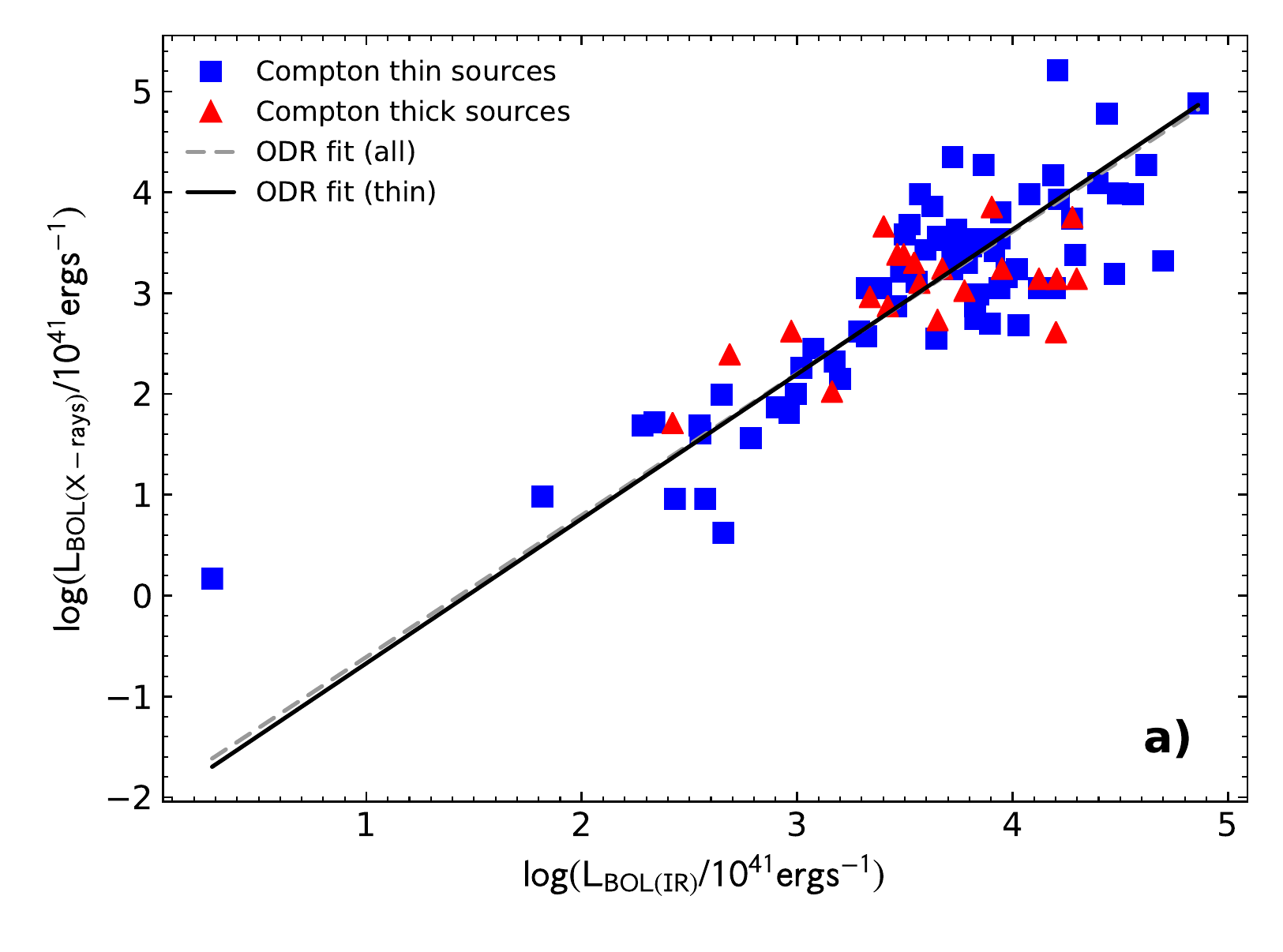}
  \includegraphics[width=0.5\textwidth]{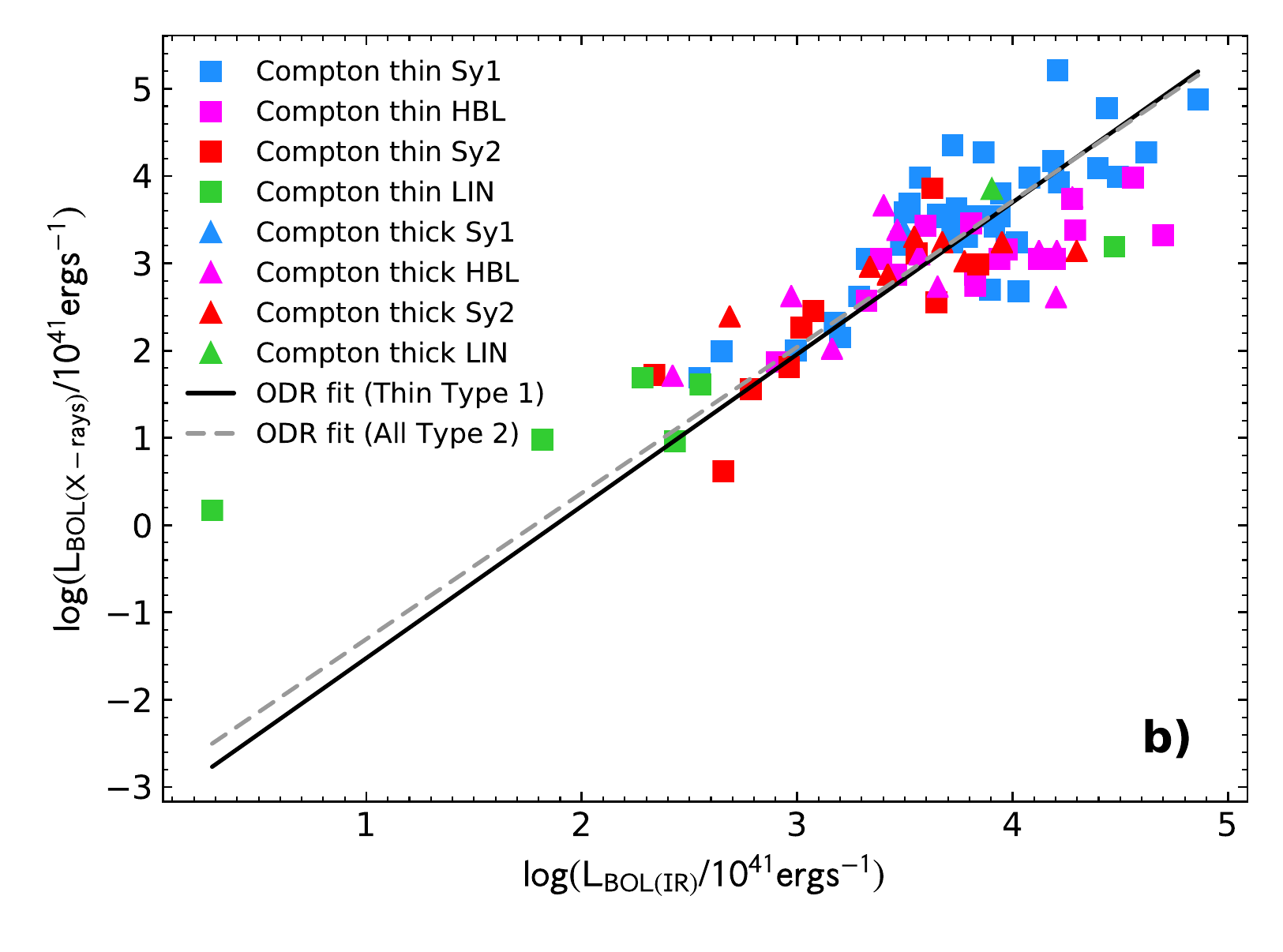}  
  \caption{{\bf a: left}  Bolometric luminosity as derived from the 2-10keV X-ray luminosity using the \citet{lusso2012} bolometric correction, as a function of the bolometric luminosity, as derived from the [OIV]26$\mu$m line luminosity and the bolometric correction of \citet{spinoglio1995} according to the formula from \citet{mordini2021}. The luminosities are expressed in logarithmic units of 10$^{41}$ \ergs. {\bf b: right} Same as {\bf a}, but with the objects color coded. 
  }
  \label{fig:LBOL_X&LBOL_IR}
  \end{figure*}

In Figure \ref{fig:LBOL_X&LBOL_IR} we show the bolometric luminosity, as computed from the corrected ($2$--$10)\, \rm{keV}$ X-ray luminosity, using the bolometric correction of \citet{lusso2012}, as a function of the bolometric luminosity as computed from the \oivp~ line luminosity with the calibration of \citet{mordini2021} and the bolometric correction of \citet{spinoglio1995}. Fig. \ref{fig:LBOL_X&LBOL_IR}(a) shows the AGN divided in Compton-thin and Compton-thick, while Fig. \ref{fig:LBOL_X&LBOL_IR}(b) shows them divided in the various nuclear activity classes.

Table \ref{tab:cor2} gives the results of the ODR fit: it appears that the correlation is very strong, having one of the highest values of the Pearson coefficient ($\rho \sim$ 0.83,  even if some caution has to be taken because these are luminosities). However the relation is not linear, with a slope of $B=1.4\pm0.1$, meaning that the X-ray estimate of the bolometric luminosity is rising faster at higher luminosities with respect to the IR estimate. This may be ascribed to differences in the large bolometric corrections involved in both cases, typically of a factor of $\sim 10$, which include non-linear dependencies between the X-ray or mid-IR line fluxes and the total luminosities derived from them. For instance, \citet{duras2020} show in their Fig.\,6 the large discrepancies that exist among different X-ray bolometric corrections in the 2-10\,keV band \citep{marconi2004}. In \citet{mordini2021}, fig.\,9b, we show that the luminosity estimates provided by the IR bolometric correction used in this work are closer to those based on the X-ray bolometric correction from \citet{lusso2012} than those obtained using the X-ray correction in \citet{marconi2004}.

\begin{figure*}
  \includegraphics[width=0.5\textwidth]{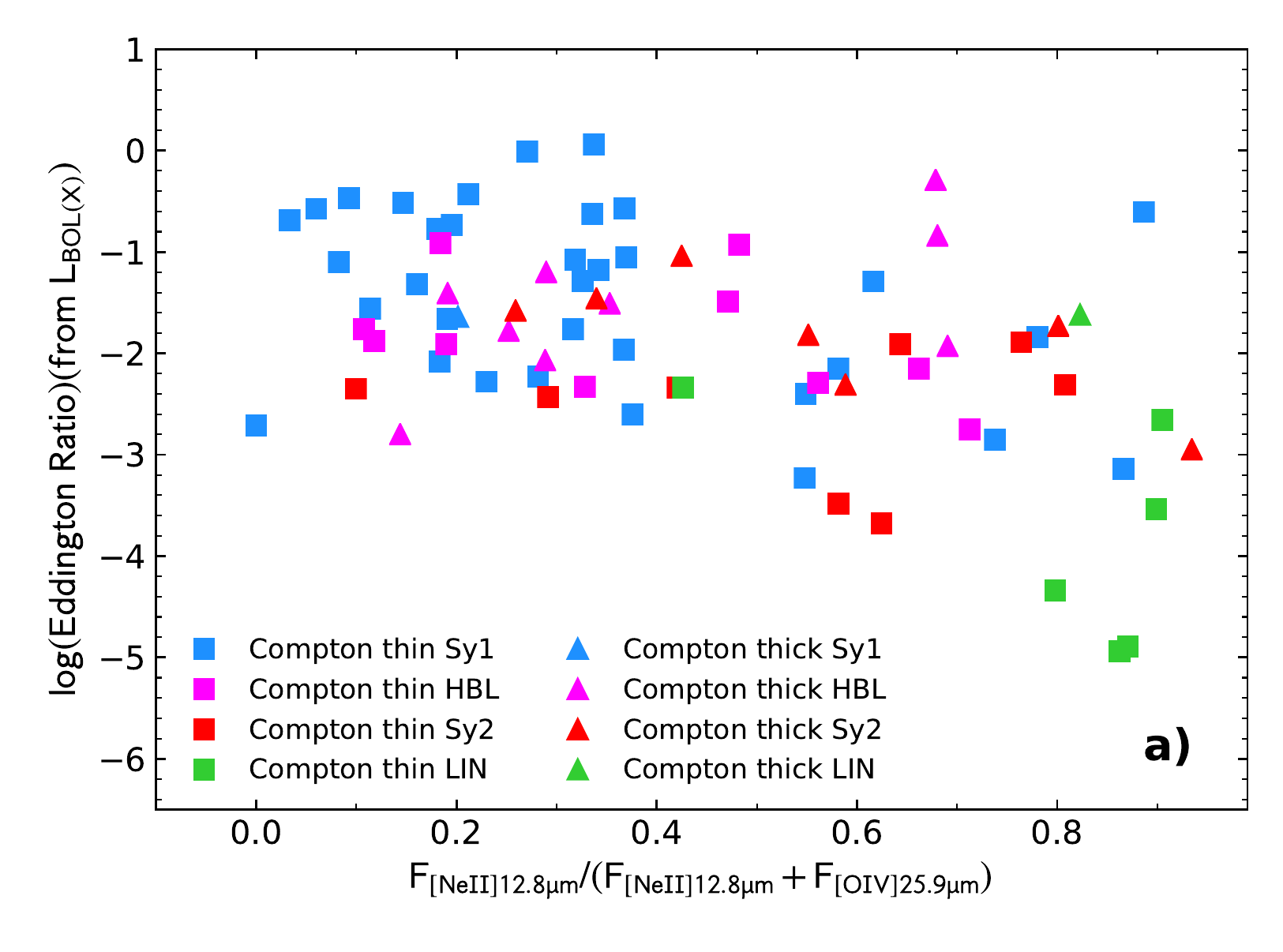}
  \includegraphics[width=0.5\textwidth]{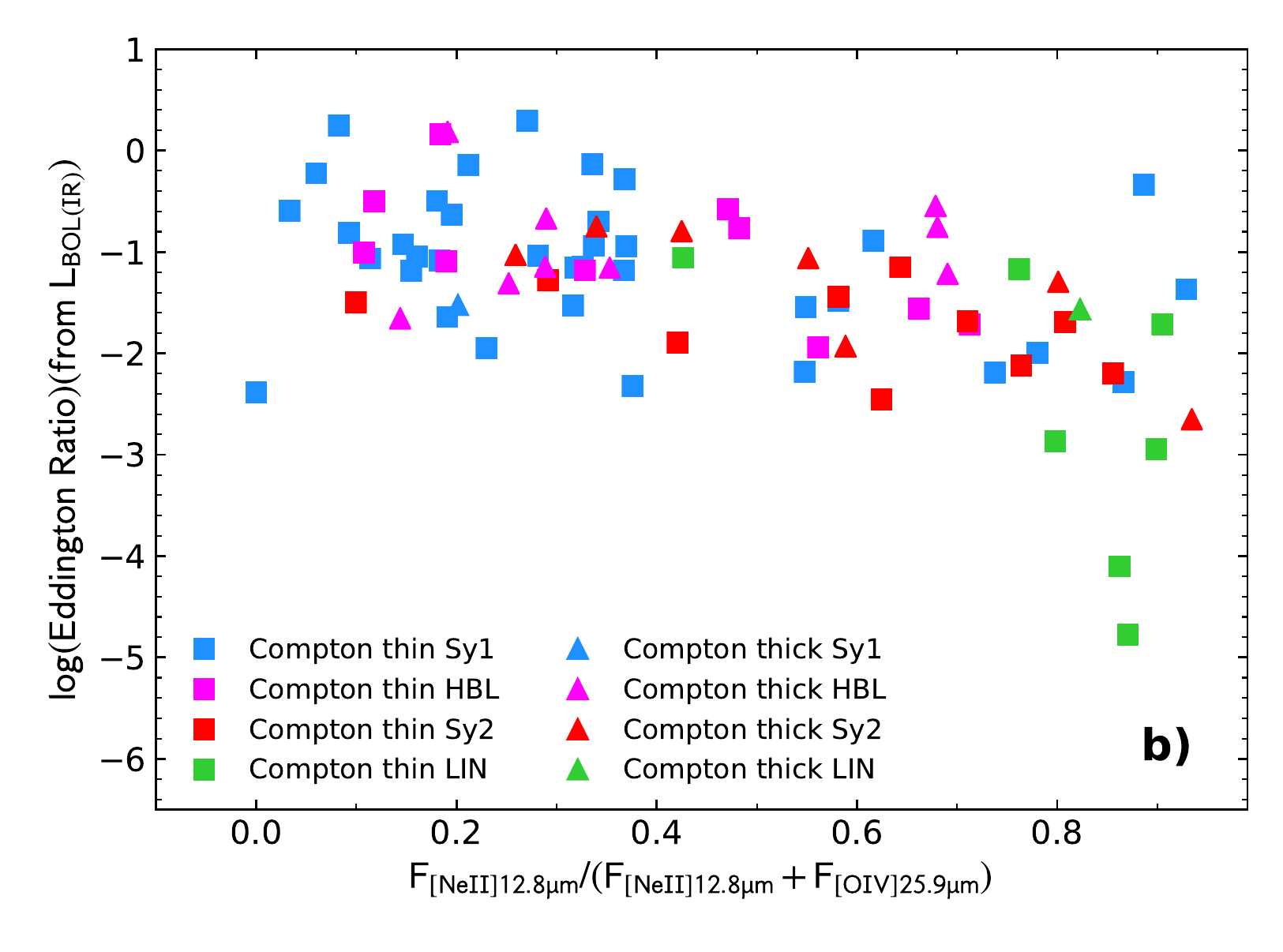}
  \caption{{\bf a: left} Luminosity-Excitation diagram (LED) \citep{fernandez2021} showing the Eddington ratio (in logarithmic form) as a function of the [NeII]/[NeII]+[OIV] line ratio, which measures the accretion disk dominance in an AGN, following the results of \citet{fernandez2021}. The Eddington ratio has been computed using the bolometric luminosity as derived from the X-ray 2-10keV luminosity. {\bf b: right} Same as {\bf a}, but with the the bolometric luminosity as derived from the \oivp~ line and the bolometric correction of \citet{spinoglio1995}, using the formula from \citet{mordini2021}. 
  }
  \label{fig:LED}
\end{figure*}  

  
\subsection{Accretion properties versus AGN types}
\subsubsection{Luminosity-Excitation Diagram}\label{sec:led}

\citet{fernandez2021} presented the luminosity-excitation diagram (LED) using the mid-IR fine structure lines of \oiv~25.9$\mu$m and \neii~in the attempt to quantify the dominance of the accretion disk in various types of AGN, as a function of the Eddington ratio. The composite line ratio of [NeII]12.8$\mu$m/(\oiv25.9$\mu$m + [NeII]12.8$\mu$m), that we define as the ``ionizing continuum softness'' (hereafter called simply {\it softness}), is a good measure of the slope of the Lyman continuum due to the difference between the ionization potential of O$^{3+}$ ($55\, \rm{eV}$) and Ne$^+$ ($22\, \rm{eV}$). The accretion disk contribution in the extreme UV boosts significantly the \oivp~ line intensity, which becomes very faint when the accretion disk cools or it is truncated (see fig.\,3 of \citealt{fernandez2021}). As a consequence, the Lyman hardness is inversely proportional to the dominance of the accretion disk in AGN, being the {\it softness}:
$${\rm [NeII]/([OIV] + [NeII])~{\sim}~1 - (L_{disk}/L_{total})}$$

The LED diagram in Fig.\,\ref{fig:LED} shows the Eddington ratio as a function of the {\it softness} for our sample of AGN. The Eddington ratio is measured using the bolometric luminosity estimated from both the ($2$--$10)\, \rm{keV}$ X-ray luminosity (see Section \ref{sec:bol}; Fig.\,\ref{fig:LED}\textbf{a}) and from the mid-IR [OIV] fine structure line  (Fig.\,\ref{fig:LED}\textbf{b}). This is divided by the estimated black hole mass, taken from the compilation in \citet{fernandez2021}.
One can see from this figure that using the X-ray determination of the bolometric luminosity has the effect of increasing the dispersion of the points in Eddington ratio, while the IR determination of the bolometric luminosity provides a better defined relation between the two axes, minimizing the spread in the Eddington ratio axis. On the other hand, IR-based luminosities tend to show slightly higher Eddington ratios, which can be caused by a small difference between the X-rays and the IR bolometric corrections. 

At low values of the Eddington ratio (log $\lambda_{\rm Edd}<-3.0$), almost only LINERs are present, and these lie at very high values of the {\it softness}, i.e. at very low excitation as traced by the nebular gas, as it is indeed expected \citep{fernandez2021}. On the other hand, most Seyfert 1 nuclei lie at low values of the {\it softness} (high gas excitation), due to the contribution of the accretion disk to the extreme UV continuum in these sources. Seyfert nuclei on the upper-right side of Fig.\,\ref{fig:LED} are compatible with a cooler or a truncated accretion disk, which does not have enough UV photons to significantly boost the \oivp~ emission line. Possible contamination in the \neii~ line, due to strong star formation activity, could in principle shift the nuclei to the right side of this diagram, however most of the objects in our sample are AGN dominated sources. In order to quantify this statement, 57 objects -- randomly selected, for which high-resolution {\it Spitzer} IRS spectra have been published  -- out of 117, i.e. $\sim50\%$ of our sample, have an AGN fraction computed \citep{tommasin2010} and the average fraction is (85$\pm$15\%),
and therefore a negligible contribution from star formation to this line is expected.

Table \ref{tab:LED_stats} gives the statistics of the various AGN populations in terms of {\it softness} and Eddington ratio ($\lambda_{\rm Edd}$). Type 1 AGN, i.e. Seyfert type 1 and Hidden Broad Line Region galaxies (HBL) together, show on average higher Eddington ratios (log($\overline{\lambda_{\rm Edd}}$)=-1.2$\pm$1.0) and lower {\it softness}
(${\rm [NeII]/([OIV]+[NeII])}=0.36{\pm}0.24$), 
while Seyfert type 2 galaxies show lower Eddington ratios (log($\overline{\lambda_{\rm Edd}}$)=-2.0$\pm$0.7) and intermediate {\it softness} (${\rm [NeII]/([OIV]+[NeII])}=0.6{\pm}0.3$) and Low Ionization Nuclear Emission Line galaxies (LINER) have even lower Eddington ratios (log($\overline{\lambda_{\rm Edd}}$) =-3.0$\pm$0.7) and higher {\it softness} (${\rm [NeII]/([OIV]+[NeII])}=0.8{\pm}0.3$), defining a decreasing trend among the three populations of AGN. 

\begin{table*}[ht!!!]
\centering
\setlength{\tabcolsep}{3.pt}
\caption{Correlation of X-ray fluxes and luminosities with IR fine-structure lines.}\label{tab:cor}
\footnotesize
\begin{tabular}{llclcc}
    \hline\\[-0.3cm]
\bf Considered variables &\bf Subset  &  \bf N &  \bf $\rho$~~~~~ (P$_{\rm null}$)  & \bf A $\pm\ \sigma$A & \bf B $\pm\ \sigma$B  \\[0.05cm]
 (1)            & (2)       & (3)           & (4)     & (5)      &  (6)     \\[0.1cm]
\hline\\[-0.25cm]
 F(2-10keV)\ vs. F(\nevp)  & all                         & 83 &  0.40 (1.9$\times 10^{-4}$) &   0.28  $\pm$ 0.16 & {\bf 1.19 $\pm$ 0.17} \\
  "~~~~~~~~~~~~~~~~~~~~"~~~~~~~~~~~~~~~~~~~~" & C-thin   & 63 &  0.37 (3.0 $\times 10^{-3}$) &  -0.23 $\pm$ 0.30 & {\bf 1.86 $\pm$ 0.36 } \\
 F(2-10keV)\ vs. F(\nevs)  & all                         & 72 &  0.46 (4.2 $\times 10^{-5}$) &  -0.14  $\pm$ 0.18 & {\bf 1.21  $\pm$ 0.18} \\
  "~~~~~~~~~~~~~~~~~~~~"~~~~~~~~~~~~~~~~~~~~" & C-thin   & 54 &  0.48 (2.7 $\times 10^{-4}$) &  -0.27 $\pm$ 0.28 & {\bf 1.67 $\pm$ 0.29} \\
 F(2-10keV)\ vs. F(\oivp)   & all                        & 94  & 0.55  (1.2 $\times 10^{-8}$) & -0.58  $\pm$ 0.21 & {\bf 1.31  $\pm$ 0.15}\\
  "~~~~~~~~~~~~~~~~~~~~"~~~~~~~~~~~~~~~~~~~~" & C-thin  & 73  &  0.58   (7.7 $\times 10^{-8}$) & -0.88 $\pm$ 0.26 & {\bf 1.54 $\pm$ 0.19}  \\
 "~~~~~~~~~~~~~~~~~~~~"~~~~~~~~~~~~~~~~~~~~" & Type~1   & 57  &  0.52  (3.1  $\times 10^{-5}$) &  -0.51 $\pm$ 0.27 & {\bf 1.28 $\pm$ 0.19}  \\
 "~~~~~~~~~~~~~~~~~~~~"~~~~~~~~~~~~~~~~~~~~" & Type~2   & 18  &  0.63  (4.8  $\times 10^{-3}$) &  -1.67 $\pm$ 0.72 & {\bf 2.03 $\pm$ 0.52}  \\
 F(2-10keV)\ vs. F(\neiii)  & all                       & 100 &  0.47  (7.0 $\times 10^{-7}$) &  -2.03  $\pm$ 0.47 & {\bf 2.78  $\pm$ 0.41} \\
  "~~~~~~~~~~~~~~~~~~~~"~~~~~~~~~~~~~~~~~~~~" & C-thin  & 79 &   0.47 (1.4 $\times 10^{-5}$) &  -2.67 $\pm$ 0.66 & {\bf 3.45 $\pm$ 0.61} \\
 F(2-10keV)\ vs. F(\neii)   & all                       & 98  &  0.15  (1.4 $\times 10^{-1}$) &  --- & {\bf ---}\\
  "~~~~~~~~~~~~~~~~~~~~"~~~~~~~~~~~~~~~~~~~~" & C-thin  & 77  &  0.07   (5.3 $\times 10^{-1}$) &  --- & {\bf ---}  \\
 F(14-195keV)\ vs. F(\nevp)  & all                      & 58  & 0.44   (4.8 $\times 10^{-4}$) &  1.01 $\pm$ 0.13 & {\bf 0.74 $\pm$ 0.12 } \\
  "~~~~~~~~~~~~~~~~~~~~"~~~~~~~~~~~~~~~~~~~~" & C-thin  & 51  & 0.40   (3.8 $\times 10^{-3}$) &  0.80   $\pm$ 0.17  & {\bf 0.99  $\pm$ 0.18}   \\
 F(14-195keV)\ vs. F(\nevs)  & all                      &  52 & 0.42   (2.2 $\times 10^{-3}$) &  0.92   $\pm$ 0.16 & {\bf 0.76  $\pm$ 0.14} \\
  "~~~~~~~~~~~~~~~~~~~~"~~~~~~~~~~~~~~~~~~~~" & C-thin  &  45 & 0.36  (1.5 $\times 10^{-2}$) &  0.64   $\pm$ 0.23  & {\bf 1.03  $\pm$ 0.22  }\\
 F(14-195keV)\ vs. F(\oivp) & all                       &  61 & 0.45  (2.3 $\times 10^{-4}$) &  0.65 $\pm$ 0.18 & {\bf 0.69  $\pm$ 0.11} \\
  "~~~~~~~~~~~~~~~~~~~~"~~~~~~~~~~~~~~~~~~~~" & C-thin  &  53 & 0.42 (1.8 $\times 10^{-3}$) &  0.45 $\pm$ 0.22  & {\bf 0.82 $\pm$ 0.15} \\
  "~~~~~~~~~~~~~~~~~~~~"~~~~~~~~~~~~~~~~~~~~" & Type~1  & 46  &  0.40  (6.1  $\times 10^{-3}$) &  0.37 $\pm$ 0.26 & {\bf 0.87 $\pm$ 0.17}  \\
 "~~~~~~~~~~~~~~~~~~~~"~~~~~~~~~~~~~~~~~~~~" & Type~2   &  9  &  0.58  (1.0  $\times 10^{-1}$) &  0.54 $\pm$ 0.52 & {\bf 0.83 $\pm$ 0.33}  \\
 F(14-195keV)\ vs. F(\neiii)  & all                     &  62 & 0.45 (2.5 $\times 10^{-4}$) &  0.75  $\pm$ 0.16  & {\bf 0.75   $\pm$ 0.12 } \\
  "~~~~~~~~~~~~~~~~~~~~"~~~~~~~~~~~~~~~~~~~~" & C-thin  &  54 & 0.41  (1.9 $\times 10^{-3}$) &  0.50   $\pm$ 0.21  & {\bf 0.97  $\pm$ 0.17}\\
 F(14-195keV)\ vs. F(\neii) & all                       &  60 & 0.29  (2.5 $\times 10^{-2}$) &  1.25 $\pm$ 0.14 & {\bf 0.44   $\pm$ 0.11} \\
  "~~~~~~~~~~~~~~~~~~~~"~~~~~~~~~~~~~~~~~~~~" & C-thin  &  52 & 0.22  (1.2 $\times 10^{-1}$) &  1.05  $\pm$ 0.17 & {\bf 0.53  $\pm$ 0.14} \\
\hline
L(2-10keV)\ vs. L(\nevp)  & all                          & 83 & 0.74  (9.6 $\times 10^{-16}$) &   2.42 $\pm$ 0.07 & {\bf 1.00 $\pm$ 0.09} \\
 "~~~~~~~~~~~~~~~~~~~~"~~~~~~~~~~~~~~~~~~~~" & C-thin    & 63 & 0.76  (7.1 $\times 10^{-13}$) &   2.45 $\pm$ 0.08 & {\bf 1.01 $\pm$ 0.10} \\
L(2-10keV)\ vs. L(\nevs)  & all                          & 72 & 0.70  (8.4 $\times 10^{-12}$) &   2.37 $\pm$ 0.08 & {\bf 1.15 $\pm$ 0.12} \\
 "~~~~~~~~~~~~~~~~~~~~"~~~~~~~~~~~~~~~~~~~~" & C-thin    & 54 & 0.72  (9.2 $\times 10^{-10}$) &   2.38 $\pm$ 0.09 & {\bf 1.23  $\pm$ 0.14} \\
L(2-10keV)\ vs. L(\oivp)   & all                         & 95 & 0.86  (5.1 $\times 10^{-29}$) &  1.76 $\pm$ 0.07 & {\bf 1.17  $\pm$ 0.07} \\
 "~~~~~~~~~~~~~~~~~~~~"~~~~~~~~~~~~~~~~~~~~" & C-thin    & 74 & 0.88  (4.0 $\times 10^{-25}$) &  1.79 $\pm$ 0.07 & {\bf 1.17 $\pm$ 0.07} \\
  "~~~~~~~~~~~~~~~~~~~~"~~~~~~~~~~~~~~~~~~~~" & Type~1  & 57  & 0.78  (5.8  $\times 10^{-13}$) & 1.82 $\pm$ 0.08 & {\bf 1.17 $\pm$ 0.11}  \\
 "~~~~~~~~~~~~~~~~~~~~"~~~~~~~~~~~~~~~~~~~~" & Type~2   & 18  & 0.77  (2.0  $\times 10^{-4}$) &  1.88 $\pm$ 0.22 & {\bf 1.49 $\pm$ 0.28}  \\
L(2-10keV)\ vs. L(\neiii)  & all                         & 101 & 0.77  (4.6 $\times 10^{-21}$) &   1.98 $\pm$ 0.08 & {\bf 1.41 $\pm$ 0.10} \\
 "~~~~~~~~~~~~~~~~~~~~"~~~~~~~~~~~~~~~~~~~~" & C-thin    & 80 & 0.78  (8.4 $\times 10^{-18}$) &   1.98 $\pm$ 0.10 & {\bf 1.47  $\pm$ 0.12} \\
L(2-10keV)\ vs. L(\neii)   & all                         & 99 & 0.61  (1.5 $\times 10^{-11}$) &   1.94 $\pm$ 0.12 & {\bf 1.65  $\pm$ 0.17 } \\
 "~~~~~~~~~~~~~~~~~~~~"~~~~~~~~~~~~~~~~~~~~" & C-thin    & 78 & 0.62  (1.8 $\times 10^{-9}$)  &   1.96 $\pm$ 0.15 & {\bf 1.81  $\pm$ 0.21 } \\
L(14-195keV)\ vs. L(\nevp)  & all                        & 58 & 0.76   (3.2 $\times 10^{-12}$) &  2.78 $\pm$ 0.08 & {\bf 1.04 $\pm$ 0.10 } \\
 "~~~~~~~~~~~~~~~~~~~~"~~~~~~~~~~~~~~~~~~~~" & C-thin    & 51 & 0.80   (1.9 $\times 10^{-12}$) &  2.80 $\pm$ 0.08 & {\bf 1.04 $\pm$ 0.10} \\
L(14-195keV)\ vs. L(\nevs)  & all                        &  52 & 0.70  (7.3 $\times 10^{-9}$) &  2.70 $\pm$ 0.09 & {\bf 1.18 $\pm$ 0.14} \\
 "~~~~~~~~~~~~~~~~~~~~"~~~~~~~~~~~~~~~~~~~~" & C-thin    &  45 & 0.74  (7.8 $\times 10^{-9}$)  &  2.69 $\pm$ 0.09 & {\bf 1.18 $\pm$ 0.14} \\
L(14-195keV)\ vs. L(\oivp)   & all                       &  61 & 0.85  (3.6 $\times 10^{-18}$) &  2.19 $\pm$ 0.07 & {\bf 1.02 $\pm$ 0.07} \\
 "~~~~~~~~~~~~~~~~~~~~"~~~~~~~~~~~~~~~~~~~~"  & C-thin   &  53 & 0.87  (2.3 $\times 10^{-17}$) &  2.18 $\pm$ 0.08 & {\bf 1.03 $\pm$ 0.08} \\
  "~~~~~~~~~~~~~~~~~~~~"~~~~~~~~~~~~~~~~~~~~" & Type~1  & 46  &  0.77  (3.3  $\times 10^{-10}$) &  2.11 $\pm$ 0.10 & {\bf 1.19 $\pm$ 0.13}  \\
 "~~~~~~~~~~~~~~~~~~~~"~~~~~~~~~~~~~~~~~~~~" & Type~2   & 9  &  0.82  (7.3  $\times 10^{-3}$) &    2.26 $\pm$ 0.17 & {\bf 0.92 $\pm$ 0.22}  \\
L(14-195keV)\ vs. L(\neiii)  & all                        &  62 & 0.85  (2.2 $\times 10^{-18}$) &  2.45 $\pm$ 0.07 & {\bf 1.15 $\pm$ 0.08} \\
 "~~~~~~~~~~~~~~~~~~~~"~~~~~~~~~~~~~~~~~~~~" & C-thin    &  54 & 0.85  (3.9 $\times 10^{-18}$)  &  2.47 $\pm$ 0.07 & {\bf 1.16 $\pm$ 0.03} \\
L(14-195keV)\ vs. L(\neii)   & all                       &  60 & 0.77  (5.2 $\times 10^{-13}$) &  2.52 $\pm$ 0.09 & {\bf 1.26 $\pm$ 0.12} \\
 "~~~~~~~~~~~~~~~~~~~~"~~~~~~~~~~~~~~~~~~~~"  & C-thin   &  52 & 0.81  (3.3 $\times 10^{-13}$) &  2.58 $\pm$ 0.10 & {\bf 1.28 $\pm$ 0.12} \\

\hline
\end{tabular}\\[0.2cm] 
\begin{tablenotes}
\footnotesize
\item \textbf{Notes.} Fit results. The $2$--$10\, \rm{keV}$ band fluxes and luminosities are intrinsic, i.e. corrected for the source intrinsic absorption measured from the X-ray spectra. The columns give for each correlation: (1) variables; (2) Subset of the entire sample on which the fit was computed: ``C-thin'' indicates the Compton-thin Sy1 and Sy2 subset, ``all'' indicates the entire sample with HX data; (3) Number of sources considered in the statistical analysis; (4) Pearson  correlation coefficient $\rho$ (1: completely correlated variables, 0: uncorrelated variables) with the relative null hypothesis (zero correlation) probability; (5) and (6): Parameters of the linear regression fit using the equation: ${\rm log(F_y) = A + B \times log(F_x)}$ or ${\rm log(L_y) = A + B \times log(L_x)}$. 
\label{tab:fit}
\end{tablenotes}
\end{table*}

\begin{table*}[ht!!!]
\centering
\setlength{\tabcolsep}{3.pt}
\caption{Other correlation among X-ray fluxes and luminosities with other AGN indicators, bolometric luminosities and \oiii~ and \oivp~ fluxes and luminosities and SFR versus BHAR. 
}\label{tab:cor2}
\footnotesize
\begin{tabular}{llclcc}
    \hline\\[-0.3cm]
\bf Considered variables &\bf Subset  &  \bf N &  \bf $\rho$~~~~~ (P$_{\rm null}$)  & \bf A $\pm\ \sigma$A & \bf B $\pm\ \sigma$B  \\[0.05cm]
 (1)            & (2)       & (3)           & (4)     & (5)      &  (6)     \\[0.1cm]
\hline\\[-0.25cm]
F(2-10keV)\ vs. F(\oiii)  & all                         & 91 &  0.59 (5.4$\times 10^{-10}$) &  -0.66 $\pm$ 0.21  & {\bf 1.49 $\pm$ 0.16} \\
  "~~~~~~~~~~~~~~~~~~~~"~~~~~~~~~~~~~~~~~~~~" & C-thin   & 73 & 0.68 (3.4$\times 10^{-11}$) &  -0.91 $\pm$ 0.22  & {\bf 1.70 $\pm$ 0.18}  \\
F(2-10keV)\ vs. F(12$\mu$m~nuclear)          & all      & 60 &  0.46 (1.9$\times 10^{-4}$) &  -2.19 $\pm$ 0.58   & {\bf 1.52 $\pm$ 0.25 } \\
  "~~~~~~~~~~~~~~~~~~~~"~~~~~~~~~~~~~~~~~~~~" & C-thin   & 47 &  0.45 (1.4 $\times 10^{-3}$) & -3.34 $\pm$ 0.94  & {\bf 2.05 $\pm$ 0.42} \\
F(14-195keV)\ vs. F(\oiii) & all                        &  59 & 0.28  (3.4 $\times 10^{-2}$) &  1.16 $\pm$ 0.14 & {\bf 0.38 $\pm$ 0.10} \\
  "~~~~~~~~~~~~~~~~~~~~"~~~~~~~~~~~~~~~~~~~~" & C-thin &  52 & 0.32 (2.2 $\times 10^{-2}$) &  0.84 $\pm$ 0.18  & {\bf 0.60 $\pm$ 0.13} \\
F(14-195keV)\ vs. F(12$\mu$m~nuclear)          & all   &  39 & 0.49  (1.7 $\times 10^{-3}$) &  -0.15  $\pm$ 0.39  & {\bf 0.81 $\pm$ 0.16} \\
  "~~~~~~~~~~~~~~~~~~~~"~~~~~~~~~~~~~~~~~~~~" & C-thin &  33 & 0.44 (1.0 $\times 10^{-2}$) &  -0.60  $\pm$ 0.54  & {\bf 1.01 $\pm$ 0.22} \\
\hline
 L(2-10keV)\ vs. L(\oiii)  & all                         & 92 &  0.76 (2.9$\times 10^{-18}$) &  1.91 $\pm$ 0.08  & {\bf 1.10 $\pm$ 0.09 } \\
  "~~~~~~~~~~~~~~~~~~~~"~~~~~~~~~~~~~~~~~~~~" & C-thin   & 74 &  0.80 (6.8 $\times 10^{-18}$) &  1.86 $\pm$ 0.09 & {\bf 1.18 $\pm$ 0.09 } \\
 L(2-10keV)\ vs. L(12$\mu$m nuclear)          & all      & 61 &  0.85 (8.1$\times 10^{-18}$) &  -0.42 $\pm$ 0.18  & {\bf 0.99 $\pm$ 0.07} \\
  "~~~~~~~~~~~~~~~~~~~~"~~~~~~~~~~~~~~~~~~~~" & C-thin   & 48 &  0.85 (1.2 $\times 10^{-14}$) & -0.48 $\pm$ 0.20  & {\bf 1.01 $\pm$ 0.08} \\
L(14-195keV)\ vs. F(\oiii)                   & all      &  59 & 0.70  (5.9 $\times 10^{-10}$) &  2.36 $\pm$ 0.07 & {\bf 0.69 $\pm$ 0.08} \\
  "~~~~~~~~~~~~~~~~~~~~"~~~~~~~~~~~~~~~~~~~~" & C-thin &  52 & 0.75 (1.7 $\times 10^{-10}$) &  2.34 $\pm$ 0.07 & {\bf 0.82 $\pm$ 0.09} \\
L(14-195keV)\ vs. F(12$\mu$m~nuclear)          & all   &  39 & 0.92  (1.7 $\times 10^{-16}$) &  -0.03 $\pm$ 0.17 & {\bf 1.00 $\pm$ 0.07} \\
  "~~~~~~~~~~~~~~~~~~~~"~~~~~~~~~~~~~~~~~~~~" & C-thin &  33 & 0.93 (3.0 $\times 10^{-15}$) &  -0.01 $\pm$ 0.17 & {\bf 1.01 $\pm$ 0.07} \\
\hline
LBOL(2-10keV)\ vs.  L(\oivp)  & all                       & 94 & 0.84  (1.0 $\times 10^{-25}$) & 3.02 $\pm$ 0.06 & {\bf 0.88 $\pm$ 0.06} \\
 "~~~~~~~~~~~~~~~~~~~~"~~~~~~~~~~~~~~~~~~~~" & C-thin    & 73 & 0.86  (3.9 $\times 10^{-22}$) &  3.06 $\pm$ 0.06 & {\bf 0.89 $\pm$ 0.06} \\
LBOL(2-10keV)\ vs.  L(\nevp)  & all                       & 82 & 0.70  (1.6 $\times 10^{-13}$) & 3.53 $\pm$ 0.07 & {\bf 0.90 $\pm$ 0.08} \\
 "~~~~~~~~~~~~~~~~~~~~"~~~~~~~~~~~~~~~~~~~~" & C-thin    & 62 & 0.74  (1.0 $\times 10^{-13}$) &  3.59 $\pm$ 0.08 & {\bf 0.86$\pm$ 0.09} \\
 LBOL(2-10keV)\ vs.  L(\nevs)  & all                     & 71 & 0.72  (1.6 $\times 10^{-12}$) & 3.50 $\pm$ 0.07 & {\bf 1.06 $\pm$ 0.10} \\
 "~~~~~~~~~~~~~~~~~~~~"~~~~~~~~~~~~~~~~~~~~" & C-thin    & 53 & 0.71  (3.5 $\times 10^{-9}$) & 3.53 $\pm$ 0.08 & {\bf 1.07$\pm$ 0.12} \\
\hline
LBOL(2-10keV)\ vs. LBOL(IR)  & all                       & 94 & 0.83  (6.6 $\times 10^{-25}$) &  -2.02 $\pm$ 0.33 & {\bf 1.41 $\pm$ 0.09} \\
 "~~~~~~~~~~~~~~~~~~~~"~~~~~~~~~~~~~~~~~~~~" & C-thin    & 73 & 0.85  (1.4 $\times 10^{-21}$) &   -2.11 $\pm$ 0.35 & {\bf 1.43$\pm$ 0.10} \\
\hline
F(\oiii)\ vs. F(\oivp)  & All.                        & 94 & 0.64  (2.5 $\times 10^{-12}$) &  -0.20 $\pm$ 0.14 & {\bf 1.04$\pm$ 0.10} \\
 "~~~~~~~~~~~~~~~~~~~~"~~~~~~~~~~~~~~~~~~~~" & C-thin & 76 & 0.62  (2.3 $\times 10^{-9}$) &   -0.05 $\pm$ 0.14 & {\bf 0.94$\pm$ 0.11} \\
F(\oiii)\ vs. F(\oivp)  & Type 1                      & 55 & 0.67  (2.1 $\times 10^{-8}$) &   0.01 $\pm$ 0.16 & {\bf 0.93$\pm$ 0.11} \\
 "~~~~~~~~~~~~~~~~~~~~"~~~~~~~~~~~~~~~~~~~~" & Type 2 & 22 & 0.43  (4.5 $\times 10^{-2}$) &   -0.51 $\pm$ 0.40 & {\bf 1.06$\pm$ 0.30} \\
 \hline
L(\oiii)\ vs. L(\oivp)  & All                      & 94 & 0.85  (1.7 $\times 10^{-27}$) &  -0.15 $\pm$ 0.05 & {\bf 1.07$\pm$ 0.06} \\
 "~~~~~~~~~~~~~~~~~~~~"~~~~~~~~~~~~~~~~~~~~" & C-thin & 76 & 0.86  (8.6 $\times 10^{-24}$) &   -0.12 $\pm$ 0.06 & {\bf 1.04$\pm$ 0.06} \\
L(\oiii)\ vs. L(\oivp)  & Type 1                      & 55 & 0.86  (4.9 $\times 10^{-17}$) &  -0.13 $\pm$ 0.06 & {\bf 1.17$\pm$ 0.09} \\
 "~~~~~~~~~~~~~~~~~~~~"~~~~~~~~~~~~~~~~~~~~" & Type 2 & 22 & 0.71  (2.3 $\times 10^{-4}$) &   -0.41 $\pm$ 0.16 & {\bf 1.09$\pm$ 0.20} \\
 \hline
{\rm SFR(M$_\odot$/yr) vs. BHAR(M$_\odot$/yr)} from [CII]158$\mu$m    & All    & 58 & 0.77  (1.2 $\times 10^{-12}$) &  1.39 $\pm$ 0.10 & {\bf 0.53$\pm$ 0.05} \\
 "~~~~~~~~~~~~~~~~~~~~"~~~~~~~~~~~~~~~~~~~~" & C-thin & 45 & 0.80  (2.7 $\times 10^{-11}$) &   1.26 $\pm$ 0.11 & {\bf 0.50$\pm$ 0.05} \\
{\rm SFR(M$_\odot$/yr) vs. BHAR(M$_\odot$/yr)} from [CII]158$\mu$m   & Type 1  & 33 & 0.73 (1.6 $\times 10^{-6}$) &   1.41 $\pm$ 0.14 & {\bf 0.70$\pm$ 0.10} \\
 "~~~~~~~~~~~~~~~~~~~~"~~~~~~~~~~~~~~~~~~~~" & Type 2 & 12 & 0.79 (2.2 $\times 10^{-3}$) &   1.76 $\pm$ 0.28 & {\bf 0.66$\pm$ 0.14} \\
{\rm SFR(M$_\odot$/yr) vs. BHAR(M$_\odot$/yr)} from PAH11.3$\mu$m    & All    & 89 & 0.62  (1.1 $\times 10^{-10}$) &  1.05 $\pm$ 0.12 & {\bf 0.53$\pm$ 0.06} \\
 "~~~~~~~~~~~~~~~~~~~~"~~~~~~~~~~~~~~~~~~~~" & C-thin & 70 & 0.60  (3.3 $\times 10^{-8}$) &   0.95 $\pm$ 0.14 & {\bf 0.48$\pm$ 0.06} \\
{\rm SFR(M$_\odot$/yr) vs. BHAR(M$_\odot$/yr)} from PAH11.3$\mu$m     & Type 1  & 46 & 0.50  (4.5 $\times 10^{-4}$) &  1.31 $\pm$ 0.23 & {\bf 0.81$\pm$ 0.14} \\
 "~~~~~~~~~~~~~~~~~~~~"~~~~~~~~~~~~~~~~~~~~" & Type 2 & 23 & 0.64  (1.0 $\times 10^{-3}$) &   1.57 $\pm$ 0.32 & {\bf 0.74$\pm$ 0.15} \\
 \hline
 {\rm sSFR(M$_\odot$/yr) vs. LBOL(IR) } from PAH11.3$\mu$m    & All    & 78 & 0.44  (6.7 $\times 10^{-5}$) &  -3.30 $\pm$ 0.36 & {\bf 0.67$\pm$ 0.10} \\
 "~~~~~~~~~~~~~~~~~~~~"~~~~~~~~~~~~~~~~~~~~"                  & C-thin & 60 & 0.42  (8.5 $\times 10^{-4}$) &  -3.10 $\pm$ 0.39 & {\bf 0.59$\pm$ 0.11} \\
{\rm sSFR(M$_\odot$/yr) vs. LBOL(IR) )} from PAH11.3$\mu$m     & Type 1  & 43 & 0.17  (2.8 $\times 10^{-1}$) & -4.70 $\pm$ 1.00 & {\bf 0.99$\pm$ 0.26} \\
 "~~~~~~~~~~~~~~~~~~~~"~~~~~~~~~~~~~~~~~~~~"                    & Type 2 & 20 & 0.46  (4.1 $\times 10^{-2}$) & -5.01 $\pm$ 1.22 & {\bf 1.21$\pm$ 0.35} \\
\hline
{\rm L(2-10keV)\ vs. L(\oivp)}   & {\bf $L_{12\mu m}<10^{44} erg s^{-1}$} &  42  &   0.83 (6.8 $\times 10^{-12}$)  &    1.83 $\pm$ 0.11  & {\bf 1.26  $\pm$ 0.12} \\
 "~~~~~~~~~~~~~~~~~~~~"~~~~~~~~~~~~~~~~~~~~" & {\bf $L_{12\mu m}>10^{44} erg s^{-1}$} & 15 &  0.63  (1.2 $\times 10^{-2}$) & 1.58 $\pm$ 0.41 & {\bf 1.76 $\pm$ 0.49} \\
  "~~~~~~~~~~~~~~~~~~~~"~~~~~~~~~~~~~~~~~~~~" & {\bf $\log(\lambda_{\rm Edd})<-1.5$}  & 47 &  0.85  (3.8  $\times 10^{-14}$) & 1.78 $\pm$ 0.11 & {\bf 1.28 $\pm$ 0.11}  \\
 "~~~~~~~~~~~~~~~~~~~~"~~~~~~~~~~~~~~~~~~~~" & {\bf $\log(\lambda_{\rm Edd})>-1.5$}   & 47 &  0.70  (3.7  $\times 10^{-8}$) &  1.70 $\pm$ 0.11 & {\bf 1.27 $\pm$ 0.16}  \\
\hline
\end{tabular}\\[0.2cm] 
\begin{tablenotes}
\footnotesize
\item \textbf{Notes.} Fit results. The $2$--$10\, \rm{keV}$ band fluxes and luminosities are intrinsic, i.e. corrected for the source intrinsic absorption measured from the X-ray spectra. The columns give for each correlation: (1) variables; (2) Subset of the entire sample on which the fit was computed: ``C-thin'' indicates the Compton-thin Sy1 and Sy2 subset, ``all'' indicates the entire sample with HX data; (3) Number of sources considered in the statistical analysis; (4) Pearson  correlation coefficient $\rho$ (1: completely correlated variables, 0: uncorrelated variables) with the relative null hypothesis (zero correlation) probability; (5) and (6): Parameters of the linear regression as in Table \ref{tab:cor}. \label{tab:fitother}
\end{tablenotes}
\end{table*}

\begin{table}[ht!!!]
\centering
\setlength{\tabcolsep}{3.pt}
\caption{Line Excitation Diagram (LED) statistics}\label{tab:LED_stats}
\begin{tabular}{llclccc}
    \hline\\[-0.3cm]
\bf L$_{BOL}$     &\bf Type & \bf N & $\overline{\rm Soft.}$ & $\sigma$ & $log(\overline{\lambda_{\rm Edd}})$ & $\sigma$\\[0.05cm]
 (1)            & (2)    &  (3)  & (4)           & (5)    & (6)           & (7)    \\[0.1cm]
 \hline\\
 X-rays         & SY1    & 36    & 0.35          & 0.25   & -1.50         & 0.94   \\
 IR             & "      & 37    & 0.34          & 0.24   & -1.10         & 0.71   \\  
 X-rays         & HBL    & 20    & 0.38          & 0.22   & -1.27         & 2.02   \\
 IR             & "      & 19    & 0.39          & 0.22   & -0.97         & 0.57   \\  
 X-rays         & Type1  & 56    & 0.36          & 0.24   & -1.41         & 1.42   \\
 IR           &(Sy1+HBL) & 56    & 0.36          & 0.24   & -1.06         & 0.67   \\   
 X-rays         & SY2    & 17    & 0.60          & 0.27   & -2.31         & 0.74   \\
 IR             & "      & 17    & 0.57          & 0.23   & -1.58         & 0.56   \\  
 X-rays         & LIN    & 8     & 0.82          & 0.17   & -3.32         & 1.28   \\
 IR             & "      & 8     & 0.79          & 0.15   & -2.52         & 1.38   \\  
\hline\\[-0.25cm]
\end{tabular}\\[0.2cm] 
\begin{tablenotes}
\footnotesize
\item \textbf{Notes.}  Column (1) gives the method for the computation of the bolometric luminosity, from which the Eddington ratio has been computed; Col.(2) the AGN type, note that Type 1 refers to the sum of the Seyfert 1's and the HBL's;  Col.(3) the number of objects;  Col.(4) and (5) the mean value of the {\it softness} and its standard deviation ($\sigma$); Col.(6) and (7) the mean value of the Eddington Ratio ${\lambda_{\rm Edd}}$ and its standard deviation ($\sigma$).
\end{tablenotes}
\end{table}

\subsubsection{Eddington ratio as a function of the bolometric luminosity}\label{sec:edd}

\begin{figure*}
  \includegraphics[width=0.5\textwidth]{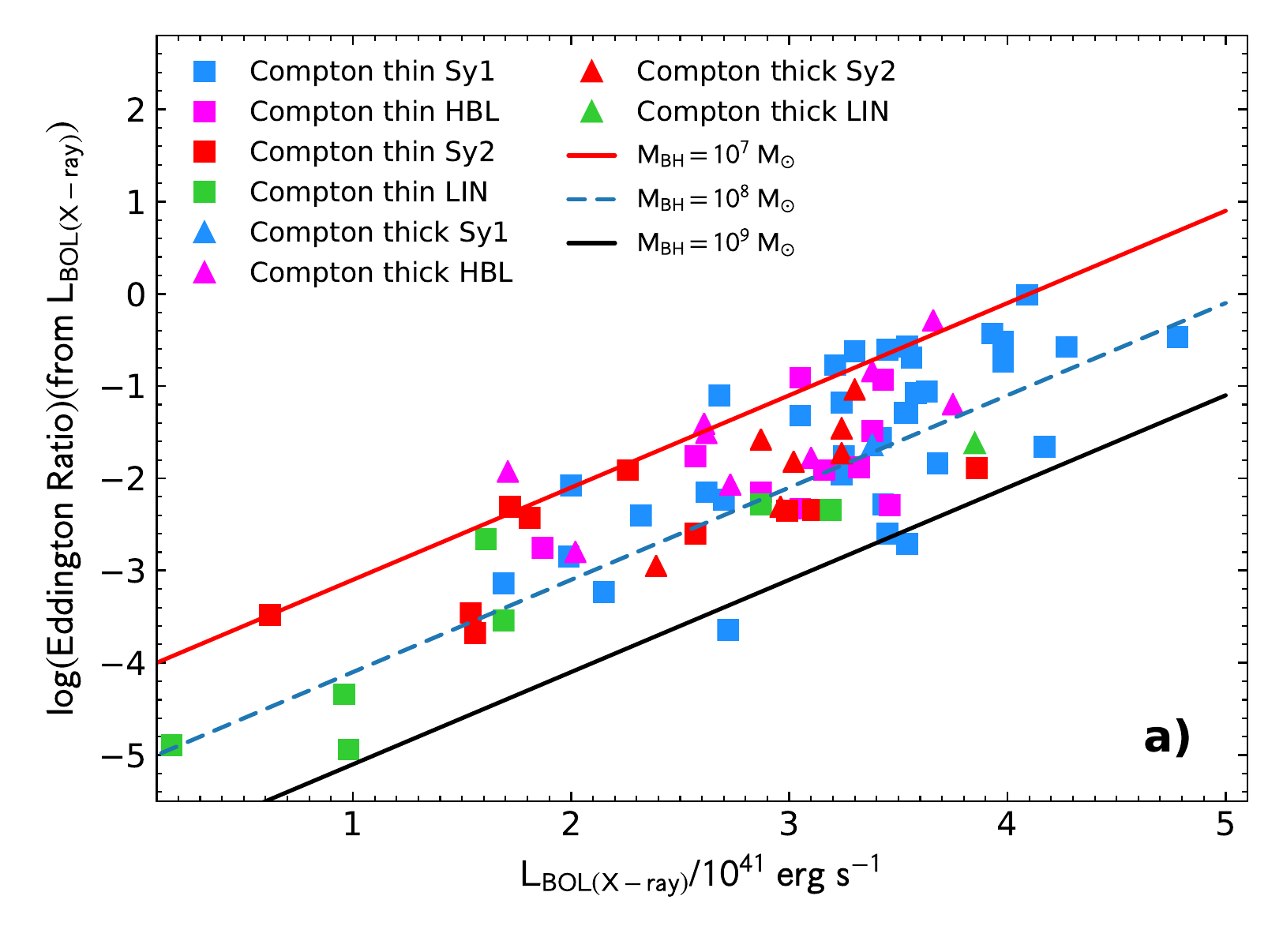}
  \includegraphics[width=0.5\textwidth]{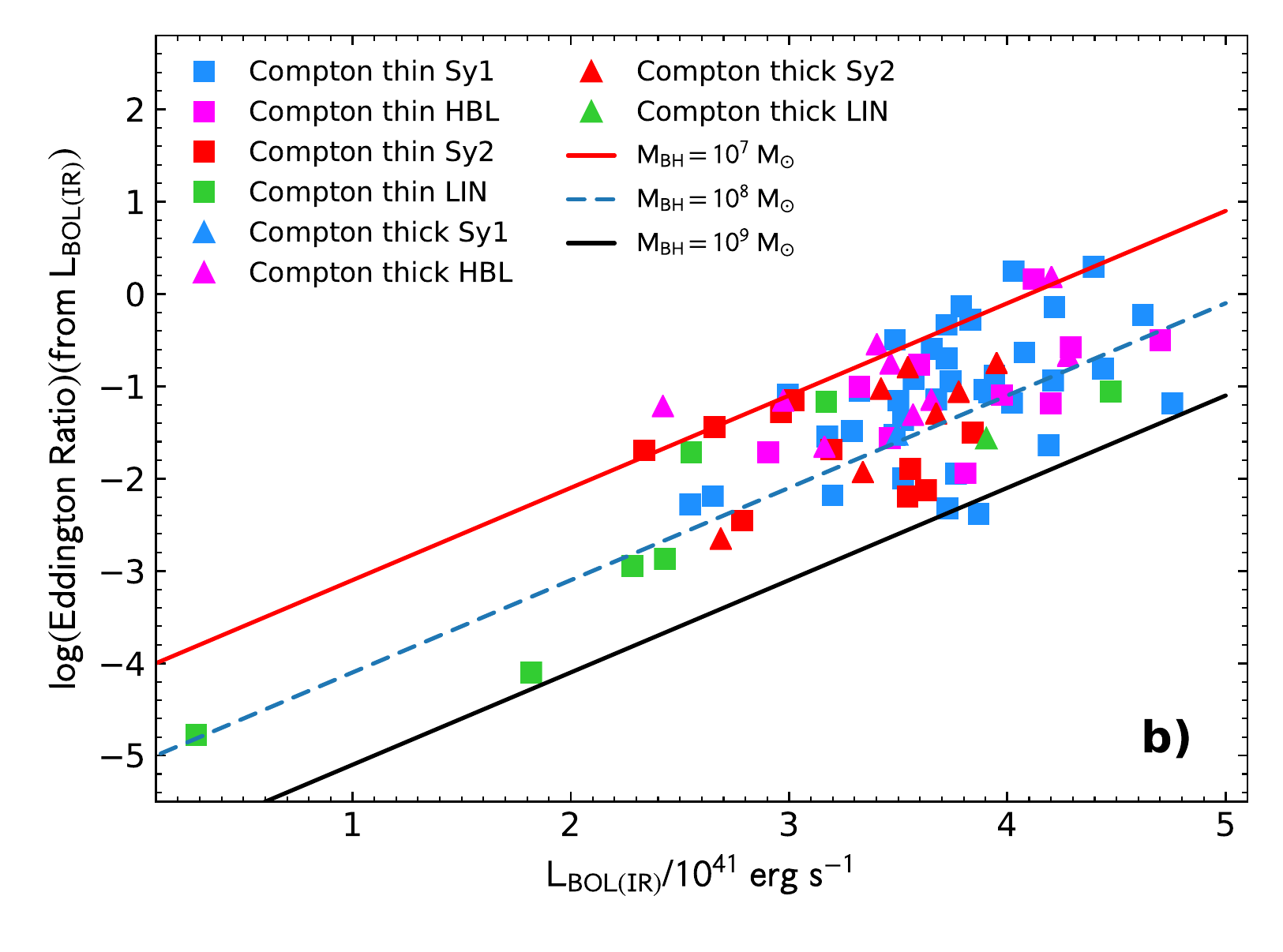}
  \caption{{\bf a: left} The Eddington-ratio as a function of the bolometric luminosity as derived from the X-ray 2-10keV luminosity for the sub-sample of 12$\mu$m AGN with a BH mass estimate. Triangles indicate the Compton-Thick sources and the lines indicate three distinct BH-masses loci at 10$^7$, 10$^8$ and 10$^9$ M$_{\odot}$. A clear separation between the HBL and the Sy2 populations is not observed either in the Eddington–ratio nor in the bolometric luminosity. {\bf b: right} Same as {\bf a:}, but with the the bolometric luminosity as derived from the IR [OIV]25.9 line and the bolometric correction of \citet{spinoglio1995}, using the formula from \citet{mordini2021}. 
  }
  \label{fig:Edd_vs_bol}
\end{figure*}

Figure \ref{fig:Edd_vs_bol} shows the Eddington ratio as a function of the bolometric luminosity as computed from the corrected ($2$--$10)\, \rm{keV}$ X-ray luminosity, using the bolometric correction of \citet{lusso2012} (panel \textbf{a}) and the Eddington ratio as a function of the bolometric luminosity as computed from the \oiv~ line luminosity with the calibration of \citet{mordini2021} and the bolometric correction of \citet{spinoglio1995} (panel \textbf{b}). The lines crossing the diagrams are the loci of the constant black-hole masses at 10$^7$, 10$^8$ and 10$^9$ M$_{\odot}$, and show that most AGN have masses of the order of 10$^8$ M$_{\odot}$. The median black hole mass is 5$\times 10^7\, \rm{M_\odot}$, spanning the range $\sim 10^7$--$10^9\, \rm{M_\odot}$ (Fig.\,\ref{fig:Edd_vs_bol}).

 \begin{figure*}
    \includegraphics[width=0.5\textwidth]{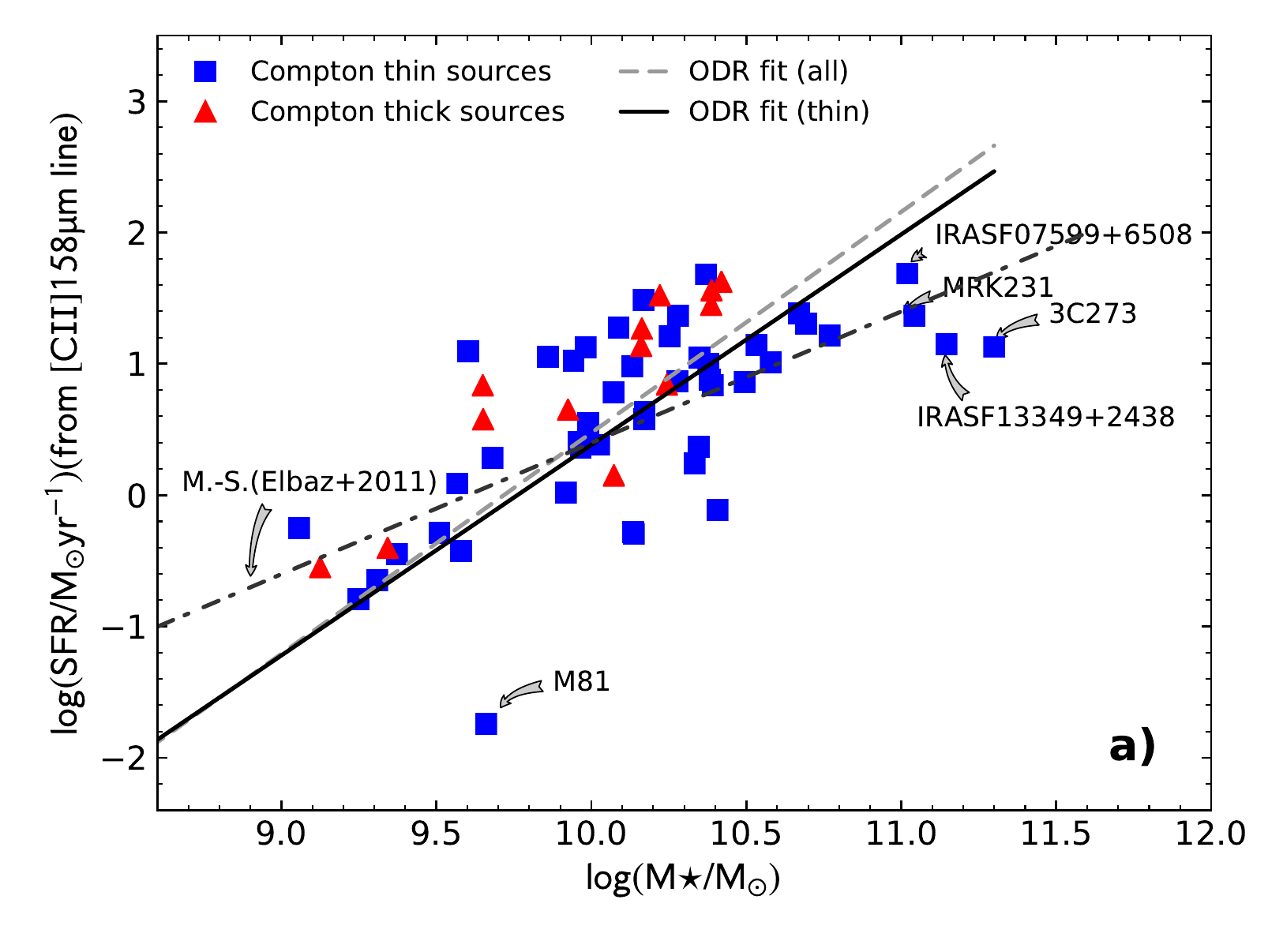}~
 \includegraphics[width=0.5\textwidth]{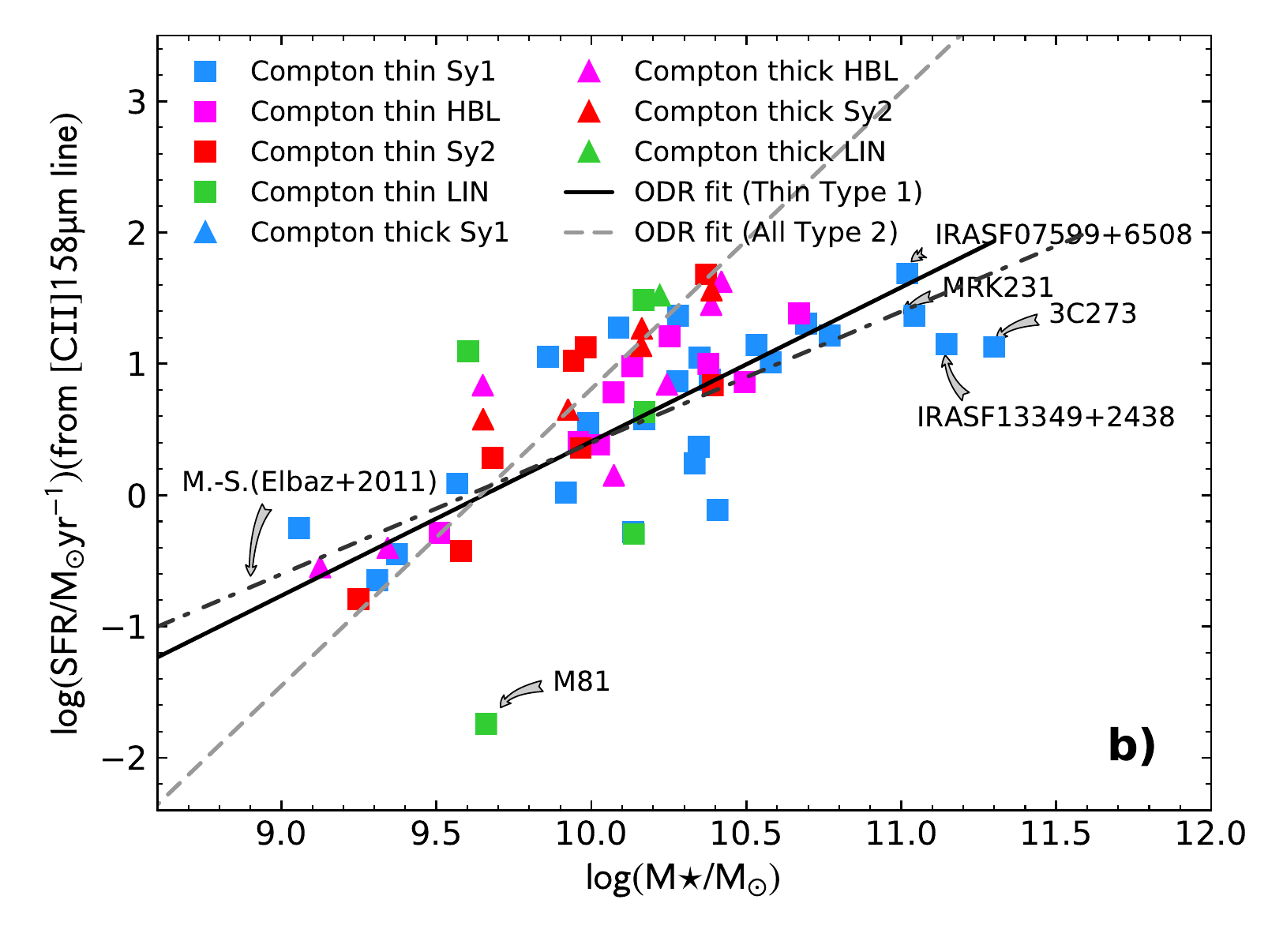} \\
     \includegraphics[width=0.5\textwidth]{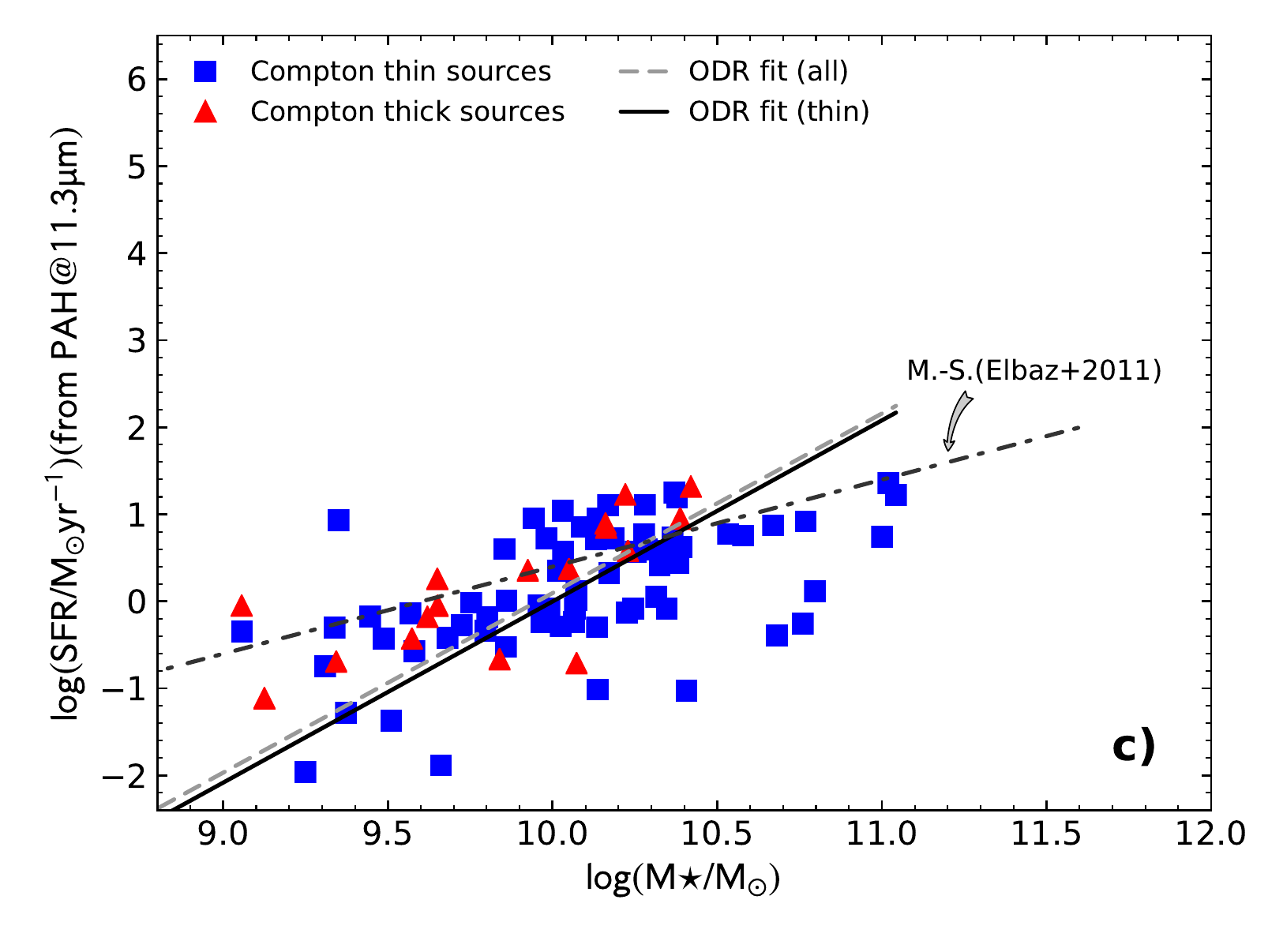}~
    \includegraphics[width=0.5\textwidth]{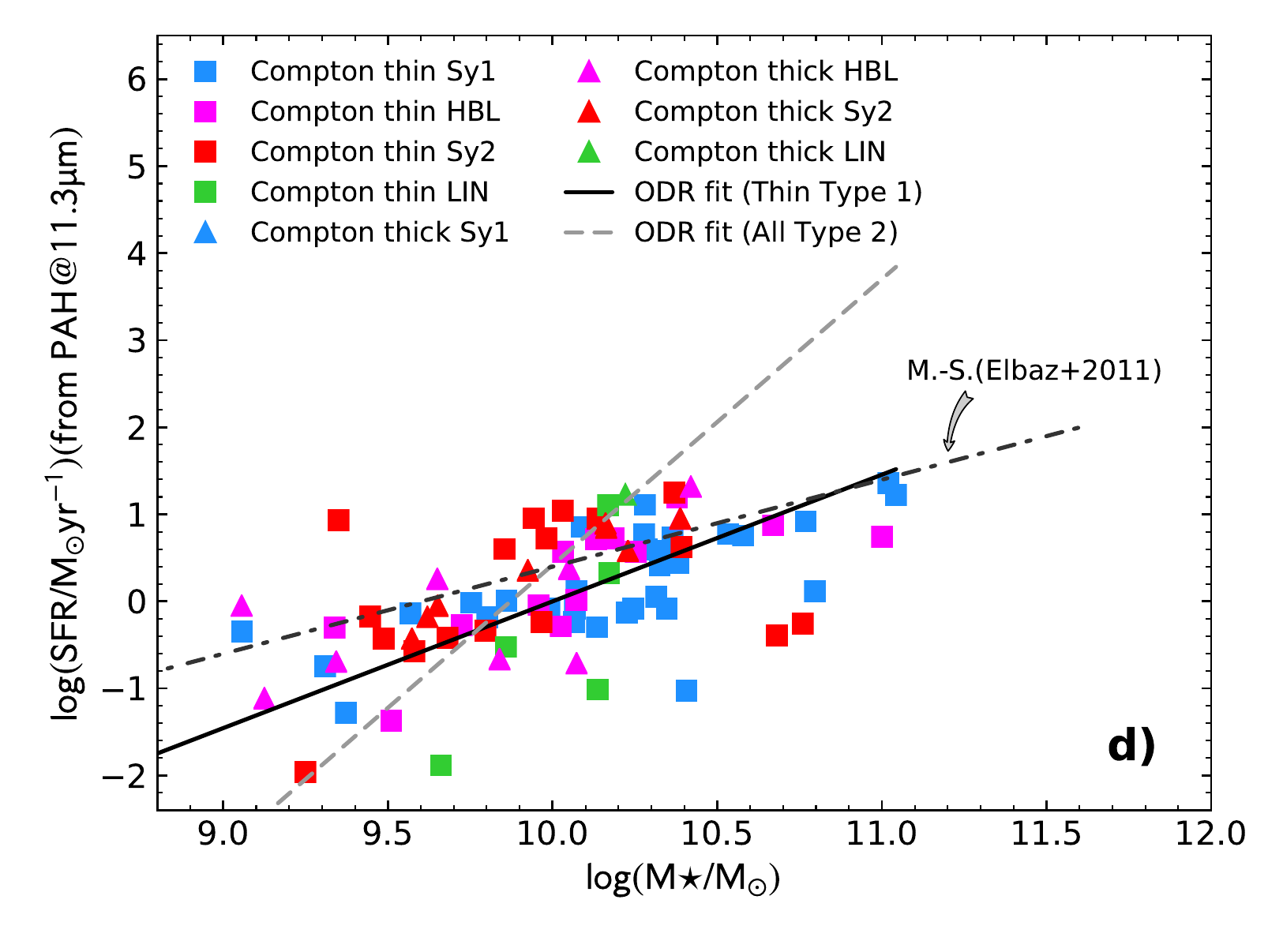}\\
  \caption{{\bf a: upper left}  The SFR, derived from the [CII]158$\mu$m fine structure line, as a function of the stellar mass of the galaxies. The galaxies are divided in Compton thick and Compton thin and the given fits are indistinguishable. The dot-dashed line represents the so called Main-Sequence for star forming galaxies \citep{elbaz2011}. {\bf b: upper right} Same as {\bf a:}, but with the color coded galaxy types. 
  {\bf c: lower left} Same as {\bf a:}, but with the SFR derived from the PAH 11.25$\mu$m emission band and the calibration from \citet{xie&ho2019}. {\bf d: lower right} Same as {\bf c:} but with the color coded galaxy types.   
  }\label{fig:SFR_MASS_CII&PAH}
  \end{figure*}

\subsection{Star formation and black hole accretion in AGN}
\subsubsection{SFR versus stellar mass of the galaxy}\label{sec:sfr&mass}

To establish how much star formation is present per unit mass of a galaxy, the SFR vs galactic stellar mass diagram has been analyzed by \citet{elbaz2011}. 

To measure the star formation rate in our AGN sample, we have used two different luminosities: the PAH\,11.3$\mu$m emission feature and the [CII]158$\mu$m emission line. 
 For the first feature we have used the calibration of \citet{xie&ho2019} and the PAH\,11.3$\mu$m  fluxes of the 12$\mu$m sample from \citet{wu2009}.
 For the [CII]158$\mu$m line we have adopted the SFR formula given in \citet{mordini2021}, which has been derived for the Star forming galaxies, and therefore corrected for the scaling factor of the difference between SF galaxies and AGN (see Table D.1 in \citet{mordini2021}). 
 The [CII]158$\mu$m line data for our sample have been taken from \citet{fernandez2016}.
 We show in Fig.\ref{fig:SFR_MASS_CII&PAH}{\bf a} and {\bf b} the diagram of the SFR, computed with the [CII] line luminosity, as a function of galaxy stellar mass and in Fig.\ref{fig:SFR_MASS_CII&PAH}{\bf c} and {\bf d} the same diagram but with the SFR computed through the PAH\,11.3$\mu$m emission feature. 
 
 The \neii~ and \neiii~ SFR determination can, in principle, be applied only to star forming galaxies \citep{mordini2021} because these lines are also produced in the AGN NLR (especially the \neiii), therefore a correction has to be applied, as e.g. given by \citet{zhuang2019}, which subtracts the AGN emission of these lines using the \nevp~ line. As a comparison, we have included this determination in the Appendix (see Fig. \ref{fig:SFR_MASS_neon}).
 
Comparing the two methods to derive the SFR in galaxies, results in a slightly different normalization: the [CII]158$\mu$m line covers a range in SFR, from $\sim$0.1 $<$ SFR $\lesssim$ 60 M$_{\odot}$ yr$^{-1}$, while the PAH 11.3$\mu$m emission feature covers the range of $\sim$0.01 $<$ SFR $\lesssim$ 20 M$_{\odot}$ yr$^{-1}$. One can argue that in galaxies dominated from an AGN, the PAH emission might be suppressed to some extent by the highly energetic AGN radiation field, while the ionic fine structure lines, once corrected for the total IR  luminosity contribution form the nucleus, can give a better estimate of the SFR. The dash-dotted line in all frames of Fig. \ref{fig:SFR_MASS_CII&PAH} shows the so-called "Main-Sequence" (M.S.) for star forming galaxies, we can see from all figures that most of the galaxies of our sample are below such a M.S. 
Moreover, while there is no difference between Compton-thin and Compton-thick objects, it appears that type 2 AGN have a steeper dependence of the SFR as a function of stellar mass and therefore have a higher SFR at a given galaxy mass. This result was already suggested by the analysis of mid-IR photometric data at 10$\mu$m from the ground, when compared to the IRAS 12$\mu$m data \citep{maiolino1995}.

\begin{figure*}
    \includegraphics[width=0.5\textwidth]{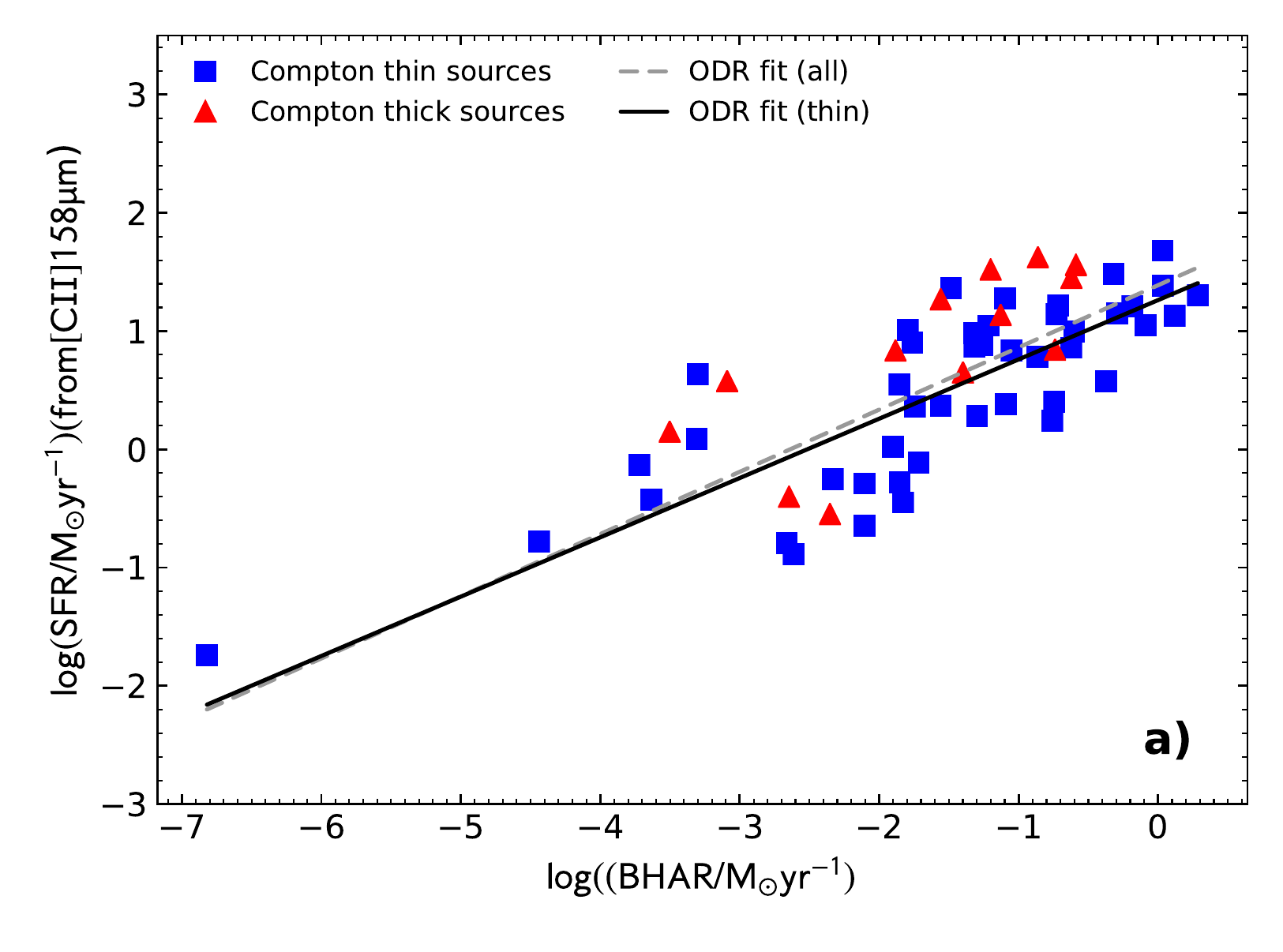}~
    \includegraphics[width=0.5\textwidth]{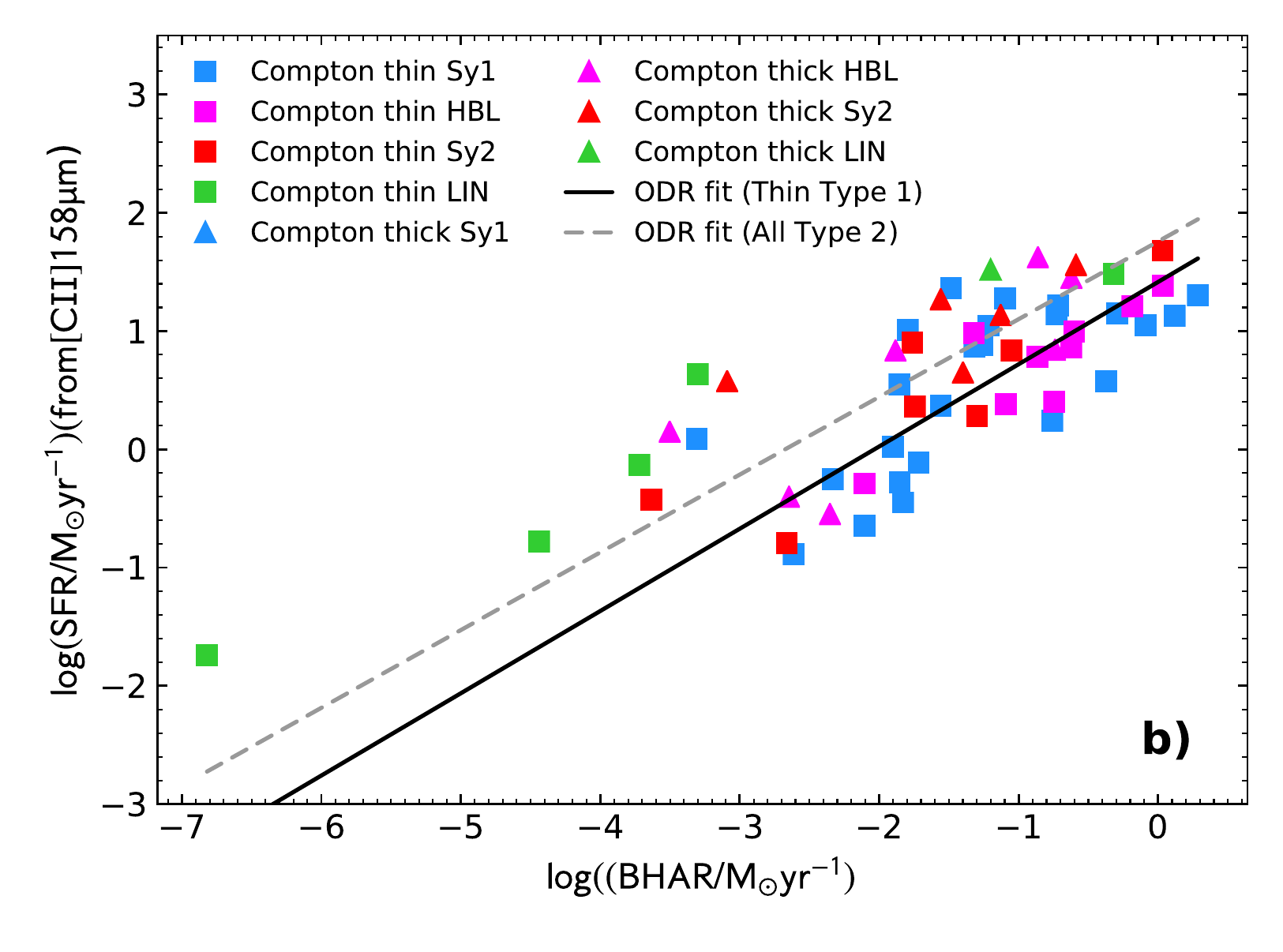}\\
      \includegraphics[width=0.5\textwidth]{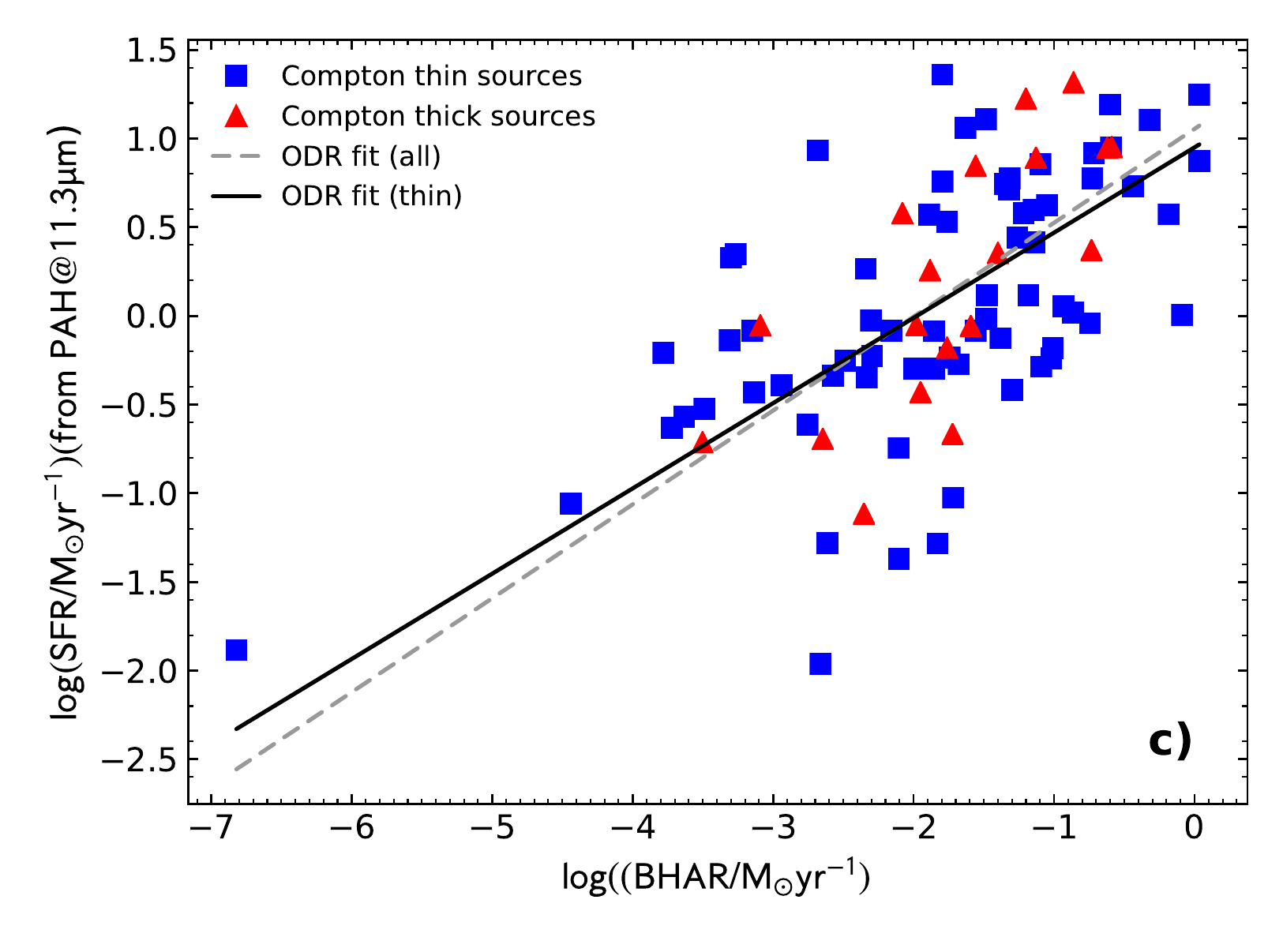}~
    \includegraphics[width=0.5\textwidth]{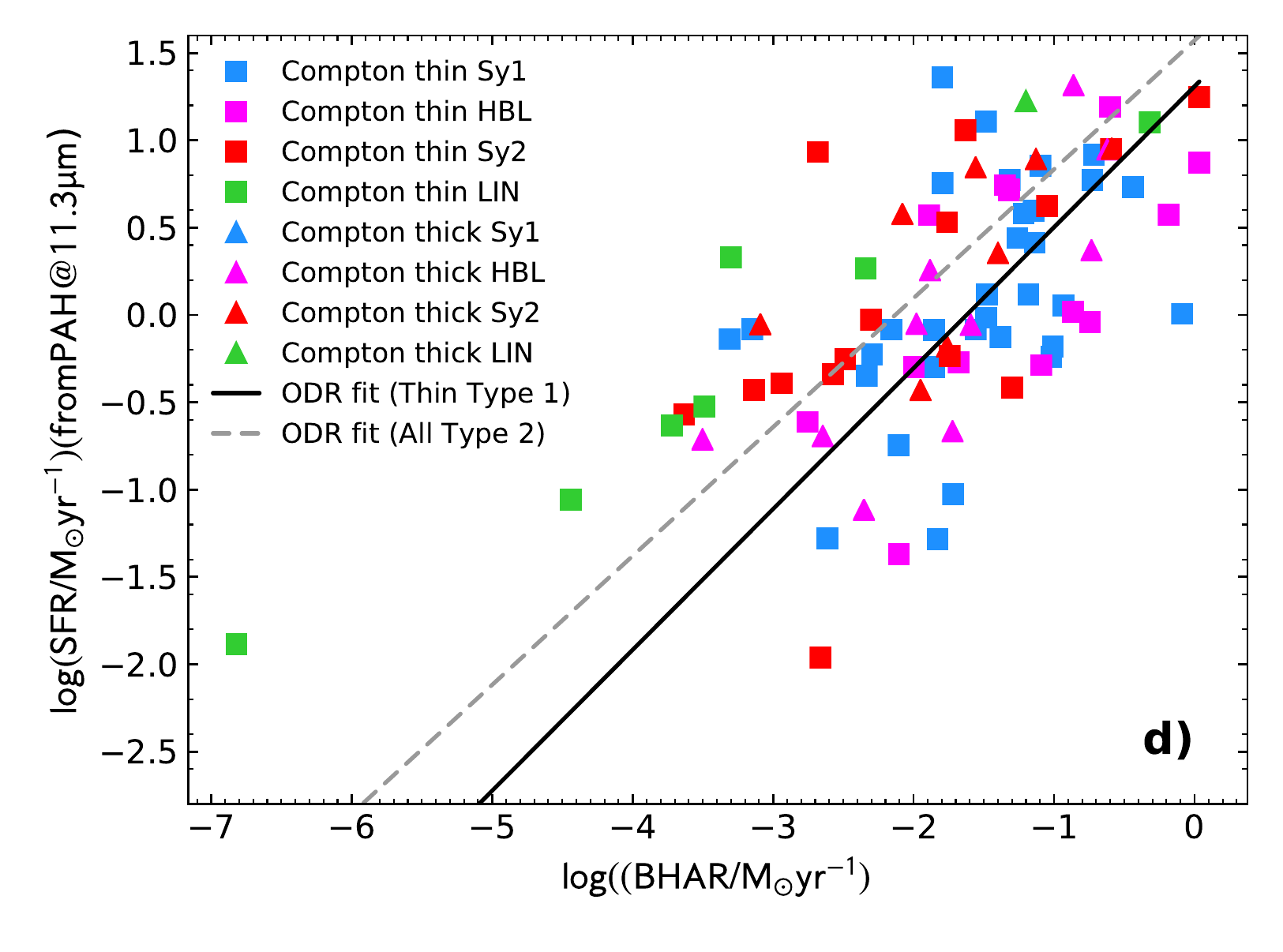}\  
  \caption{{\bf a: upper left} Star Formation rate, as derived from the [CII]158$\mu$m line luminosity and the calibration of \citet{mordini2021}, as a function of the Black-hole accretion rate as derived from the  \oivp\ line luminosity and the calibration from \citet{mordini2021}, adopting the bolometric correction of \citet{lusso2012} for all objects and Compton thin only galaxies. Blue squares indicate Compton thin objects, while red triangles Compton thick. The solid line shows the fit for all Compton thin AGN, while the broken line the fit for all objects. {\bf b: upper right} Same as in {\bf a} but with the color coded objects and the fit of Compton thin type 1 galaxies and all type 2 galaxies. The value of the correlations is given in Table \ref{tab:cor2}.
  {\bf c: lower left} Same as in {\bf a}, but with the Star Formation rate, as derived from the PAH11.3$\mu$m feature, as measured from low-resolution \textit{Spitzer} data \citep{wu2009} and the calibration from \citet{xie&ho2019}. {\bf d: lower right} Same as in {\bf c}, but with the color coded objects and the fit of Compton thin type 1 galaxies and all type 2 galaxies.
  }\label{fig:BHAR_SFR_CII&PAH}
  \end{figure*}

\begin{figure*}
    \includegraphics[width=0.5\textwidth]{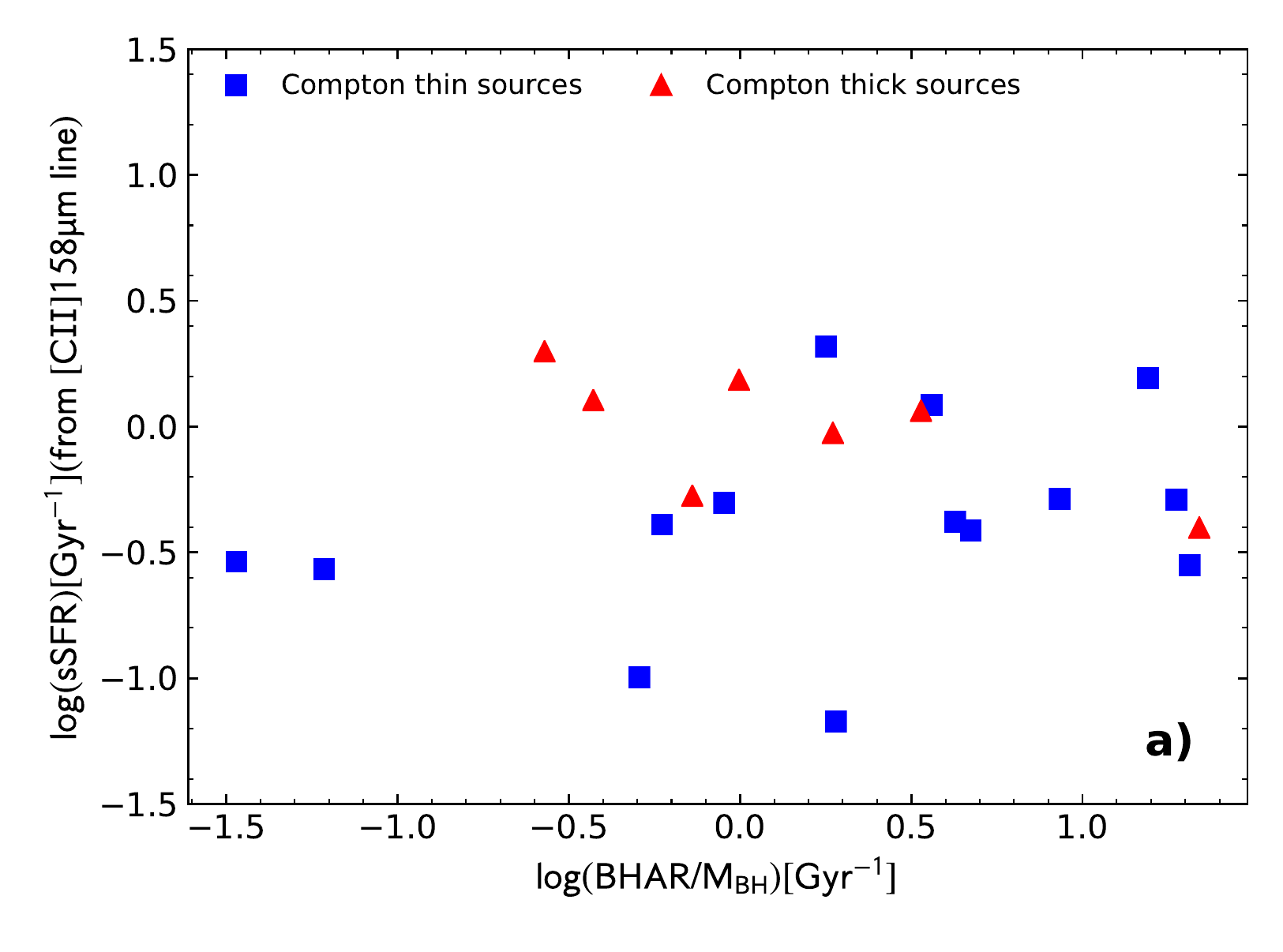}~
    \includegraphics[width=0.5\textwidth]{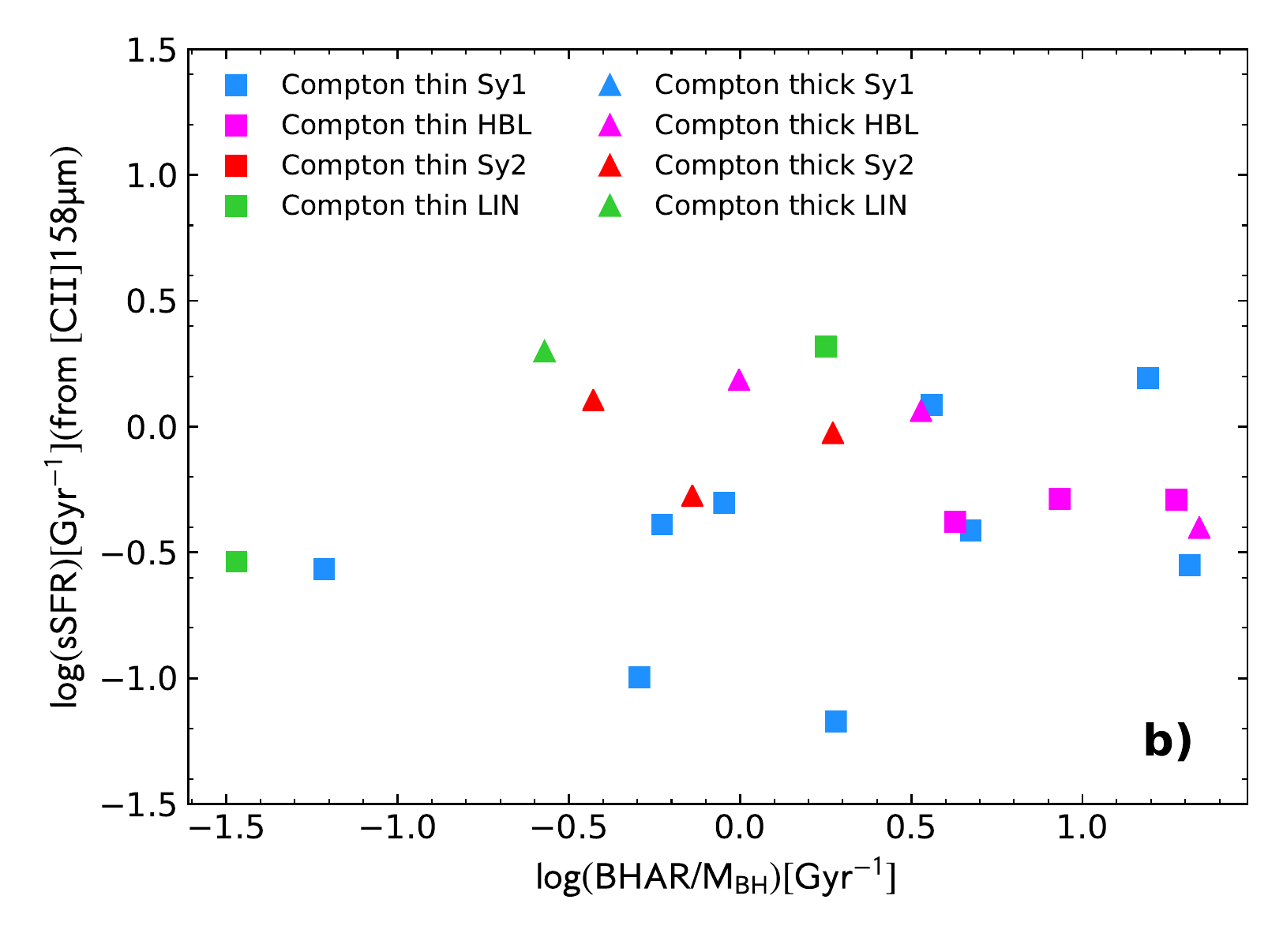}\\
    \includegraphics[width=0.5\textwidth]{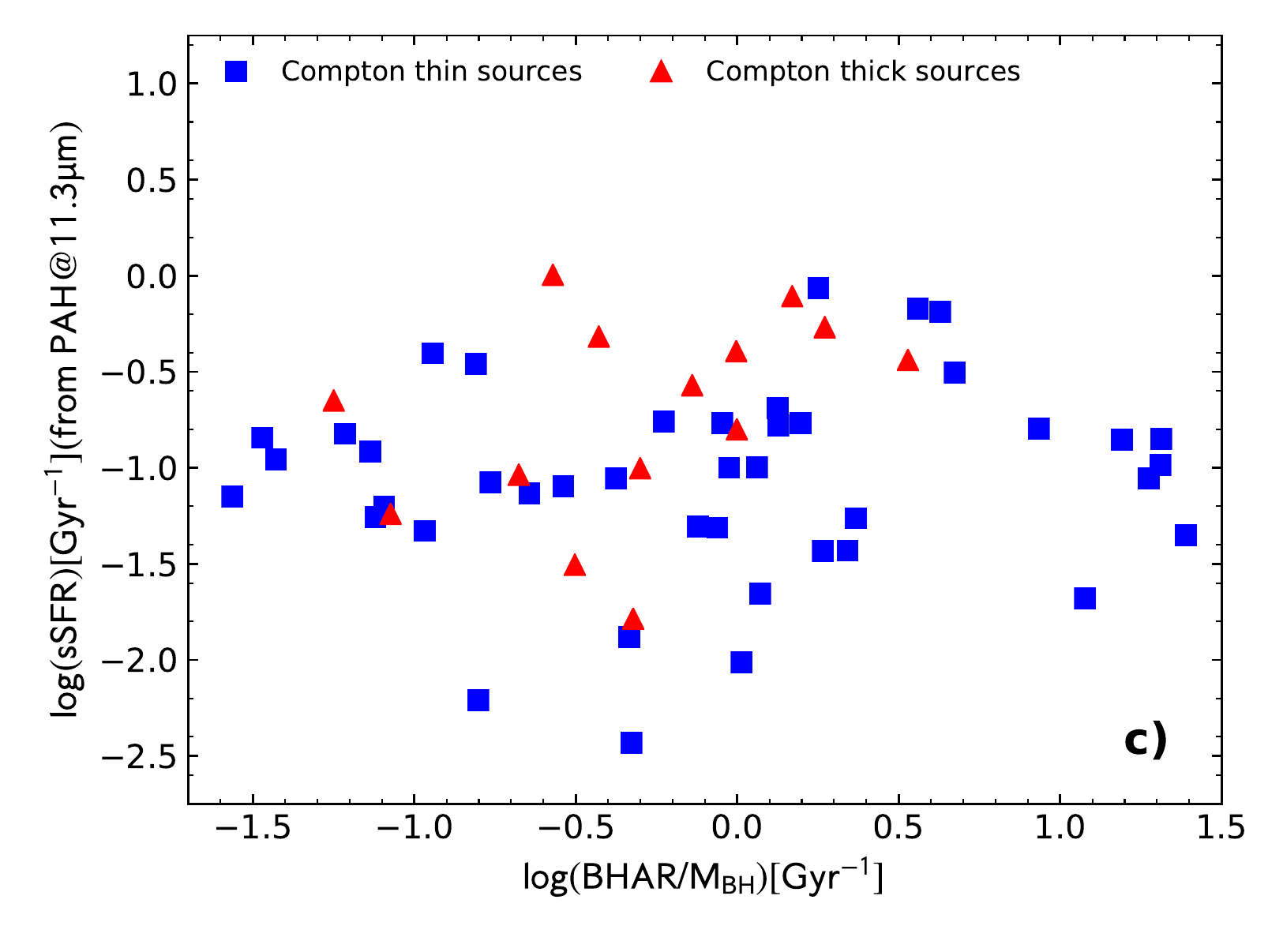}~
    \includegraphics[width=0.5\textwidth]{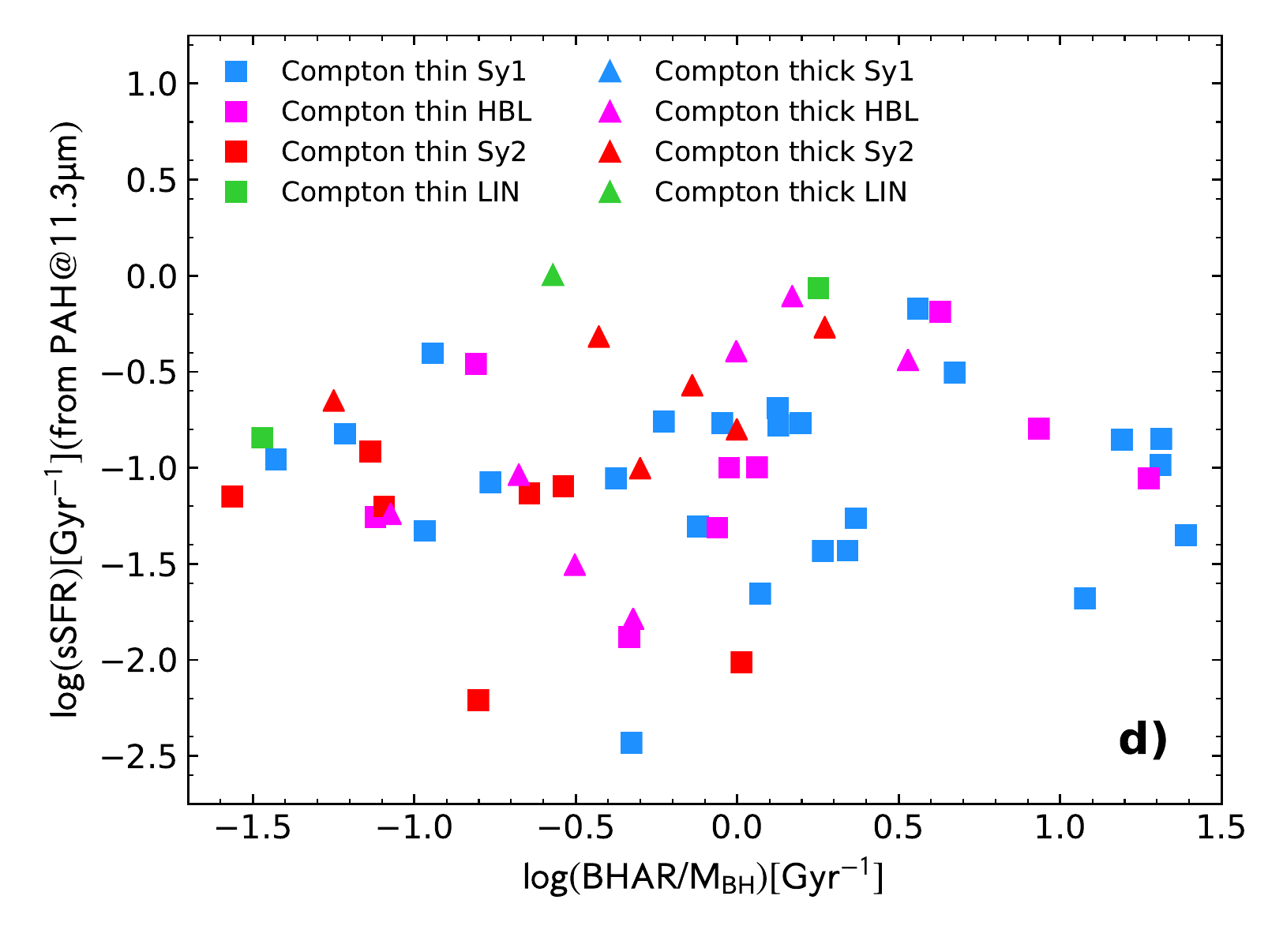}\\    
   \caption{{\bf a: upper left} Specific Star Formation rate, i.e. sSFR\,=\,SFR/M$_{\star}$, as derived from the [CII]158$\mu$m luminosities and the calibration of \citet{mordini2021} (see text), as a function of the {\it specific} Black-hole accretion rate, i.e  sBHAR\,=\,BHAR/M$_{BH}$ as derived from the \oivp\ line luminosity and the calibration from \citet{mordini2021}, adopting the bolometric correction of \citet{lusso2012} for all objects and Compton thin only galaxies. Blue squares indicate Compton thin objects, while red triangles Compton thick. The solid line shows the fit for all Compton thin AGN, while the broken line the fit for all objects. {\bf b: upper right} Same as in {\bf a} but with the color coded objects and the fit of Compton thin type 1 galaxies and all type 2 galaxies. The value of the correlations is given in Table \ref{tab:cor2}.
  {\bf c: lower left} Same as in {\bf a}, but with the Star Formation rate, as derived from the PAH11.3$\mu$m feature, as measured from low-resolution \textit{Spitzer} data \citep{wu2009} and the calibration from \citet{xie&ho2019}.
  {\bf d: lower right} Same as in {\bf c}, but with the color coded objects and the fit of Compton thin type 1 galaxies and all type 2 galaxies.}\label{fig:sBHAR_sSFR}
  \end{figure*}

\subsubsection{Correlation between the SFR and the BHAR}\label{sec:sfr&BH}

 We have derived the correlation between SFR and BHAR, as can be seen from Fig. \ref{fig:BHAR_SFR_CII&PAH} and in Table \ref{tab:cor2}, measuring the BHAR with the \oivp~ line and the calibration from \citet{mordini2021}, with the bolometric correction from \citet{lusso2012} and computing the SFR in the two different ways described above (see section \ref{sec:sfr&mass}). 

For the sub-sample of 58 AGN, for which the [CII]158$\mu$m line has been observed, using this SFR determination, we can derive the following formula (see Table \ref{tab:cor2}):
\begin{equation}
     {\rm SFR(M_{\odot}/yr) = (24.5\pm5.5) \cdot BHAR(M_{\odot}/yr)^{0.53} }
\end{equation}

Using the PAH 11.3$\mu$m feature, for the sample of 89 galaxies for which we have measured this feature, we derive (see Table \ref{tab:cor2}:
\begin{equation}
     {\rm SFR(M_{\odot}/yr) = (11.2\pm3.2) \cdot BHAR(M_{\odot}/yr)^{0.53} }
\end{equation}

\citet{diamond2012} used the same RSA Seyfert sample of 89 galaxies already discussed in \citet{Diamond-Stanic2009} to study the relationship between black-hole growth and star formation in Seyfert galaxies and found with the PAH11.3$\mu$m feature a steeper slope of $B\sim 0.8$, with a lower scaling factor of 7.6, but consistent within the errors.

As can be seen in Table \ref{tab:cor2}, there is no significant difference between CT and Compton thin AGN. The small apparent difference between Type 1 and Type 2 AGN (see Table  \ref{tab:cor2}) is not statistically significant. 


\begin{figure*}
    \includegraphics[width=0.5\textwidth]{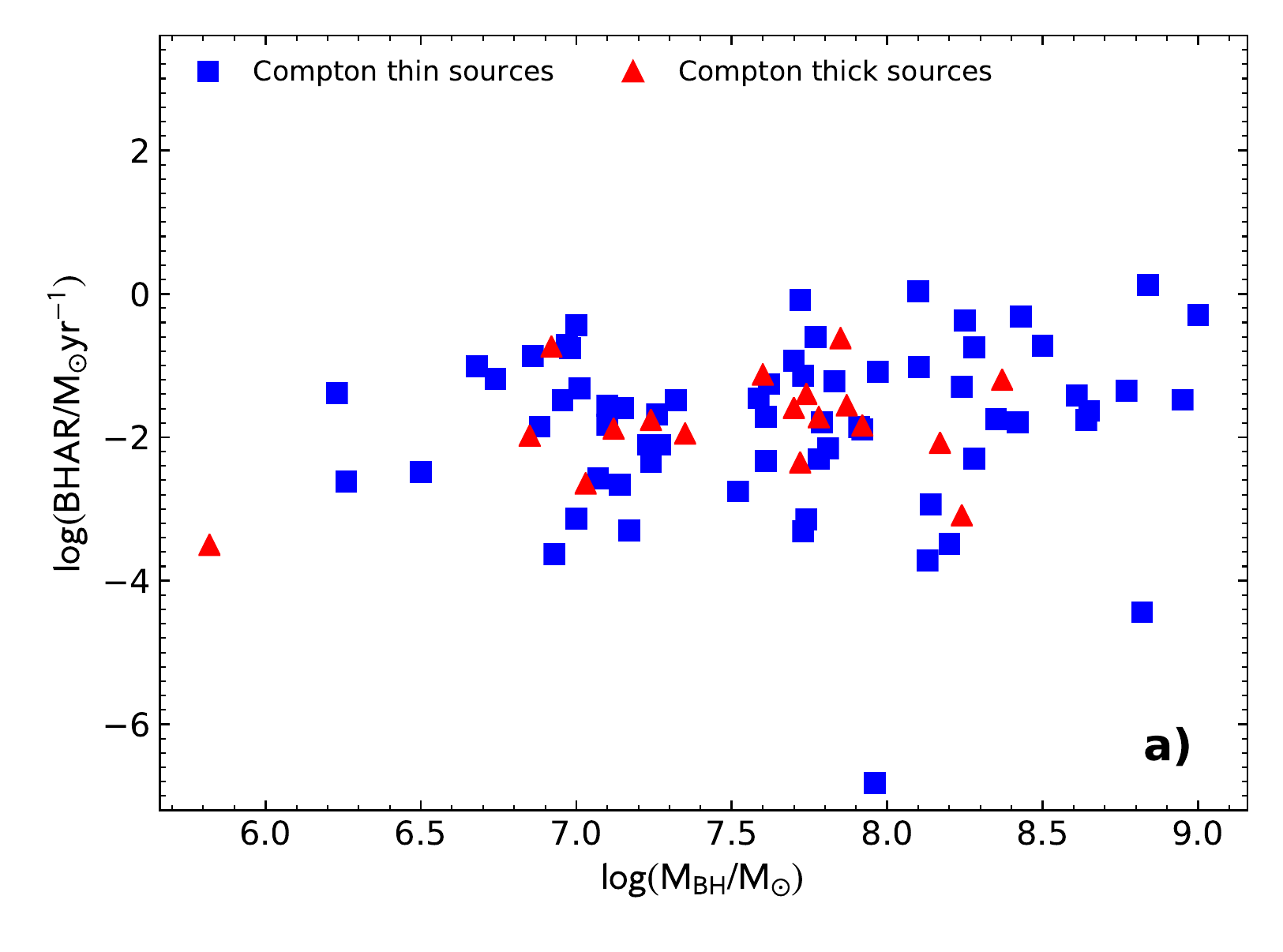}~
    \includegraphics[width=0.5\textwidth]{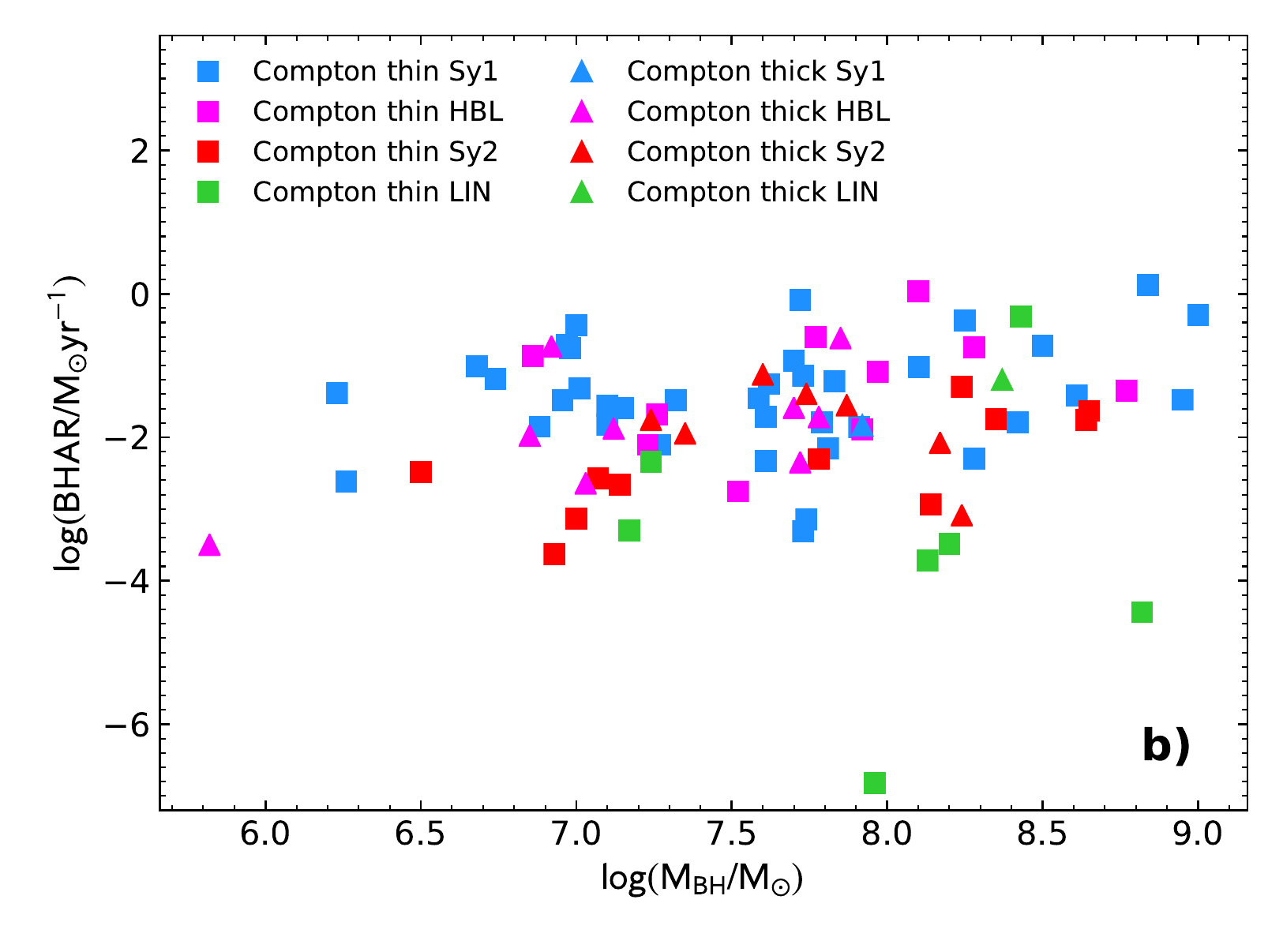}\\
   \caption{{\bf a: left} Black hole accretion rate (BHAR), as derived from the  \oivp\ line luminosity and the calibration from \citet{mordini2021}, adopting the bolometric correction of \citet{lusso2012} as a function of the black-hole mass for all objects divided in Compton thin (blue squares) and Compton thick (red triangles).  
   {\bf b:  right} Same as in {\bf a} but with the color coded objects and the fit of Compton thin type 1 galaxies and all type 2 galaxies. Excluding the the low ionization emission line objects (LINER) an increasing trend of the BHAR ads a function of the BH mass is apparent in the data. 
  }\label{fig:BHAR_MBH}
  \end{figure*}

\subsubsection{Correlation between specific SFR and ``specific'' BHAR}\label{sec:ssfr&sBH}

In Fig.\,\ref{fig:sBHAR_sSFR}, we show the specific SFR, i.e. the star formation rate divided by the stellar mass of the galaxy, sSFR\,=\,SFR\,/\,M${\star}$(M$_{\odot}$), as a function of what we name "specific" BHAR, i.e. the black hole accretion rate divided by the black hole mass, sBHAR\,=\,BHAR\,/\,M$_{\rm BH}$(M$_{\odot}$). Again, as in Section \ref{sec:sfr&mass}, we have determined the SFR using the two methods: in Fig.\,\ref{fig:sBHAR_sSFR}\,{\bf a, b} using the neon lines and in   Fig.\,\ref{fig:sBHAR_sSFR}\,{\bf c, d} using the PAH. There is no correlation between the two quantities.  

\subsubsection{Black hole accretion rate as a function of the Black hole mass}

In analogy with the SFR versus stellar mass diagram, we have plotted in Fig. \ref{fig:BHAR_MBH} the BHAR as a function of the black hole mass. The lack of any correlation testifies that there is no dependence of the BHAR with the black hole mass.

\section{Discussion}\label{sec:discuss}

\subsection{Correlations between NLR lines and X-rays}\label{subsec:discuss-corr}

Bright forbidden emission lines from the NLR are usually adopted as unbiased orientation-independent tracers of the AGN bolometric power \citep{kauffmann2003,heckman2004,brinchmann2004}. In particular, the correlation between the \oiii~ line and the hard X-ray luminosity has been exploited in the past to derive bolometric corrections to the narrow emission line fluxes \citep[e.g.][]{lamassa2009,lamassa2010,lamastra2009}. Nevertheless, the optical narrow line fluxes can be affected by dust extinction, and thus a correction is required to derive reliable bolometric luminosities. Another caveat is the relatively low ionization potential of the bright narrow lines in the optical range. This is $35\, \rm{eV}$ in the case of \oiii, i.e. below the double ionization edge of helium ($54\, \rm{eV}$), meaning that non-negligible contamination from star formation activity is possible, especially in low-metallicity environments \citep[e.g.][]{izotov2018}. In this regard, the main advantages of using high-excitation lines in the mid-IR range is that these tracers are essentially unaffected by dust extinction and they probe unambiguously the accretion power in AGN \citep{genzel1998,melendez2008,rigby2009}.

In an X-ray selected sample of 40 galaxies, \citet{melendez2008} found a marginally stronger correlation between the 2-10\,keV intrinsic luminosity and the \oiii~ luminosity when compared to that obtained using the \oivp~ line instead. This has been interpreted as uncorrected absorption affecting both the \oiii~ and the 2-10\,keV luminosities \citep{melendez2008,Diamond-Stanic2009}. In our 12$\mu$m-selected sample of 116 galaxies, we have analyzed both the flux and the luminosity correlations between the hard X-ray emission (2-10\,keV and 14-195\,keV) and the emission lines of \oivp~ and the two \nevp~ and \nevs~ in the mid-IR, and the \oiii~ optical line. In terms of flux, the \oivp~ and the \oiii~ lines show a similar correlation coefficient when compared with the 2-10\,keV band for both the Compton-thin sub-sample and for the all the nuclei (Pearson coeff. $\rho \sim 0.6$; Tables\,\ref{tab:cor} and \ref{tab:cor2}). The correlations are weaker when the 14-195\,keV band is considered ($\rho \sim 0.4$ for \oivp~ and $\rho \sim 0.3$ for \oiii). These values are similar to those found by \citet{berney2015} for the correlations between the optical lines and the X-ray fluxes in the AGN BAT sample. As expected, the distance factor boosts significantly all the correlations when the luminosities are considered ($\rho \gtrsim 0.7$). We conclude that no significant differences between the \oivp~ and the \oiii~ lines are found for our 12$\mu$m-selected sample. Thus, the selection criteria and the smaller number statistics in \citet{melendez2008} could explain the differences with our results.

When compared with the \oivp~ line, the higher excitation lines of \nevp~ show a weaker correlation with the 2-10\,keV fluxes ($\rho \sim 0.4$), but comparable results for the 14-195\,keV band. Overall, there is a correlation between the high excitation lines and the X-ray bands (Figs.\,\ref{fig:HX-OIVF&L}, \ref{fig:HX-NeV14F&L} and \ref{fig:HX-NeV24F&L}) with a large scatter in its flux version, which becomes significantly stronger only when the luminosities are considered.

The \neiii~ line requires a similar ionization potential as \oiii~ ($41\, \rm{eV}$) and shows a weak correlation with both X-ray bands ($\rho \sim 0.45$; Fig.\,\ref{fig:HX-NeIIIF&L}). In contrast, the lower excitation line of \neii~ ($21\,\rm{eV}$) shows no correlation with either of the X-ray bands ($\rho \sim 0.2$ with p-null values of $\sim 10\%$; Fig.\,\ref{fig:HX-NeIIF&L}). Still, the corresponding luminosity correlations in both cases are strong ($\rho \gtrsim 0.6$), suggesting that the results obtained for the \neii~ line are spurious.


We explore here if there is a possible dependence of the mid-IR high ionization lines with the covering factor (CF), i.e. the fraction of the sky seen from the AGN that is blocked by obscuring material. This could in principle block part of the NLR clouds from the primary excitation and ionization originating in the central supermassive black hole, and introduce systematic differences between sources with different covering factors. This could affect the NLR line to X-ray correlation and therefore introduce additional uncertainties in our results.

This effect can be serious for the so-called {\it elusive} AGN, where the nuclear UV source is completely embedded and the ionizing photons cannot escape, which prevents the formation of a classical narrow-line region \citep{maiolino2003} and for buried AGN found in LINER ULIRGs without optical Seyfert signatures \citep{imanishi2006}.

An attempt to measure the covering factors of the mid-IR NLR lines of our sample of AGN is beyond the scope of the present paper, however, we have used an indirect method to test this scenario. 
The CF should be mainly driven by the AGN luminosity, because the obscuring material surrounding the nucleus recedes at higher AGN luminosities. This is shown by e.g. \citet{toba2014} on the dependence of the CF from the mid-IR AGN luminosity, and by \citet{ricci2017b} on the dependence of the CF from the Eddington ratio. These two tests demonstrate statistically that, for our sample, the mid-IR NLR lines are not affected by variations in the covering factor and therefore the results of this study, that the mid-IR NLR lines are a good indicator of the AGN intrinsic luminosity, are indeed robust.

\begin{figure*}
    \includegraphics[width=0.5\textwidth]{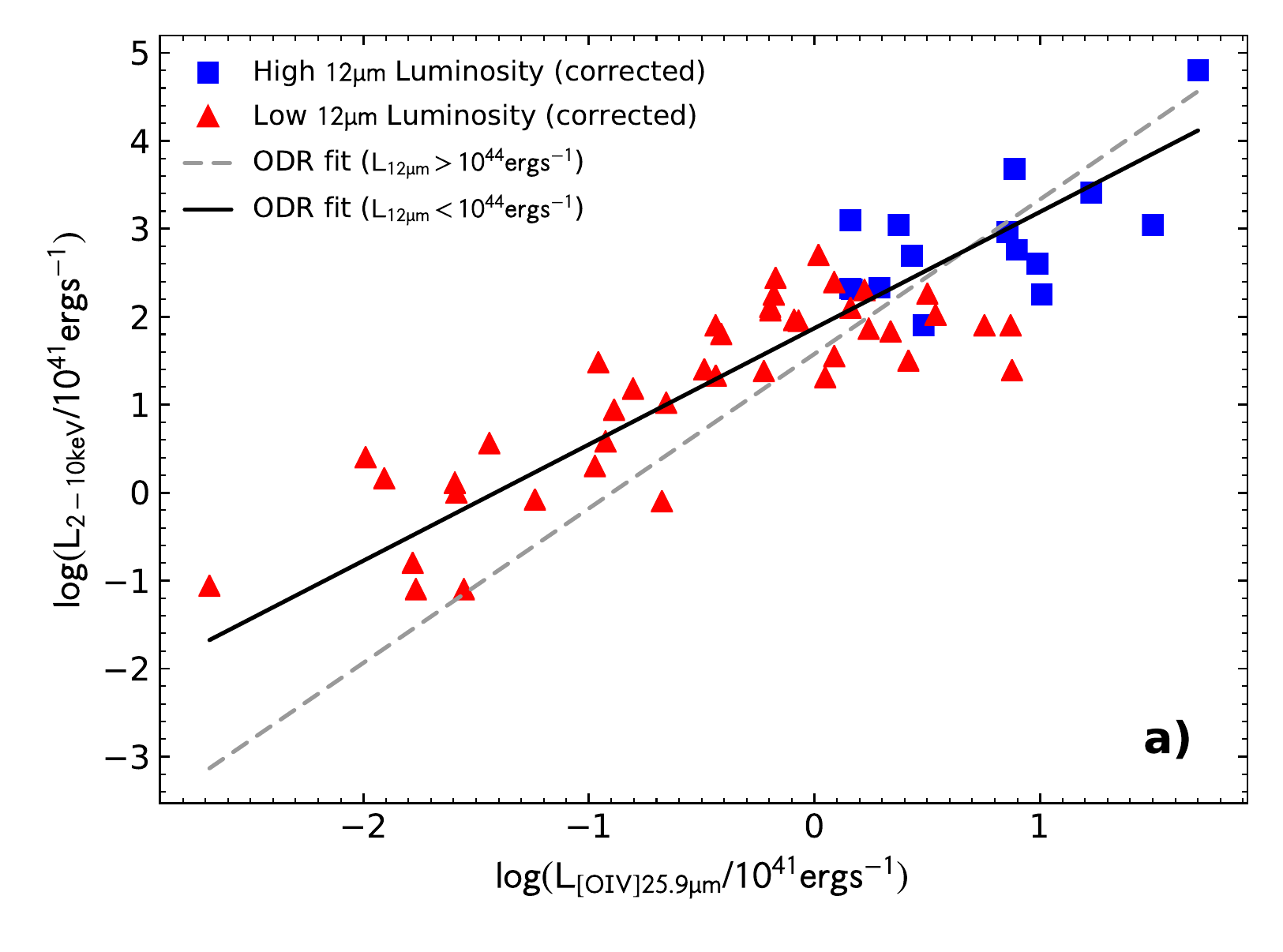}~
    \includegraphics[width=0.5\textwidth]{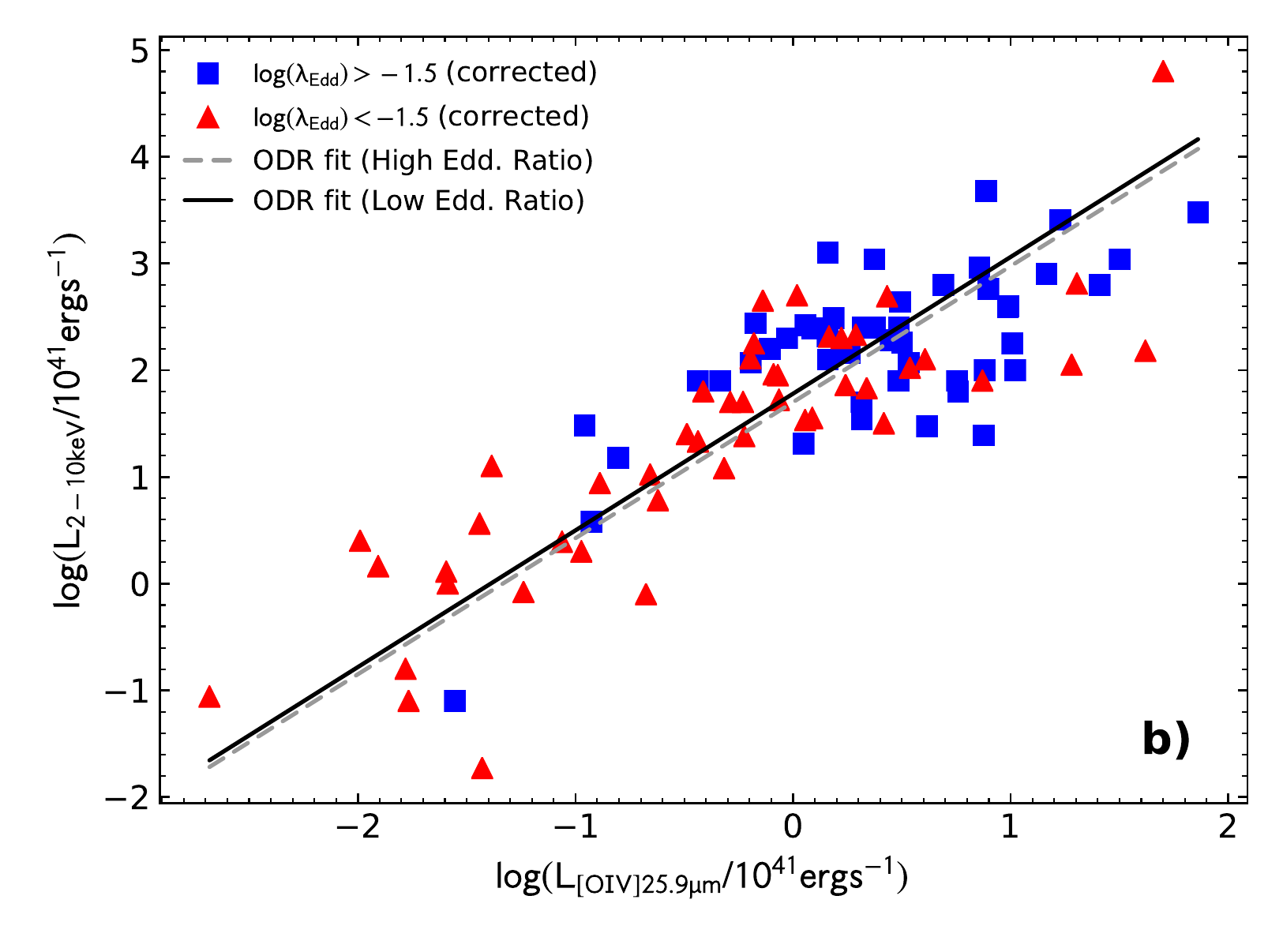}\\
   \caption{{\bf a: left} X-ray (2--10) keV intrinsic luminosity as a function of the \oiv\ 25.9$\mu$m line luminosity for all sources with a measure of the nuclear 12$\mu$m luminosity. The blue squares indicate the AGN with $L_{12\mu m}>10^{44} erg s^{-1}$, while red triangles show objects with $L_{12\mu m}<10^{44} erg s^{-1}$. The solid lines represent the linear regression fit with low luminosity objects, while the broken line shows the fit to the high luminosity objects. 
   {\bf b:  right} Same as in {\bf a} but with the AGN divided by their Eddington ratio:  The blue squares indicate the AGN with $\log(\lambda_{\rm Edd})>-1.5$, while red triangles show objects with $\log(\lambda_{\rm Edd})<-1.5$.
  }\label{fig:covering_fac}
  \end{figure*}

\citet{toba2014}, 
using the Wide-field Infrared Survey Explorer \citep[WISE,][]{wright2010} MIR all-sky survey and complete optical spectroscopy of emission lines for $\sim$ 3000 AGN-dominated MIR sources, found that the CF decreased with increasing MIR luminosity. In their Fig. 28, they show that the CF can be as high as 0.6$<$CF$\lesssim$0.8 for luminosities $L_{22\mu m} \lesssim 10^{44} erg s^{-1}$ and decreases to 0.2$<$CF$\lesssim$0.4 at $L_{22\mu m} \gtrsim 10^{44} erg s^{-1}$, thus defining a turnover luminosity at 
$L_{22\mu m} = 10^{44} erg s^{-1}$.
Following the arguments above, we have plotted in Fig.\ref{fig:covering_fac}{\bf a} the X-ray (2--10) keV intrinsic luminosity as a function of the \oiv\ 25.9$\mu$m line luminosity dividing the AGN with the nuclear 12$\mu$m luminosity above and below the turnover luminosity of $L_{12\mu m} = 10^{44} erg s^{-1}$. We have adopted the nuclear 12$\mu$m luminosity, instead of the 22$\mu$m WISE luminosity used in \citet{toba2014}, because the nuclear 12$\mu$m luminosity is not affected by the galaxy emission. The result between the two fits, of high and low luminosity AGN, does not show any significant difference (see the fit results in Table \ref{tab:cor2}). This indicates that, for our sample of AGN, there is no apparent effect of the covering factor affecting the mid-IR line luminosities. This is in line with the isotropic properties of the NLR lines found by \citet{mulchaey1994}.

\citet{ricci2017b} confirmed the decrease in the covering factor of the circumnuclear material with increasing accretion rates, already found previously \citep{ueda2003,maiolino2007,treister2008, burlon2011}, using a systematic multi-wavelength survey of hard X-ray-selected black holes. Their result can be summarized by a relation between the covering factor of dusty gas and the Eddington ratio, inferred for both Compton-thick and Compton-thin AGN, that shows that the covering factor is significantly larger  at $\log(\lambda_{\rm Edd})<-1.5$ ($\overline{CF} \sim 0.85$), and lower for $\log(\lambda_{\rm Edd})>-1.5$ ($\overline{CF}\sim 0.40$). We have therefore plotted in Fig.\ref{fig:covering_fac}{\bf b} the X-ray (2--10) keV intrinsic luminosity as a function of the \oiv\ 25.9$\mu$m line luminosity dividing the AGN by their Eddington ratio. The figure shows that there is no distinction between the objects with low and high Eddington ratio (see the fit results in Table \ref{tab:cor2}).

In conclusion, we can affirm that for our sample of AGN, there is no apparent dependence of the NLR mid-IR lines from the covering factor of circumnuclear material around the nuclei.

\subsection{Line to X-ray flux ratios in Seyferts 1 and 2}
\citet{melendez2008} also found that the observed hard X-rays/[\textsc{O\,iv}] ratio is systematically lower in Seyfert 2 galaxies when compared to type 1 nuclei. This difference further increases for CT Seyfert 2's. They interpret this result as absorption affecting the observed $2$--$10\, \rm{keV}$ spectra (corrected by Galactic N$_H$), which would even depress the observed luminosities in 
the hardest X-ray band ($14$--$195\, \rm{keV}$). In our Fig.\,\ref{fig:X-OIVF&L}(d) and Fig.\,\ref{fig:HX-OIVF&L}(d), and in Table \ref{tab:cor} we do not find any statistically significant difference, either comparing the whole sample of AGN with the Compton-thin objects only and dividing them in the two broad classes of type 1's and type 2's. 

\citet{Diamond-Stanic2009} and \citet{rigby2009} have tried to overcome the problems encountered by \citet{melendez2008} by selecting a complete sample of 89 optically selected Seyfert galaxies from the Revised Shapley-Ames sample \citep{maiolino-rieke1995,ho1997}, containing 18 Seyfert 1's and 71 Seyfert 2's. \citet{Diamond-Stanic2009} have found that the \oivp~ luminosity distributions of Seyfert 1's and 2's are indistinguishable and also that both Compton-thin and Compton-thick Seyfert 2's are statistically indistinguishable from Seyfert 1's. The same authors found that the correlation between the \oiii~ luminosity and the \oivp~ luminosity shows a significant difference between Seyfert 1's and Seyfert 2's with type 1's having a \oivp/\oiii~ ratio of order unity, while type 2's have on average \oivp/\oiii $\sim 5$. 
To the contrary, we do not find any significant difference in this correlation for our sample, as can be seen either in the flux and in the luminosity correlations (see Sect. \ref{sec:oxygen}, Figs. \ref{fig:OIII-OIVF&L} a--d) and Table \ref{tab:cor2}).

Finally \citet{Diamond-Stanic2009} showed that the relation between the {\it observed} 2-10 keV X-ray luminosity and the \oivp~ line luminosity of Seyfert 2's (which include Compton thick objects) is significantly displaced compared to Seyfert 1's. This is not a surprise because the {\it observed} 2-10 keV X-ray luminosities need to be corrected for absorption (see our Fig. \ref{fig:X-OIVF&L}(b,d) and Table \ref{tab:cor}). When this correction is applied, there is no statistical difference between Seyfert 1's and 2's.

\citet{rigby2009} have used the same optically selected sample of \citet{Diamond-Stanic2009} to study the correlation between the \oivp~ luminosity and the hard (10–200 keV) X-rays from {\it Swift}, INTEGRAL, and {\it BeppoSAX}. They found that Seyfert 2's have systematically lower X-ray/\oiv~ luminosity ratios than Seyfert 1's. In particular, for their sample, Seyfert 2's have an average hard X-ray to \oiv~ ratio that is 3.1$\pm$0.8 times lower than Seyfert 1's. their Compton-thin Seyfert 2's show a decrement of 1.9 $\pm$0.5, and known Compton-thick Seyfert 2's show a decrement of 5.0$\pm$2.7. To the contrary, with the complete mid-IR selected sample, we do not find any statistically significant difference among the various populations (all vs Compton-thin and AGN type 1's vs type 2's) either in flux or in luminosity (see Fig. \ref{fig:HX-OIVF&L} and Table \ref{tab:cor}). We ascribe their result to the heterogeneous hard X-ray coverage of their sample, being referred to different instruments with different observed energy ranges, which might cause significant systematic errors in the inferred X-ray luminosity of the considered AGN.

To establish the best isotropic indicators of intrinsic AGN luminosity, \citet{lamassa2010} have identified five different proxies, namely the \oiii~, the \oivp~ line luminosities, the mid-IR, the radio and the hard X-ray (E $>$ 10 keV) continuum emission, using two complete samples of low-redshift Type 2 AGN: an optically selected \oiii~ sample of 20 Seyfert 2's galaxies and the original 12$\mu$m sample of 32 Seyfert 2's \citep{spinoglio1989}. When they compared the Seyfert 2's with the typical values of the Seyfert 1's, they found that the mean ratio of the observed \oiii~ flux to the \oivp~ flux is lower in Seyfert 2's than in Seyfert 1's by a factor of 1.5–2. 
Analyzing the correlations in Table \ref{tab:cor2}, shown in Figs. \ref{fig:OIII-OIVF&L}, as discussed already above and in Sect. \ref{sec:oxygen}, we are unable to confirm their result: the slopes of the correlations of Type 1's and Type 2's are well within the errors, and the intercepts have very large errors, therefore no statistically significant difference is apparent in our data. 

\citet{lamassa2010} have also found that the 14-195 keV hard X-ray flux (relative to the both \oiii~ and \oiv) is suppressed by about an order of magnitude in mid-IR selected Seyfert 2's compared to Seyfert 1's (consistent with \citealt{rigby2009}). This is not the case for our larger complete 12$\mu$m selected sample of AGN, where the hard X-ray 14-195 keV X-ray data are more homogeneous and up-to-date, being collected mostly from the latest BAT survey \citep{oh2018}, as can be seen in Fig. \ref{fig:HX-OIVF&L} and Table \ref{tab:cor}.

\subsection{Accretion properties in a complete AGN sample}

The luminosity-excitation diagram (LED; \citealt{fernandez2021}) for the AGN in the 12\,$\mu$m sample, already introduced in Sect. \ref{sec:led}, is shown in Fig. \ref{fig:LED}. It includes all galaxies of the sample with observations of the \oivp~ and \neii~ lines, a measure of the bolometric luminosity either from the X-ray emission or from the IR (see Section \ref{sec:bol}), and an estimate of the black hole mass (Section\,\ref{sec:led}). In spite of the smaller number statistics, when compared to the heterogeneous sample analyzed in \citet{fernandez2021}, especially among the LINER class, the 12\,$\mu$m sample is complete and not biased toward type 1 or type 2 sources \citep{rush1993}. Thus, it is a better sample to investigate intrinsic differences in the accretion properties of the various classes of AGN.

 The clear distinction between type 1 and type 2 nuclei, which appears in the {\it softness} axis (see Sect. \ref{sec:led}), with the majority of the Seyfert 1 population showing values of {\it softness}  $<0.5$, caused by strong emission of the \oivp~ line, is due to the intense UV radiation able to reach the high ionization potential of O$^{3+}$ ($\sim 55\, \rm{eV}$), and is associated with the presence of a prominent accretion disk in these nuclei. This is also true for the hidden broad line Seyferts (HBL). Nevertheless, a certain fraction of the Seyfert 1 and HBL population (19\% of Sy1 and 20-30\% of HBL), depending on the method with which the bolometric luminosity is computed) is still found at {\it softness} values $> 0.5$. The average value of the {\it softness} for type 1 AGN (Seyfert 1's and HBL) is $\sim$0.36  (see table \ref{tab:LED_stats}).

On the other hand, the Seyfert 2 population shows a wider distribution in excitation properties, with about half of the nuclei closer to the high excitation Seyfert 1's and the other half at low excitation values, in agreement with the bimodal distribution found by \citet{fernandez2021} for the type 2 nuclei. The average value of the {\it softness} for type 2 nuclei is $\sim$0.6  (see Table \ref{tab:LED_stats}).

In spite of the low statistics for the LINER in our sample, only 8 objects, 
the average value of the {\it softness} is $\sim$ 0.8, (see table \ref{tab:LED_stats}), higher than any other AGN types, 
as expected for AGN lacking a significant contribution from the accretion disk in the UV range.

As already shown in Table \ref{tab:LED_stats}, Type 1 nuclei (i.e. Seyfert 1's and HBL galaxies) show on average higher Eddington ratios ($log(\overline{\lambda_{\rm Edd}})= -1.2\pm0.2$) with respect to type 2 objects ($log(\overline{\lambda_{\rm Edd}})= -2.0\pm0.3$), i.e. on average they appear to accrete more efficiently. As a matter of fact, very few Seyfert 2s are above $\sim 10\%$ Eddington, and most of the type 1s above this limit show prominent \oivp~ emission and therefore high excitation.  As expected, the lower Eddington luminosities observed correspond to LINER nuclei ($log(\overline{\lambda_{\rm Edd}})= -2.9\pm0.4$).
Two of the type 1 nuclei seem to be accreting at their Eddington limits in Fig.\,\ref{fig:LED}a, while Fig.\,\ref{fig:LED}b shows values slightly over Eddington. This is likely caused by an offset introduced by the different bolometric corrections applied to the X-ray and IR luminosities, although there is a good overall agreement between the two calibrations.

These results suggest that the accretion properties of type 1 and type 2 nuclei are intrinsically different. Most of the galaxies above $\sim 10\%$ Eddington are type 1s, which tend to show higher excitation values possibly due to the presence of a more dominant accretion disk in these nuclei. Half of the Seyfert 2s share similar excitation properties at lower luminosities, and could be consistent with the obscured Seyfert 1s in agreement with the unified model. The other half of the Seyfert 2 population and some of the type 1 nuclei show a relatively weak \oivp~ emission, which could be explained by the lack of a prominent accretion disk in their ionizing continuum emission.

\subsection{Host galaxy versus black hole properties}

The result presented in Section \ref{sec:sfr&BH} that a strong correlation (see the correlation parameters in Table \ref{tab:cor2}) is found between the SFR and the BHAR should be analyzed carefully. This correlation does not necessarily mean that the two processes have a physical and causal link, but they may both depend on the fundamental physical quantity. In order to investigate the origin of this correlation we normalize both quantities by their related masses, the stellar mass of the galaxy for the SFR and the black hole mass for the BHAR, i.e. looking for a correlation between the specific SFR (sSFR) and the ratio of BHAR to the black hole mass. As shown in Section \ref{sec:ssfr&sBH}, no relation is found between the sSFR and the specific BHAR (Fig.\,\ref{fig:sBHAR_sSFR}), meaning that galaxies that are more efficient at forming stars do not necessarily feed their central black holes at a higher specific rate. Therefore, the correlation found between SFR and BHAR might be driven by the mass. That is, a massive galaxy would likely form more stars, it would host a heavier black hole at its center, which would be then more capable at producing energy through accretion processes. On the other hand, we cannot rule out the possibility that indeed a physical link exists among the two processes.

\begin{figure*}
    \includegraphics[width=0.5\textwidth]{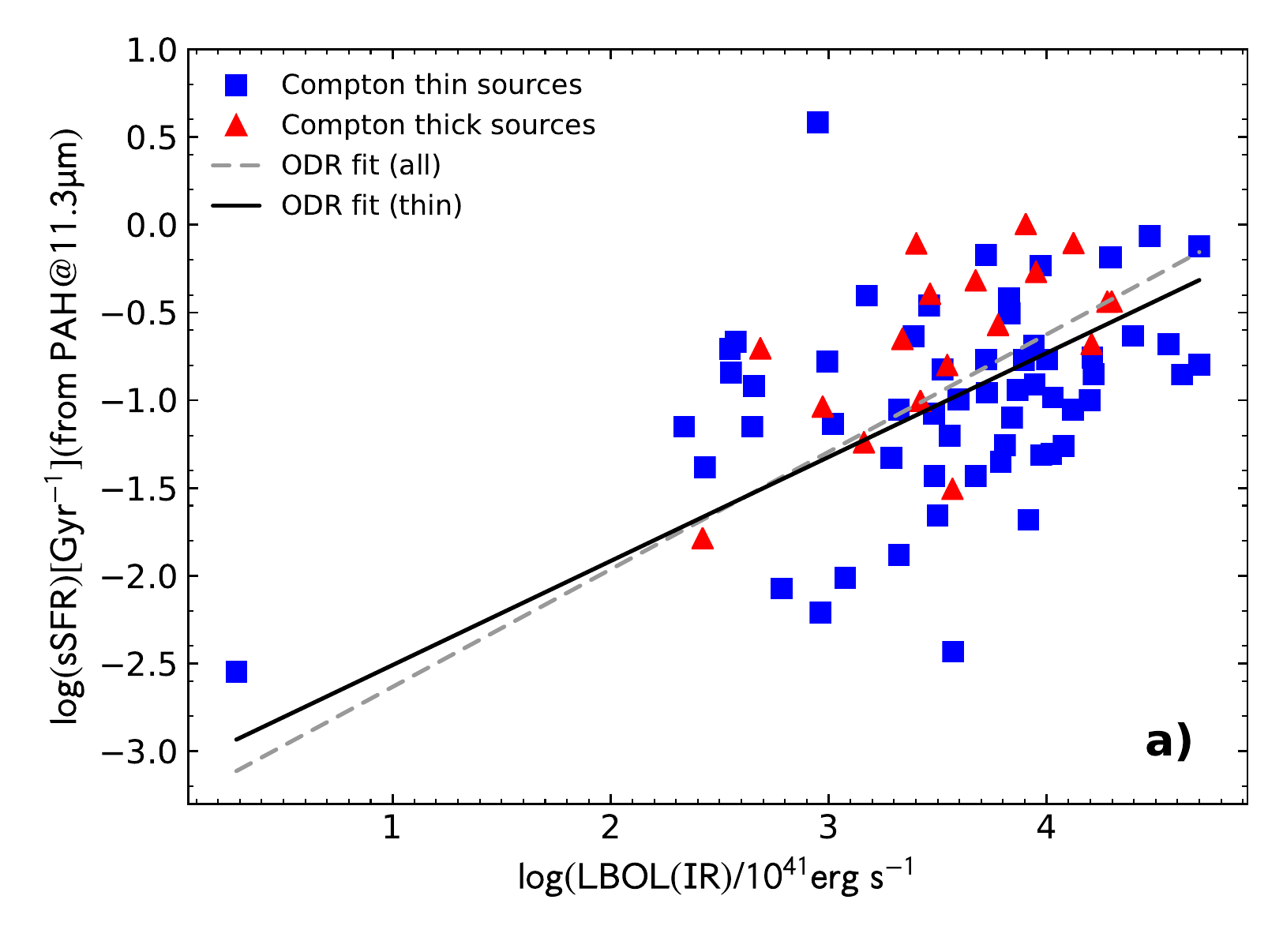}~
    \includegraphics[width=0.5\textwidth]{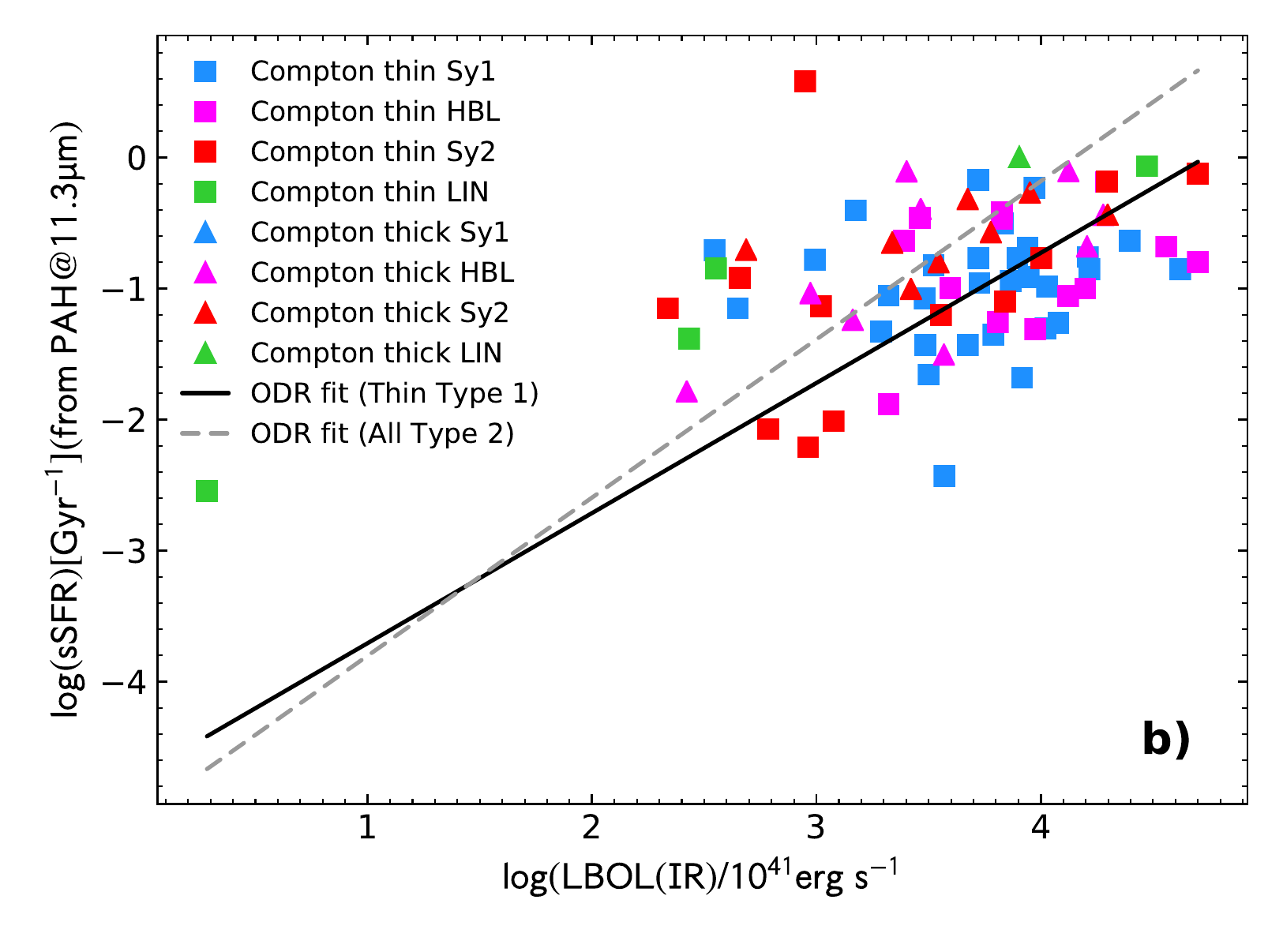}\\
   \caption{{\bf a: left} Specific star formation rate (sSFR), as derived from the  PAH11.3$\mu$m feature \citep{wu2009} and the calibration from \citet{xie&ho2019}, as a function of the bolometric luminosity, as derived from the IR, for all objects divided in Compton thin (blue squares) and Compton thick (red triangles).  
   {\bf b:  right} Same as in {\bf a} but with the color coded objects and the fit of Compton thin type 1 galaxies and all type 2 galaxies. Excluding the the low ionization emission line objects (LINER) an increasing trend of the BHAR ads a function of the BH mass is apparent in the data. 
  }\label{fig:sSFRvsLbol}
  \end{figure*}

Our results presented in Section \ref{sec:sfr&BH}, showing a correlation between SFR and BHAR, can be compared with results of \citet{zhuang2020} who analyzed a large sample of optically selected type 1 and type 2 AGN and showed: {\it i}) a linear relation between AGN luminosity (accretion rate) and SFR and {\it ii}) that type-2 AGN, independent of stellar mass, Eddington ratio, redshift or molecular gas mass, exhibit intrinsically stronger star formation activity than type-1 AGNs. Both results are consistent with our findings. This is in apparent disagreement with the simplest AGN unified model in which there are no intrinsic differences between type 1 and type 2 Seyfert galaxies. 
 
Moreover, our finding that normalizing to the mass makes the correlation disappear, is consistent with the results of \citet{zhuang2021}, who showed that the strong correlation between SFR and BHAR in their optically selected sample of 453 z$\sim$0.3 type 1 AGNs is driven by the mutual dependence of these parameters on the molecular gas mass. They also report that the SFR and BHAR are still weakly correlated after removing the dependence of star formation rate on molecular gas mass. However we do not see this effect using our -- different -- normalization.

\citet{ward2022} have compared the results of simulations (TNG, EAGLE and SIMBA), also for the local Universe (epoch of redshift $z$ = 0), invoking AGN (negative) feedback to suppress star formation, with the observations searching signatures of AGN feedback by looking at trends between the bolometric luminosity and galaxy properties, or comparing AGN to non-AGN. The lack of negative trends between sSFR and AGN luminosity could indicate that AGN feedback may be ineffective. We have looked for correlations between the sSFR and the galaxy stellar mass, the AGN bolometric luminosity and the strength of the AGN as measured by the \oivp  /\neii~ line ratio (measuring the accretion disk dominance in AGN), but we did not find any {\it negative} tend. To the contrary, we find a positive correlation between the sSFR and the bolometric luminosity of the AGN, as can be seen in Fig. \ref{fig:sSFRvsLbol}, that might indicate a {\it positive} AGN role in boosting the star formation. The results of the fit are given in Table \ref{tab:cor2}.


\subsection{What can be done by the JWST}

The new James Webb Space Telescope (JWST; \citealt{gardner2006}), launched on the 25th of December 2021, thanks to its mid-IR spectrometer MIRI \citep{rieke2015}, can obtain spectra across the whole mid-IR range, for AGN and galaxies in the local Universe. This will provide local samples  to study the co-existence of star formation and black hole accretion.
Further, JWST will be able to observe the shorter wavelength high-ionization line of \nevp~ up to a redshift of z$\sim$0.9, i.e. at the {\it beginning} of the so-called {\it cosmic noon} of galaxy evolution. The other high-ionization lines discussed in this paper, the \nevs~ and the \oivp~ lines, can be observed only at lower redshifts, namely at z$\lesssim$0.13, i.e. in the local Universe. However, other near-IR lines can be used to trace BHAR and SFR. 
The fine structure lines of [MgIV]4.49$\mu$m (IP = 80.1eV) and [ArVI]4.53$\mu$m (IP = 75eV) can be used as BHAR tracers for the 1 $<$ z $<$ 3 range, and the [NeVI]7.65$\mu$m line (IP = 126.2eV) as the best tracer for z $<$ 1.5 \citep{satyapal2021,mordini2022}. The [ArII]6.98$\mu$m (IP = 15.7eV) and [ArIII]8.99$\mu$m (IP = 27.6eV) lines can be used to measure the SFR at z $<$ 3 and z $<$ 2, respectively, while the stronger [NeII]12.8$\mu$m line is
within the JWST-MIRI spectral range above z $<$ 1.2. At higher redshifts, the PAH features at 6.2 and 7.7$\mu$m can be observed at z $<$ 3 and z $<$ 2.7, respectively. 
We refer to the work of \citet{mordini2022} for the assessment of these spectral lines and features, that can be detected by JWST-MIRI in galaxies and active galactic nuclei up to redshift z$\sim$3. 

\section{Summary}

Unbiased tracers of the AGN power, detected even in dust-embedded regions,
are essential to understand the co-evolution of supermassive black holes and their host galaxies. In this work we have demonstrated the potential of high-excitation emission lines in the mid-IR range to achieve this goal. Mid-IR transitions of ionic species with ionization potentials above the double ionization edge of helium ($> 54\, \rm{eV}$) show a negligible contribution from stellar populations, and therefore are excellent isotropic tracers of the AGN contribution, unaffected by dust obscuration and temperature effects as opposed to the optical lines. This study is based on the 12$\mu$m AGN sample, a bolometric flux-limited sample unbiased by selection effects (e.g. dust obscuration, \oiii~ luminosity), and the legacy of \textit{Spitzer}/IRS for the spectroscopic observations of these galaxies. We compare the brightest high-excitation mid-IR lines (\oivp~ and \nevp, \nevs) with the hard X-ray continuum and the \oiii~ emission-line luminosities. Accretion power measurements derived from both X-ray continuum and the \oivp~ line are used to investigate intrinsic differences among the various AGN types. Finally, we compare the main properties of the AGN host galaxies with those of the accreting supermassive black hole at their centers. The main conclusions of this study are:

\begin{itemize}
    \item The 2-10keV X-ray fluxes and luminosities, only when corrected for absorption, give a robust measure of the AGN strength and these correlate with the high ionization mid-IR lines. The 14-195\,keV hard X-ray fluxes also correlate with the mid-IR high ionization lines.
    
    \item The high ionization mid-IR fine structure lines (\nevp~, \nevs~ and \oivp) are a very powerful measure of the AGN strength and in particular the BHAR.
    
    \item Simple formulae are given to derive the AGN bolometric luminosities from the luminosity of one of the three high-ionization lines of \oivp, \nevp~ and \nevs.
    
    \item The high-ionization mid-IR emission lines (\nevp, \nevs, \oivp) linearly correlate with other AGN bolometric indicators (compact nuclear $12\, \rm{\micron}$ emission, as well as [OIII] $\lambda$5007\AA \ line emission), both in terms of flux and luminosity.
    
    \item No evidence of systematic differences in these correlations is found among the various Seyfert populations, including type 1 and type 2, and Compton-thick and -thin AGN. 
    
    \item Differences are found between Seyfert type 1 and type 2 in the Eddington ratios and the excitation properties of the narrow line gas. Type 1 AGN, i.e. Seyfert type 1 and Hidden Broad Line Region galaxies (HBL) together, show on average higher Eddington ratios and stronger excitation, as traced by the \oiv/[Ne\,II] line ratio, while Seyfert type 2 show lower Eddington ratios and lower excitation. Low Ionization Nuclear Emission Line galaxies (LINER) have even lower Eddington ratios  and lower excitation, defining a decreasing trend among the three populations of AGN.
    
    \item In the mid-IR, the SFR in AGN can be measured by the PAH 11.3$\mu$m feature and in the 
    far-IR with the fine structure line of [CII]158$\mu$m, once corrected for the offset between the luminosity of the AGN and the star forming galaxies. 
    
    
    \item A positive correlation is found between the Black Hole Accretion Rate (BHAR) and the Star Formation Rate (SFR) (SFR$\propto$ $\sqrt{BHAR}$), measured through the [CII]158$\mu$m line and the PAH 11.3$\mu$m feature, but no correlation between the {\it specific} SFR (sSFR) and the {\it specific} BHAR, i.e. the BHAR/M$_{BH}$.
    
    \item The JWST will be able to measure these lines in many galaxies in the local Universe with the MIRI spectrograph, and assess the co-existence of star formation and black hole accretion. Moreover, the \nevp~ line can be detected in AGN up to a redshift of z $\sim$0.9, allowing an accurate measure of the AGN bolometric luminosity, to be compared e.g., with the results of hard X-ray observations.
\end{itemize}

\begin{acknowledgements}
JAFO and LS acknowledge financial support by the Agenzia Spaziale Italiana (ASI) under the research contract 2018-31-HH.0. JAFO acknowledges the financial support from the Spanish Ministry of Science and Innovation and the European Union -- NextGenerationEU through the Recovery and Resilience Facility project ICTS-MRR-2021-03-CEFCA. We thank the anonymous referee who helped improving this work.
\end{acknowledgements}

\clearpage

\bibliography{12mgsX}{}

\begin{thebibliography}{}
\expandafter\ifx\csname natexlab\endcsname\relax\def\natexlab#1{#1}\fi
\providecommand{\url}[1]{\href{#1}{#1}}
\providecommand{\dodoi}[1]{doi:~\href{http://doi.org/#1}{\nolinkurl{#1}}}
\providecommand{\doeprint}[1]{\href{http://ascl.net/#1}{\nolinkurl{http://ascl.net/#1}}}
\providecommand{\doarXiv}[1]{\href{https://arxiv.org/abs/#1}{\nolinkurl{https://arxiv.org/abs/#1}}}

\bibitem[{{Akylas} \& {Georgantopoulos}(2009)}]{akylas2009}
{Akylas}, A., \& {Georgantopoulos}, I. 2009, \aap, 500, 999,
  \dodoi{10.1051/0004-6361/200811371}

\bibitem[{{Asmus} {et~al.}(2015){Asmus}, {Gandhi}, {H{\"o}nig}, {Smette}, \&
  {Duschl}}]{asmus2015}
{Asmus}, D., {Gandhi}, P., {H{\"o}nig}, S.~F., {Smette}, A., \& {Duschl}, W.~J.
  2015, \mnras, 454, 766, \dodoi{10.1093/mnras/stv1950}

\bibitem[{{Asmus} {et~al.}(2014){Asmus}, {H{\"o}nig}, {Gandhi}, {Smette}, \&
  {Duschl}}]{asmus2014}
{Asmus}, D., {H{\"o}nig}, S.~F., {Gandhi}, P., {Smette}, A., \& {Duschl}, W.~J.
  2014, \mnras, 439, 1648, \dodoi{10.1093/mnras/stu041}

\bibitem[{{Bassani} {et~al.}(1999){Bassani}, {Dadina}, {Maiolino}, {Salvati},
  {Risaliti}, {Della Ceca}, {Matt}, \& {Zamorani}}]{bassani1999}
{Bassani}, L., {Dadina}, M., {Maiolino}, R., {et~al.} 1999, \apjs, 121, 473,
  \dodoi{10.1086/313202}

\bibitem[{{Baum} {et~al.}(2010){Baum}, {Gallimore}, {O'Dea}, {Buchanan},
  {Noel-Storr}, {Axon}, {Robinson}, {Elitzur}, {Dorn}, \&
  {Staudaher}}]{baum2010}
{Baum}, S.~A., {Gallimore}, J.~F., {O'Dea}, C.~P., {et~al.} 2010, \apj, 710,
  289, \dodoi{10.1088/0004-637X/710/1/289}

\bibitem[{{Berney} {et~al.}(2015){Berney}, {Koss}, {Trakhtenbrot}, {Ricci},
  {Lamperti}, {Schawinski}, {Balokovi{\'c}}, {Crenshaw}, {Fischer}, {Gehrels},
  {Harrison}, {Hashimoto}, {Ichikawa}, {Mushotzky}, {Oh}, {Stern}, {Treister},
  {Ueda}, {Veilleux}, \& {Winter}}]{berney2015}
{Berney}, S., {Koss}, M., {Trakhtenbrot}, B., {et~al.} 2015, \mnras, 454, 3622,
  \dodoi{10.1093/mnras/stv2181}

\bibitem[{{Bi} {et~al.}(2020){Bi}, {Feng}, \& {Ho}}]{bi2020}
{Bi}, S., {Feng}, H., \& {Ho}, L.~C. 2020, \apj, 900, 124,
  \dodoi{10.3847/1538-4357/aba761}

\bibitem[{{Bianchi} {et~al.}(2008){Bianchi}, {Chiaberge}, {Piconcelli},
  {Guainazzi}, \& {Matt}}]{bianchi2008}
{Bianchi}, S., {Chiaberge}, M., {Piconcelli}, E., {Guainazzi}, M., \& {Matt},
  G. 2008, \mnras, 386, 105, \dodoi{10.1111/j.1365-2966.2008.13078.x}

\bibitem[{{Bianchi} {et~al.}(2005){Bianchi}, {Guainazzi}, {Matt}, {Chiaberge},
  {Iwasawa}, {Fiore}, \& {Maiolino}}]{bianchi2005}
{Bianchi}, S., {Guainazzi}, M., {Matt}, G., {et~al.} 2005, \aap, 442, 185,
  \dodoi{10.1051/0004-6361:20053389}

\bibitem[{{Bianchi} {et~al.}(2009){Bianchi}, {Guainazzi}, {Matt}, {Fonseca
  Bonilla}, \& {Ponti}}]{bianchi2009}
{Bianchi}, S., {Guainazzi}, M., {Matt}, G., {Fonseca Bonilla}, N., \& {Ponti},
  G. 2009, \aap, 495, 421, \dodoi{10.1051/0004-6361:200810620}

\bibitem[{{Brightman} \& {Nandra}(2008)}]{brightman2008}
{Brightman}, M., \& {Nandra}, K. 2008, \mnras, 390, 1241,
  \dodoi{10.1111/j.1365-2966.2008.13841.x}

\bibitem[{{Brightman} \& {Nandra}(2011)}]{brightman2011}
---. 2011, \mnras, 413, 1206, \dodoi{10.1111/j.1365-2966.2011.18207.x}

\bibitem[{{Brinchmann} {et~al.}(2004){Brinchmann}, {Charlot}, {White},
  {Tremonti}, {Kauffmann}, {Heckman}, \& {Brinkmann}}]{brinchmann2004}
{Brinchmann}, J., {Charlot}, S., {White}, S.~D.~M., {et~al.} 2004, \mnras, 351,
  1151, \dodoi{10.1111/j.1365-2966.2004.07881.x}

\bibitem[{{Burlon} {et~al.}(2011){Burlon}, {Ajello}, {Greiner}, {Comastri},
  {Merloni}, \& {Gehrels}}]{burlon2011}
{Burlon}, D., {Ajello}, M., {Greiner}, J., {et~al.} 2011, \apj, 728, 58,
  \dodoi{10.1088/0004-637X/728/1/58}

\bibitem[{{Cardamone} {et~al.}(2007){Cardamone}, {Moran}, \&
  {Kay}}]{cardamone2007}
{Cardamone}, C.~N., {Moran}, E.~C., \& {Kay}, L.~E. 2007, \aj, 134, 1263,
  \dodoi{10.1086/520801}

\bibitem[{{Cusumano} {et~al.}(2010){Cusumano}, {La Parola}, {Segreto},
  {Ferrigno}, {Maselli}, {Sbarufatti}, {Romano}, {Chincarini}, {Giommi},
  {Masetti}, {Moretti}, {Parisi}, \& {Tagliaferri}}]{cusumano2010}
{Cusumano}, G., {La Parola}, V., {Segreto}, A., {et~al.} 2010, \aap, 524, A64,
  \dodoi{10.1051/0004-6361/201015249}

\bibitem[{{da Cunha} {et~al.}(2008){da Cunha}, {Charlot}, \&
  {Elbaz}}]{dacunha2008}
{da Cunha}, E., {Charlot}, S., \& {Elbaz}, D. 2008, \mnras, 388, 1595,
  \dodoi{10.1111/j.1365-2966.2008.13535.x}

\bibitem[{{De Looze} {et~al.}(2011){De Looze}, {Baes}, {Bendo}, {Cortese}, \&
  {Fritz}}]{delooze2011}
{De Looze}, I., {Baes}, M., {Bendo}, G.~J., {Cortese}, L., \& {Fritz}, J. 2011,
  \mnras, 416, 2712, \dodoi{10.1111/j.1365-2966.2011.19223.x}

\bibitem[{{De Looze} {et~al.}(2014){De Looze}, {Cormier}, {Lebouteiller},
  {Madden}, {Baes}, {Bendo}, {Boquien}, {Boselli}, {Clements}, {Cortese},
  {Cooray}, {Galametz}, {Galliano}, {Graci{\'a}-Carpio}, {Isaak}, {Karczewski},
  {Parkin}, {Pellegrini}, {R{\'e}my-Ruyer}, {Spinoglio}, {Smith}, \&
  {Sturm}}]{delooze2014}
{De Looze}, I., {Cormier}, D., {Lebouteiller}, V., {et~al.} 2014, \aap, 568,
  A62, \dodoi{10.1051/0004-6361/201322489}

\bibitem[{{Del Moro} {et~al.}(2016){Del Moro}, {Alexander}, {Bauer}, {Daddi},
  {Kocevski}, {McIntosh}, {Stanley}, {Brandt}, {Elbaz}, {Harrison}, {Luo},
  {Mullaney}, \& {Xue}}]{delmoro2016}
{Del Moro}, A., {Alexander}, D.~M., {Bauer}, F.~E., {et~al.} 2016, \mnras, 456,
  2105, \dodoi{10.1093/mnras/stv2748}

\bibitem[{{Deluit} \& {Courvoisier}(2003)}]{deluit2003}
{Deluit}, S., \& {Courvoisier}, T.~J.~L. 2003, \aap, 399, 77,
  \dodoi{10.1051/0004-6361:20021794}

\bibitem[{{Diamond-Stanic} \& {Rieke}(2012)}]{diamond2012}
{Diamond-Stanic}, A.~M., \& {Rieke}, G.~H. 2012, \apj, 746, 168,
  \dodoi{10.1088/0004-637X/746/2/168}

\bibitem[{{Diamond-Stanic} {et~al.}(2009){Diamond-Stanic}, {Rieke}, \&
  {Rigby}}]{Diamond-Stanic2009}
{Diamond-Stanic}, A.~M., {Rieke}, G.~H., \& {Rigby}, J.~R. 2009, \apj, 698,
  623, \dodoi{10.1088/0004-637X/698/1/623}

\bibitem[{{Driver} {et~al.}(2016){Driver}, {Wright}, {Andrews}, {Davies},
  {Kafle}, {Lange}, {Moffett}, {Mannering}, {Robotham}, {Vinsen}, {Alpaslan},
  {Andrae}, {Baldry}, {Bauer}, {Bamford}, {Bland-Hawthorn}, {Bourne}, {Brough},
  {Brown}, {Cluver}, {Croom}, {Colless}, {Conselice}, {da Cunha}, {De Propris},
  {Drinkwater}, {Dunne}, {Eales}, {Edge}, {Frenk}, {Graham}, {Grootes},
  {Holwerda}, {Hopkins}, {Ibar}, {van Kampen}, {Kelvin}, {Jarrett}, {Jones},
  {Lara-Lopez}, {Liske}, {Lopez-Sanchez}, {Loveday}, {Maddox}, {Madore},
  {Mahajan}, {Meyer}, {Norberg}, {Penny}, {Phillipps}, {Popescu}, {Tuffs},
  {Peacock}, {Pimbblet}, {Prescott}, {Rowlands}, {Sansom}, {Seibert}, {Smith},
  {Sutherland}, {Taylor}, {Valiante}, {Vazquez-Mata}, {Wang}, {Wilkins}, \&
  {Williams}}]{driver2016}
{Driver}, S.~P., {Wright}, A.~H., {Andrews}, S.~K., {et~al.} 2016, \mnras, 455,
  3911, \dodoi{10.1093/mnras/stv2505}

\bibitem[{{Duras} {et~al.}(2020){Duras}, {Bongiorno}, {Ricci}, {Piconcelli},
  {Shankar}, {Lusso}, {Bianchi}, {Fiore}, {Maiolino}, {Marconi}, {Onori},
  {Sani}, {Schneider}, {Vignali}, \& {La Franca}}]{duras2020}
{Duras}, F., {Bongiorno}, A., {Ricci}, F., {et~al.} 2020, \aap, 636, A73,
  \dodoi{10.1051/0004-6361/201936817}

\bibitem[{{Efstathiou} {et~al.}(2014){Efstathiou}, {Pearson}, {Farrah},
  {Rigopoulou}, {Graci{\'a}-Carpio}, {Verma}, {Spoon}, {Afonso},
  {Bernard-Salas}, {Clements}, {Cooray}, {Cormier}, {Etxaluze}, {Fischer},
  {Gonz{\'a}lez-Alfonso}, {Hurley}, {Lebouteiller}, {Oliver}, {Rowan-Robinson},
  \& {Sturm}}]{efstathiou2014}
{Efstathiou}, A., {Pearson}, C., {Farrah}, D., {et~al.} 2014, \mnras, 437, L16,
  \dodoi{10.1093/mnrasl/slt131}

\bibitem[{{Elbaz} {et~al.}(2011){Elbaz}, {Dickinson}, {Hwang},
  {D{\'\i}az-Santos}, {Magdis}, {Magnelli}, {Le Borgne}, {Galliano},
  {Pannella}, {Chanial}, {Armus}, {Charmandaris}, {Daddi}, {Aussel}, {Popesso},
  {Kartaltepe}, {Altieri}, {Valtchanov}, {Coia}, {Dannerbauer}, {Dasyra},
  {Leiton}, {Mazzarella}, {Alexander}, {Buat}, {Burgarella}, {Chary}, {Gilli},
  {Ivison}, {Juneau}, {Le Floc'h}, {Lutz}, {Morrison}, {Mullaney}, {Murphy},
  {Pope}, {Scott}, {Brodwin}, {Calzetti}, {Cesarsky}, {Charlot}, {Dole},
  {Eisenhardt}, {Ferguson}, {F{\"o}rster Schreiber}, {Frayer}, {Giavalisco},
  {Huynh}, {Koekemoer}, {Papovich}, {Reddy}, {Surace}, {Teplitz}, {Yun}, \&
  {Wilson}}]{elbaz2011}
{Elbaz}, D., {Dickinson}, M., {Hwang}, H.~S., {et~al.} 2011, \aap, 533, A119,
  \dodoi{10.1051/0004-6361/201117239}

\bibitem[{{Elitzur} \& {Shlosman}(2006)}]{elitzur2006}
{Elitzur}, M., \& {Shlosman}, I. 2006, \apjl, 648, L101, \dodoi{10.1086/508158}

\bibitem[{{Fern{\'a}ndez-Ontiveros} \&
  {Mu{\~n}oz-Darias}(2021)}]{fernandez2021}
{Fern{\'a}ndez-Ontiveros}, J.~A., \& {Mu{\~n}oz-Darias}, T. 2021, \mnras, 504,
  5726, \dodoi{10.1093/mnras/stab1108}

\bibitem[{{Fern{\'a}ndez-Ontiveros} {et~al.}(2016){Fern{\'a}ndez-Ontiveros},
  {Spinoglio}, {Pereira-Santaella}, {Malkan}, {Andreani}, \&
  {Dasyra}}]{fernandez2016}
{Fern{\'a}ndez-Ontiveros}, J.~A., {Spinoglio}, L., {Pereira-Santaella}, M.,
  {et~al.} 2016, \apjs, 226, 19, \dodoi{10.3847/0067-0049/226/2/19}

\bibitem[{{Gallimore} {et~al.}(2010){Gallimore}, {Yzaguirre}, {Jakoboski},
  {Stevenosky}, {Axon}, {Baum}, {Buchanan}, {Elitzur}, {Elvis}, {O'Dea}, \&
  {Robinson}}]{gallimore2010}
{Gallimore}, J.~F., {Yzaguirre}, A., {Jakoboski}, J., {et~al.} 2010, \apjs,
  187, 172, \dodoi{10.1088/0067-0049/187/1/172}

\bibitem[{{Gardner} {et~al.}(2006){Gardner}, {Mather}, {Clampin}, {Doyon},
  {Greenhouse}, {Hammel}, {Hutchings}, {Jakobsen}, {Lilly}, {Long}, {Lunine},
  {McCaughrean}, {Mountain}, {Nella}, {Rieke}, {Rieke}, {Rix}, {Smith},
  {Sonneborn}, {Stiavelli}, {Stockman}, {Windhorst}, \& {Wright}}]{gardner2006}
{Gardner}, J.~P., {Mather}, J.~C., {Clampin}, M., {et~al.} 2006, \ssr, 123,
  485, \dodoi{10.1007/s11214-006-8315-7}

\bibitem[{{Genzel} {et~al.}(1998){Genzel}, {Lutz}, {Sturm}, {Egami}, {Kunze},
  {Moorwood}, {Rigopoulou}, {Spoon}, {Sternberg}, {Tacconi-Garman}, {Tacconi},
  \& {Thatte}}]{genzel1998}
{Genzel}, R., {Lutz}, D., {Sturm}, E., {et~al.} 1998, \apj, 498, 579,
  \dodoi{10.1086/305576}

\bibitem[{{Ghosh} \& {Soundararajaperumal}(1992)}]{ghosh1992}
{Ghosh}, K.~K., \& {Soundararajaperumal}, S. 1992, \mnras, 259, 175,
  \dodoi{10.1093/mnras/259.1.175}

\bibitem[{{G{\'o}mez-Guijarro} {et~al.}(2017){G{\'o}mez-Guijarro},
  {Gonz{\'a}lez-Mart{\'\i}n}, {Ramos Almeida}, {Rodr{\'\i}guez-Espinosa}, \&
  {Gallego}}]{gomez2017}
{G{\'o}mez-Guijarro}, C., {Gonz{\'a}lez-Mart{\'\i}n}, O., {Ramos Almeida}, C.,
  {Rodr{\'\i}guez-Espinosa}, J.~M., \& {Gallego}, J. 2017, \mnras, 469, 2720,
  \dodoi{10.1093/mnras/stx1037}

\bibitem[{{Gonz{\'a}lez-Mart{\'\i}n} {et~al.}(2009){Gonz{\'a}lez-Mart{\'\i}n},
  {Masegosa}, {M{\'a}rquez}, {Guainazzi}, \&
  {Jim{\'e}nez-Bail{\'o}n}}]{gonzalez-martin2009}
{Gonz{\'a}lez-Mart{\'\i}n}, O., {Masegosa}, J., {M{\'a}rquez}, I., {Guainazzi},
  M., \& {Jim{\'e}nez-Bail{\'o}n}, E. 2009, \aap, 506, 1107,
  \dodoi{10.1051/0004-6361/200912288}

\bibitem[{{Goto} {et~al.}(2019){Goto}, {Oi}, {Utsumi}, {Momose}, {Matsuhara},
  {Hashimoto}, {Toba}, {Ohyama}, {Takagi}, {Chiang}, {Kim}, {Kilerci Eser},
  {Malkan}, {Kim}, {Miyaji}, {Im}, {Nakagawa}, {Jeong}, {Pearson}, {Barrufet},
  {Sedgwick}, {Burgarella}, {Buat}, \& {Ikeda}}]{goto2019}
{Goto}, T., {Oi}, N., {Utsumi}, Y., {et~al.} 2019, \pasj, 71, 30,
  \dodoi{10.1093/pasj/psz009}

\bibitem[{{Grupe} {et~al.}(2004){Grupe}, {Mathur}, \& {Komossa}}]{grupe2004}
{Grupe}, D., {Mathur}, S., \& {Komossa}, S. 2004, \aj, 127, 3161,
  \dodoi{10.1086/421002}

\bibitem[{{Gruppioni} {et~al.}(2016){Gruppioni}, {Berta}, {Spinoglio},
  {Pereira-Santaella}, {Pozzi}, {Andreani}, {Bonato}, {De Zotti}, {Malkan},
  {Negrello}, {Vallini}, \& {Vignali}}]{gruppioni2016}
{Gruppioni}, C., {Berta}, S., {Spinoglio}, L., {et~al.} 2016, \mnras, 458,
  4297, \dodoi{10.1093/mnras/stw577}

\bibitem[{{Guainazzi} {et~al.}(2005){Guainazzi}, {Matt}, \&
  {Perola}}]{guainazzi2005}
{Guainazzi}, M., {Matt}, G., \& {Perola}, G.~C. 2005, \aap, 444, 119,
  \dodoi{10.1051/0004-6361:20053643}

\bibitem[{{Heckman} {et~al.}(2004){Heckman}, {Kauffmann}, {Brinchmann},
  {Charlot}, {Tremonti}, \& {White}}]{heckman2004}
{Heckman}, T.~M., {Kauffmann}, G., {Brinchmann}, J., {et~al.} 2004, \apj, 613,
  109, \dodoi{10.1086/422872}

\bibitem[{{Ho} {et~al.}(1997){Ho}, {Filippenko}, \& {Sargent}}]{ho1997}
{Ho}, L.~C., {Filippenko}, A.~V., \& {Sargent}, W. L.~W. 1997, \apjs, 112, 315,
  \dodoi{10.1086/313041}

\bibitem[{{Huang} {et~al.}(2019){Huang}, {Hu}, {Zhao}, {Zhang}, {Lu}, {Wang},
  {Zhang}, {Du}, {Li}, {Bai}, {Ho}, {Bian}, {Yuan}, \& {Wang}}]{huang2019}
{Huang}, Y.-K., {Hu}, C., {Zhao}, Y.-L., {et~al.} 2019, \apj, 876, 102,
  \dodoi{10.3847/1538-4357/ab16ef}

\bibitem[{{Ichikawa} {et~al.}(2019){Ichikawa}, {Ricci}, {Ueda}, {Bauer},
  {Kawamuro}, {Koss}, {Oh}, {Rosario}, {Shimizu}, {Stalevski}, {Fuller},
  {Packham}, \& {Trakhtenbrot}}]{ichikawa2019}
{Ichikawa}, K., {Ricci}, C., {Ueda}, Y., {et~al.} 2019, \apj, 870, 31,
  \dodoi{10.3847/1538-4357/aaef8f}

\bibitem[{{Imanishi} {et~al.}(2006){Imanishi}, {Dudley}, \&
  {Maloney}}]{imanishi2006}
{Imanishi}, M., {Dudley}, C.~C., \& {Maloney}, P.~R. 2006, \apj, 637, 114,
  \dodoi{10.1086/498391}

\bibitem[{{Iwasawa} {et~al.}(2011){Iwasawa}, {Sanders}, {Teng}, {U}, {Armus},
  {Evans}, {Howell}, {Komossa}, {Mazzarella}, {Petric}, {Surace}, {Vavilkin},
  {Veilleux}, \& {Trentham}}]{iwasawa2011}
{Iwasawa}, K., {Sanders}, D.~B., {Teng}, S.~H., {et~al.} 2011, \aap, 529, A106,
  \dodoi{10.1051/0004-6361/201015264}

\bibitem[{{Iyomoto} {et~al.}(1996){Iyomoto}, {Makishima}, {Fukazawa},
  {Tashiro}, {Ishisaki}, {Nakai}, \& {Taniguchi}}]{iyomoto1996}
{Iyomoto}, N., {Makishima}, K., {Fukazawa}, Y., {et~al.} 1996, \pasj, 48, 231,
  \dodoi{10.1093/pasj/48.2.231}

\bibitem[{{Izotov} {et~al.}(2018){Izotov}, {Worseck}, {Schaerer}, {Guseva},
  {Thuan}, {Fricke}, \& {Orlitov{\'a}}}]{izotov2018}
{Izotov}, Y.~I., {Worseck}, G., {Schaerer}, D., {et~al.} 2018, \mnras, 478,
  4851, \dodoi{10.1093/mnras/sty1378}

\bibitem[{{Jaffe} {et~al.}(2004){Jaffe}, {Meisenheimer}, {R{\"o}ttgering},
  {Leinert}, {Richichi}, {Chesneau}, {Fraix-Burnet}, {Glazenborg-Kluttig},
  {Granato}, {Graser}, {Heijligers}, {K{\"o}hler}, {Malbet}, {Miley},
  {Paresce}, {Pel}, {Perrin}, {Przygodda}, {Schoeller}, {Sol}, {Waters},
  {Weigelt}, {Woillez}, \& {de Zeeuw}}]{jaffe2004}
{Jaffe}, W., {Meisenheimer}, K., {R{\"o}ttgering}, H.~J.~A., {et~al.} 2004,
  \nat, 429, 47, \dodoi{10.1038/nature02531}

\bibitem[{{Kauffmann} {et~al.}(2003){Kauffmann}, {Heckman}, {Tremonti},
  {Brinchmann}, {Charlot}, {White}, {Ridgway}, {Brinkmann}, {Fukugita}, {Hall},
  {Ivezi{\'c}}, {Richards}, \& {Schneider}}]{kauffmann2003}
{Kauffmann}, G., {Heckman}, T.~M., {Tremonti}, C., {et~al.} 2003, \mnras, 346,
  1055, \dodoi{10.1111/j.1365-2966.2003.07154.x}

\bibitem[{{Kim} {et~al.}(2017){Kim}, {Ho}, {Peng}, {Barth}, \& {Im}}]{kim2017}
{Kim}, M., {Ho}, L.~C., {Peng}, C.~Y., {Barth}, A.~J., \& {Im}, M. 2017, \apjs,
  232, 21, \dodoi{10.3847/1538-4365/aa8a75}

\bibitem[{{Koss} {et~al.}(2017){Koss}, {Trakhtenbrot}, {Ricci}, {Lamperti},
  {Oh}, {Berney}, {Schawinski}, {Balokovi{\'c}}, {Baronchelli}, {Crenshaw},
  {Fischer}, {Gehrels}, {Harrison}, {Hashimoto}, {Hogg}, {Ichikawa}, {Masetti},
  {Mushotzky}, {Sartori}, {Stern}, {Treister}, {Ueda}, {Veilleux}, \&
  {Winter}}]{koss2017}
{Koss}, M., {Trakhtenbrot}, B., {Ricci}, C., {et~al.} 2017, \apj, 850, 74,
  \dodoi{10.3847/1538-4357/aa8ec9}

\bibitem[{{La Caria} {et~al.}(2019){La Caria}, {Vignali}, {Lanzuisi},
  {Gruppioni}, \& {Pozzi}}]{lacaria2019}
{La Caria}, M.~M., {Vignali}, C., {Lanzuisi}, G., {Gruppioni}, C., \& {Pozzi},
  F. 2019, \mnras, 487, 1662, \dodoi{10.1093/mnras/stz1381}

\bibitem[{{LaMassa} {et~al.}(2009){LaMassa}, {Heckman}, {Ptak},
  {Hornschemeier}, {Martins}, {Sonnentrucker}, \& {Tremonti}}]{lamassa2009}
{LaMassa}, S.~M., {Heckman}, T.~M., {Ptak}, A., {et~al.} 2009, \apj, 705, 568,
  \dodoi{10.1088/0004-637X/705/1/568}

\bibitem[{{LaMassa} {et~al.}(2010){LaMassa}, {Heckman}, {Ptak}, {Martins},
  {Wild}, \& {Sonnentrucker}}]{lamassa2010}
---. 2010, \apj, 720, 786, \dodoi{10.1088/0004-637X/720/1/786}

\bibitem[{{LaMassa} {et~al.}(2011){LaMassa}, {Heckman}, {Ptak}, {Martins},
  {Wild}, {Sonnentrucker}, \& {Hornschemeier}}]{lamassa2011}
---. 2011, \apj, 729, 52, \dodoi{10.1088/0004-637X/729/1/52}

\bibitem[{{Lamastra} {et~al.}(2009){Lamastra}, {Bianchi}, {Matt}, {Perola},
  {Barcons}, \& {Carrera}}]{lamastra2009}
{Lamastra}, A., {Bianchi}, S., {Matt}, G., {et~al.} 2009, \aap, 504, 73,
  \dodoi{10.1051/0004-6361/200912023}

\bibitem[{{Lee} {et~al.}(2013){Lee}, {Kriss}, {Chakravorty}, {Rahoui}, {Young},
  {Brandt}, {Hines}, {Ogle}, \& {Reynolds}}]{lee2013}
{Lee}, J.~C., {Kriss}, G.~A., {Chakravorty}, S., {et~al.} 2013, \mnras, 430,
  2650, \dodoi{10.1093/mnras/stt050}

\bibitem[{{Levenson} {et~al.}(2006){Levenson}, {Heckman}, {Krolik}, {Weaver},
  \& {{\.Z}ycki}}]{levenson2006}
{Levenson}, N.~A., {Heckman}, T.~M., {Krolik}, J.~H., {Weaver}, K.~A., \&
  {{\.Z}ycki}, P.~T. 2006, \apj, 648, 111, \dodoi{10.1086/505735}

\bibitem[{{Liu} {et~al.}(2014){Liu}, {Wang}, {Yang}, {Zhu}, \&
  {Zhou}}]{liu2014}
{Liu}, T., {Wang}, J.-X., {Yang}, H., {Zhu}, F.-F., \& {Zhou}, Y.-Y. 2014,
  \apj, 783, 106, \dodoi{10.1088/0004-637X/783/2/106}

\bibitem[{{Lusso} {et~al.}(2012){Lusso}, {Comastri}, {Simmons}, {Mignoli},
  {Zamorani}, {Vignali}, {Brusa}, {Shankar}, {Lutz}, {Trump}, {Maiolino},
  {Gilli}, {Bolzonella}, {Puccetti}, {Salvato}, {Impey}, {Civano}, {Elvis},
  {Mainieri}, {Silverman}, {Koekemoer}, {Bongiorno}, {Merloni}, {Berta}, {Le
  Floc'h}, {Magnelli}, {Pozzi}, \& {Riguccini}}]{lusso2012}
{Lusso}, E., {Comastri}, A., {Simmons}, B.~D., {et~al.} 2012, \mnras, 425, 623,
  \dodoi{10.1111/j.1365-2966.2012.21513.x}

\bibitem[{{Lutz} {et~al.}(2004){Lutz}, {Maiolino}, {Spoon}, \&
  {Moorwood}}]{lutz2004}
{Lutz}, D., {Maiolino}, R., {Spoon}, H.~W.~W., \& {Moorwood}, A.~F.~M. 2004,
  \aap, 418, 465, \dodoi{10.1051/0004-6361:20035838}

\bibitem[{{Madau} \& {Dickinson}(2014)}]{madau2014}
{Madau}, P., \& {Dickinson}, M. 2014, \araa, 52, 415,
  \dodoi{10.1146/annurev-astro-081811-125615}

\bibitem[{{Mahajan} {et~al.}(2018){Mahajan}, {Drinkwater}, {Driver}, {Hopkins},
  {Graham}, {Brough}, {Brown}, {Holwerda}, {Owers}, \&
  {Pimbblet}}]{mahajan2018}
{Mahajan}, S., {Drinkwater}, M.~J., {Driver}, S., {et~al.} 2018, \mnras, 475,
  788, \dodoi{10.1093/mnras/stx3202}

\bibitem[{{Maiolino} \& {Rieke}(1995)}]{maiolino-rieke1995}
{Maiolino}, R., \& {Rieke}, G.~H. 1995, \apj, 454, 95, \dodoi{10.1086/176468}

\bibitem[{{Maiolino} {et~al.}(1995){Maiolino}, {Ruiz}, {Rieke}, \&
  {Keller}}]{maiolino1995}
{Maiolino}, R., {Ruiz}, M., {Rieke}, G.~H., \& {Keller}, L.~D. 1995, \apj, 446,
  561, \dodoi{10.1086/175815}

\bibitem[{{Maiolino} {et~al.}(2007){Maiolino}, {Shemmer}, {Imanishi}, {Netzer},
  {Oliva}, {Lutz}, \& {Sturm}}]{maiolino2007}
{Maiolino}, R., {Shemmer}, O., {Imanishi}, M., {et~al.} 2007, \aap, 468, 979,
  \dodoi{10.1051/0004-6361:20077252}

\bibitem[{{Maiolino} {et~al.}(2003){Maiolino}, {Comastri}, {Gilli}, {Nagar},
  {Bianchi}, {B{\"o}ker}, {Colbert}, {Krabbe}, {Marconi}, {Matt}, \&
  {Salvati}}]{maiolino2003}
{Maiolino}, R., {Comastri}, A., {Gilli}, R., {et~al.} 2003, \mnras, 344, L59,
  \dodoi{10.1046/j.1365-8711.2003.07036.x}

\bibitem[{{Malkan} {et~al.}(2017){Malkan}, {Jensen}, {Rodriguez}, {Spinoglio},
  \& {Rush}}]{malkan2017}
{Malkan}, M.~A., {Jensen}, L.~D., {Rodriguez}, D.~R., {Spinoglio}, L., \&
  {Rush}, B. 2017, \apj, 846, 102, \dodoi{10.3847/1538-4357/aa8302}

\bibitem[{{Marconi} {et~al.}(2004){Marconi}, {Risaliti}, {Gilli}, {Hunt},
  {Maiolino}, \& {Salvati}}]{marconi2004}
{Marconi}, A., {Risaliti}, G., {Gilli}, R., {et~al.} 2004, \mnras, 351, 169,
  \dodoi{10.1111/j.1365-2966.2004.07765.x}

\bibitem[{{Mel{\'e}ndez} {et~al.}(2008){Mel{\'e}ndez}, {Kraemer}, {Armentrout},
  {Deo}, {Crenshaw}, {Schmitt}, {Mushotzky}, {Tueller}, {Markwardt}, \&
  {Winter}}]{melendez2008}
{Mel{\'e}ndez}, M., {Kraemer}, S.~B., {Armentrout}, B.~K., {et~al.} 2008, \apj,
  682, 94, \dodoi{10.1086/588807}

\bibitem[{{Mordini} {et~al.}(2021){Mordini}, {Spinoglio}, \&
  {Fern{\'a}ndez-Ontiveros}}]{mordini2021}
{Mordini}, S., {Spinoglio}, L., \& {Fern{\'a}ndez-Ontiveros}, J.~A. 2021, \aap,
  653, A36, \dodoi{10.1051/0004-6361/202140696}

\bibitem[{{Mordini} {et~al.}(2022){Mordini}, {Spinoglio}, \&
  {Fern{\'a}ndez-Ontiveros}}]{mordini2022}
---. 2022, \pasa, 39, e012, \dodoi{10.1017/pasa.2022.10}

\bibitem[{{Mulchaey} {et~al.}(1994){Mulchaey}, {Koratkar}, {Ward}, {Wilson},
  {Whittle}, {Antonucci}, {Kinney}, \& {Hurt}}]{mulchaey1994}
{Mulchaey}, J.~S., {Koratkar}, A., {Ward}, M.~J., {et~al.} 1994, \apj, 436,
  586, \dodoi{10.1086/174933}

\bibitem[{{M{\"u}ller-S{\'a}nchez} {et~al.}(2018){M{\"u}ller-S{\'a}nchez},
  {Hicks}, {Malkan}, {Davies}, {Yu}, {Shaver}, \& {Davis}}]{muller-sanchez2018}
{M{\"u}ller-S{\'a}nchez}, F., {Hicks}, E.~K.~S., {Malkan}, M., {et~al.} 2018,
  \apj, 858, 48, \dodoi{10.3847/1538-4357/aab9ad}

\bibitem[{{Oh} {et~al.}(2018){Oh}, {Koss}, {Markwardt}, {Schawinski},
  {Baumgartner}, {Barthelmy}, {Cenko}, {Gehrels}, {Mushotzky}, {Petulante},
  {Ricci}, {Lien}, \& {Trakhtenbrot}}]{oh2018}
{Oh}, K., {Koss}, M., {Markwardt}, C.~B., {et~al.} 2018, \apjs, 235, 4,
  \dodoi{10.3847/1538-4365/aaa7fd}

\bibitem[{{Peng} {et~al.}(2002){Peng}, {Ho}, {Impey}, \& {Rix}}]{peng2002}
{Peng}, C.~Y., {Ho}, L.~C., {Impey}, C.~D., \& {Rix}, H.-W. 2002, \aj, 124,
  266, \dodoi{10.1086/340952}

\bibitem[{{Pereira-Santaella} {et~al.}(2013){Pereira-Santaella}, {Spinoglio},
  {Busquet}, {Wilson}, {Glenn}, {Isaak}, {Kamenetzky}, {Rangwala}, {Schirm},
  {Baes}, {Barlow}, {Boselli}, {Cooray}, \& {Cormier}}]{pereira2013}
{Pereira-Santaella}, M., {Spinoglio}, L., {Busquet}, G., {et~al.} 2013, \apj,
  768, 55, \dodoi{10.1088/0004-637X/768/1/55}

\bibitem[{{Polletta} {et~al.}(1996){Polletta}, {Bassani}, {Malaguti},
  {Palumbo}, \& {Caroli}}]{polletta1996}
{Polletta}, M., {Bassani}, L., {Malaguti}, G., {Palumbo}, G.~G.~C., \&
  {Caroli}, E. 1996, \apjs, 106, 399, \dodoi{10.1086/192342}

\bibitem[{{Pott} {et~al.}(2010){Pott}, {Malkan}, {Elitzur}, {Ghez}, {Herbst},
  {Sch{\"o}del}, \& {Woillez}}]{pott2010}
{Pott}, J.-U., {Malkan}, M.~A., {Elitzur}, M., {et~al.} 2010, \apj, 715, 736,
  \dodoi{10.1088/0004-637X/715/2/736}

\bibitem[{{Ricci} {et~al.}(2015){Ricci}, {Ueda}, {Koss}, {Trakhtenbrot},
  {Bauer}, \& {Gandhi}}]{ricci2015}
{Ricci}, C., {Ueda}, Y., {Koss}, M.~J., {et~al.} 2015, \apjl, 815, L13,
  \dodoi{10.1088/2041-8205/815/1/L13}

\bibitem[{{Ricci} {et~al.}(2017{\natexlab{a}}){Ricci}, {Trakhtenbrot}, {Koss},
  {Ueda}, {Del Vecchio}, {Treister}, {Schawinski}, {Paltani}, {Oh}, {Lamperti},
  {Berney}, {Gandhi}, {Ichikawa}, {Bauer}, {Ho}, {Asmus}, {Beckmann}, {Soldi},
  {Balokovi{\'c}}, {Gehrels}, \& {Markwardt}}]{ricci2017}
{Ricci}, C., {Trakhtenbrot}, B., {Koss}, M.~J., {et~al.} 2017{\natexlab{a}},
  \apjs, 233, 17, \dodoi{10.3847/1538-4365/aa96ad}

\bibitem[{{Ricci} {et~al.}(2017{\natexlab{b}}){Ricci}, {Trakhtenbrot}, {Koss},
  {Ueda}, {Schawinski}, {Oh}, {Lamperti}, {Mushotzky}, {Treister}, {Ho},
  {Weigel}, {Bauer}, {Paltani}, {Fabian}, {Xie}, \& {Gehrels}}]{ricci2017b}
---. 2017{\natexlab{b}}, \nat, 549, 488, \dodoi{10.1038/nature23906}

\bibitem[{{Rieke} {et~al.}(2015){Rieke}, {Wright}, {B{\"o}ker}, {Bouwman},
  {Colina}, {Glasse}, {Gordon}, {Greene}, {G{\"u}del}, {Henning}, {Justtanont},
  {Lagage}, {Meixner}, {N{\o}rgaard-Nielsen}, {Ray}, {Ressler}, {van Dishoeck},
  \& {Waelkens}}]{rieke2015}
{Rieke}, G.~H., {Wright}, G.~S., {B{\"o}ker}, T., {et~al.} 2015, \pasp, 127,
  584, \dodoi{10.1086/682252}

\bibitem[{{Rigby} {et~al.}(2009){Rigby}, {Diamond-Stanic}, \&
  {Aniano}}]{rigby2009}
{Rigby}, J.~R., {Diamond-Stanic}, A.~M., \& {Aniano}, G. 2009, \apj, 700, 1878,
  \dodoi{10.1088/0004-637X/700/2/1878}

\bibitem[{{Risaliti} {et~al.}(2000){Risaliti}, {Gilli}, {Maiolino}, \&
  {Salvati}}]{risaliti2000}
{Risaliti}, G., {Gilli}, R., {Maiolino}, R., \& {Salvati}, M. 2000, \aap, 357,
  13.
\newblock \doarXiv{astro-ph/0002460}

\bibitem[{{Risaliti} {et~al.}(1999){Risaliti}, {Maiolino}, \&
  {Salvati}}]{risaliti1999}
{Risaliti}, G., {Maiolino}, R., \& {Salvati}, M. 1999, \apj, 522, 157,
  \dodoi{10.1086/307623}

\bibitem[{{Rush} {et~al.}(1993){Rush}, {Malkan}, \& {Spinoglio}}]{rush1993}
{Rush}, B., {Malkan}, M.~A., \& {Spinoglio}, L. 1993, \apjs, 89, 1,
  \dodoi{10.1086/191837}

\bibitem[{{Saade} {et~al.}(2022){Saade}, {Brightman}, {Stern}, {Malkan}, \&
  {Garcia}}]{saade2022}
{Saade}, M.~L., {Brightman}, M., {Stern}, D., {Malkan}, M.~A., \& {Garcia},
  J.~A. 2022, arXiv e-prints, arXiv:2205.14216.
\newblock \doarXiv{2205.14216}

\bibitem[{{Satyapal} {et~al.}(2021){Satyapal}, {Kamal}, {Cann}, {Secrest}, \&
  {Abel}}]{satyapal2021}
{Satyapal}, S., {Kamal}, L., {Cann}, J.~M., {Secrest}, N.~J., \& {Abel}, N.~P.
  2021, \apj, 906, 35, \dodoi{10.3847/1538-4357/abbfaf}

\bibitem[{{Schmitt} \& {Kinney}(1996)}]{schmitt1996}
{Schmitt}, H.~R., \& {Kinney}, A.~L. 1996, \apj, 463, 498,
  \dodoi{10.1086/177264}

\bibitem[{{Shu} {et~al.}(2010){Shu}, {Liu}, \& {Wang}}]{shu2010}
{Shu}, X.~W., {Liu}, T., \& {Wang}, J.~X. 2010, \apj, 722, 96,
  \dodoi{10.1088/0004-637X/722/1/96}

\bibitem[{{Singh} {et~al.}(2011){Singh}, {Shastri}, \& {Risaliti}}]{singh2011}
{Singh}, V., {Shastri}, P., \& {Risaliti}, G. 2011, \aap, 532, A84,
  \dodoi{10.1051/0004-6361/201016387}

\bibitem[{{Spinoglio} \& {Malkan}(1989)}]{spinoglio1989}
{Spinoglio}, L., \& {Malkan}, M.~A. 1989, \apj, 342, 83, \dodoi{10.1086/167577}

\bibitem[{{Spinoglio} \& {Malkan}(1992)}]{spinoglio1992}
---. 1992, \apj, 399, 504, \dodoi{10.1086/171943}

\bibitem[{{Spinoglio} {et~al.}(1995){Spinoglio}, {Malkan}, {Rush}, {Carrasco},
  \& {Recillas-Cruz}}]{spinoglio1995}
{Spinoglio}, L., {Malkan}, M.~A., {Rush}, B., {Carrasco}, L., \&
  {Recillas-Cruz}, E. 1995, \apj, 453, 616, \dodoi{10.1086/176425}

\bibitem[{{Spinoglio} {et~al.}(2015){Spinoglio}, {Pereira-Santaella}, {Dasyra},
  {Calzoletti}, {Malkan}, {Tommasin}, \& {Busquet}}]{spinoglio2015}
{Spinoglio}, L., {Pereira-Santaella}, M., {Dasyra}, K.~M., {et~al.} 2015, \apj,
  799, 21, \dodoi{10.1088/0004-637X/799/1/21}

\bibitem[{{Strickland}(2007)}]{strickland2007}
{Strickland}, D.~K. 2007, \mnras, 376, 523,
  \dodoi{10.1111/j.1365-2966.2007.11478.x}

\bibitem[{{Sturm} {et~al.}(2002){Sturm}, {Lutz}, {Verma}, {Netzer},
  {Sternberg}, {Moorwood}, {Oliva}, \& {Genzel}}]{sturm2002}
{Sturm}, E., {Lutz}, D., {Verma}, A., {et~al.} 2002, \aap, 393, 821,
  \dodoi{10.1051/0004-6361:20021043}

\bibitem[{{Tan} {et~al.}(2012){Tan}, {Wang}, {Shu}, \& {Zhou}}]{tan2012}
{Tan}, Y., {Wang}, J.~X., {Shu}, X.~W., \& {Zhou}, Y. 2012, \apjl, 747, L11,
  \dodoi{10.1088/2041-8205/747/1/L11}

\bibitem[{{Teng} {et~al.}(2009){Teng}, {Veilleux}, {Anabuki}, {Dermer},
  {Gallo}, {Nakagawa}, {Reynolds}, {Sanders}, {Terashima}, \&
  {Wilson}}]{teng2009}
{Teng}, S.~H., {Veilleux}, S., {Anabuki}, N., {et~al.} 2009, \apj, 691, 261,
  \dodoi{10.1088/0004-637X/691/1/261}

\bibitem[{{Theios} {et~al.}(2016){Theios}, {Malkan}, \& {Ross}}]{theios2016}
{Theios}, R.~L., {Malkan}, M.~A., \& {Ross}, N.~R. 2016, \apj, 822, 45,
  \dodoi{10.3847/0004-637X/822/1/45}

\bibitem[{{Toba} {et~al.}(2014){Toba}, {Oyabu}, {Matsuhara}, {Malkan},
  {Gandhi}, {Nakagawa}, {Isobe}, {Shirahata}, {Oi}, {Ohyama}, {Takita},
  {Yamauchi}, \& {Yano}}]{toba2014}
{Toba}, Y., {Oyabu}, S., {Matsuhara}, H., {et~al.} 2014, \apj, 788, 45,
  \dodoi{10.1088/0004-637X/788/1/45}

\bibitem[{{Tommasin} {et~al.}(2010){Tommasin}, {Spinoglio}, {Malkan}, \&
  {Fazio}}]{tommasin2010}
{Tommasin}, S., {Spinoglio}, L., {Malkan}, M.~A., \& {Fazio}, G. 2010, \apj,
  709, 1257, \dodoi{10.1088/0004-637X/709/2/1257}

\bibitem[{{Tommasin} {et~al.}(2008){Tommasin}, {Spinoglio}, {Malkan}, {Smith},
  {Gonz{\'a}lez-Alfonso}, \& {Charmandaris}}]{tommasin2008}
{Tommasin}, S., {Spinoglio}, L., {Malkan}, M.~A., {et~al.} 2008, \apj, 676,
  836, \dodoi{10.1086/527290}

\bibitem[{{Tran}(2003)}]{tran2003}
{Tran}, H.~D. 2003, \apj, 583, 632, \dodoi{10.1086/345473}

\bibitem[{{Treister} {et~al.}(2008){Treister}, {Krolik}, \&
  {Dullemond}}]{treister2008}
{Treister}, E., {Krolik}, J.~H., \& {Dullemond}, C. 2008, \apj, 679, 140,
  \dodoi{10.1086/586698}

\bibitem[{{Tristram} {et~al.}(2007){Tristram}, {Meisenheimer}, {Jaffe},
  {Schartmann}, {Rix}, {Leinert}, {Morel}, {Wittkowski}, {R{\"o}ttgering},
  {Perrin}, {Lopez}, {Raban}, {Cotton}, {Graser}, {Paresce}, \&
  {Henning}}]{tristram2007}
{Tristram}, K.~R.~W., {Meisenheimer}, K., {Jaffe}, W., {et~al.} 2007, \aap,
  474, 837, \dodoi{10.1051/0004-6361:20078369}

\bibitem[{{Ueda} {et~al.}(2003){Ueda}, {Akiyama}, {Ohta}, \&
  {Miyaji}}]{ueda2003}
{Ueda}, Y., {Akiyama}, M., {Ohta}, K., \& {Miyaji}, T. 2003, \apj, 598, 886,
  \dodoi{10.1086/378940}

\bibitem[{{Ueda} {et~al.}(2001){Ueda}, {Ishisaki}, {Takahashi}, {Makishima}, \&
  {Ohashi}}]{ueda2001}
{Ueda}, Y., {Ishisaki}, Y., {Takahashi}, T., {Makishima}, K., \& {Ohashi}, T.
  2001, \apjs, 133, 1, \dodoi{10.1086/319189}

\bibitem[{{Ueda} {et~al.}(2005){Ueda}, {Ishisaki}, {Takahashi}, {Makishima}, \&
  {Ohashi}}]{ueda2005}
---. 2005, \apjs, 161, 185, \dodoi{10.1086/468187}

\bibitem[{{Ward} {et~al.}(2022){Ward}, {Harrison}, {Costa}, \&
  {Mainieri}}]{ward2022}
{Ward}, S.~R., {Harrison}, C.~M., {Costa}, T., \& {Mainieri}, V. 2022, \mnras,
  514, 2936, \dodoi{10.1093/mnras/stac1219}

\bibitem[{{Wechsler} \& {Tinker}(2018)}]{wechsler2018}
{Wechsler}, R.~H., \& {Tinker}, J.~L. 2018, \araa, 56, 435,
  \dodoi{10.1146/annurev-astro-081817-051756}

\bibitem[{{Winter} {et~al.}(2008){Winter}, {Mushotzky}, {Tueller}, \&
  {Markwardt}}]{winter2008}
{Winter}, L.~M., {Mushotzky}, R.~F., {Tueller}, J., \& {Markwardt}, C. 2008,
  \apj, 674, 686, \dodoi{10.1086/525274}

\bibitem[{{Wright} {et~al.}(2010){Wright}, {Eisenhardt}, {Mainzer}, {Ressler},
  {Cutri}, {Jarrett}, {Kirkpatrick}, {Padgett}, {McMillan}, {Skrutskie},
  {Stanford}, {Cohen}, {Walker}, {Mather}, {Leisawitz}, {Gautier}, {McLean},
  {Benford}, {Lonsdale}, {Blain}, {Mendez}, {Irace}, {Duval}, {Liu}, {Royer},
  {Heinrichsen}, {Howard}, {Shannon}, {Kendall}, {Walsh}, {Larsen}, {Cardon},
  {Schick}, {Schwalm}, {Abid}, {Fabinsky}, {Naes}, \& {Tsai}}]{wright2010}
{Wright}, E.~L., {Eisenhardt}, P. R.~M., {Mainzer}, A.~K., {et~al.} 2010, \aj,
  140, 1868, \dodoi{10.1088/0004-6256/140/6/1868}

\bibitem[{{Wu} {et~al.}(2009){Wu}, {Charmandaris}, {Huang}, {Spinoglio}, \&
  {Tommasin}}]{wu2009}
{Wu}, Y., {Charmandaris}, V., {Huang}, J., {Spinoglio}, L., \& {Tommasin}, S.
  2009, \apj, 701, 658, \dodoi{10.1088/0004-637X/701/1/658}

\bibitem[{{Xia} {et~al.}(2018){Xia}, {Malkan}, {Ross}, \& {Ancheta}}]{xia2018}
{Xia}, J., {Malkan}, M.~A., {Ross}, N.~R., \& {Ancheta}, A.~J. 2018, \apj, 869,
  138, \dodoi{10.3847/1538-4357/aaedc2}

\bibitem[{{Xie} \& {Ho}(2019)}]{xie&ho2019}
{Xie}, Y., \& {Ho}, L.~C. 2019, \apj, 884, 136,
  \dodoi{10.3847/1538-4357/ab4200}

\bibitem[{{Yamada} {et~al.}(2020){Yamada}, {Ueda}, {Tanimoto}, {Oda},
  {Imanishi}, {Toba}, \& {Ricci}}]{yamada2020}
{Yamada}, S., {Ueda}, Y., {Tanimoto}, A., {et~al.} 2020, \apj, 897, 107,
  \dodoi{10.3847/1538-4357/ab94b1}

\bibitem[{{York} {et~al.}(2000){York}, {Adelman}, {Anderson}, {Anderson},
  {Annis}, {Bahcall}, {Bakken}, {Barkhouser}, {Bastian}, {Berman}, {Boroski},
  {Bracker}, {Briegel}, {Briggs}, {Brinkmann}, {Brunner}, {Burles}, {Carey},
  {Carr}, {Castander}, {Chen}, {Colestock}, {Connolly}, {Crocker}, {Csabai},
  {Czarapata}, {Davis}, {Doi}, {Dombeck}, {Eisenstein}, {Ellman}, {Elms},
  {Evans}, {Fan}, {Federwitz}, {Fiscelli}, {Friedman}, {Frieman}, {Fukugita},
  {Gillespie}, {Gunn}, {Gurbani}, {de Haas}, {Haldeman}, {Harris}, {Hayes},
  {Heckman}, {Hennessy}, {Hindsley}, {Holm}, {Holmgren}, {Huang}, {Hull},
  {Husby}, {Ichikawa}, {Ichikawa}, {Ivezi{\'c}}, {Kent}, {Kim}, {Kinney},
  {Klaene}, {Kleinman}, {Kleinman}, {Knapp}, {Korienek}, {Kron}, {Kunszt},
  {Lamb}, {Lee}, {Leger}, {Limmongkol}, {Lindenmeyer}, {Long}, {Loomis},
  {Loveday}, {Lucinio}, {Lupton}, {MacKinnon}, {Mannery}, {Mantsch}, {Margon},
  {McGehee}, {McKay}, {Meiksin}, {Merelli}, {Monet}, {Munn}, {Narayanan},
  {Nash}, {Neilsen}, {Neswold}, {Newberg}, {Nichol}, {Nicinski}, {Nonino},
  {Okada}, {Okamura}, {Ostriker}, {Owen}, {Pauls}, {Peoples}, {Peterson},
  {Petravick}, {Pier}, {Pope}, {Pordes}, {Prosapio}, {Rechenmacher}, {Quinn},
  {Richards}, {Richmond}, {Rivetta}, {Rockosi}, {Ruthmansdorfer}, {Sandford},
  {Schlegel}, {Schneider}, {Sekiguchi}, {Sergey}, {Shimasaku}, {Siegmund},
  {Smee}, {Smith}, {Snedden}, {Stone}, {Stoughton}, {Strauss}, {Stubbs},
  {SubbaRao}, {Szalay}, {Szapudi}, {Szokoly}, {Thakar}, {Tremonti}, {Tucker},
  {Uomoto}, {Vanden Berk}, {Vogeley}, {Waddell}, {Wang}, {Watanabe},
  {Weinberg}, {Yanny}, {Yasuda}, \& {SDSS Collaboration}}]{york2000}
{York}, D.~G., {Adelman}, J., {Anderson}, John~E., J., {et~al.} 2000, \aj, 120,
  1579, \dodoi{10.1086/301513}

\bibitem[{{Zhang} {et~al.}(2019){Zhang}, {Du}, {Smith}, {Zhao}, {Hu}, {Xiao},
  {Li}, {Huang}, {Wang}, {Bai}, {Ho}, \& {Wang}}]{zhang2019}
{Zhang}, Z.-X., {Du}, P., {Smith}, P.~S., {et~al.} 2019, \apj, 876, 49,
  \dodoi{10.3847/1538-4357/ab1099}

\bibitem[{{Zhuang} \& {Ho}(2020)}]{zhuang2020}
{Zhuang}, M.-Y., \& {Ho}, L.~C. 2020, \apj, 896, 108,
  \dodoi{10.3847/1538-4357/ab8f2e}

\bibitem[{{Zhuang} {et~al.}(2019){Zhuang}, {Ho}, \& {Shangguan}}]{zhuang2019}
{Zhuang}, M.-Y., {Ho}, L.~C., \& {Shangguan}, J. 2019, \apj, 873, 103,
  \dodoi{10.3847/1538-4357/ab0650}

\bibitem[{{Zhuang} {et~al.}(2021){Zhuang}, {Ho}, \& {Shangguan}}]{zhuang2021}
---. 2021, \apj, 906, 38, \dodoi{10.3847/1538-4357/abc94d}

\end{thebibliography}
\bibliographystyle{aasjournal}




\clearpage

\appendix

\section{Other figures with correlations}
 
\begin{figure*}[h!]
  \includegraphics[width=0.5\textwidth]{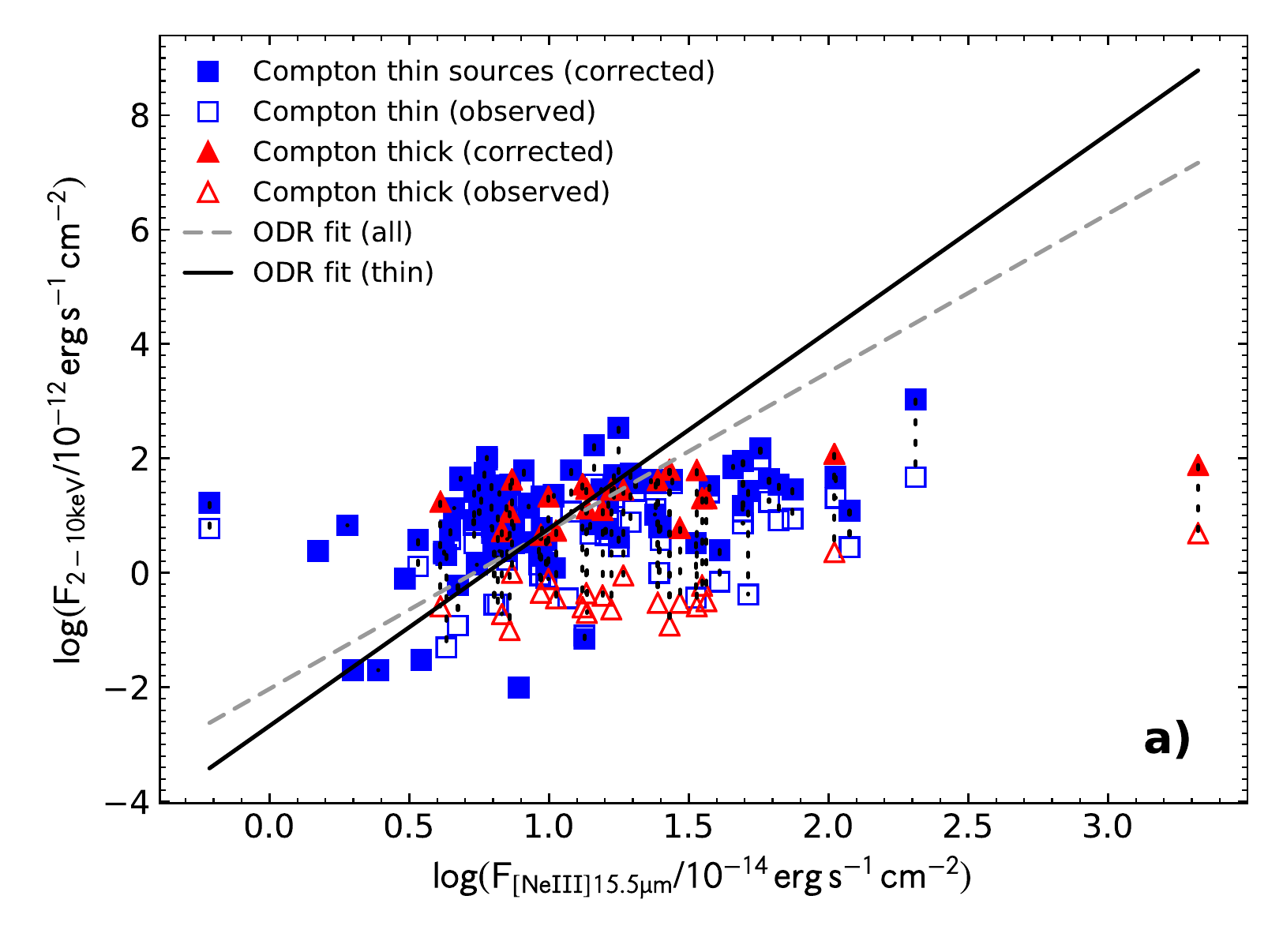}
  \includegraphics[width=0.5\textwidth]{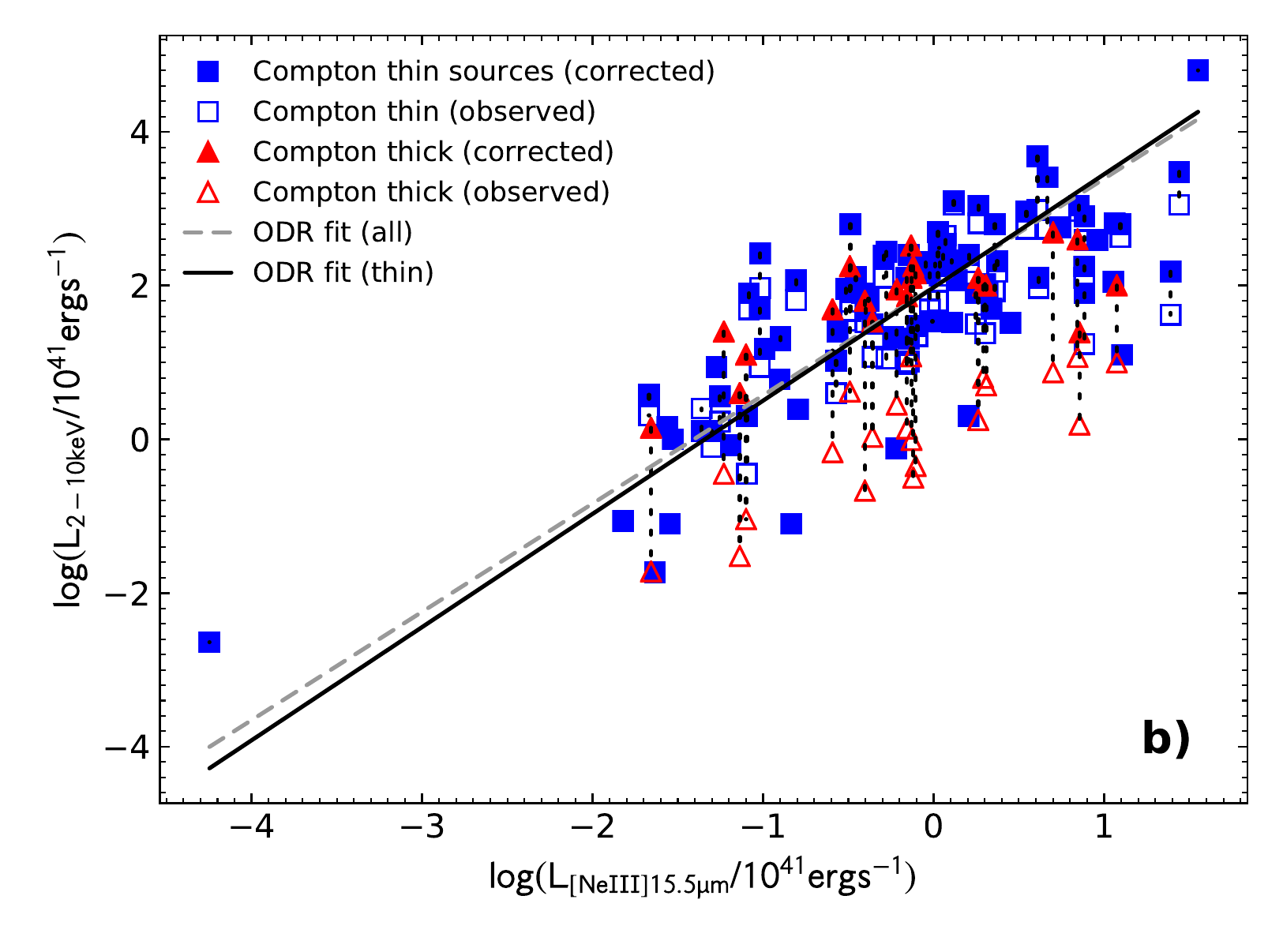}
  \caption{{\bf a: left} X-ray (2--10) keV intrinsic flux as a function of the \neiii\ line flux for all sources with reliable hard-X data and an optical spectroscopic classification. The blue squares indicate the Compton thin AGN, while Compton thick objects are denoted by red triangles. The solid lines represent the linear regression fit with all objects, while the dotted line the fit with only Compton thin objects. {\bf b: right} X-ray (2-10) keV intrinsic luminosity vs. \neiii\ line luminosity (both in logarithmic units of \ergs). 
  }
  \label{fig:X-NeIIIF&L}
  \end{figure*}

\begin{figure*}
  \includegraphics[width=0.5\textwidth]{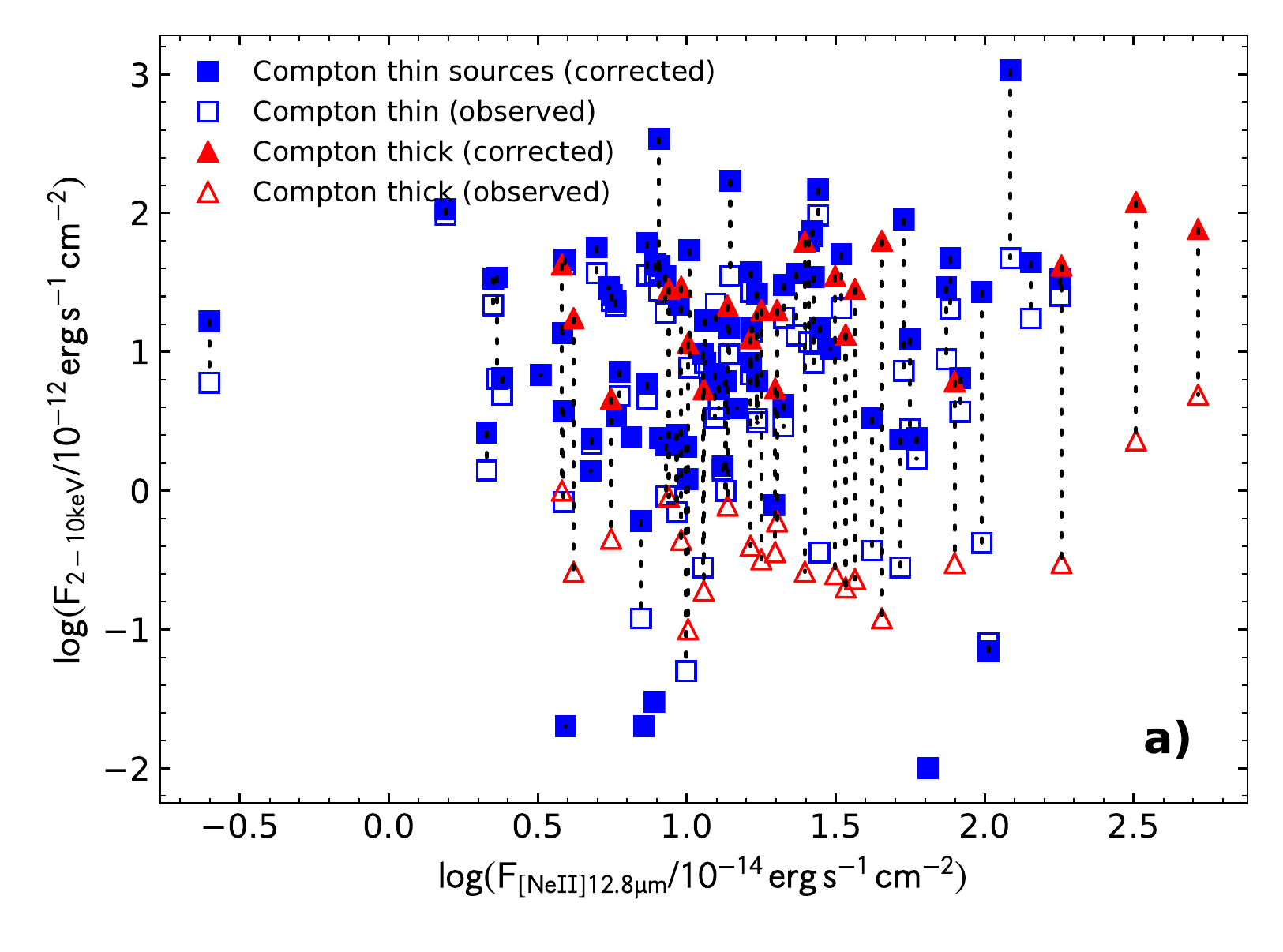}
  \includegraphics[width=0.5\textwidth]{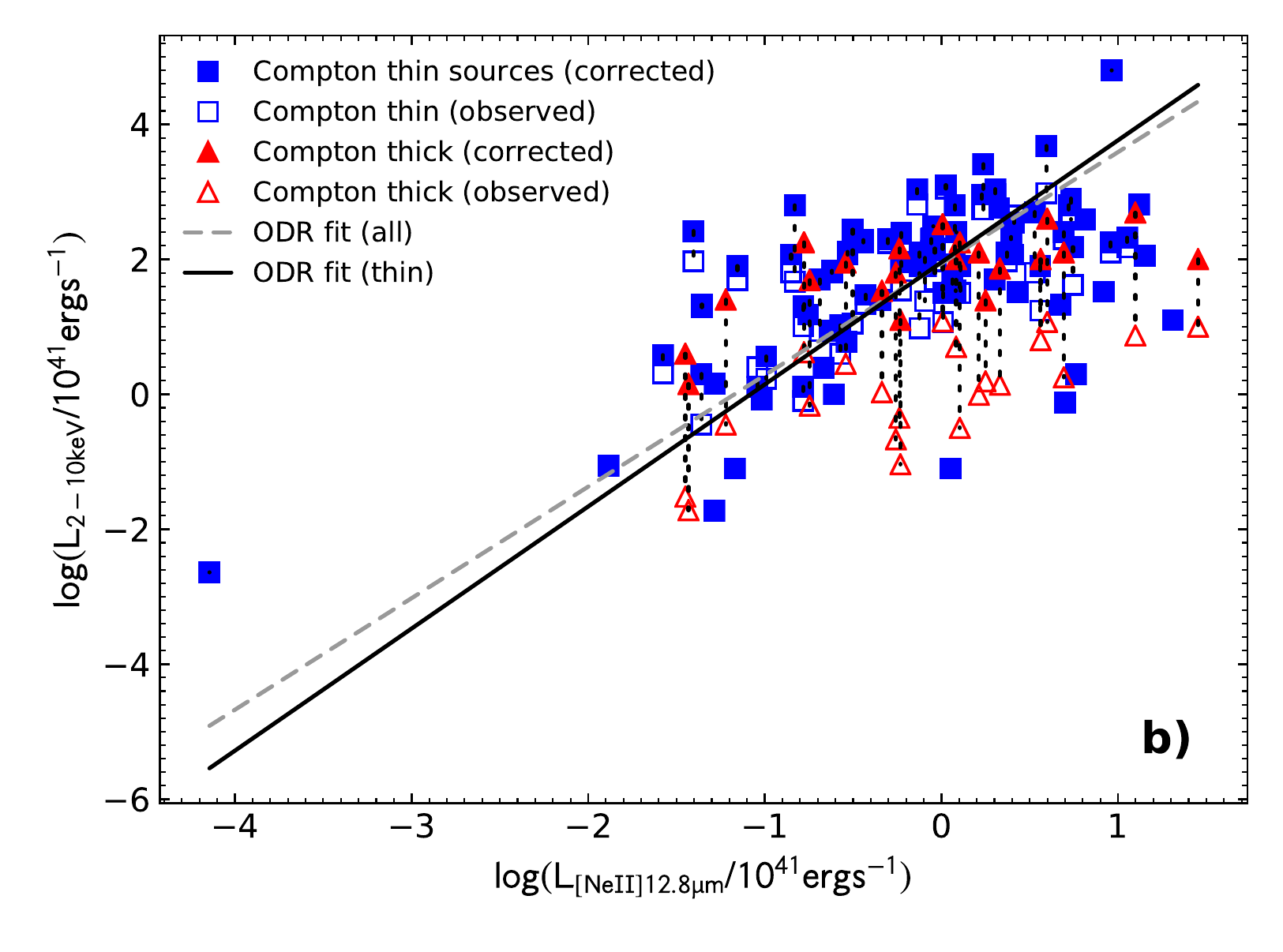}
  \caption{{\bf a: left} X-ray (2--10) keV intrinsic flux as a function of the \neii\  line flux for all sources with reliable hard-X data and an optical spectroscopic classification. The blue squares indicate the Compton thin AGN, while Compton thick objects are denoted by red triangles. The solid lines represent the linear regression fit with all objects, while the dotted line the fit with only Compton thin objects . {\bf b: right} X-ray (2-10) keV intrinsic luminosity vs. \neii\  line luminosity (both in logarithmic units of \ergs). 
  }
  \label{fig:X-NeIIF&L}
  \end{figure*}

\begin{figure*}
  \includegraphics[width=0.5\textwidth]{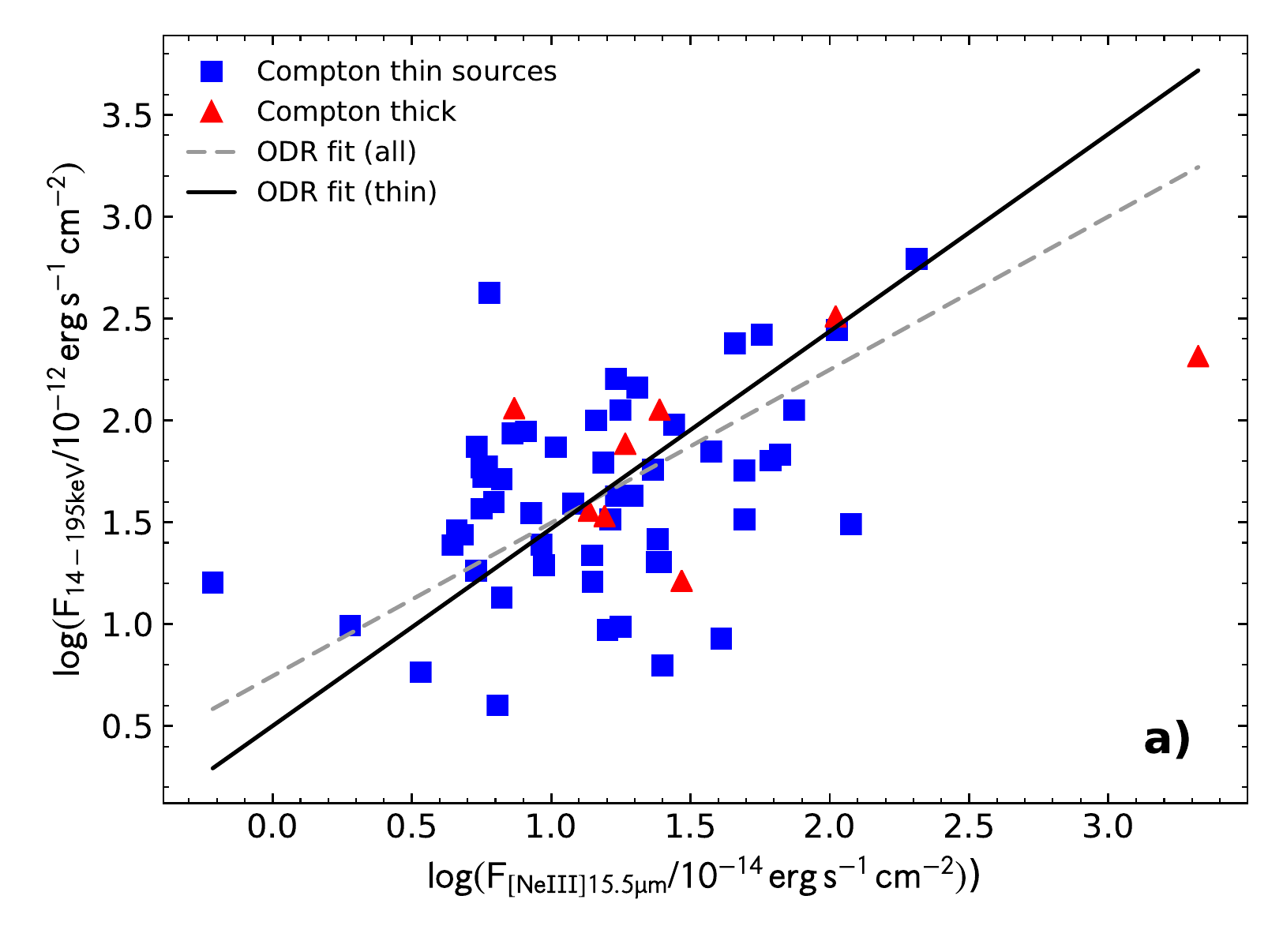}
  \includegraphics[width=0.5\textwidth]{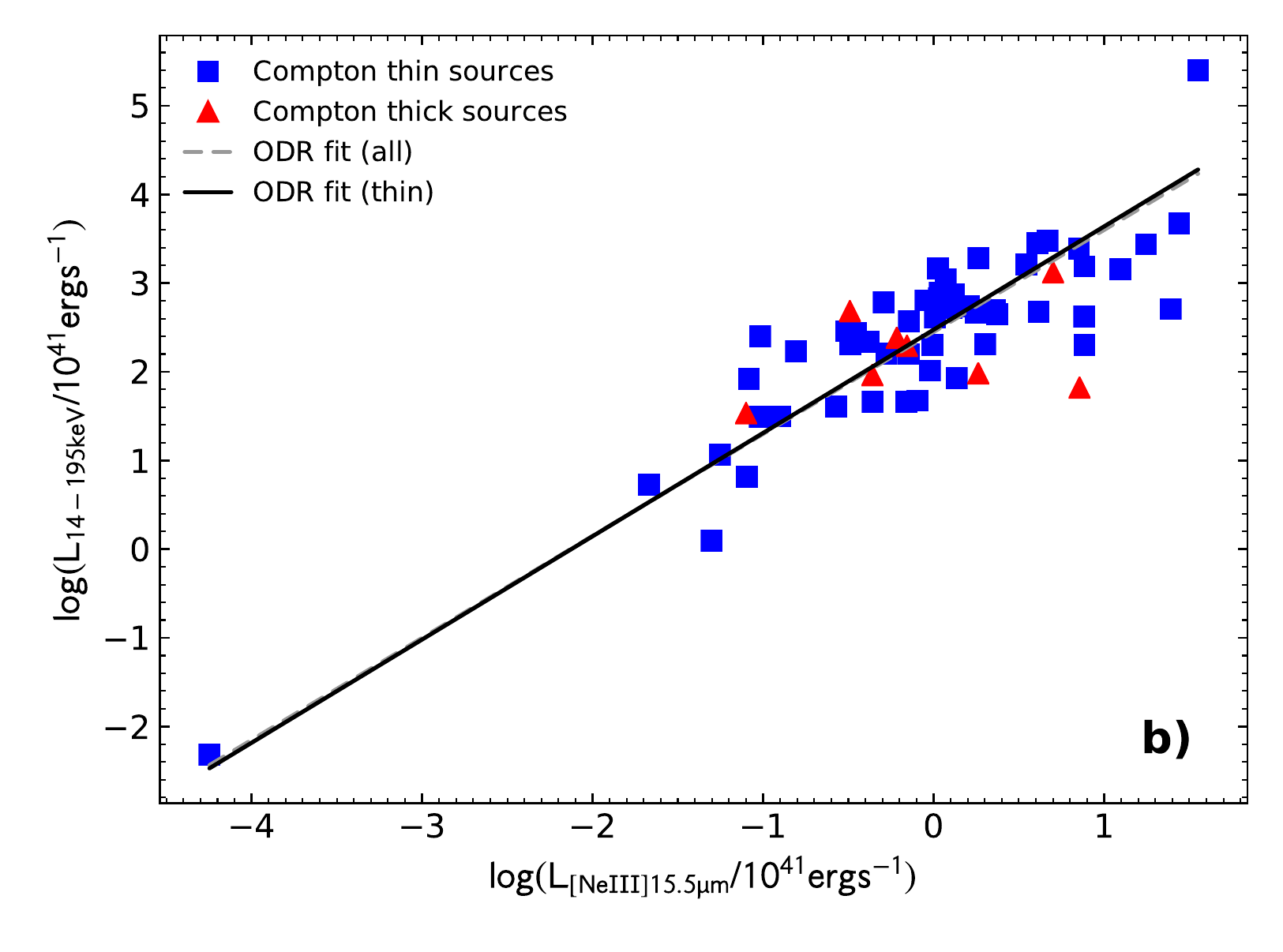}
  \caption{{\bf a: left} X-ray (14-195) keV intrinsic flux as a function of the \neiii~ line flux for all sources with reliable hard-X data and an optical spectroscopic classification. The blue squares indicate the Compton thin AGN, while Compton thick objects are denoted by red triangles. The solid lines represent the linear regression fit with all objects, while the dotted line the fit with only Compton thin objects. {\bf b: right} X-ray (14-195) keV intrinsic luminosity vs. \neiii\  line luminosity (both in logarithmic units of \ergs). 
  }
  \label{fig:HX-NeIIIF&L}
  \end{figure*}

\begin{figure*}
  \includegraphics[width=0.5\textwidth]{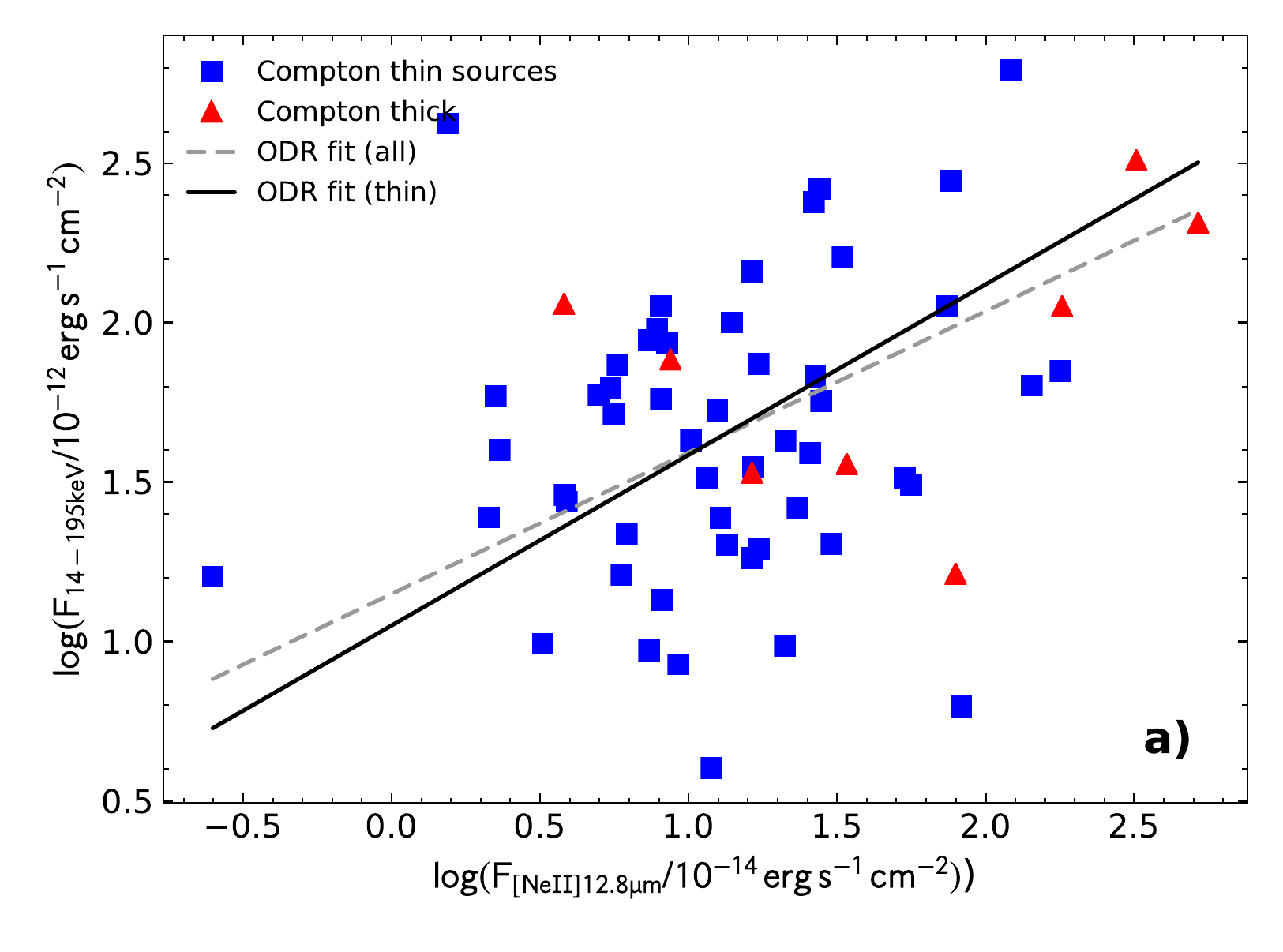}
  \includegraphics[width=0.5\textwidth]{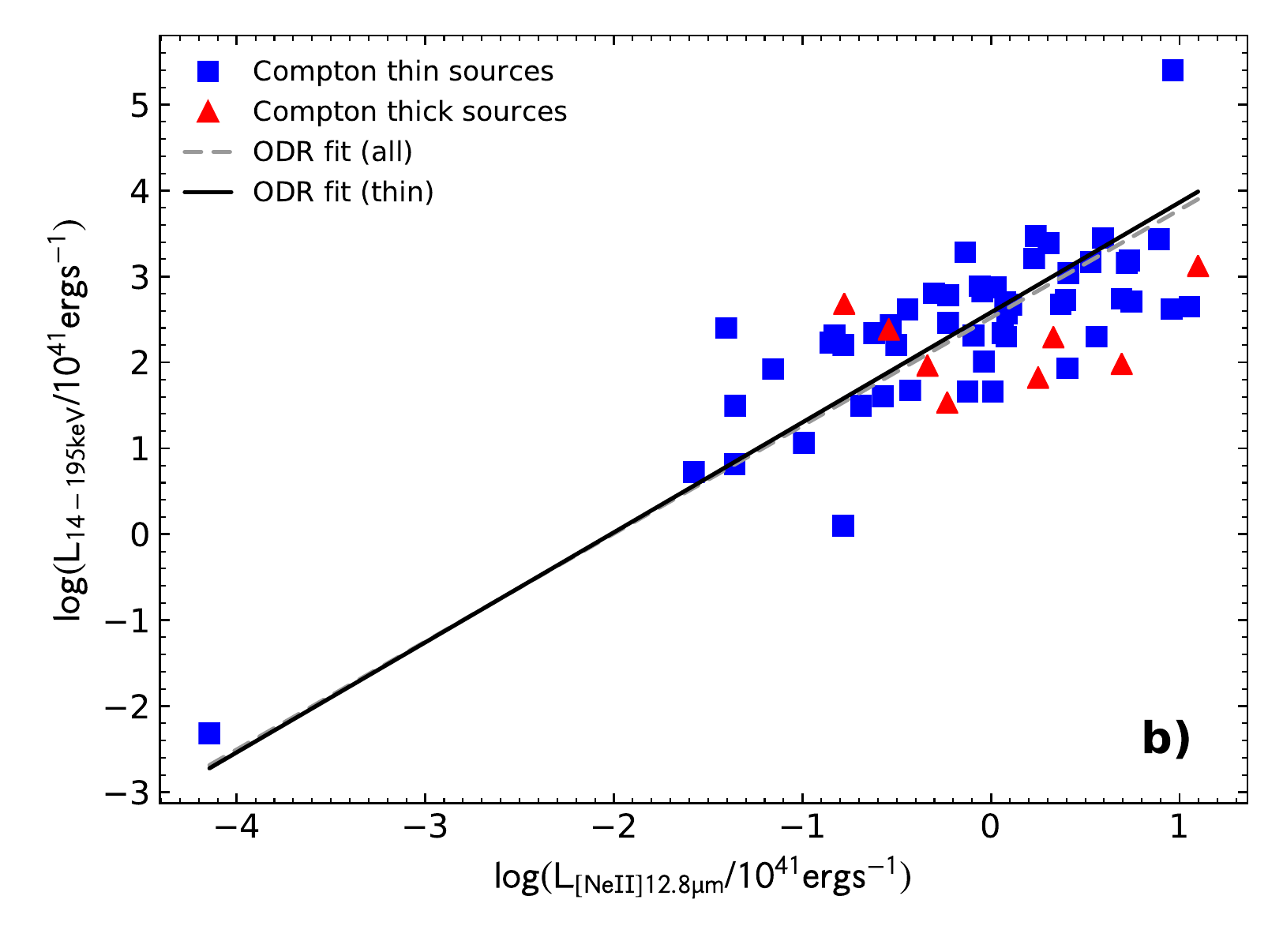}
  \caption{{\bf a: left} X-ray (14-195) keV intrinsic flux as a function of the \neii~ line flux for all sources with reliable hard-X data and an optical spectroscopic classification. The blue squares indicate the Compton thin AGN, while Compton thick objects are denoted by red triangles. The solid lines represent the linear regression fit with all objects, while the dotted line the fit with only Compton thin objects. {\bf b: right} X-ray (14-195) keV intrinsic luminosity vs. \neii\ line luminosity (both in logarithmic units of \ergs). 
  }
  \label{fig:HX-NeIIF&L}
  \end{figure*}

We present here the other correlations discussed in the main text.
 Fig.\ref{fig:X-NeIIIF&L} shows the (2-10)keV X-ray flux and luminosity as a function of the \neiii~ line flux and luminosity and Fig.\ref{fig:X-NeIIF&L}, the same for \neii. 
 Fig.\ref{fig:HX-NeIIIF&L} shows the (14-195)keV X-ray flux and luminosity as a function of the \neiii~ line flux and luminosity and Fig. \ref{fig:HX-NeIIF&L} the same for the \neii~ line.

\begin{figure*}
  \includegraphics[width=0.5\textwidth]{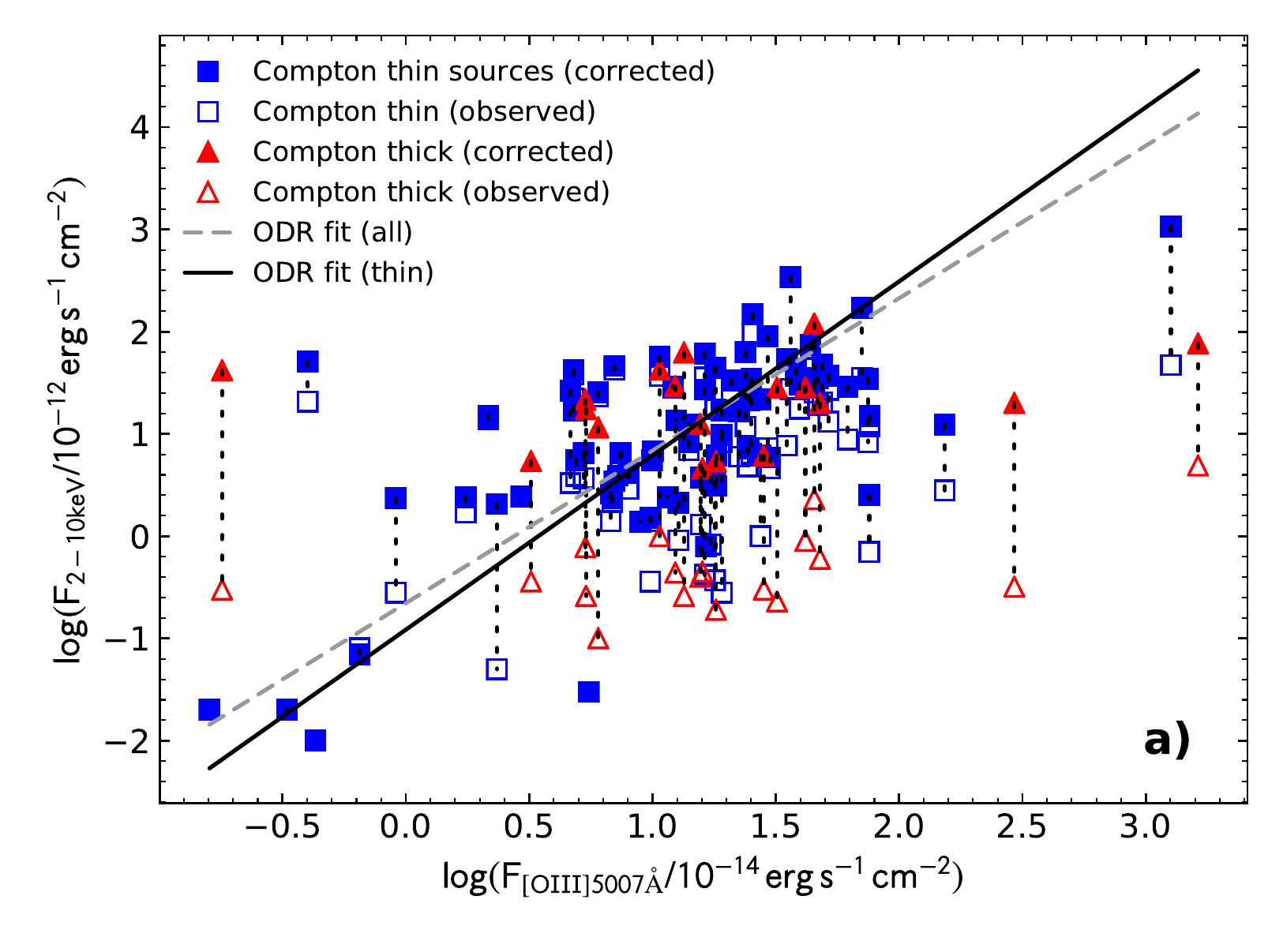}
  \includegraphics[width=0.5\textwidth]{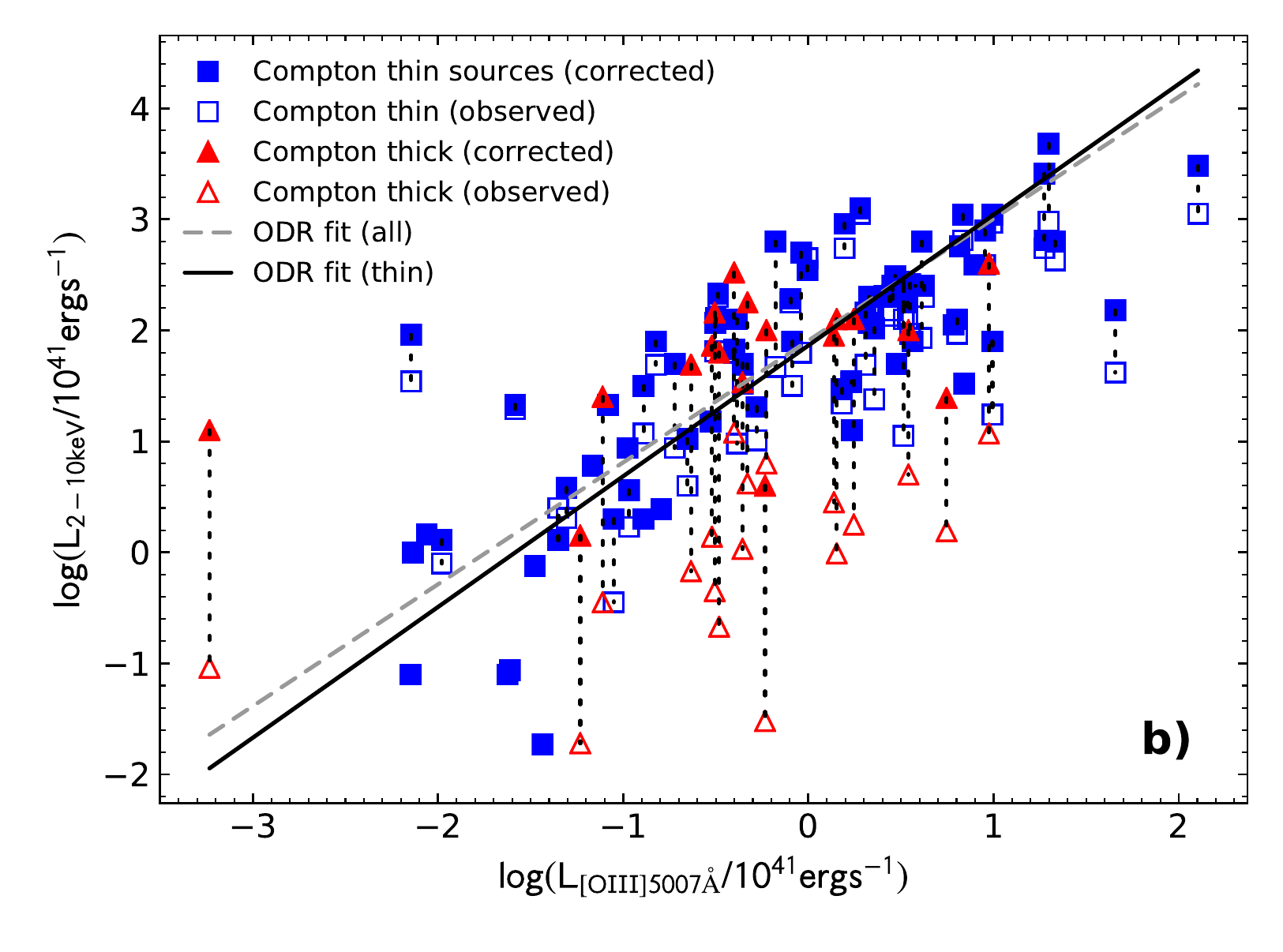}
  \caption{{\bf a: left} X-ray (2-10) keV intrinsic flux as a function of the \oiii~line flux for all sources with reliable hard-X data and an optical spectroscopic classification. The blue squares indicate the Compton thin AGN, while Compton thick objects are denoted by red triangles. The solid lines represent the linear regression fit with all objects, while the dotted line the fit with only Compton thin objects . {\bf b: right} X-ray (2-10) keV intrinsic luminosity vs. \oiii~ line luminosity (both in logarithmic units of \ergs). 
  }
  \label{fig:X-OIII5007F&L}
  \end{figure*}

\begin{figure*}
  \includegraphics[width=0.5\textwidth]{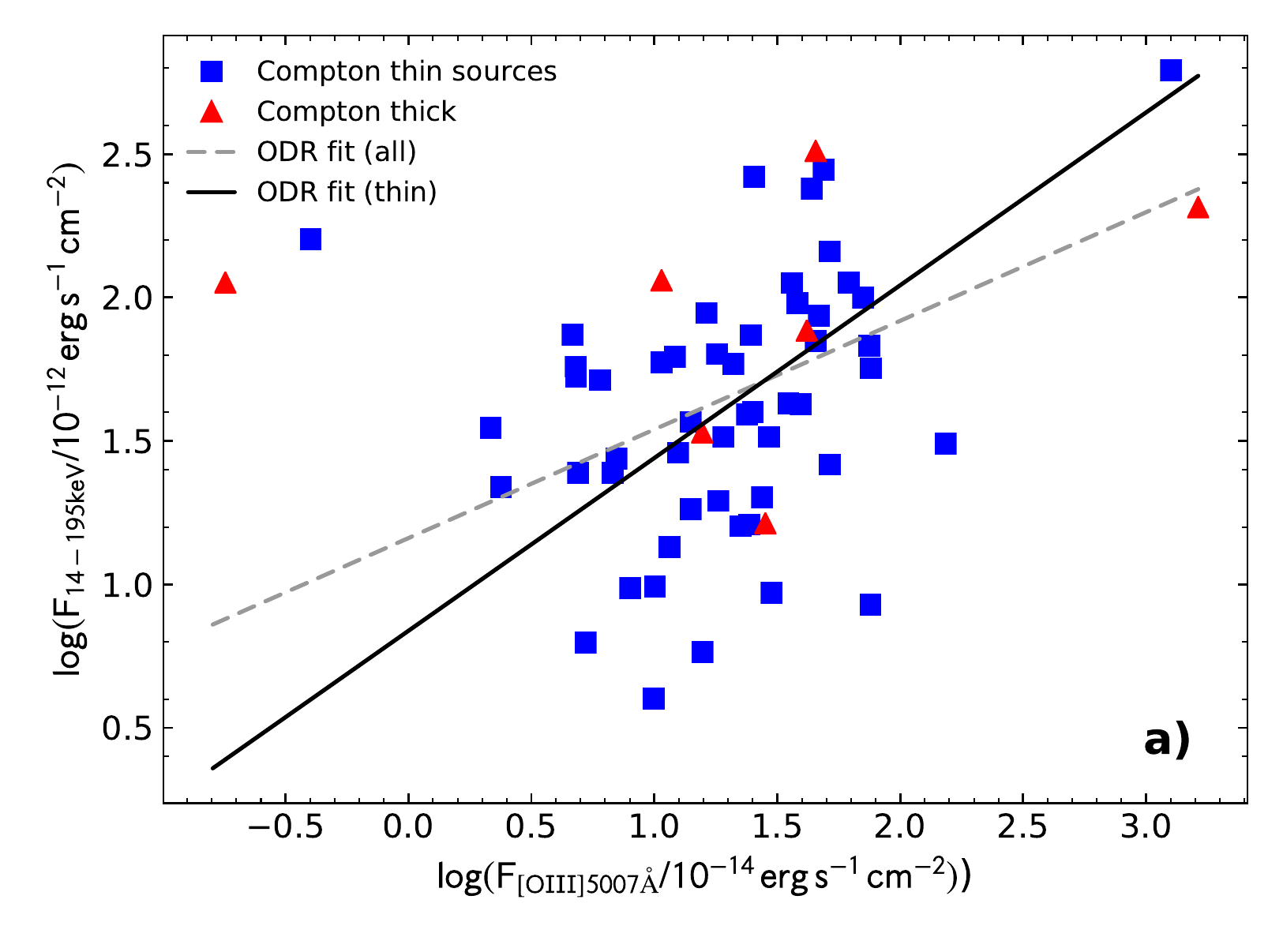}
  \includegraphics[width=0.5\textwidth]{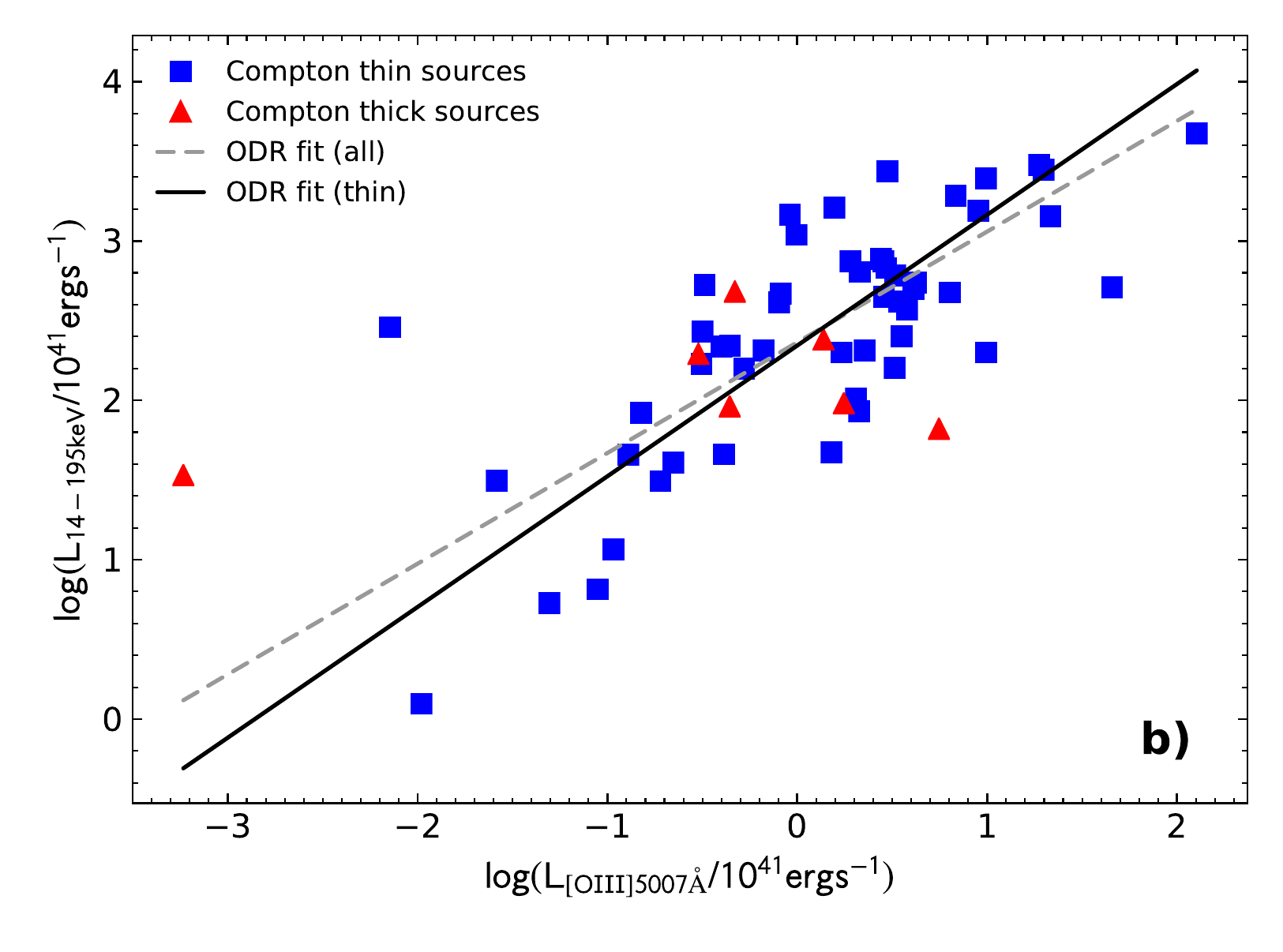}
  \caption{{\bf a: left} X-ray (14-195) keV intrinsic flux as a function of the \oiii~line flux for all sources with reliable hard-X data and an optical spectroscopic classification. The blue squares indicate the Compton thin AGN, while Compton thick objects are denoted by red triangles. The solid lines represent the linear regression fit with all objects, while the dotted line the fit with only Compton thin objects. {\bf b: right} X-ray (14-195) keV intrinsic luminosity vs. \oiii~ line luminosity (both in logarithmic units of \ergs). 
  }
  \label{fig:HX-OIII5007F&L}
  \end{figure*}

\begin{figure*}
  \includegraphics[width=0.5\textwidth]{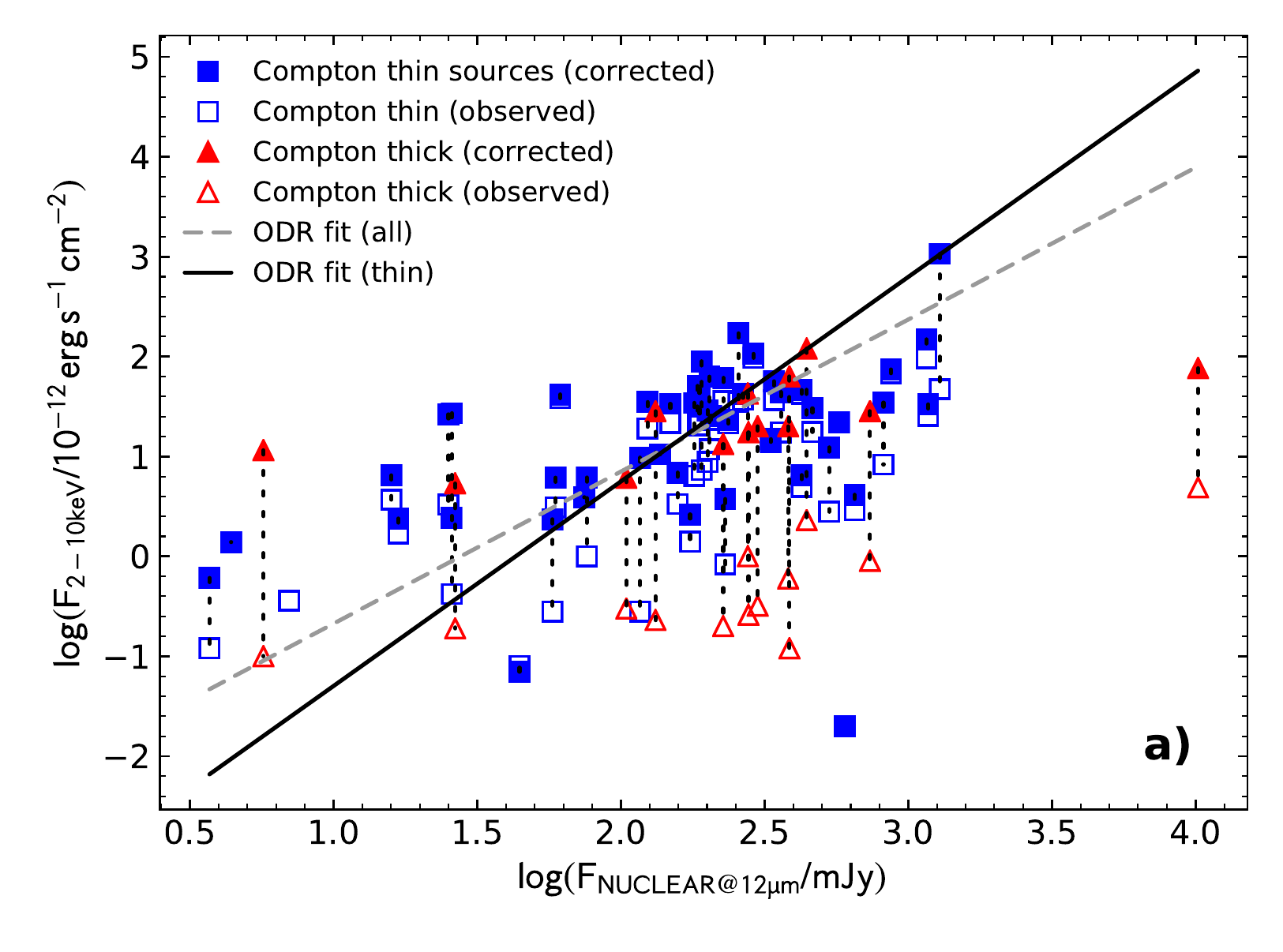}
  \includegraphics[width=0.5\textwidth]{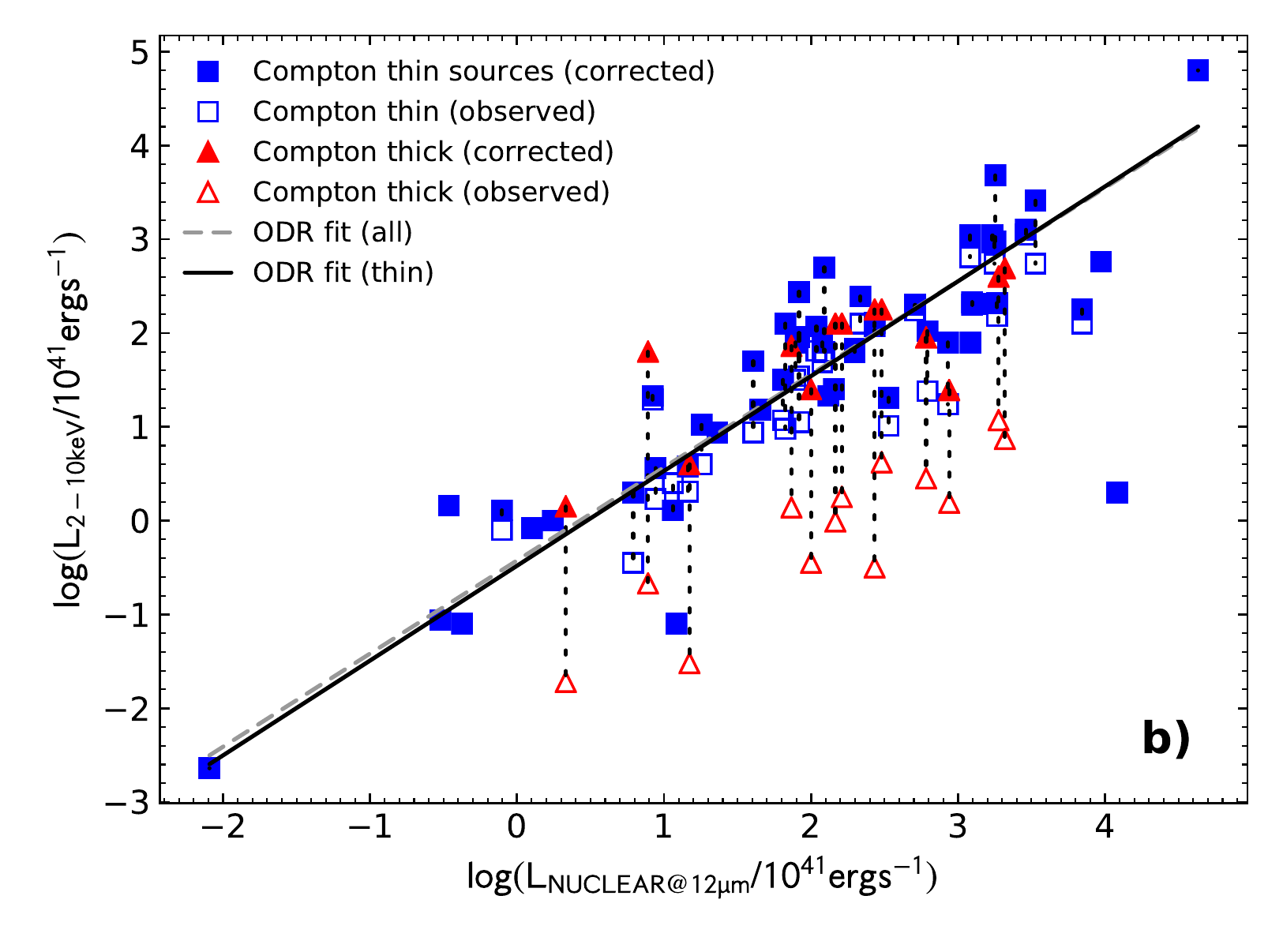}
  \caption{{\bf a: left} X-ray (2-10) keV intrinsic flux as a function of the nuclear 12$\mu$m flux for all sources with reliable hard-X data and an optical spectroscopic classification. The blue squares indicate the Compton thin AGN, while Compton thick objects are denoted by red triangles. The solid lines represent the linear regression fit with all objects, while the dotted line the fit with only Compton thin objects. {\bf b: right} X-ray (2-10) keV intrinsic luminosity vs. 12$\mu$m luminosity (both in logarithmic units of \ergs). 
  }
  \label{fig:X-12umF&L}
  \end{figure*}

\begin{figure*}
  \includegraphics[width=0.5\textwidth]{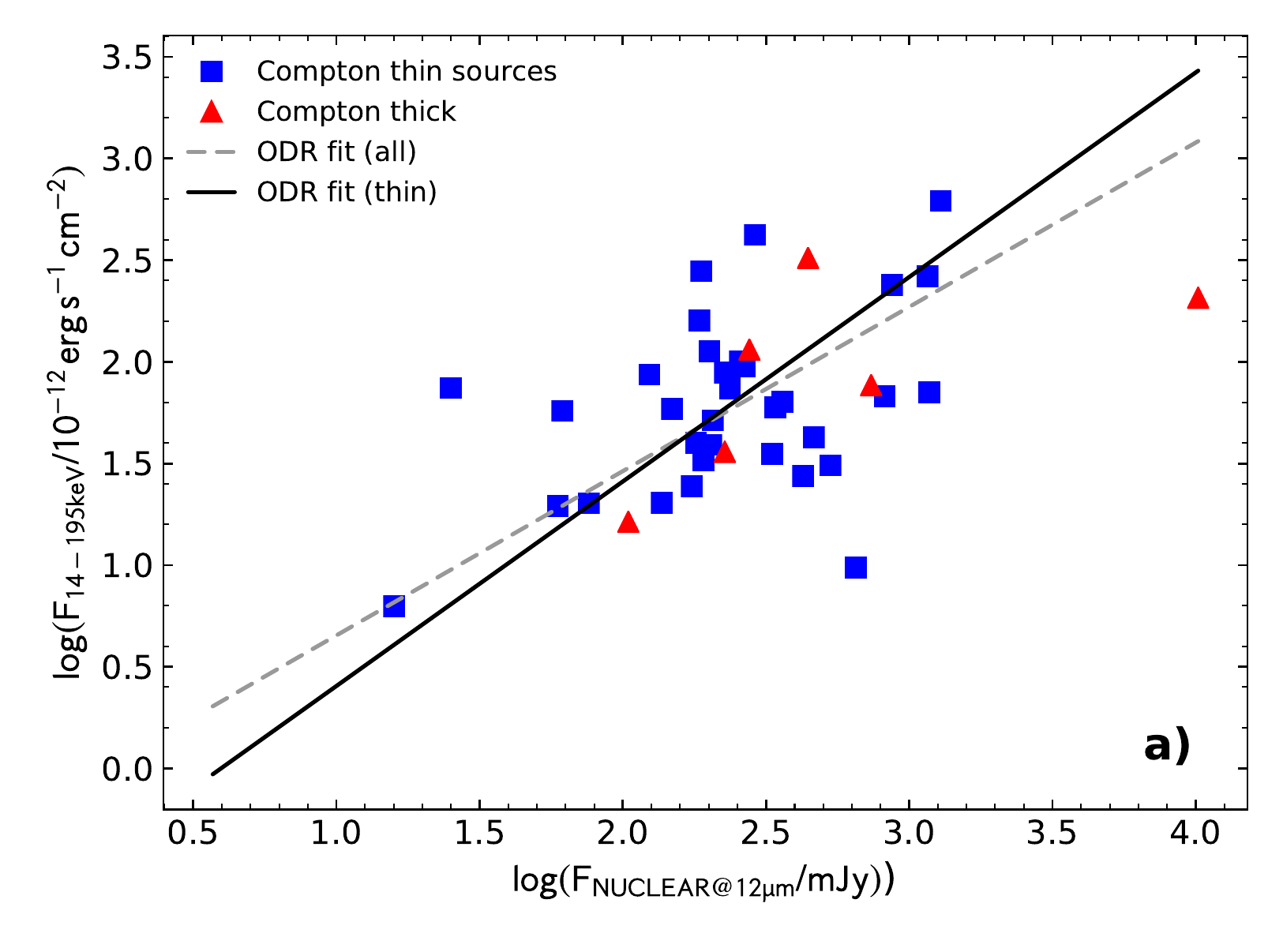}
  \includegraphics[width=0.5\textwidth]{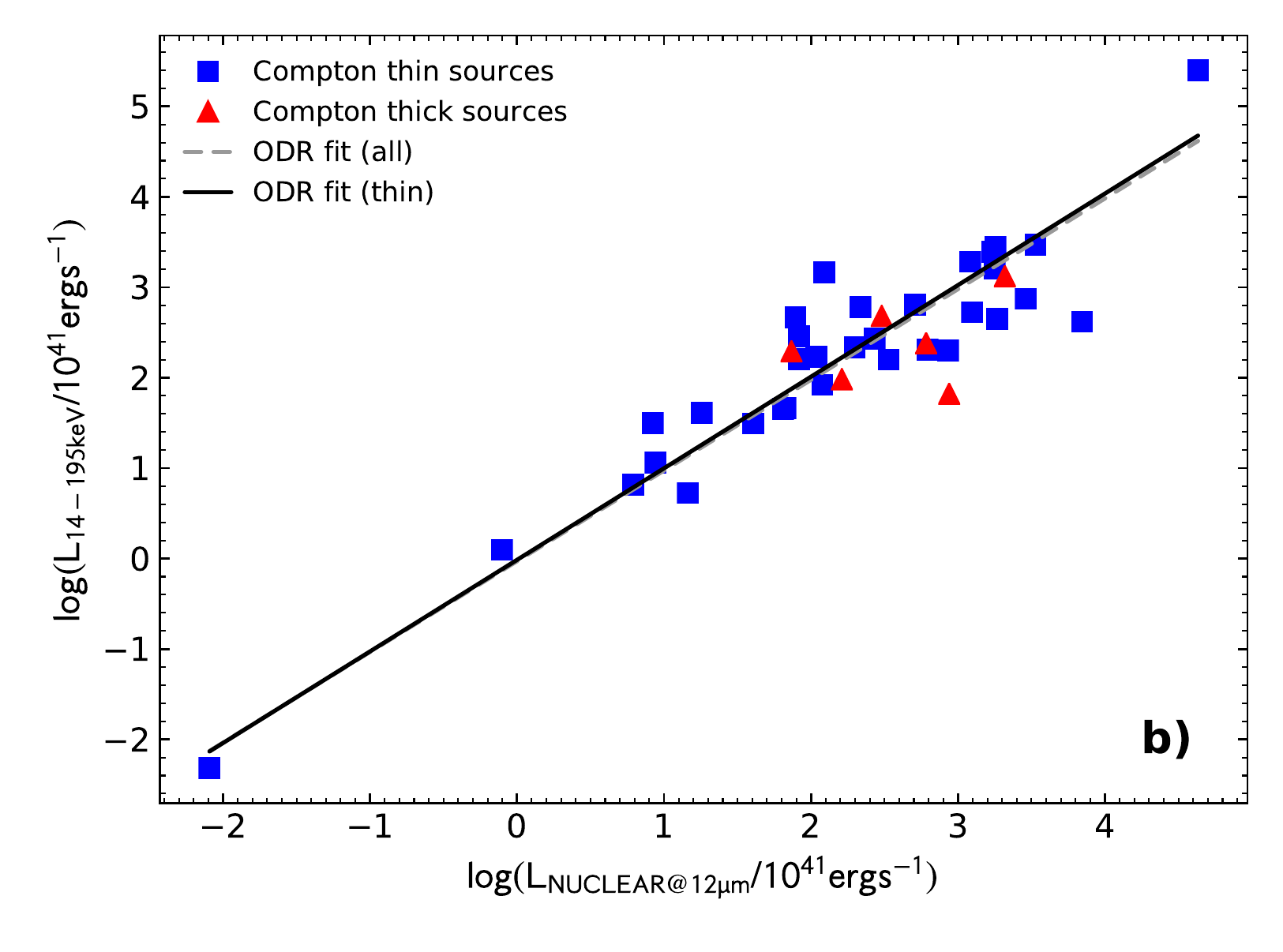}
  \caption{{\bf a: left} X-ray (14-195) keV intrinsic flux as a function of the nuclear 12$\mu$m flux for all sources with reliable hard-X data and an optical spectroscopic classification. The blue squares indicate the Compton thin AGN, while Compton thick objects are denoted by red triangles. The solid lines represent the linear regression fit with all objects, while the dotted line the fit with only Compton thin objects. {\bf b: right} X-ray (14-195) keV intrinsic luminosity vs. 12$\mu$m luminosity (both in logarithmic units of \ergs). 
  }
  \label{fig:HX-12umF&L}
  \end{figure*}

 Fig.\ref{fig:X-OIII5007F&L} shows the (2-10)keV X-ray flux and luminosity as a function of the [OIII]$\lambda$5007$\AA$ line flux and luminosity and Fig. \ref{fig:HX-OIII5007F&L} the same for the (14-195)keV X-ray flux and luminosity. 
 Fig.\ref{fig:X-12umF&L} presents the (2-10)keV X-ray flux and luminosity as a function of the nuclear 12$\mu$m flux and luminosity from \citet{asmus2014} and Fig. \ref{fig:HX-12umF&L} the same for the (14-195)keV X-ray flux and luminosity.
  
\begin{figure*}
    \includegraphics[width=0.5\textwidth]{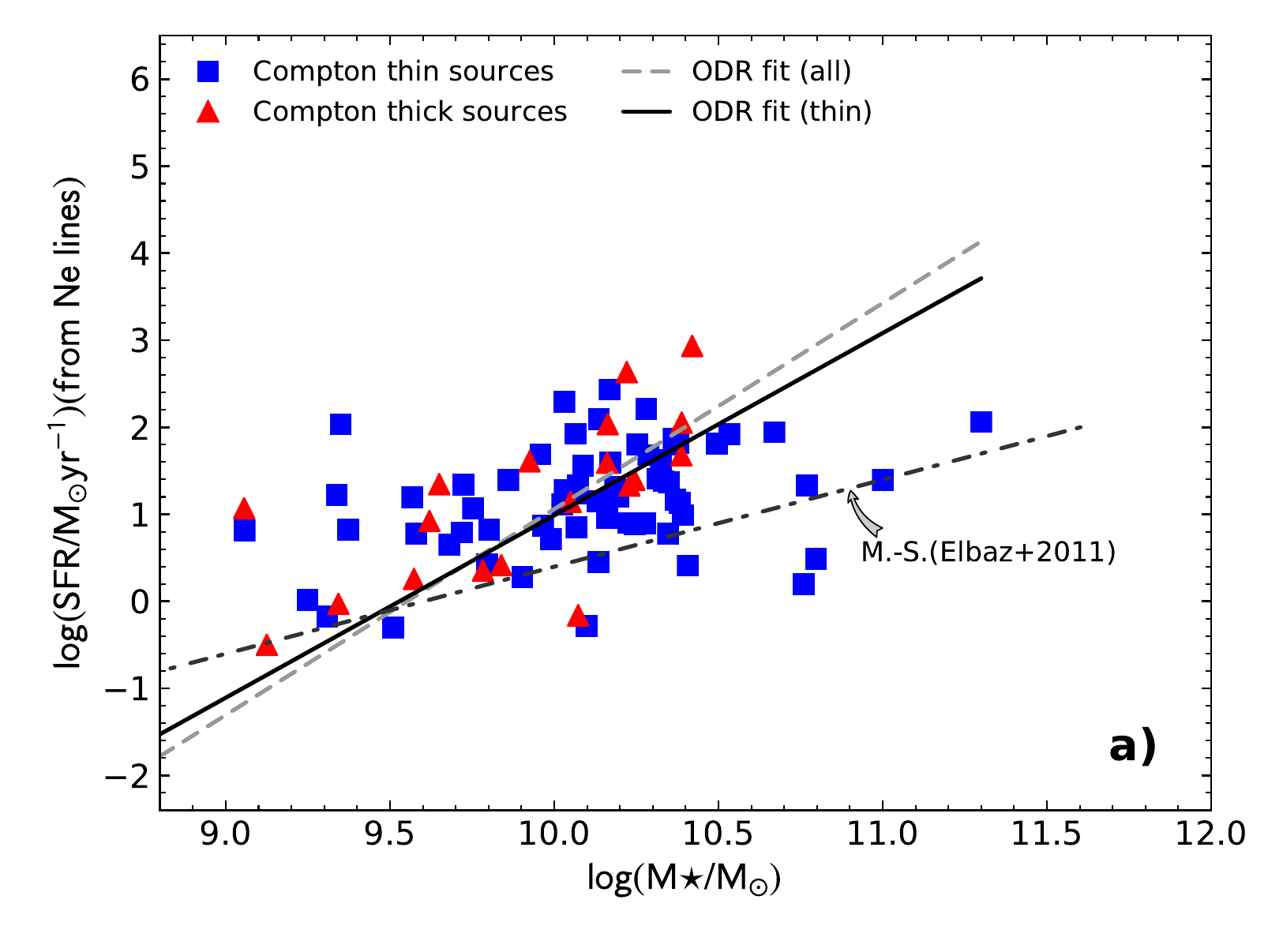}~
 \includegraphics[width=0.5\textwidth]{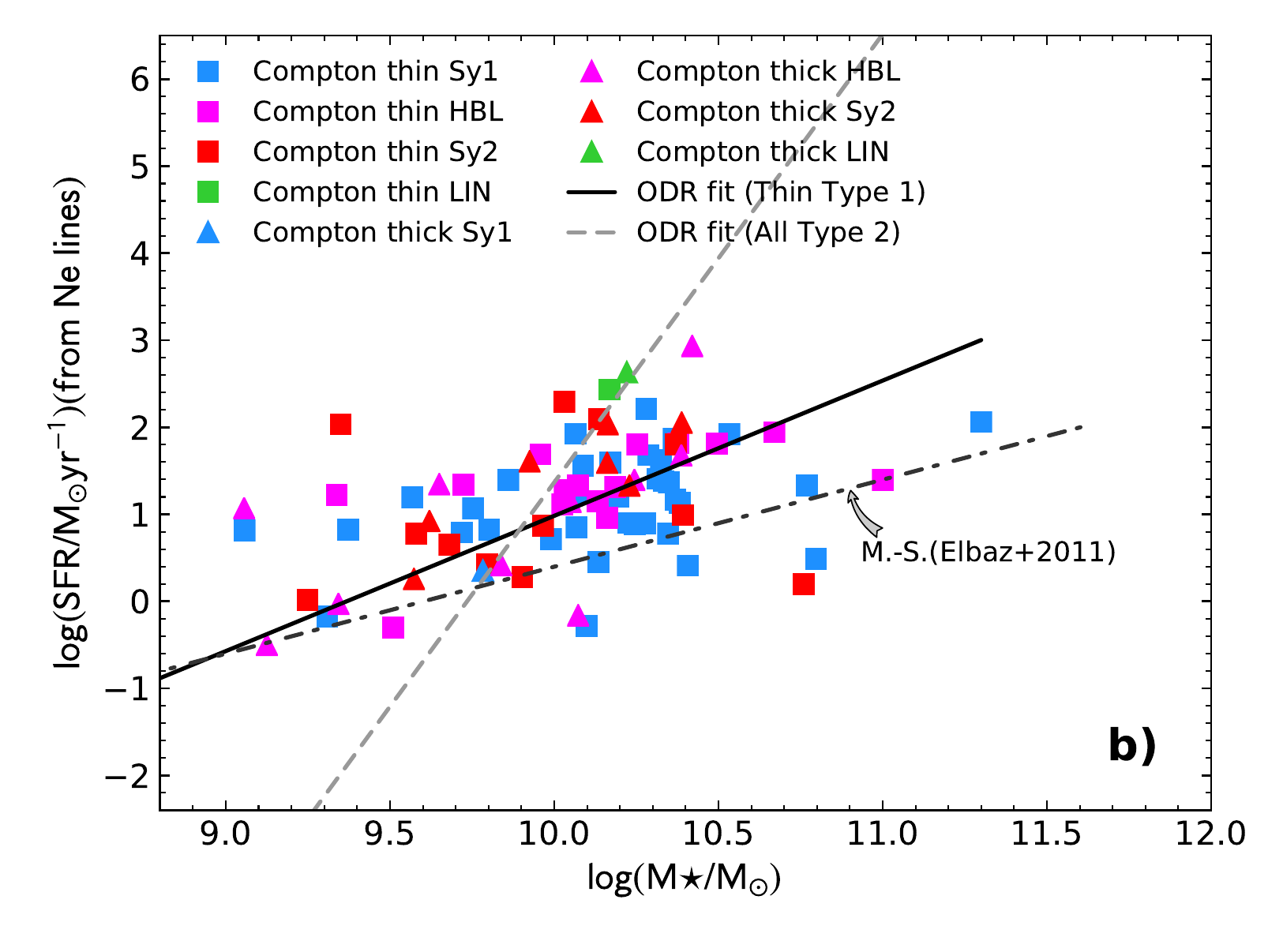} \\
 
  \caption{{\bf a: left}  The SFR, derived from the neon mid-IR fine structure lines, as a function of the stellar mass of the galaxies. The galaxies are divided in Compton thick and Compton thin and the given fits are indistinguishable. The dot-dashed line represents the so called Main-Sequence for star forming galaxies \citep{elbaz2011}. {\bf b: right} Same as {\bf a:}, but with the color coded galaxy types. 
  }\label{fig:SFR_MASS_neon}
  \end{figure*}

\begin{figure*}
    \includegraphics[width=0.5\textwidth]{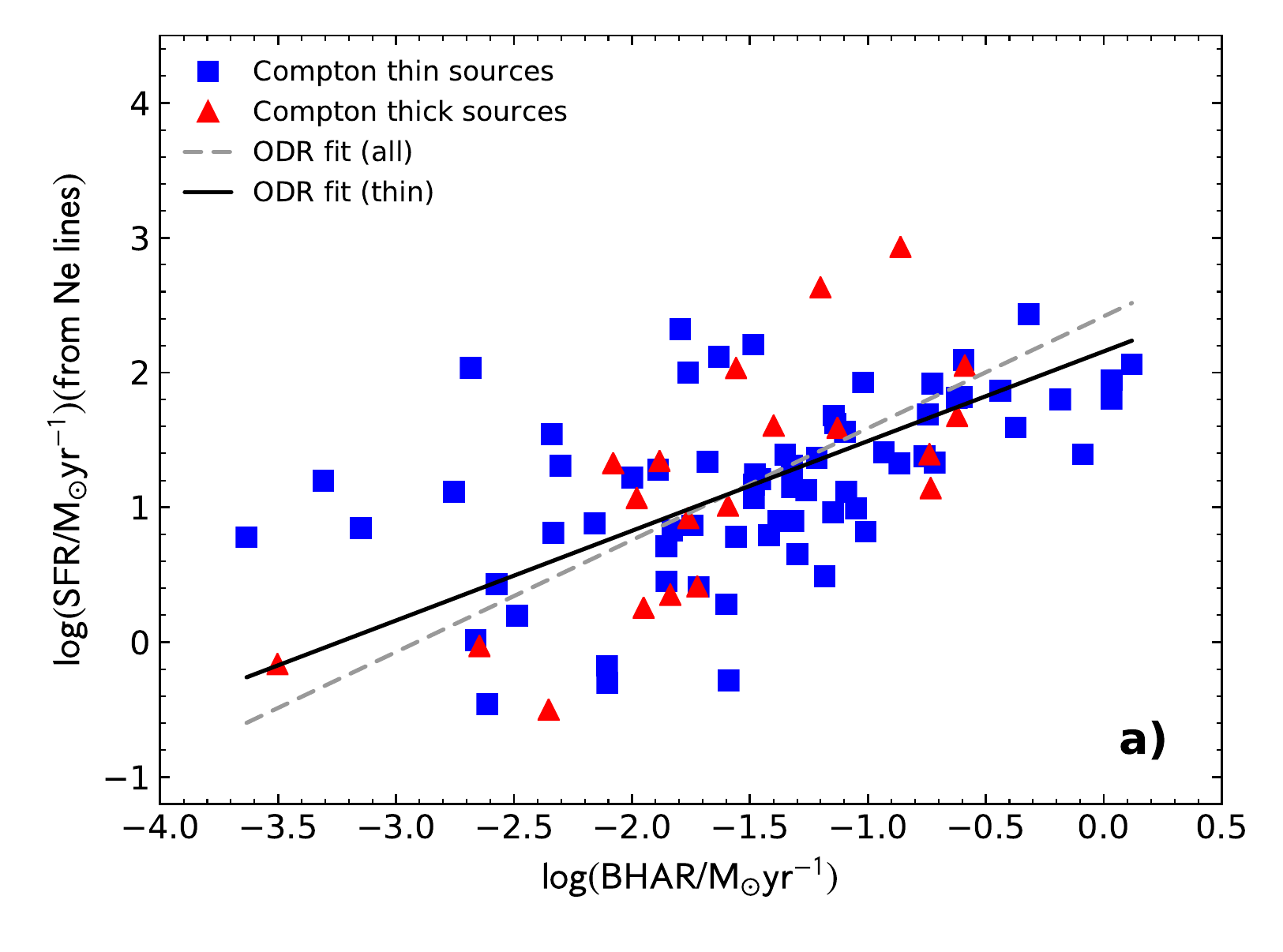}~
    \includegraphics[width=0.5\textwidth]{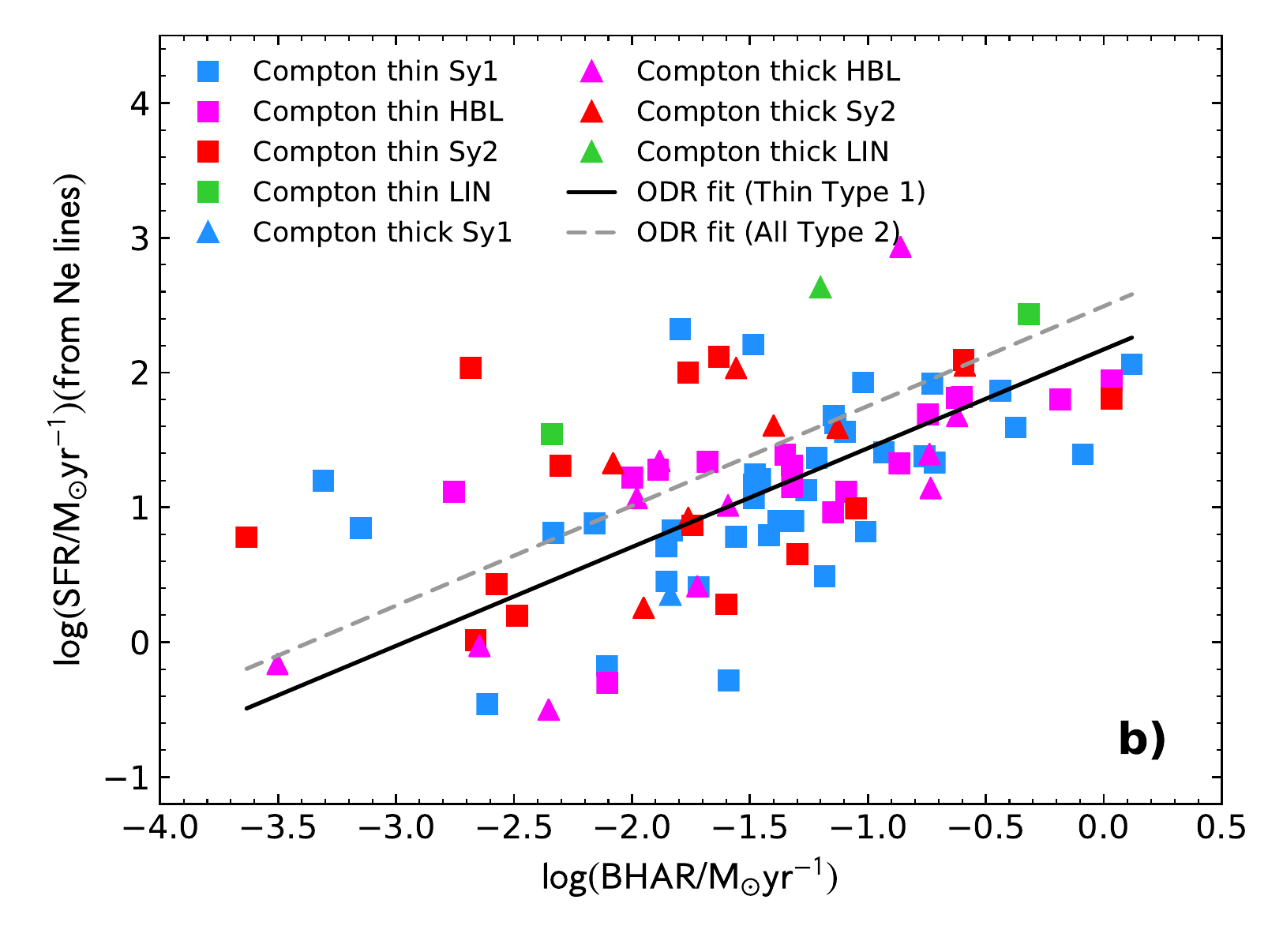}\\
  \caption{{\bf a: left} Star Formation rate, as derived from the sum of the \neiii~ and \neii~ luminosities and the calibration of \citet{mordini2021} with the correction for AGN from \citet{zhuang2019}, as a function of the Black-hole accretion rate as derived from the  \oivp\ line luminosity and the calibration from \citet{mordini2021}, adopting the bolometric correction of \citet{lusso2012} for all objects and Compton thin only galaxies. Blue squares indicate Compton thin objects, while red triangles Compton thick. The solid line shows the fit for all Compton thin AGN, while the broken line the fit for all objects. {\bf b: right} Same as in {\bf a} but with the color coded objects and the fit of Compton thin type 1 galaxies and all type 2 galaxies. The value of the correlations is given in Table \ref{tab:cor2}.
  }\label{fig:BHAR_SFR_neon}
  \end{figure*}

Fig.\ref{fig:SFR_MASS_neon} shows the SFR-Galaxy mass diagram from the SFR determination derived from the \neii~ and \neiii~ lines.
Finally Fig.\ref{fig:BHAR_SFR_neon} shows the SFR, as derived from the neon lines, as a function of the BHAR.

\clearpage
  
\section{Sample}

We present here all the tables with the data of the sample.

\begin{ThreePartTable}
\setlength{\tabcolsep}{2.pt}
\setlength{\LTcapwidth}{\textwidth}
\scriptsize
\begin{longtable}{rlcccc}
\caption{The AGN sample: coordinates, redshift, types}\label{tab:sample0}
\\ \hline\\[-0.3cm]
\bf n. & \bf Name         & \bf RA (J2000.0)  & \bf dec (J2000.0) & \bf $z$     & \bf Type \\
       & (1)             & (2)               & (3)               & (4)         & (5)       \\
\hline\\[-0.4cm]
\endfirsthead
\caption{continued.}\\
\hline\\[-0.3cm]
\bf n. &\bf Name        & \bf RA (J2000.0)  & \bf dec (J2000.0) & \bf $z$     & \bf Type \\
       & (1)            & (2)               & (3)               & (4)         & (5)       \\
\hline\\[-0.4cm]
\endhead
\endfoot
\hline\\
\endlastfoot
  1 &  MRK0335          &  00:06:19.5  & +20:12:10  &  0.0258  & Sy1   \\  
  2 &  NGC34=Mrk938=N17 &  00:11:06.5  & -12:06:26  &  0.0196  & Sy2   \\  
  3 &  IRASF00198-7926  &  00:21:57.0  & -79:10:14  &  0.0728  & Sy2   \\  
  4 &  ESO012-G021      &  00:40:47.8  & -79:14:27  &  0.0300  & Sy1   \\  
  5 &  NGC0262=MRK348   &  00:48:47.1  & +31:57:25  &  0.0150  & HBL   \\  
  6 &  Izw001=UGC00545  &  00:53:34.9  & +12:41:36  &  0.0611  & Sy1   \\  
  7 &  IRASF00521-7054  &  00:53:56.2  & -70:38:03  &  0.0689  & HBL   \\  
  8 &  ESO541-IG012     &  01:02:17.5  & -19:40:09  &  0.0566  & Sy2   \\  
  9 &  NGC0424          &  01:11:27.5  & -38:05:01  &  0.0118  & HBL CT \\ 
 10 &  NGC0526A         &  01:23:54.2  & -35:03:56  &  0.0191  & Sy1   \\  
 11 &  NGC0513          &  01:24:26.8  & +33:47:58  &  0.0195  & HBL   \\  
 12 &  IRASF01475-0740  &  01:50:02.7  & -07:25:48  &  0.0177  & HBL   \\  
 13 &  MRK1034NED02     &  02:23:22.0  & +32:11:50  &  0.0338  & Sy1   \\  
 14 &  ESO545-G013      &  02:24:40.2  & -19:08:27  &  0.0337  & Sy1   \\  
 15 &  NGC0931=Mrk1040  &  02:28:14.5  & +31:18:42  &  0.0167  & Sy1   \\  
 16 &  NGC1068          &  02:42:40.7  & -00:00:48  &  0.0038  & HBL CT \\ 
 17 &  NGC1056          &  02:42:48.5  & +28:34:29  &  0.0052  & Sy2   \\  
 18 &  NGC1097          &  02:46:19.1  & -30:16:28  &  0.0042  & LIN   \\  
 19 &  NGC1125          &  02:51:40.4  & -16:39:02  &  0.0109  & HBL CT \\ 
 20 &  NGC1144          &  02:55:12.2  & -00:11:01  &  0.0288  & Sy2   \\  
 21 &  MCG-02-08-039    &  03:00:29.8  & -11:24:59  &  0.0299  & HBL   \\  
 22 &  NGC1194          &  03:03:49.2  & -01:06:12  &  0.0136  & Sy1 CT \\ 
 23 &  NGC1241          &  03:11:14.7  & -08:55:20  &  0.0135  & Sy2   \\  
 24 &  NGC1320          &  03:24:48.7  & -03:02:33  &  0.0089  & Sy2   \\  
 25 &  NGC1365          &  03:33:36.4  & -36:08:25  &  0.0055  & Sy1   \\  
 26 &  NGC1386          &  03:36:45.4  & -35:59:57  &  0.0029  & HBL CT \\ 
 27 &  IRASF03362-1642  &  03:38:34.5  & -16:32:16  &  0.0369  & Sy2   \\  
 28 &  IRASF03450+0055  &  03:47:40.2  & +01:05:14  &  0.0310  & Sy1   \\  
 29 &  NGC1566          &  04:20:00.6  & -54:56:17  &  0.0050  & Sy1   \\  
 30 &  3C120            &  04:33:11.1  & +05:21:16  &  0.0330  & Sy1   \\  
 31 &  MRK0618          &  04:36:22.2  & -10:22:34  &  0.0356  & Sy1   \\  
 32 &  IRASF04385-0828  &  04:40:54.9  & -08:22:22  &  0.0151  & HBL   \\  
 33 &  NGC1667          &  04:48:37.1  & -06:19:12  &  0.0152  & Sy2   \\  
 34 &  ESO033-G002      &  04:55:59.6  & -75:32:27  &  0.0181  & Sy2   \\  
 35 &  ESO362-G018      &  05:19:35.5  & -32:39:30  &  0.0124  & Sy1   \\  
 36 &  IRASF05189-2524  &  05:21:01.4  & -25:21:45  &  0.0426  & HBL   \\  
 37 &  ESO253-G003      &  05:25:18.3  & -46:00:20  &  0.0425  & Sy2   \\  
 38 &  IRASF05563-3820  &  05:58:02.0  & -38:20:05  &  0.0339  & Sy1   \\  
 39 &  MRK0006          &  06:52:12.2  & +74:25:37  &  0.0188  & Sy1   \\  
 40 &  MRK0009          &  07:36:57.0  & +58:46:13  &  0.0399  & Sy1   \\  
 41 &  MRK0079          &  07:42:32.8  & +49:48:35  &  0.0222  & Sy1   \\  
 42 &  IRASF07599+6508  &  08:04:30.5  & +64:59:53  &  0.1483  & Sy1   \\  
 43 &  NGC2639          &  08:43:38.0  & +50:12:20  &  0.0111  & Sy1   \\  
 44 &  IRASF08572+3915  &  09:00:25.3  & +39:03:54  &  0.0583  & Sy2/LIN \\
 45 &  MRK0704          &  09:18:26.0  & +16:18:19  &  0.0292  & Sy1   \\  
 46 &  UGC05101         &  09:35:51.6  & +61:21:11  &  0.0394  & LIN CT \\ 
 47 &  NGC2992          &  09:45:42.0  & -14:19:35  &  0.0077  & Sy1   \\  
 48 &  MRK1239          &  09:52:19.1  & -01:36:43  &  0.0199  & Sy1   \\  
 49 &  M81              &  09:55:33.2  & +69:03:55  &  -0.00013 & LIN   \\  
 50 &  3C234            &  10:01:49.5  & +28:47:09  &  0.1849  & Sy1   \\  
 51 &  NGC3079          &  10:01:57.8  & +55:40:47  &  0.0037  & Sy2 CT \\ 
 52 &  NGC3227          &  10:23:30.6  & +19:51:54  &  0.0039  & Sy1   \\  
 53 &  NGC3511          &  11:03:23.7  & -23:05:11  &  0.0037  & Sy1   \\  
 54 &  NGC3516          &  11:06:47.5  & +72:34:07  &  0.0088  & Sy1   \\  
 55 &  MCG+00-29-023    &  11:21:12.2  & -02:59:03  &  0.0249  & Sy2   \\  
 56 &  NGC3660          &  11:23:32.2  & -08:39:30  &  0.0123  & Sy2   \\  
 57 &  NGC3982          &  11:56:28.1  & +55:07:31  &  0.0037  & HBL   \\  
 58 &  NGC4051          &  12:03:09.6  & +44:31:53  &  0.0023  & Sy1   \\  
 59 &  UGC07064         &  12:04:43.3  & +31:10:38  &  0.0250  & HBL   \\  
 60 &  NGC4151          &  12:10:32.6  & +39:24:21  &  0.0033  & Sy1   \\  
 61 &  NGC4253=MRK766   &  12:18:26.5  & +29:48:46  &  0.0129  & Sy1   \\  
 62 &  NGC4388          &  12:25:46.7  & +12:39:41  &  0.0084  & HBL   \\  
 63 &  3C273            &  12:29:06.7  & +02:03:09  &  0.1583  & Sy1   \\  
 64 &  NGC4501=M88      &  12:31:59.0  & +14:25:10  &  0.0076  & Sy2   \\  
 65 &  NGC4579=M58      &  12:37:43.5  & +11:49:05  &  0.0051  & LIN   \\  
 66 &  NGC4593          &  12:39:39.4  & -05:20:39  &  0.0090  & Sy1   \\  
 67 &  NGC4594=M104     &  12:39:58.8  & -11:37:28  &  0.0034  & LIN   \\  
 68 &  NGC4602          &  12:40:36.5  & -05:07:55  &  0.0085  & Sy1   \\  
 69 &  TO1238-36=I3639  &  12:40:52.9  & -36:45:22  &  0.0109  & HBL CT \\ 
 70 &  M-2-33-34=N4748  &  12:52:12.4  & -13:24:54  &  0.0146  & Sy1   \\  
 71 &  MRK0231          &  12:56:14.2  & +56:52:25  &  0.0422  & Sy1   \\  
 72 &  NGC4922          &  13:01:24.5  & +29:18:30  &  0.0236  & Sy2   \\  
 73 &  NGC4941          &  13:04:13.1  & -05:33:06  &  0.0037  & Sy2   \\  
 74 &  NGC4968          &  13:07:06.0  & -23:40:43  &  0.0099  & Sy2   \\  
 75 &  NGC5005          &  13:10:56.2  & +37:03:33  &  0.0032  & LIN   \\  
 76 &  NGC5033          &  13:13:27.5  & +36:35:38  &  0.0029  & Sy1   \\  
 77 &  MCG-03-34-064    &  13:22:24.4  & -16:43:43  &  0.0165  & HBL   \\  
 78 &  NGC5135          &  13:25:44.0  & -29:50:02  &  0.0137  & Sy2 CT \\ 
 79 &  NGC5194=M51A     &  13:29:52.3  & +47:11:54  &  0.0015  & Sy2   \\  
 80 &  MCG-06-30-015    &  13:35:53.7  & -34:17:45  &  0.0077  & Sy1   \\  
 81 &  IRASF13349+2438  &  13:37:18.7  & +24:23:03  &  0.1076  & Sy1   \\  
 82 &  NGC5256=Mrk266   &  13:38:17.8  & +48:16:35  &  0.0279  & Sy2   \\  
 83 &  MRK0273          &  13:44:42.1  & +55:53:13  &  0.0378  & LIN   \\  
 84 &  IC4329A          &  13:49:19.3  & -30:18:34  &  0.0161  & Sy1   \\  
 85 &  NGC5347          &  13:53:17.8  & +33:29:27  &  0.0078  & HBL CT \\ 
 86 &  MRK0463=UGC8850  &  13:56:02.9  & +18:22:19  &  0.0504  & HBL   \\  
 87 &  NGC5506          &  14:13:14.8  & -03:12:27  &  0.0062  & Sy2   \\  
 88 &  NGC5548          &  14:17:59.5  & +25:08:12  &  0.0172  & Sy1   \\  
 89 &  MRK0817          &  14:36:22.1  & +58:47:39  &  0.0315  & Sy1   \\  
 90 &  IRASF15091-2107  &  15:11:59.8  & -21:19:02  &  0.0446  & Sy1   \\  
 91 &  NGC5929          &  15:26:06.1  & +41:40:14  &  0.0083  & HBL   \\  
 92 &  NGC5953          &  15:34:32.3  & +15:11:42  &  0.0066  & Sy2   \\  
 93 &  ARP220=UGC9913   &  15:34:57.3  & +23:30:12  &  0.0181  & Sy2   \\  
 94 &  N5995=M-2-40-4   &  15:48:24.9  & -13:45:28  &  0.0252  & HBL   \\  
 95 &  IRASF15480-0344  &  15:50:41.5  & -03:53:18  &  0.0303  & HBL   \\  
 96 &  ESO141-G055      &  19:21:14.3  & -58:40:13  &  0.0371  & Sy1   \\  
 97 &  IRASF19254-7245  &  19:31:22.5  & -72:39:20  &  0.0617  & HBL CT \\ 
 98 &  NGC6810          &  19:43:34.1  & -58:39:21  &  0.0068  & LIN   \\  
 99 &  NGC6860          &  20:08:46.1  & -61:05:56  &  0.0149  & Sy1   \\  
100 &  NGC6890          &  20:18:18.1  & -44:48:23  &  0.0081  & Sy2   \\  
101 &  MRK0509          &  20:44:09.7  & -10:43:25  &  0.0344  & Sy1   \\  
102 &  IC5063           &  20:52:02.0  & -57:04:09  &  0.0113  & HBL   \\  
103 &  MRK0897          &  21:07:45.8  & +03:52:40  &  0.0263  & Sy2   \\  
104 &  NGC7130          &  21:48:19.5  & -34:57:09  &  0.0162  & Sy2 CT \\ 
105 &  NGC7172          &  22:02:01.7  & -31:52:18  &  0.0087  & Sy2 CT \\ 
106 &  IRASF22017+0319  &  22:04:19.2  & +03:33:50  &  0.0611  & HBL   \\  
107 &  NGC7213          &  22:09:16.2  & -47:10:00  &  0.0058  & LIN   \\  
108 &  3C445            &  22:23:49.6  & -02:06:12  &  0.0562  & Sy1   \\  
109 &  NGC7314          &  22:35:46.0  & -26:03:02  &  0.0048  & HBL   \\  
110 &  MCG+03-58-007    &  22:49:36.9  & -19:16:24  &  0.0315  & HBL   \\  
111 &  NGC7469          &  23:03:15.6  & +08:52:26  &  0.0163  & Sy1   \\  
112 &  NGC7496          &  23:09:47.2  & -43:25:40  &  0.0055  & Sy2   \\  
113 &  NGC7582          &  23:18:23.5  & -42:22:14  &  0.0053  & Sy2 CT \\ 
114 &  NGC7590          &  23:18:55.0  & -42:14:17  &  0.0053  & Sy2   \\  
115 &  NGC7603          &  23:18:56.6  & +00:14:38  &  0.0295  & Sy1   \\  
116 &  NGC7674          &  23:27:56.7  & +08:46:45  &  0.0289  & HBL CT \\ 
117 &  CGCG381-051      &  23:48:41.3  & +02:14:21  &  0.0307  & Sy2   \\  

\end{longtable}
\end{ThreePartTable}

\clearpage
\begin{ThreePartTable}
\begin{TableNotes}
\footnotesize
\item \textbf{Notes.} The columns give for each AGN in the sample: (1) name; (2) AGN type: Sy1 = Seyfert type 1; Sy2 = Seyfert type 2; HBL = Hidden Broad Line Region AGN; LIN = LINER galaxy; (3) Hydrogen column density from the literature (see col. 11);  
(4) whether the galaxy is a Compton Thick object; (5) SWIFT name from \citet{ricci2017}; or ${\dag}$ from \citet{cusumano2010} (6) intrinsic 2-10 keV X-ray flux; (7) observed 14-195 keV X-ray flux, from \citet{oh2018};  (8) logarithm of the 2-10 keV corrected X-ray luminosity; (9) logarithm of the 14-195 keV corrected X-ray luminosity; (10) Bolometric luminosity, as derived from the 2-10 keV X-ray luminosity, following \citet{lusso2012}; (11) references for the X-ray observations and hydrogen column densities, no X: means that no x-rays observations are available; 1: \citet{ichikawa2019}; 2: \citet{ghosh1992}; 3: \citet{ricci2017}; 4:\citet{brightman2011}; 5: \citet{brightman2008}; 6: \citet{bassani1999}; 7: \citet{grupe2004}; 8 :\citet{akylas2009}; 9 :\citet{cardamone2007};  10: \citet{risaliti2000};  11 :\citet{risaliti1999};  12 :\citet{guainazzi2005};  13 :\citet{strickland2007};  14 :\citet{gonzalez-martin2009};  15 :\citet{winter2008};  16:  \citet{lee2013}; 17 :\citet{lutz2004} ; 18: \citet{deluit2003}; 19: \citet{cusumano2010}; 20: \citet{tan2012}; 21: \citet{iyomoto1996}; 22: \citet{polletta1996}; 23: \citet{iwasawa2011}; 24: \citet{liu2014}; 25: \citet{singh2011}; 26: \citet{ueda2001}; 27: \citet{bianchi2009}; 28: \citet{ueda2005}; 29: \citet{bianchi2008}; 30: \citet{teng2009}; 31: \citet{shu2010}; 32: \citet{bianchi2005}; 33: \citet{bi2020}. 
\end{TableNotes}
\setlength{\tabcolsep}{1.pt}
\setlength{\LTcapwidth}{\textwidth}
\scriptsize
\begin{longtable}{rlcccccccccc}
\caption{X-ray properties of the AGN sample: \& SWIFT ID., X-ray fluxes and luminosities.}\label{tab:sample1} 
\\ \hline\\[-0.2cm]    
\bf n. &\bf Name       & \bf Type &  \bf log(N$_H$) & CT & \bf SWIFT Name    & \bf F$_{\rm 2-10keV}$    & \bf F$_{\rm 14-195keV}$ & \bf log(L$_{\rm 2-10keV}$) & \bf log(L$_{\rm 14-195keV}$) & \bf log(L$_{\rm BOL(X)}$) & refs. \\[0.05cm]
       &               &          &  (cm$^{-2}$)    &       &             & \multicolumn{2}{c}{($10^{-12} \rm{erg s^{-1} cm^{-2}}$)}     &  \multicolumn{3}{c}{($\rm{erg\,s^{-1}}$)}  \\[0.05cm]
 & (1)             & (2)            & (3)         & (4)     & (5)      &  (6)    & (7)             & (8)                & (9)        & (10)      & (11)         \\[0.1cm]
\hline\\[-0.2cm]
\endfirsthead
\caption{continued.}\\
\hline\\[-0.2cm]
\bf n. &\bf Name       & \bf Type &  \bf log(N$_H$) & CT & \bf SWIFT Name    & \bf F$_{\rm 2-10keV}$    & \bf F$_{\rm 14-195keV}$ & \bf log(L$_{\rm 2-10keV}$) & \bf log(L$_{\rm 14-195keV}$) & \bf log(L$_{\rm BOL(X)}$) & refs. \\[0.05cm]
       &               &          &  (cm$^{-2}$)    &       &             & \multicolumn{2}{c}{($10^{-12} \rm{erg s^{-1} cm^{-2}}$)}     &  \multicolumn{3}{c}{($\rm{erg\,s^{-1}}$)}  \\[0.05cm]
 & (1)             & (2)            & (3)         & (4)     & (5)      &  (6)    & (7)             & (8)                & (9)        & (10)      & (11)         \\[0.1cm]
\hline\\[-0.2cm]
\endhead
\endfoot
\hline\\
\insertTableNotes
\endlastfoot
  1 &  MRK0335          &  Sy1   & 20.48 &   & J0006.2+2012  & 16.69  & 15.97  & 43.42  & 43.40 & 44.56 & 1,3\\ 
  2 &  NGC34=Mrk938=N17 &  Sy2   & 23.67 &   &    ---        &  2.35  &  ---   & 42.33  &  ---  & 43.57 & 12,4\\ 
  3 &  IRASF00198-7926  &  Sy2   &       &   &    ---        &  ---   & 21.8   &  ---   & 44.44 &  ---  & 18\\ 
  4 &  ESO012-G021      & Sy1    & 19.70 & & J0040.9-7915~${\dag}$ & 5.51 & 4.0 & 43.07 & 42.93 & 44.21 & 2,19\\ 
  5 &  NGC0262=MRK348   &  HBL   & 23.12 &   & J0048.8+3155  &  37.45 & 144.8  & 43.3   & 43.89 & 44.43 & 3,4 \\ 
  6 &  Izw001=UGC00545  &  Sy1   & 20.48 &   &    ---        &  6.51  &  ---   & 43.76  &  ---  & 44.93 & 4,17\\ 
  7 &  IRASF00521-7054  &  HBL   &       &   &    ---        &  3.46  &  ---   & 43.59  &  ---  & 44.74 & 20\\ 
  8 &  ESO541-IG012     &  Sy2   &       &   &    ---        &  ---   &  ---   &  ---   &  ---  &  ---  & no X\\ 
  9 &  NGC0424          &  HBL   & 24.33 & Y & J0111.4-3808  &  28.4  & 21.51  & 42.95  & 42.85 & 44.10 & 3,4 \\
 10 &  NGC0526A         &  Sy1   & 22.30 &   & J0123.8-3504  &  23.1  & 73.91  & 43.3   & 43.80 & 44.43 & 3,4,17\\ 
 11 &  NGC0513          &  HBL   & 22.78 &   & J0124.5+3350  &  5.57  & 24.47  & 42.7   & 43.34 & 43.87 & 1,3,4 \\ 
 12 &  IRASF01475-0740  &  HBL   & 24.30 & Y &    ---        &  21.5  &  ---   & 42.08  &  ---  & 43.37 & 5,4\\ 
 13 &  MRK1034NED02     &  Sy1   &       &   &    ---        &  ---   &  ---   &  ---   &  ---  &  ---  & no X\\ 
 14 &  ESO545-G013      &  Sy1   &       &   &    ---        &  4.19  &  ---   & 43.64  &  ---  & 44.79 & 33 \\ 
 15 &  NGC0931=Mrk1040  &  Sy1   & 21.09 &   & J0228.1+3118  & 29.19  & 62.22  & 43.28  & 43.61 & 44.42 & 1,3\\ 
 16 &  NGC1068          &  HBL &$>$25.00 & Y & J0242.6+0000 & 76.4   & 37.90  & 42.39   & 42.0  & 43.61 & 3,6 \\
 17 &  NGC1056          &  Sy2   &       &   &    ---        &  ---   &  ---   &  ---   &  ---  &  ---  & no X\\ 
 18 &  NGC1097          &  LIN   & 20.78 &   &    ---        &  2.39  &  ---   & 41.00  &  ---  & 42.61 & 17,21\\ 
 19 &  NGC1125          &  HBL   & 24.27 & Y & J0251.6-1639  & 12.5   & 16.23  & 42.53  & 42.66 & 43.73 & 3\\
 20 &  NGC1144          &  Sy2   & 23.76 &   & J0255.2-0011  & 26.4   & 74.31  & 43.70  & 44.16 & 44.86 & 1,3\\ 
 21 &  MCG-02-08-039    &  HBL   & 23.67 &   &    ---        &  3.75  &  ---   & 42.9   &  ---  & 44.05 & 4\\ 
 22 &  NGC1194          &  Sy1   & 24.33 & Y & J0304.1-0108  &  42.5  & 36.22  & 43.25  & 43.20 & 44.38 & 3\\
 23 &  NGC1241          &  Sy2   &       &   &    ---        &  ---   &   ---  &  ---   &  ---  &  ---  & no X\\ 
 24 &  NGC1320          &  Sy2   & 24.31 & Y &    ---        &  29.3  &  ---   & 42.7   &  ---  & 43.87 & 4\\ 
 25 &  NGC1365          &  Sy1   & 23.60 &   & J0333.6-3607  &  44.15 & 63.52  & 42.5   & 42.66 & 43.70 & 3,4,17\\ 
 26 &  NGC1386          &  HBL   & 20.66 & Y &    ---        &  20.0  &  ---   & 40.9   &  ---  & 42.55 &  4\\
 27 &  IRASF03362-1642  &  Sy2   &       &   &    ---        &  ---   &  ---   &  ---   &  ---  &  ---  & no X\\ 
 28 &  IRASF03450+0055  &  Sy1   &       &   &    ---        &  ---   &  ---   &  ---   &  ---  &  ---  & no X\\ 
 29 &  NGC1566          &  Sy1   & 20.00 &   & J0420.0-5457  &  6.13  & 19.54  & 41.56  & 42.06 & 42.99 & 1,3\\ 
 30 &  3C120            &  Sy1   & 21.45 &   & J0433.0+0521  & 42.52  & 95.38  & 44.04  & 44.39 & 45.27 & 3\\ 
 31 &  MRK0618          &  Sy1   & 20.95 &   & J0436.3-1022  &  8.37  & 18.30  & 43.4   & 43.74 & 44.54 & 1,3,4\\ 
 32 &  IRASF04385-0828  &  HBL   &       &   &    ---        & 14.71  &  ---   & 42.9   &  ---  & 44.05 & 4,22\\ 
 33 &  NGC1667          &  Sy2 &$>$24.00 & Y &    ---        & 11.53  &  ---   & 42.8   &  ---  & 43.96 & 6,4\\ 
 34 &  ESO033-G002      &  Sy2   & 22.02 &   & J0456.3-7532  &  2.63  & 24.49  & 42.31  & 43.2  & 43.55 & 3,6\\ 
 35 &  ESO362-G018      &  Sy1   & 23.43 & & J0519.4-3240~${\dag}$ & 6.90 & ---& 42.4   & ---   & 43.62 & 4,15\\ 
 36 &  IRASF05189-2524  &  HBL   & 22.92 &   & J0521.0-2522  &  4.14   &  9.71 & 43.25  & 43.62 & 44.38 & 1,3,4\\ 
 37 &  ESO253-G003      &  Sy2   &       &   &    ---        &  ---    &  ---  &  ---   &  ---  &  ---  & no X\\ 
 38 &  IRASF05563-3820  &  Sy1   & 20.00 &   & J0557.9-3822  & 46.26   & 27.45 & 44.1   & 43.87 & 45.35 & 1,3,4\\ 
 39 &  MRK0006          &  Sy1   & 23.00 &   & J0651.9+7426  & 15.04   & 56.70 & 43.1   & 43.68 & 44.24 & 3,4,17\\ 
 40 &  MRK0009          &  Sy1   & 20.46 &   & J0736.9+5846  &    6.8  &  9.84 & 43.40  & 43.57 & 44.54 & 1,3\\ 
 41 &  MRK0079          &  Sy1   & 20.00 &   & J0742.5+4948  & 54.07   & 42.72 & 43.8   & 43.70 & 44.98 & 1,3,4\\ 
 42 &  IRASF07599+6508  &  Sy1   &       &   &    ---        &  0.02   &  ---  & 42.1   &  ---  & 43.39 & 17\\ 
 43 &  NGC2639          &  Sy1   & 23.62 &   &    ---        &  2.06   &  ---  & 41.78  &  ---  & 43.15 & 6,4\\ 
 44 &  IRASF08572+3915 & Sy2/LIN &       &   &    ---        &  0.02   &  ---  & 41.30  &  ---  & 42.81 & 23 \\
 45 &  MRK0704          &  Sy1   & 20.00 &   & J0918.5+1618  & 12.44   & 36.84 & 43.4   & 43.87 & 44.54 & 1,3,4\\ 
 46 &  UGC05101         &  LIN   & 24.35 & Y & J0935.9+6120  & 13.30   &  7.13 & 43.69  & 43.42 & 44.85 & 3,4\\
 47 &  NGC2992          &  Sy1   & 21.72 &   & J0945.6-1420  & 89.67   & 32.65 & 43.1   & 42.66 & 44.24 & 1,3,4\\ 
 48 &  MRK1239          &  Sy1   & 23.48 &   &    ---        & 21.87   &  ---  & 43.312 &  ---  & 44.45 & 7,24\\ 
 49 &  M81              &  LIN   & 20.97 &   & J0955.5+6907  &  10.5   & 20.26 & 38.36  & 38.68 & 41.17 & 3,4,17\\ 
 50 &  3C234            &  Sy1   & 23.51 &   & J1001.8+2848  &  3.73   & 5.82  & 44.48  & 44.67 & 45.88 & 1,3\\ 
 51 &  NGC3079          &  Sy2   & 25.10 & Y & J1001.7+5543  &  41.7   & 36.7  & 42.10  & 42.08 & 43.39 & 3,4 \\
 52 &  NGC3227          &  Sy1   & 22.81 &   & J1023.5+1952  & 29.07   &112.5  & 42.02  & 42.61 & 43.32 & 3,4,17\\ 
 53 &  NGC3511          &  Sy1   &       &   &    ---        &  ---    &  ---  &  ---   &  ---  &  ---  & no X\\ 
 54 &  NGC3516          &  Sy1   & 20.60 &   & J1106.5+7234  &  30.7   & 112.4 & 42.72  & 43.31 & 43.89 & 3,17\\ 
 55 &  MCG+00-29-023    &  Sy2   &       &   &    ---        &  ---    &  ---  &  ---   &  ---  &  ---  & no X\\ 
 56 &  NGC3660          &  Sy2   & 20.48 &   &    ---        &  2.43   &  ---  & 41.94  &  ---  & 43.26 & 5\\ 
 57 &  NGC3982          &  HBL   & 23.63 & Y &    ---        &  5.36   &  ---  & 40.2   &  ---  & 42.13 & 4,8\\ 
 58 &  NGC4051          &  Sy1   & 20.00 &   & J1203.0+4433  & 30.35   & 42.49 & 41.58  & 41.73 & 43.00 & 1,3,4\\ 
 59 &  UGC07064         &  HBL   & 22.59 &   & J1204.9+3105  &  2.4    & 13.48 & 42.54  & 43.30 & 43.74 & 1,3\\ 
 60 &  NGC4151          &  Sy1   & 22.48 &   & J1210.5+3924  & 1068.   & 618.9 & 43.44  & 43.20 & 44.58 & 3,4\\ 
 61 &  NGC4253=MRK766   &  Sy1   & 21.90 &   & J1218.5+2952  & 36.68   & 26.17 & 43.16  & 43.01 & 44.30 & 3,4\\ 
 62 &  NGC4388          &  HBL   & 23.63 &   & J1225.8+1240  & 47.54   & 278.9 & 42.9   & 43.67 & 44.05 & 3,4\\ 
 63 &  3C273            &  Sy1   & 20.00 &   & J1229.1+0202  & 106.3   & 421.6 & 45.8   & 46.40 &  ---  & 3,4\\ 
 64 &  NGC4501=M88      & Sy2 & $>$24.00 &  &    ---        &  0.61   &  ---   & 40.92  &  ---  & 42.56 & 4,5\\ 
 65 &  NGC4579=M58      &  LIN  & 22.38 & & J1237.7+1150~${\dag}$ & 3.91 & --- & 41.11  & ---   & 42.69 & 4,17\\ 
 66 &  NGC4593          &  Sy1   & 20.30 &   & J1239.6-0519  &  61.26  & 88.30 & 43.07  & 43.23 & 44.21 & 3,4,17\\ 
 67 &  NGC4594=M104     &  LIN   & 21.74 &   &    ---        &  1.39   &  ---  & 39.94  &  ---  & 41.98 & 4,6\\ 
 68 &  NGC4602          &  Sy1   &       &   &    ---        &  ---    &  ---  &  ---   &  ---  &  ---  & no X\\ 
 69 &  TOL1238-36=IC3639 & HBL   &  24.2 & Y &    ---        &  63.2   &  ---  & 42.38  &  ---  & 43.61 & 12\\
 70 &  M-2-33-34=NGC4748 & Sy1   & 20.00 &   & J1252.3-1323  &  5.89   & 9.35  & 42.47  & 42.67 & 43.68 & 1,3\\ 
 71 &  MRK0231          &  Sy1   &       &   &    ---        &  0.79   &  ---  & 42.52  &  ---  & 43.72 & 4,25\\ 
 72 &  NGC4922          &  Sy2   &       &   &    ---        &  ---    & --- & $<$37.64 & ---   &  ---  & no X, 33\\ 
 73 &  NGC4941          &  Sy2   & 23.72 &   & J1304.3-0532  &  6.16   & 20.16 & 41.30  & 41.81 & 42.81 & 3 \\ 
 74 &  NGC4968          & Sy2 & $>$24.00 & Y &    ---        &  62.3   &  ---  & 43.2   &  ---  & 44.34 & 4,6\\ 
 75 &  NGC5005          &  LIN   & 20.00 &   &    ---        &  ---    &  ---  & 39.9   &  ---  & 41.96 & 4,14\\ 
 76 &  NGC5033          &  Sy1   & 22.46 &   & J1313.6+3650B &  6.47   &  6.26 & 41.11  & 41.10 & 42.69 & 3,17\\ 
 77 &  MCG-03-34-064    &  HBL   & 23.80 &   & J1322.2-1641  & 12.32   & 30.98 & 42.9   & 43.30 & 44.05 & 1,3,4\\ 
 78 &  NGC5135          & Sy2 & $>$24.00 & Y &    ---        & 42.8    &  ---  & 43.26  &  ---  & 44.39 & 6,26\\
 79 &  NGC5194=M51A     &  Sy2   & 24.00 &   &    ---        & 27.13   &  ---  & 41.16  &  ---  & 42.72 & 4,6,14\\ 
 80 &  MCG-06-30-015    &  Sy1   & 20.85 &   & J1335.8-3416  & 56.58   & 59.53 & 42.90  & 42.92 & 44.05 & 1,3,4\\ 
 81 &  IRASF13349+2438  &  Sy1   & 21.66 &   &    ---        &  2.36   &  ---  & 43.81  &  ---  & 44.99 & 16,27\\ 
 82 &  NGC5256=Mrk266   & Sy2 & $>$25.00 & Y &    ---        &  5.43   &  ---  & 43.0   &  ---  & 44.14 & 10,12\\ 
 83 &  MRK0273          &  LIN   & 23.69 &  & J1344.2+5551~${\dag}$ & 3.32 & ---& 43.05 &  ---  & 44.19 & 4,6\\ 
 84 &  IC4329A          &  Sy1   & 21.62 &   & J1349.3-3018  & 148.6   & 263.3 & 43.96  & 44.21 & 45.17 & 3,4,17\\ 
 85 &  NGC5347          & HBL & $>$24.00 & Y &    ---        & 22.0   &  ---  & 42.48   &  ---  & 43.69 & 11,28 \\
 86 &  MRK0463=UGC8850  &  HBL   & 23.57 &   & J1355.9+1822  &  2.52   &  8.48 & 43.18  & 43.71 & 44.32 & 1,3,17,28\\ 
 87 &  NGC5506          &  Sy2   & 22.53 &   & J1413.2-0312  & 74.28   & 239.4 & 42.83  & 43.34 & 43.99 & 3,4,17\\ 
 88 &  NGC5548          &  Sy1   & 21.71 &   & J1417.9+2507  & 35.04   & 86.47 & 43.39  & 43.78 & 44.53 & 3,4,17\\ 
 89 &  MRK0817          &  Sy1   & 20.00 &   & J1436.4+5846  &   13.6  & 28.77 & 43.49  & 43.83 & 44.63 & 1,3\\ 
 90 &  IRASF15091-2107  &  Sy1   & 21.28 &   & J1512.0-2119  & 16.86   & 32.67 & 43.9   & 44.19 & 45.09 & 1,3,4\\ 
 91 &  NGC5929          &  HBL   & 23.44 &   &    ---        &  1.50   &  ---  & 41.39  &  ---  & 42.87 & 9,25\\ 
 92 &  NGC5953          &  Sy2   &       &   &    ---        & $<$0.005 & --- & $<$39.73 & ---  &  ---  & 12\\ 
 93 &  ARP220=UGC9913   &  Sy2   &       &   &    ---        &  0.01   &  ---  & 40.88  &  ---  & 42.54 & 30\\ 
 94 &  N5995=MCG-2-40-4 &  HBL   & 21.97 &   & J1548.5-1344  &  14.9   & 35.21 & 43.33  & 43.72 & 44.46 & 1,3,28\\ 
 95 &  IRASF15480-0344  & HBL & $>$24.20 & Y &    ---        &  4.60  &  ---   & 43.0   &  ---  & 44.14 & 4,12\\ 
 96 &  ESO141-G055      &  Sy1   & 20.74 &   & J1921.1-5842  & 33.64   & 58.77 & 44.04  & 44.28 & 45.27 & 3,17\\ 
 97 &  IRASF19254-7245  & HBL & $>$24.00 & Y &    ---        &  35.1   &  ---  & 42.8   &  ---  & 43.96 & 4,17,26 \\
 98 &  NGC6810          & LIN    & 22.00 &   &    ---        &  0.08   &  ---  & 39.9   &  ---  & 41.96 & 4,13\\ 
 99 &  NGC6860          & Sy1    & 21.08 &   & J2009.0-6103  &  25.6   & 51.52 & 43.11  & 43.43 & 44.25 & 1,3,15\\ 
100 &  NGC6890          & Sy2    & 23.38 &   &    ---        &  9.74   &  ---  & 42.18  &  ---  & 43.45 & 4,9,28\\ 
101 &  MRK0509          & Sy1    & 20.00 &   & J2044.2-1045  & 170.8   &100.14 & 44.68  & 44.45 & 46.21 & 1,3,4,27\\ 
102 &  IC5063           &  HBL   & 23.56 &   & J2052.0-5704  & 34.63   & 67.76 & 43.02  & 43.31 & 44.16 & 1,3\\ 
103 &  MRK0897          &  Sy2   &       &   &    ---        &  ---    &  ---  &  ---   &  ---  &  ---  & no X\\ 
104 &  NGC7130          &  Sy2   & 24.00 & Y & J2148.3-3454  &  6.10   & 17.41 & 42.55  & 43.03 & 43.75 & 3,14,26\\
105 &  NGC7172          &  Sy2   & 22.91 &   & J2201.9-3152  & 50.89   &160.02 & 42.96  & 43.46 & 44.11 & 1,3,4 \\
106 &  IRASF22017+0319  &  HBL   & 22.83 &   & J2204.7+0337  &   7.2   & 16.16 & 43.80  & 44.15 & 44.98 & 1,3\\ 
107 &  NGC7213          &  LIN   &       &   & J2209.4-4711  &  62.92  & 39.04 & 42.7   & 42.49 & 43.87 & 3,4\\ 
108 &  3C445            &  Sy1   & 23.43 &   & J2223.9-0207  &  34.37  & 39.82 & 44.41  & 44.47 & 45.78 & 1,3,4\\ 
109 &  NGC7314          &  HBL   & 22.08 &   & J2235.9-2602  &   41.8  & 57.42 & 42.33  & 42.50 & 43.57 & 3,4,17\\ 
110 &  MCG+03-58-007    &  HBL   & 23.44 &   &    ---        &  2.13   &  ---  & 42.7   &  ---  & 43.87 & 4\\ 
111 &  NGC7469          &  Sy1   & 20.53 &   & J2303.3+0852  &  33.21  & 70.63 & 43.32  & 43.65 & 44.45 & 1,3,4\\ 
112 &  NGC7496          &  Sy2   &       &   &    ---        &  ---    &  ---  &  ---   &  ---  &  ---  & no X\\ 
113 &  NGC7582          &  Sy2   & 24.33 & Y & J2318.4-4223  &  120.   & 82.28 & 42.86  & 42.74 & 44.02 & 3,4\\
114 &  NGC7590          &  Sy2   &       &   &    ---        &  0.03   &  ---  & 39.27  &  ---  & 41.62 & 26,31\\ 
115 &  NGC7603          &  Sy1   & 20.00 &   & J2318.9+0013  &  22.04  & 52.96 & 43.65  & 44.04 & 44.68 & 1,3,25\\ 
116 &  NGC7674          & HBL & $>$25.00 & Y &    ---       &  20.     &  ---  & 43.6   &  ---  & 44.75 & 17,32\\
117 &  CGCG381-051      &  Sy2   &       &   &    ---       &  ---     &  ---  &  ---   &  ---  &  ---  & no X\\ 

\end{longtable}
\end{ThreePartTable}

\normalsize

\begin{table*}[ht!!!]
\centering
\setlength{\tabcolsep}{3.pt}
\caption{Observed and absorption corrected (2-10) keV fluxes and luminosities for Compton thick objects.}\label{tab:CT}
\footnotesize
\begin{tabular}{rlccccccc}
    \hline\\[-0.2cm]
\bf n. &\bf Name        &  \bf Type &  \bf {\rm log(N$_H$)}  & \bf {\rm F$_{X}$(OBS)} & \bf {\rm F$_{X}$(COR)}       &  \bf {\rm log(L$_{X}$)(OBS)} & \bf {\rm log(L$_{X}$)(COR)} & \bf (ref.) \\[0.05cm]
       &                &           & (cm$^{-2}$)   & \multicolumn{2}{c}{($10^{-12}\, \rm{erg\,s^{-1}\,cm^{-2}}$)}  & \multicolumn{2}{c}{($\rm{erg\,s^{-1}}$)}  & \\[0.05cm]
      & (1)            & (2)       & (3)           & (4)     & (5)      &  (6)    & (7) & (8)    \\[0.1cm]
\hline\\[-0.25cm]
  9& NGC0424         & HBL & 24.33  &    0.9   &   28.4  & 41.45   & 42.95  & 1 \\ 
 12& IRASF01475-0740 & HBL & 24.30  &   0.82   &   21.5  & 40.66    & 42.08  & 2,5    \\ 
 16& NGC1068         & HBL & $>$25  &    4.9   &   76.4  & 41.19   & 42.39  & 1 \\ 
 19& NGC1125         & HBL & 24.27  &    0.4   &   12.5  & 41.03   & 42.53  & 1 \\ 
 22& NGC1194         & Sy1 & 24.33  &    1.0   &   42.5  & 41.62   & 43.25  & 1 \\ 
 24& NGC1320         & Sy2 & 24.00  &    0.48  &   29.3  & 40.83   & 42.7   & 3,5 \\
 26& NGC1386         & HBL & 23.55  &    0.32   &  20.0  & 39.48   & 40.9   & 3,5 \\ 
 33& NGC1667         & Sy2 & $>$24  &    0.10   &   11.5  & 40.33  & 42.80  & 3,5 \\ 
 46& UGC05101        & LIN & 24.35  &    0.2   &   13.3  & 41.87   & 43.69  & 1 \\ 
 51& NGC3079         & Sy2 & 25.10  &    0.3   &   41.7  & 39.96   & 42.10  & 1 \\ 
 57& NGC3982         & HBL & 23.63  &   0.19   &   5.36  & 39.28   & 40.2   & 3,4 \\
 69& TOL1238-36=I3639 & HBL & 24.2  &   0.08   &   63.2  & 40.50   & 42.38  & 2,3 \\ 
 74& NGC4968         & Sy2 & $>$24.  &   0.26   &   62.3  & 40.65   & 43.2  & 3,5 \\ 
 78& NGC5135         & Sy2 & $>$24.  &   0.23   &   42.8  & 40.99   & 43.26   & 3,6 \\ 
 82& NGC5256=Mrk266  & Sy2 & $>$25.  &   0.36   &   5.43  & 41.8    & 43.0   & 4 \\ 
 85& NGC5347         & HBL & $>$24.  &   0.26   &   22.0  & 40.55   & 42.48   & 3,7 \\ 
 95& IRASF15480-0344 & HBL & 24.2   &   0.45   &   4.60  &  41.7   & 43.0    & 5 \\
 97& IRASF19254-7245 & HBL & $>$24. &   0.25   &   35.1  &  42.0   &  42.8   & 5 \\ 
104& NGC7130         & Sy2 & $>$24.0  &   0.3  &    6.1  & 41.25   & 42.55  & 1 \\ 
113& NGC7582         & Sy2 & 24.33  &    2.3   &   120.  & 41.14   & 42.86  & 1 \\ 
116& NGC7674         & HBL & $>$25.  &   0.60  &   20.   & 42.07   & 43.6   &  5  \\ 

\hline
\end{tabular}\\[0.2cm] 
\begin{tablenotes}
\footnotesize
\item \textbf{Notes.} The columns give for each AGN in the sample: (1) name; (2) AGN type (see Table \ref{tab:sample1}) ; (3) logarithm of Hydrogen intrinsic absorption column density (see Table \ref{tab:sample1}); (4) observed F$_{\rm 2-10keV}$ flux; (5) absorption corrected F$_{\rm 2-10keV}$ flux; (6) observed L$_{\rm 2-10keV}$; (7) absorption corrected L$_{\rm 2-10keV}$; (8) reference:  1: observed and corrected 2-10\,keV fluxes and luminosities from extrapolated 14-150keV data by \citet{ricci2017}; 2: \citet{guainazzi2005}; 3: observed 2-10\,keV fluxes and luminosities from \citet{lamassa2011}; 4: \citet{ueda2005}; 5: \citet{brightman2011}; 6: \citet{yamada2020}; 7: \citet{levenson2006}.
\end{tablenotes}
\end{table*}

\clearpage

\begin{ThreePartTable}
\begin{TableNotes}
\footnotesize
\item \textbf{Notes.} The columns give for each AGN in the sample: (1) name; (2) redshift; (3) AGN type (see Table \ref{tab:sample1}) ; (4) $\rm [NeII]12.8\mu m$ flux; (5) $\rm [NeIII]15.5\mu m$ flux; (6) $\rm [NeV]14.3 \mu m$ flux; (7) $\rm [NeV]24.3 \mu m$ flux; (8) $\rm [OIV]26 \mu m$ flux; (9) $\rm PAH\,11.3\mu m$ flux from \citet{wu2009}, where available, otherwise from \citet{tommasin2008, tommasin2010}; (10) $\rm [CII]158 \mu m$ flux from \citet{fernandez2016} (11) $\rm [OIII]\lambda 5007$ flux from \citet{malkan2017}, $^{\dag}$ from \citet{lamassa2010}, $^{\ddag}$ from \citet{koss2017} ; (11) nuclear 12${\mu}m$ flux from \citet{asmus2014}. The IR line fluxes have been taken from \citet{tommasin2008, tommasin2010}. Note that $\rm [NeV]1$ refers to the $\rm [NeV]14.3 \mu m$ line, $\rm [NeV]2$ refers to the $\rm [NeV]24.3 \mu m$ line. Other notes: $^1$ data fom \citet{efstathiou2014}
\end{TableNotes}
\setlength{\tabcolsep}{2.pt}
\setlength{\LTcapwidth}{\textwidth}
\scriptsize
\begin{longtable}{rlcccccccccccc}
\caption{Observed properties of the local galaxy sample:  AGN types, IR lines and continuum fluxes.}\label{tab:sample2}\\
\hline\\[-0.2cm]
       &                  &              &           & \multicolumn{8}{c}{\bf Line and continuum fluxes} &      &  \\
\bf n. & \bf  Name        & \bf  $z$     & \bf  Type & \bf  F$_{\rm [NeII]}$ & \bf  F$_{\rm [NeV]1}$ & \bf  F$_{\rm [NeIII]}$ & \bf  F$_{\rm [NeV]2}$ & \bf F$_{\rm [OIV]}$ &  \bf F$_{\rm PAH(11.3)}$ & \bf F$_{\rm [CII]}$ & \bf F$_{\rm [OIII]\lambda 5007}$ & \bf F$_{\rm 12{\mu}m}$  \\[0.05cm]
                &          &          &  &   \multicolumn{8}{c}{($10^{-14}\, \rm{erg\,s^{-1}\,cm^{-2}}$)}  & (mJy)  \\[0.05cm]
& (1)             &  (2)     & (3)      &  (4)    & (5)             & (6)      & (7)    & (8)              & (9)   & (10) & (11) & (12)  \\[0.1cm]
\hline\\[-0.2cm]
\endfirsthead
\caption{continued.}\\
\hline\\[-0.2cm]
       &                  &              &           & \multicolumn{8}{c}{\bf Line and continuum fluxes} &      &  \\
\bf n. & \bf  Name        & \bf  $z$     & \bf  Type & \bf  F$_{\rm [NeII]}$ & \bf  F$_{\rm [NeV]1}$ & \bf  F$_{\rm [NeIII]}$ & \bf  F$_{\rm [NeV]2}$& \bf F$_{\rm [OIV]}$ & \bf F$_{\rm PAH(11.3)}$ & \bf F$_{\rm [CII]}$ & \bf F$_{\rm [OIII]\lambda 5007}$ & \bf F$_{\rm 12{\mu}m}$  \\[0.05cm]
                &          &          &  &   \multicolumn{8}{c}{($10^{-14}\, \rm{erg\,s^{-1}\,cm^{-2}}$)}  & (mJy)  \\[0.05cm]
& (1)             &  (2)     & (3)      &  (4)    & (5)             & (6)      & (7)    & (8)              & (9)   & (10) & (11)  & (12) \\[0.1cm]
\hline\\[-0.2cm]
\endhead
\endfoot
\hline\\
\insertTableNotes
\endlastfoot
  1 &  MRK0335          &  0.0258  & Sy1   &  0.25 &  0.38  &  0.61 &  1.97  &  7.24 &$<$13.3 & ---   & 22.45 &  ---   \\
  2 &  NGC34=Mrk938=N17 &  0.0196  & Sy2   & 52.10 &$<$2.19 & 6.37 &$<$0.37 &$<$0.66 &  221.  &82.15  &  0.91 &  57.8    \\
  3 &  IRASF00198-7926  &  0.0728  & Sy2   &  6.19 & 12.27  & 14.03 & 11.38  & 33.03 &  31.2  &32.64  &  2.38 &  ---    \\ 
  4 &  ESO012-G021      &  0.0300  & Sy1   & 11.95 &  3.19  &  6.42 &  4.61  & 15.98 &  75.3  &67.52  &  9.95 &  ---     \\ 
  5 &  NGC0262=MRK348   &  0.0150  & HBL   & 16.40 &  5.82  & 20.40 &  4.95  & 17.60 &  23.2  & ---   & 51.68 &  ---     \\
  6 &  Izw001=UGC00545  &  0.0611  & Sy1   &  2.40 &  4.56  &  6.26 &$<$1.35 &  8.92 &  21.0  &13.93  &  7.43 &  425.9   \\ 
  7 &  IRASF00521-7054  &  0.0689  & HBL   &  5.80 &  5.78  &  8.13 &  2.42  &  8.63 &  ---   & 4.34  &  7.02 &   ---     \\
  8 &  ESO541-IG012     &  0.0566  & Sy2   &  1.87 &  2.21  &  2.02 &  1.16  &  4.98 &  12.5  & 6.05  &  2.12 &  ---     \\
  9 &  NGC0424          &  0.0118  & HBL   &  8.70 & 16.10  & 18.45 &  6.37  & 25.80 &  15.3  & ---   &  41.7 &  736.2   \\
 10 &  NGC0526A         &  0.0191  & Sy1   &  5.77 &  6.35  & 10.40 &  5.92  & 19.30 &$<$6.2  & ---   &  24.7 &  236.4   \\
 11 &  NGC0513          &  0.0195  & HBL   & 12.80 &  1.91  &  4.43 &  1.09  &  6.54 &  93.5  & ---   &   4.9 &  ---     \\  
 12 &  IRASF01475-0740  &  0.0177  & HBL   & 13.70 &  6.38  &  9.95 &  1.87  &  6.49 &  27.5  & ---   &  5.35 &  ---     \\ 
 13 &  MRK1034NED02     &  0.0338  & Sy1   & 34.99 &  1.06  &  3.61 &$<$0.60 &  2.68 &  187.  & ---   &   1.1 &  ---     \\ 
 14 &  ESO545-G013      &  0.0337  & Sy1   & 10.05 &  4.80  & 10.50 &  3.22  & 11.54 &  21.9  & ---   &  ---  &  ---    \\ 
 15 &  NGC0931=Mrk1040  &  0.0167  & Sy1   &  5.47 & 14.30  & 15.41 & 13.67  & 42.60 &  45.3  & ---   & 12.13 &  ---     \\
 16 &  NGC1068          &  0.0038  & HBL   & 520.0 & 1110.  & 2100. & 800.0  & 2200. &  ---   &1360.7 & 1629.8 &10205.8   \\
 17 &  NGC1056          &  0.0052  & Sy2   & 33.60 &$<$1.80 & 10.40 &$<$1.23 &  1.40 &  224.  & ---   &  ---  &  ---     \\ 
 18 &  NGC1097          &  0.0042  & LIN   & 59.10 &$<$0.92 &  7.08 &$<$2.31 &  6.20 &  1160. &653.26 &  1.75 &  16.8    \\
 19 &  NGC1125          &  0.0109  & HBL   & 16.37 &  5.09  & 15.55 &  9.69  & 40.36 &  71.6  & ---   & 15.65 &  ---      \\
 20 &  NGC1144          &  0.0288  & Sy2   & 17.20 &  0.92  &  5.40 &  1.82  &  5.31 &  130.  & ---   &  4.65 &  25.1    \\
 21 &  MCG-02-08-039    &  0.0299  & HBL   &  3.86 &  6.59  &  9.79 &  5.24  & 14.40 &$<$11.9 & ---   & 17.25 & 230.7  \\
 22 &  NGC1194          &  0.0136  & Sy1   &  3.81 &  4.28  &  7.37 &  3.76  & 15.10 &$<$24.7 & ---   &  10.7 & 276.6   \\
 23 &  NGC1241          &  0.0135  & Sy2   & 13.40 &  1.61  &  8.08 &  1.23  &  5.46 &  50.1  & ---   &  1.56 &  ---     \\
 24 &  NGC1320          &  0.0089  & Sy2   &  9.58 & 10.70  & 13.60 &  7.46  & 27.40 &  46.0  & ---   & 12.38 &  ---    \\
 25 &  NGC1365          &  0.0055  & Sy1   & 143.0 & 19.10  & 61.30 & 97.50  & 365.0 & 1200.  &1104.5 &  18.  &  360.7   \\
 26 &  NGC1386          &  0.0029  & HBL   & 17.80 & 34.50  & 36.60 & 35.00  & 106.0 & 91.3   &64.05  & 293.2 &  299.3   \\
 27 &  IRASF03362-1642  &  0.0369  & Sy2   &  0.00 &  0.00  &  0.00 &  0.00  &  0.00 & ---    & ---   &  2.59 &  ---      \\
 28 &  IRASF03450+0055  &  0.0310  & Sy1   &  1.09 &$<$1.48 &  1.82 &$<$1.88 &  2.52 &$<$16.2 & 2.45  &  6.86 &  ---    \\
 29 &  NGC1566          &  0.0050  & Sy1   & 17.22 &  0.97  &  9.42 &  0.00  &  6.12 & 322.   & ---   &  18.2 &  59.3    \\
 30 &  3C120            &  0.0330  & Sy1   &  7.84 & 16.60  & 27.60 & 29.00  & 123.0 & 9.04   &31.20  &  38.2 &  266.0   \\
 31 &  MRK0618          &  0.0356  & Sy1   & 16.40 &  3.89  &  5.38 &$<$1.29 & 10.20 & 29.7   & ---   & 14.05 &  ---      \\
 32 &  IRASF04385-0828  &  0.0151  & HBL   & 13.90 &  2.28  &  7.06 &$<$1.44 &  8.56 & 21.6   & ---   &  ---  &  ---     \\ 
 33 &  NGC1667          &  0.0152  & Sy2   & 10.10 &  1.32  &  7.23 &  1.22  &  7.06 & 156.   & ---   &  6.03 &  5.7     \\
 34 &  ESO033-G002      &  0.0181  & Sy2   &  2.13 &  6.34  &  9.22 &  5.27  & 14.39 &$<$27.  & ---   &  6.77 & 174.2   \\
 35 &  ESO362-G018      &  0.0124  & Sy1   & 12.40 &  3.27  &  7.49 &  2.62  &  8.93 & 52.2   & ---   &  ---  &  157.5   \\
 36 &  IRASF05189-2524  &  0.0426  & HBL   & 21.12 & 17.53  & 17.76 & 11.73  & 23.71 & 80.2   &16.29  &  7.99 &  650.9   \\
 37 &  ESO253-G003      &  0.0425  & Sy2   & 16.00 &  6.81  & 18.00 &  7.20  & 24.20 & 46.7   & ---   &  ---  &   ---    \\ 
 38 &  IRASF05563-3820  &  0.0339  & Sy1   &  3.89 &  2.54  &  4.82 &$<$0.79 &  5.31 &  ---   & ---   &  7.01 &  426.5   \\
 39 &  MRK0006          &  0.0188  & Sy1   & 28.00 &  9.39  & 49.34 & 10.43  & 48.24 & 15.9   & ---   &  76.  &  ---     \\ 
 40 &  MRK0009          &  0.0399  & Sy1   &  3.23 &  2.21  &  1.90 &  2.24  &  5.55 & 35.4   &13.16  & 10.05 &  ---     \\
 41 &  MRK0079          &  0.0222  & Sy1   & 10.20 &  6.55  & 19.60 & 12.70  & 42.00 & 22.2   & ---   &  35.1 &  ---     \\ 
 42 &  IRASF07599+6508  &  0.1483  & Sy1   &  3.92 &$<$0.75 & 2.45 &$<$3.00 &$<$1.80 & 9.7   & 8.01  &  0.33 &  ---      \\
 43 &  NGC2639          &  0.0111  & Sy1   &  9.95 &$<$0.64 &  4.29 &$<$1.70 &  8.21 & 47.0   & ---   &  2.34 &  ---     \\ 
 44 &  IRASF08572+3915  &  0.0583  & Sy2/LIN & 7.18 &$<$0.75 & 1.99 &$<$5.40 &$<$6.00 &$<$23.6& 11.2$^1$   &  0.16 & 602.1   \\
 45 &  MRK0704          &  0.0292  & Sy1   &$<$3.30 &  3.93 &  5.63 &$<$3.75 & 11.80 &$<$15.8 & ---   & 14.06 &  ---    \\
 46 &  UGC05101         &  0.0394  & LIN   & 34.13 &  2.57  & 13.66 &  2.82  &  7.35 & 101.   &73.18  &  ---  &  227.0   \\
 47 &  NGC2992          &  0.0077  & Sy1   & 53.50 & 23.30  & 49.40 & 23.20  & 103.0 & 158.   & ---   &  29.3 &  191.2   \\
 48 &  MRK1239          &  0.0199  & Sy1   &  9.40 &  3.40  &  9.38 &  3.22  & 15.60 & 31.9   & ---   &  27.5 &  574.3   \\
 49 &  M81              &  -0.00013  & LIN   & 30.31 &$<$0.73 & 23.93 &  0.00  &  4.51 & 232.   &42.84  &  ---  &  136.9   \\
 50 &  3C234            &  0.1849  & Sy1   &$<$1.29 &  2.35 &  3.40 &  2.91  &  8.97 &$<$4.5  & 1.91  &  15.7 &  ---    \\ 
 51 &  NGC3079          &  0.0037  & Sy2   & 181.0 &$<$2.57 & 24.50 &$<$2.66 & 12.70 & 626.   &722.4  &  0.18 &  ---     \\ 
 52 &  NGC3227          &  0.0039  & Sy1   & 74.30 & 22.60  & 74.40 & 16.00  & 61.10 & 290.   &76.46  &  61.8 &  200.5   \\
 53 &  NGC3511          &  0.0037  & Sy1   &  8.75 &$<$0.49 & 1.00 &$<$1.66 &$<$1.78 & 127.   & ---   &  0.72 &  ---     \\
 54 &  NGC3516          &  0.0088  & Sy1   &  8.07 &  7.88  & 17.72 & 10.39  & 46.92 & 12.2   &21.62  & 36.31 &  ---     \\
 55 &  MCG+00-29-023    &  0.0249  & Sy2   & 47.10 &  1.01  & 4.43 &$<$5.42 &$<$6.01 & 167.   & ---   &  0.78 &  ---     \\
 56 &  NGC3660          &  0.0123  & Sy2   &  6.51 &  0.98  &  1.49 &  1.66  &  3.61 & 29.7   & ---   & 	2.94 &  25.7    \\
 57 &  NGC3982          &  0.0037  & HBL   & 11.40 &  2.89  &  6.79 &  1.62  &  5.11 & 140.   &239.3  &  18.1 &  26.5    \\
 58 &  NGC4051          &  0.0023  & Sy1   & 21.20 & 10.70  & 17.10 & 32.20  & 94.60 & 99.9   &42.50  & 39.45 &  464.0   \\
 59 &  UGC07064         &  0.0250  & HBL   &  8.15 &  4.16  &  6.65 &  5.17  & 14.00 & 78.7   &45.18  & 11.54 &  ---     \\
 60 &  NGC4151          &  0.0033  & Sy1   & 122.0 & 71.70  & 205.0 & 81.70  & 261.0 & 48.2   &64.02  & 1264.1 & 1287.4   \\
 61 &  NGC4253=MRK766   &  0.0129  & Sy1   & 23.30 & 21.00  & 24.10 & 18.50  & 46.10 & 43.8   & ---   & 51.75 &  ---     \\
 62 &  NGC4388          &  0.0084  & HBL   & 76.60 & 46.10  & 106.0 & 73.00  & 340.0 & 143.   &239.05 & 48.87 &  187.8   \\
 63 &  3C273            &  0.1583  & Sy1   &  1.55 &  3.38  &  6.00 &  2.94  &  8.47 &$<$10.3 & 1.65  & ---  & 289.5   \\
 64 &  NGC4501=M88      &  0.0076  & Sy2   &  7.02 &$<$1.50 &  4.72 &$<$3.60 &  4.22 & 69.1   & ---   &   --- &   3.7    \\
 65 &  NGC4579=M58      &  0.0051  & LIN   & 14.75 &$<$0.44 &  7.09 &  0.00  &  1.66 & 88.6   &60.86  &  7.24 &  74.6    \\
 66 &  NGC4593          &  0.0090  & Sy1   &  7.34 &  3.09  &  8.13 &$<$4.32 & 33.30 & 61.0   &13.35  &  16.3 &  227.4   \\
 67 &  NGC4594=M104     &  0.0034  & LIN   &  4.76 &$<$0.52 &  5.52 &  0.00  &  0.76 & 75.6   &25.85  &  8.93 &   4.4    \\
 68 &  NGC4602          &  0.0085  & Sy1   &  7.57 &  0.82  & 0.63 &$<$1.20 &$<$2.30 & 52.1   & ---   &  ---  &  ---     \\ 
 69 &  TO1238-36=I3639  &  0.0109  & HBL   & 45.15 & 11.15  & 27.00 &  5.35  & 21.21 & 146.   &162.8  &   --- &  386.1   \\ 
 70 &  M-2-33-34=N4748  &  0.0146  & Sy1   &  7.37 &  6.75  & 15.93 & 20.00  & 81.93 & 30.1   & ---   & 29.97 &  ---     \\
 71 &  MRK0231          &  0.0422  & Sy1   & 19.67 &$<$3.00 & 3.05 &$<$18.0 &$<$9.50 & 87.7   &42.95  &  16.5 &  ---     \\
 72 &  NGC4922          &  0.0236  & Sy2   & 35.10 &  2.77  &  8.86 &$<$2.61 &  5.92 & 58.2   &41.38  &  3.47 &  ---     \\
 73 &  NGC4941          &  0.0037  & Sy2   & 13.50 &  8.21  & 24.80 &  7.27  & 32.80 & 8.20   &21.03  & 27.45 &  76.1    \\
 74 &  NGC4968          &  0.0099  & Sy2   & 24.90 & 17.57  & 33.80 & 10.60  & 33.70 & 65.6   & ---   & 13.43 &  0.      \\
 75 &  NGC5005          &  0.0032  & LIN   & 27.90 &$<$0.92 & 11.70 &$<$1.64 &  7.06 & 288.   & ---   &  9.79 &  7.0     \\
 76 &  NGC5033          &  0.0029  & Sy1   & 82.79 &  1.79  & 25.08 &  0.00  & 12.78 & 844.   &333.53 &  5.26 &  15.9    \\
 77 &  MCG-03-34-064    &  0.0165  & HBL   & 56.20 & 62.90  & 119.0 & 38.10  & 115.0 & 32.4   &23.35  &  153. &  530.6   \\
 78 &  NGC5135          &  0.0137  & Sy2   & 36.70 &  4.88  & 16.70 & 15.20  & 71.30 & 392.   &223.88 &  32.  &  132.0   \\
 79 &  NGC5194=M51A     &  0.0015  & Sy2   & 97.76 &  3.15  & 51.62 &  0.00  & 23.25 & 1180.  &329.25 &  16.3 &  25.8    \\
 80 &  MCG-06-30-015    &  0.0077  & Sy1   &  4.98 &  5.01  &  5.88 &  7.37  & 26.00 & 30.1   & 7.05  &  10.7 &  340.8   \\
 81 &  IRASF13349+2438  &  0.1076  & Sy1   &  4.80 &$<$0.87 &  4.21 &$<$1.47 &  7.35 &$<$16.9 & 3.76  &  6.88 &   0.    \\ 
 82 &  NGC5256=Mrk266   &  0.0279  & Sy2   & 19.80 &  2.31  & 10.60 & 11.90  & 56.80 & 108.   &161.53 &  3.22 &   0.     \\ 
 83 &  MRK0273          &  0.0378  & LIN   & 41.90 & 11.68  & 33.57 & 15.38  & 56.36 & 83.5   &72.94  & 17.96 &   0.     \\
 84 &  IC4329A          &  0.0161  & Sy1   & 27.60 & 29.30  & 57.00 & 34.60  & 117.0 &$<$28.5 &16.16  &  25.5 &1157.7   \\
 85 &  NGC5347          &  0.0078  & HBL   &  4.17 &  2.08  &  4.09 &$<$1.74 &  7.64 & 32.9   &12.93  &  5.38 &  278.0   \\
 86 &  MRK0463=UGC8850  &  0.0504  & HBL   &  9.25 & 18.25  & 40.78 & 19.93  & 69.17 & 27.8   &31.39  &  75.69&  ---     \\ 
 87 &  NGC5506          &  0.0062  & Sy2   & 26.40 & 18.50  & 45.60 & 56.50  & 239.0 & 98.0   &120.80 &  43.73&  870.8   \\
 88 &  NGC5548          &  0.0172  & Sy1   &  8.47 &  5.40  &  7.27 &  3.89  & 17.50 & 27.1   &19.61  &  46.77&  123.8   \\
 89 &  MRK0817          &  0.0315  & Sy1   &  3.83 &  1.86  &  4.58 &  3.60  &  6.53 &$<$17.2 & ---   &  12.49&  ---    \\ 
 90 &  IRASF15091-2107  &  0.0446  & Sy1   & 11.52 &  8.48  & 16.29 &  8.12  & 31.03 & 25.7   & ---   &  19.1 &  ---     \\ 
 91 &  NGC5929          &  0.0083  & HBL   & 13.20 &  1.14  &  9.83 &  2.17  &  5.32 & 35.0   & ---   &  9.83 &  ---     \\ 
 92 &  NGC5953          &  0.0066  & Sy2   & 67.20 &  2.24  & 16.70 &  6.44  & 21.00 & 407.   & ---   &  12.31&$<$29.5   \\ 
 93 &  ARP220=UGC9913   &  0.0181  & Sy2   & 64.54 &$<$2.90 & 7.80 &$<$14.0 &$<$21.0 & 154.   &125.77 &   0.43&  ---     \\ 
 94 &  N5995=M-2-40-4   &  0.0252  & HBL   & 16.50 &  6.13  &  8.47 &  3.11  & 12.90 & 82.9   & ---   &   2.16&  332.4   \\
 95 &  IRASF15480-0344  &  0.0303  & HBL   &  5.57 &  6.08  &  9.35 &  8.90  & 35.00 & 24.5   & ---   &  15.95&  ---     \\
 96 &  ESO141-G055      &  0.0371  & Sy1   &  2.24 &  2.25  &  5.62 &  1.62  &  7.26 & 19.2   &15.84  &  20.95&  148.7   \\
 97 &  IRASF19254-7245  &  0.0617  & HBL   & 31.48 &  2.77  & 13.19 &$<$1.60 &  6.35 & 50.6   &39.12  &  ---  &  ---     \\
 98 &  NGC6810          &  0.0068  & LIN   & 103.0 &$<$1.09 & 13.40 &$<$2.35 &  2.55 & 464.   & ---   &  0.65 &  44.4    \\
 99 &  NGC6860          &  0.0149  & Sy1   &  5.60 &  2.85  &  6.65 &  2.41  & 12.10 & 35.9   &41.60  &  6.02 &  206.1   \\
100 &  NGC6890          &  0.0081  & Sy2   & 11.32 &  5.77  &  6.57 &  3.77  & 10.10 & 83.3   & ---   &  19.1 &  116.6   \\
101 &  MRK0509          &  0.0344  & Sy1   & 14.00 &  4.74  & 14.50 &  6.82  & 27.50 & 47.7   &36.32  & 70.73 &  256.4   \\
102 &  IC5063           &  0.0113  & HBL   & 26.70 & 30.30  & 66.30 & 23.60  & 114.0 & 39.6   &47.16  & 74.85 &  820.6   \\
103 &  MRK0897          &  0.0263  & Sy2   & 24.03 &  1.06  &  4.38 &$<$0.80 &  0.62 & 117.   & ---   &  3.71 &   8.2    \\
104 &  NGC7130          &  0.0162  & Sy2   & 79.30 &  9.09  & 29.40 &  5.22  & 19.70 &  252.  &225.63 &  28.3 &  104.5   \\
105 &  NGC7172          &  0.0087  & Sy2   & 33.00 & 10.20  & 17.10 & 13.80  & 45.40 &  75.2  &74.79  &  0.4  &  185.0   \\
106 &  IRASF22017+0319  &  0.0611  & HBL   &  5.95 &  8.33  & 14.07 &  9.40  & 29.04 &  9.56  &13.62  &  24.33&  ---     \\ 
107 &  NGC7213          &  0.0058  & LIN   & 25.70 &$<$1.85 &12.00 &$<$0.89 &$<$13.5 &  28.8  &30.87  &  23.85&  203.3   \\
108 &  3C445            &  0.0562  & Sy1   &  2.31 &  1.95  &  6.23 &  5.84  & 22.60 &  ---   & 3.15  &  25.15&  180.0   \\
109 &  NGC7314          &  0.0048  & HBL   &  8.08 & 16.90  & 23.20 & 21.50  & 67.00 &  18.7  &45.91  &  4.79 &  61.5    \\
110 &  MCG+03-58-007    &  0.0315  & HBL   &  8.52 &  6.63  &  9.29 &  3.91  &  8.80 &  50.3  & ---   &  12.74&  ---     \\
111 &  NGC7469          &  0.0163  & Sy1   & 179.0 & 17.50  & 37.50 & 17.10  & 22.90 &  454.  &287.29 &  45.38& 1173.8   \\
112 &  NGC7496          &  0.0055  & Sy2   & 48.08 &$<$1.80 & 6.67 &$<$2.40 &$<$2.40 &  216.  & ---   &   4.89&  169.5   \\
113 &  NGC7582          &  0.0053  & Sy2   & 322.0 & 38.80  & 105.0 & 63.60  & 262.0 &  771.  &425.93 &   45.3&  443.2   \\
114 &  NGC7590          &  0.0053  & Sy2   &  7.78 &$<$1.50 &  3.49 &$<$1.20 &  5.60 &  130.  & ---   &   5.52&$<$10.6   \\
115 &  NGC7603          &  0.0295  & Sy1   & 12.50 &$<$0.75 &  5.71 &$<$0.93 &  3.52 &  62.2  &35.06  &   4.8 &  ---     \\ 
116 &  NGC7674          &  0.0289  & HBL   & 20.10 & 21.20  & 35.30 & 16.50  & 49.30 &  100.  &113.26 &  47.78&  382.2   \\
117 &  CGCG381-051      &  0.0307  & Sy2   & 19.10 &$<$0.70 & 1.35 &$<$1.18 &$<$1.26 &  40.4  & ---  &   0.51&   ---   \\ 
\end{longtable}
\end{ThreePartTable}

\normalsize

\clearpage

\begin{ThreePartTable}
\begin{TableNotes}
\footnotesize
\item \textbf{Notes.} The columns give for each AGN in the sample: (1) name; (2) redshift; (3) AGN type (see Table \ref{tab:sample1}) ; (4) K band galaxy flux density: F$_{\rm K}(GAL)$ flux; (5) K band total flux density from the 2MASS Extended Catalog (ref.)  F$_{\rm K}(TOT)$; (6) K band nuclear flux density F$_{\rm K}(NUC)$ (refs.); (7) Galaxy mass, estimated from the K-band galaxy luminosity (see text), or $\ddag$ from HST observations of \citet{kim2017}, or $^1$: \citep{zhang2019} ; (13) bolometric luminosity, as derived from \citet{mordini2021} using the [OIV]26$\mu$m line and the bolometric correction of \citet{lusso2012}; (14) Black hole mass, as derived from the compilation of \citet{fernandez2021} or $\dag$: from \citet{kim2017};  $^2$: \citet{huang2019}; $^3$: \citet{lee2013}.
\end{TableNotes}
\setlength{\tabcolsep}{2.pt}
\setlength{\LTcapwidth}{\textwidth}
\scriptsize
\begin{longtable}{rlcccccccc}
\caption{Observed properties of the local galaxy sample:  AGN types, K-band fluxes, galaxy and BH masses and bolometric luminosities.}\label{tab:sample3}\\
\hline\\[-0.2cm]
       &                  &              &           & \multicolumn{3}{c}{\bf K-band Continuum fluxes}                        &                     &                              &     \\
\bf n. & \bf  Name        & \bf  $z$     & \bf  Type & \bf  F$_{\rm K}(GAL)$  & \bf  F$_{\rm K}(TOT)$ & \bf  F$_{\rm K}(NUC)$ &  \bf log(M$_{GAL}$) &  \bf  log(L$_{\rm BOL(IR)}$) & \bf log(M{$_{\rm BH}$}) \\
       &                  &              &           &   \multicolumn{3}{c}{(Jy)}  & (M$_{\odot}$) & ($\rm{erg\,s^{-1}}$) &(M$_{\odot}$)  \\
       & (1)              &  (2)         & (3)       &  (4)    & (5)    & (6)           & (7)    & (8)              & (9)   \\
\hline\\[-0.2cm]
\endfirsthead
\caption{continued.}\\
\hline\\[-0.2cm]
       &                  &              &           & \multicolumn{3}{c}{\bf K-band Continuum fluxes} &      &  &    \\
\bf n. & \bf  Name        & \bf  $z$     & \bf  Type & \bf  F$_{\rm K}(GAL)$  & \bf  F$_{\rm K}(TOT)$ & \bf  F$_{\rm K}(NUC)$ &  \bf log(M$_{GAL}$) &  \bf  log(L$_{\rm BOL(IR)}$) & \bf log(M{$_{\rm BH}$}) \\
       &                  &              &           &   \multicolumn{3}{c}{(Jy)}  & (M$_{\odot}$) & ($\rm{erg\,s^{-1}}$) &(M$_{\odot}$)  \\
       & (1)              &  (2)         & (3)       &  (4)    & (5)    & (6)           & (7)    & (8)              & (9)   \\
\hline\\[-0.2cm]
\endhead
\endfoot
\hline\\
\insertTableNotes
\endlastfoot
  1 &  MRK0335          &  0.0258  & Sy1    &  4.72e-02  &  6.31e-02  &  1.60e-02  &    10.10  & 44.65 & 7.15    \\
  2 &  NGC34=Mrk938=N17 &  0.0196  & Sy2    &  5.70e-02  &  6.23e-02  &  5.25e-03  &     9.94  & ---   & 8.07    \\
  3 &  IRASF00198-7926  &  0.0728  & Sy2    &  1.10e-02  &  1.32e-02  &  2.15e-03  &    10.37  & 45.70 & ---    \\ 
  4 &  ESO012-G021      &  0.0300  & Sy1    &  3.39e-02  &  3.90e-02  &  5.10e-03  &    10.09  & 44.97 & ---     \\ 
  5 &  NGC0262=MRK348   &  0.0150  & HBL    &  5.88e-02  &  6.10e-02  &  2.17e-03  &     9.72  & 44.60 & 7.26    \\
  6 &  Izw001=UGC00545  &  0.0611  & Sy1    &  3.93e-02  &  5.76e-02  &  1.84e-02  &    10.77  & 45.22 & 6.97$^2$    \\ 
  7 &  IRASF00521-7054  &  0.0689  & HBL    &  1.64e-02  &  2.37e-02  &  7.24e-03  &    10.49  & 45.28 & ---    \\
  8 &  ESO541-IG012     &  0.0566  & Sy2    &  1.92e-02  &  2.61e-02  &  6.85e-03  &    10.39  & 45.00 &  ---   \\
  9 &  NGC0424          &  0.0118  & HBL    &  1.24e-01  &  1.49e-01  &  2.49e-02  &     9.84  & 44.57 & 7.78    \\
 10 &  NGC0526A         &  0.0191  & Sy1    &  3.58e-02  &  4.46e-02  &  8.79e-03  &     9.72  & 44.76 & 8.61    \\
 11 &  NGC0513          &  0.0195  & HBL    &  7.07e-02  &  7.22e-02  &  1.51e-03  &    10.03  & 44.46 & 7.92    \\  
 12 &  IRASF01475-0740  &  0.0177  & HBL    &  9.07e-03  &  9.73e-03  &  6.55e-04  &     9.06  & 44.40 & 6.85    \\ 
 13 &  MRK1034NED02     &  0.0338  & Sy1    &  ---       &   4.59e-02 &  ---       &   ---     & 44.52 & 7.79    \\ 
 14 &  ESO545-G013      &  0.0337  & Sy1    &  4.65e-02  &  4.80e-02  &  1.53e-03  &    10.32  & 44.94 &  ---   \\ 
 15 &  NGC0931=Mrk1040  &  0.0167  & Sy1    &  1.23e-01  &  1.30e-01  &  7.11e-03  &    10.80  & 44.92 & 6.74    \\
 16 &  NGC1068          &  0.0038  & HBL    &  3.03e+00  &  3.23e+00  &  1.96e-01  &    10.24  & 45.20 & 6.92    \\
 17 &  NGC1056          &  0.0052  & Sy2    &  ---       &   1.50e-01 &   ---      &    ---    & 43.24 & ---    \\ 
 18 &  NGC1097          &  0.0042  & LIN    &  2.10e+00  &  2.10e+00  &  1.22e-03  &    10.17  & 43.55 & 7.17    \\
 19 &  NGC1125          &  0.0109  & HBL    &  ---       &  ---       &  ---       &    ---    & 44.65 & 7.70    \\
 20 &  NGC1144          &  0.0288  & Sy2    &  ---       &   1.65e-01 &   ---      &    ---    & 44.63 & 8.65    \\
 21 &  MCG-02-08-039    &  0.0299  & HBL    &  4.05e-02  &  4.08e-02  &  2.91e-04  &    10.16  & 44.94 &  ---   \\
 22 &  NGC1194          &  0.0136  & Sy1    &  8.19e-02  &  8.33e-02  &  1.46e-03  &     9.78  & 44.49 & 7.92    \\
 23 &  NGC1241          &  0.0135  & Sy2    &  ---       &   2.32e-01 &   ---      &   ---     & 44.19 & 7.78    \\
 24 &  NGC1320          &  0.0089  & Sy2    &  1.18e-01  &  1.20e-01  &  2.40e-03  &     9.57  & 44.42 & 7.35    \\
 25 &  NGC1365          &  0.0055  & Sy1    &  1.84e+00  &  1.88e+00  &  4.02e-02  &    10.35  & 44.89 & 7.83    \\
 26 &  NGC1386          &  0.0029  & HBL    &  3.95e-01  &  3.96e-01  &  3.87e-04  &     9.12  & 44.16 & 7.72    \\
 27 &  IRASF03362-1642  &  0.0369  & Sy2    &  ---       &   2.44e-02 &   ---      &   ---     & ---   &  ---   \\
 28 &  IRASF03450+0055  &  0.0310  & Sy1    &  2.15e-02  &  3.14e-02  &  9.91e-03  &     9.92  & 44.43 &         \\
 29 &  NGC1566          &  0.0050  & Sy1    &  1.17e+00  &  1.17e+00  &  4.61e-03  &    10.07  & 43.65 & 7.74    \\
 30 &  3C120            &  0.0330  & Sy1    &  4.28e-02  &  5.58e-02  &  1.29e-02  &    9.86$\ddag$  & 45.62 &7.72$\dag$    \\
 31 &  MRK0618          &  0.0356  & Sy1    &  3.81e-02  &  4.73e-02  &  9.20e-03  &    10.29  &  44.94 & 7.73    \\
 32 &  IRASF04385-0828  &  0.0151  & HBL    &  2.39e-02  &  2.86e-02  &  4.70e-03  &     9.34  & 44.39 &         \\ 
 33 &  NGC1667          &  0.0152  & Sy2    &  1.83e-01  &  1.84e-01  &  8.87e-04  &    10.23  & 44.34 & 8.17    \\
 34 &  ESO033-G002      &  0.0181  & Sy2    &  6.09e-02  &  6.50e-02  &  4.05e-03  &     9.90  & 44.65 &         \\
 35 &  ESO362-G018      &  0.0124  & Sy1    &  6.32e-02  &  6.52e-02  &  1.94e-03  &    10.24$\ddag$  & 44.29 & 7.81$\dag$    \\
 36 &  IRASF05189-2524  &  0.0426  & HBL    &  3.28e-02  &  4.72e-02  &  1.43e-02  &    10.38  & 45.29 & 7.77    \\
 37 &  ESO253-G003      &  0.0425  & Sy2    &  1.89e-02  &  2.20e-02  &  3.08e-03  &    10.14  & 45.29 &         \\ 
 38 &  IRASF05563-3820  &  0.0339  & Sy1    &  5.08e-02  &  7.24e-02  &  2.17e-02  &    10.37  & 44.72 &         \\
 39 &  MRK0006          &  0.0188  & Sy1    &  8.20e-02  &  1.00e-01  &  1.80e-02  &    10.06  & 45.01 & 8.102   \\ 
 40 &  MRK0009          &  0.0399  & Sy1    &  2.98e-02  &  3.87e-02  &  8.95e-03  &    10.28  & 44.83 & 7.01    \\
 41 &  MRK0079          &  0.0222  & Sy1    &  7.40e-02  &  8.04e-02  &  6.37e-03  &    10.31$\ddag$  & 45.08 & 7.70$\dag$   \\ 
 42 &  IRASF07599+6508  &  0.1483  & Sy1    &  3.16e-02  &  4.47e-02  &  1.32e-02  &    11.02$\ddag$  &        &  8.47$\dag$  \\
 43 &  NGC2639          &  0.0111  & Sy1    &  ---    &   2.92e-01 &   --- &   ---     & 44.20 & 8.28    \\ 
 44 &  IRASF08572+3915  & 0.0583 & Sy2/LIN  &  2.95e-03  &  3.90e-03  &  9.55e-04  &     9.60  &       &         \\
 45 &  MRK0704          &  0.0292  & Sy1    &  3.83e-02  &  4.83e-02  &  9.91e-03  &    10.52$\ddag$  & 44.87 & 7.62$\dag$    \\
 46 &  UGC05101         &  0.0394  & LIN    &  2.67e-02  &  3.30e-02  &  6.25e-03  &    10.22  & 44.90 & 8.37    \\
 47 &  NGC2992          &  0.0077  & Sy1    &  2.39e-01  &  2.43e-01  &  4.36e-03  &     9.75  & 44.72 & 7.32    \\
 48 &  MRK1239          &  0.0199  & Sy1    &  7.46e-02  &  9.61e-02  &  2.15e-02  &    10.07  & 44.73 & 8.95    \\
 49 &  M81              & -0.00013 & LIN    &  1.96e+01  &  1.96e+01  &  6.31e-03  &     9.66  & 41.28 & 7.96    \\
 50 &  3C234            &  0.1849  & Sy1    &  3.60e-03  &  5.33e-03  &  1.72e-03  &    10.69  & 45.86 & ---     \\ 
 51 &  NGC3079          &  0.0037  & Sy2    &  8.16e-01  &  8.30e-01  &  1.37e-02  &     9.65  & 43.69 & 8.24    \\ 
 52 &  NGC3227          &  0.0039  & Sy1    &  5.75e-01  &  5.87e-01  &  1.13e-02  &     9.06$\ddag$  & 44.17 & 7.61$\dag$    \\
 53 &  NGC3511          &  0.0037  & Sy1    &  ---       &   3.95e-01 &   ---      &   ---     &  ---  &  ---    \\
 54 &  NGC3516          &  0.0088  & Sy1    &  2.56e-01  &  2.63e-01  &  6.02e-03  &    10.41$\ddag$  & 44.57 &7.61$\dag$    \\
 55 &  MCG+00-29-023    &  0.0249  & Sy2    &  4.33e-02  &  4.52e-02  &  1.98e-03  &    10.03  &  ---  &  ---    \\
 56 &  NGC3660          &  0.0123  & Sy2    &  1.03e-01  &  1.03e-01  &  2.70e-04  &     9.80  & 44.02 & 7.07    \\
 57 &  NGC3982          &  0.0037  & HBL    &  ----      &   1.92e-01 &   ---      &   10.07$\ddag$  & 43.42 & 5.82$\dag$    \\
 58 &  NGC4051          &  0.0023  & Sy1    &  5.63e-01  &  5.70e-01  &  6.85e-03  &   8.50$\ddag$  & 43.99 & 6.26$\dag$    \\
 59 &  UGC07064         &  0.0250  & HBL    &  5.41e-02  &  5.58e-02  &  1.72e-03  &    10.13  & 44.83 & ---    \\
 60 &  NGC4151          &  0.0033  & Sy1    &  7.23e-01  &  7.44e-01  &  2.11e-02  &     9.37$\ddag$  & 44.50 &7.10$\dag$    \\
 61 &  NGC4253=MRK766   &  0.0129  & Sy1    &  6.61e-02  &  7.73e-02  &  1.13e-02  &    10.23$\ddag$  & 44.79 &6.23$\dag$    \\
 62 &  NGC4388          &  0.0084  & HBL    &  4.17e-01  &  4.19e-01  &  1.69e-03  &    10.07  & 45.12 & 6.86    \\
 63 &  3C273            &  0.1583  & Sy1    &  4.69e-02  &  7.07e-02  &  2.38e-02  &    11.3$^1$  & 45.75 &8.839    \\
 64 &  NGC4501=M88      &  0.0076  & Sy2    &  2.08e+00  &  2.08e+00  &  4.66e-04  &    10.68  & 43.78 & 8.14    \\
 65 &  NGC4579=M58      &  0.0051  & LIN    &  ----      &   1.70e+00 &   ---      &   ---     & 43.28 & 8.13    \\
 66 &  NGC4593          &  0.0090  & Sy1    &  4.20e-01  &  4.27e-01  &  6.08e-03  &    10.13  & 44.48 &6.882    \\
 67 &  NGC4594=M104     &  0.0034  & LIN    &  ---       &   6.90e+00 &   ---      &   ---     & 42.82 & 8.82    \\
 68 &  NGC4602          &  0.0085  & Sy1    &  ---       &   2.56e-01 &   ---      &   ---     &  ---  & ---     \\ 
 69 &  TO1238-36=I3639  &  0.0109  & HBL    &  9.39e-02  &  9.65e-02  &  2.51e-03  &     9.65  & 44.46 & 7.12    \\ 
 70 &  M-2-33-34=N4748  &  0.0146  & Sy1    &  7.42e-02  &  7.66e-02  &  2.44e-03  &     9.80  & 45.03 & 6.68    \\
 71 &  MRK0231          &  0.0422  & Sy1    &  1.52e-01  &  1.99e-01  &  4.66e-02  &    11.04$\ddag$  &  ---  & 8.50$\ddag$    \\
 72 &  NGC4922          &  0.0236  & Sy2    &  ---       &   7.76e-02 &   ---      &   ---     & 44.54 & 8.64    \\
 73 &  NGC4941          &  0.0037  & Sy2    &  3.23e-01  &  3.44e-01  &  2.11e-02  &     9.25  & 43.96 & 7.14    \\
 74 &  NGC4968          &  0.0099  & Sy2    &  1.06e-01  &  1.08e-01  &  1.38e-03  &     9.62  & 44.54 & 7.24    \\
 75 &  NGC5005          &  0.0032  & LIN    &  1.76e+00  &  1.76e+00  &  6.79e-03  &     9.86  & 43.43 & 8.20    \\
 76 &  NGC5033          &  0.0029  & Sy1    &  1.10e+00  &  1.10e+00  &  3.60e-03  &     9.57  & 43.55 & 7.73    \\
 77 &  MCG-03-34-064    &  0.0165  & HBL    &  8.33e-02  &  8.40e-02  &  6.61e-04  &     9.96  & 45.20 & 8.28    \\
 78 &  NGC5135          &  0.0137  & Sy2    &  1.93e-01  &  1.96e-01  &  3.19e-03  &    10.16  & 44.95 & 7.60    \\
 79 &  NGC5194=M51A     &  0.0015  & Sy2    &  4.22e+00  &  4.22e+00  &  4.25e-03  &     9.58  & 43.34 & 6.93    \\
 80 &  MCG-06-30-015    &  0.0077  & Sy1    &  8.57e-02  &  9.80e-02  &  1.22e-02  &     9.31  & 44.32 & 7.27    \\
 81 &  IRASF13349+2438  &  0.1076  & Sy1    &  3.02e-02  &  4.41e-02  &  1.39e-02  &    11.15  & 45.49 & 9.0$^3$     \\ 
 82 &  NGC5256=Mrk266   &  0.0279  & Sy2    &  7.83e-02  &  8.02e-02  &  1.91e-03  &    10.39  & 45.30 & ---     \\ 
 83 &  MRK0273          &  0.0378  & LIN    &  2.58e-02  &  2.93e-02  &  3.53e-03  &    10.17  & 45.47 & 8.43    \\
 84 &  IC4329A          &  0.0161  & Sy1    &  1.80e-01  &  2.00e-01  &  2.01e-02  &    10.33$\ddag$  & 45.19 & 6.98$\dag$    \\
 85 &  NGC5347          &  0.0078  & HBL    &  9.05e-02  &  9.25e-02  &  1.96e-03  &     9.34  & 43.97 & 7.03    \\
 86 &  MRK0463=UGC8850  &  0.0504  & HBL    &  4.60e-02  &  6.20e-02  &  1.60e-02  &    10.67  & 45.70 & 8.10    \\ 
 87 &  NGC5506          &  0.0062  & Sy2    &  3.12e-01  &  3.54e-01  &  4.21e-02  &     9.68  & 44.84 & 8.24    \\
 88 &  NGC5548          &  0.0172  & Sy1    &  1.14e-01  &  1.17e-01  &  3.22e-03  &    10.35$\ddag$  & 44.67 &7.89$\dag$    \\
 89 &  MRK0817          &  0.0315  & Sy1    &  3.94e-02  &  4.85e-02  &  9.04e-03  &    10.20  & 44.74 &7.586    \\ 
 90 &  IRASF15091-2107  &  0.0446  & Sy1    &  2.90e-02  &  3.77e-02  &  8.63e-03  &    10.36  & 45.40 & 7.0     \\ 
 91 &  NGC5929          &  0.0083  & HBL    &  ---       &   1.77e-01 &   ---      &   ---     & 43.90 & 7.52    \\ 
 92 &  NGC5953          &  0.0066  & Sy2    &  ---       &   1.45e-01 &   ---      &   ---     & 44.17 & 7.24    \\ 
 93 &  ARP220=UGC9913   &  0.0181  & Sy2    &  7.32e-02  &  7.64e-02  &  3.25e-03  &     9.98  & ---   & 7.90    \\ 
 94 &  N5995=M-2-40-4   &  0.0252  & HBL    &  1.03e-01  &  1.17e-01  &  1.39e-02  &    11.00$\ddag$  & 44.81 &6.77$\dag$    \\
 95 &  IRASF15480-0344  &  0.0303  & HBL    &  3.05e-02  &  3.24e-02  &  1.87e-03  &    10.05  & 45.21 &  ---    \\
 96 &  ESO141-G055      &  0.0371  & Sy1    &  4.37e-02  &  5.48e-02  &  1.12e-02  &    10.38  & 44.87 &  ---    \\
 97 &  IRASF19254-7245  &  0.0617  & HBL    &  1.73e-02  &  1.92e-02  &  1.94e-03  &    10.42  & 45.12 &  ---    \\
 98 &  NGC6810          &  0.0068  & LIN    &  5.62e-01  &  5.65e-01  &  2.96e-03  &    10.02  & 43.57 &  ---    \\
 99 &  NGC6860          &  0.0149  & Sy1    &  1.10e-01  &  1.17e-01  &  6.98e-03  &     9.99  & 44.48 & 7.91    \\
100 &  NGC6890          &  0.0081  & Sy2    &  1.62e-01  &  1.63e-01  &  1.74e-03  &    10.76$\ddag$  & 44.08 &  6.50$\dag$    \\
101 &  MRK0509          &  0.0344  & Sy1    &  4.95e-02  &  6.64e-02  &  1.69e-02  &    10.53$\ddag$  & 45.21 &8.14$\dag$    \\
102 &  IC5063           &  0.0113  & HBL    &  2.07e-01  &  2.11e-01  &  3.63e-03  &    10.03  & 44.97 & 7.97    \\
103 &  MRK0897          &  0.0263  & Sy2    &  8.07e-03  &  1.66e-02  &  8.55e-03  &     9.35  & 43.95 &  ---    \\
104 &  NGC7130          &  0.0162  & Sy2    &  1.39e-01  &  1.40e-01  &  9.55e-04  &    10.16  & 44.67 & 7.87    \\
105 &  NGC7172          &  0.0087  & Sy2    &  3.05e-01  &  3.14e-01  &  8.87e-03  &     9.97  & 44.55 & 8.35    \\
106 &  IRASF22017+0319  &  0.0611  & HBL    &  1.20e-02  &  1.59e-02  &  3.87e-03  &    10.25  & 45.56 &  ---    \\ 
107 &  NGC7213          &  0.0058  & LIN    &  1.02e+00  &  1.02e+00  &  6.19e-03  &    10.14  &  ---  & 8.05    \\
108 &  3C445            &  0.0562  & Sy1    &  ---       &  ---       &  ---       &  10.17$\ddag$  & 45.44 & 8.25$\dag$    \\
109 &  NGC7314          &  0.0048  & HBL    &  3.51e-01  &  3.54e-01  &  3.02e-03  &     9.51  & 44.32 & 7.23    \\
110 &  MCG+03-58-007    &  0.0315  & HBL    &  3.87e-02  &  4.58e-02  &  7.18e-03  &    10.19  & 44.83 & ---     \\
111 &  NGC7469          &  0.0163  & Sy1    &  1.79e-01  &  1.93e-01  &  1.39e-02  &    10.28  & 44.72 &6.956    \\
112 &  NGC7496          &  0.0055  & Sy2    &  2.30e-01  &  2.31e-01  &  7.11e-04  &     9.45  &  ---  & 6.98    \\
113 &  NGC7582          &  0.0053  & Sy2    &  7.47e-01  &  7.90e-01  &  4.25e-02  &     9.92  & 44.78 & 7.74    \\
114 &  NGC7590          &  0.0053  & Sy2    &  2.73e-01  &  2.89e-01  &  1.63e-02  &     9.49  & 43.66 & 7.00    \\
115 &  NGC7603          &  0.0295  & Sy1    &  1.09e-01  &  1.24e-01  &  1.54e-02  &    10.58  & 44.52 & 8.42    \\ 
116 &  NGC7674          &  0.0289  & HBL    &  7.28e-02  &  8.12e-02  &  8.39e-03  &    10.39  & 45.28 & 7.85    \\
117 &  CGCG381-051      &  0.0307  & Sy2    &  1.89e-02  &  1.96e-02  &  6.67e-04  &     9.85  &  ---  &  ---    \\
\end{longtable}
\end{ThreePartTable}
\normalsize

\end{document}